\documentclass[a4paper,11pt,english]{article}
\usepackage[utf8]{inputenc}
\usepackage[T1]{fontenc}
\usepackage{lmodern, amsmath, adjustbox, a4wide, multirow, xcolor, graphicx, xfrac, slashed, cancel,mathtools}
\usepackage[numbers, sort&compress, elide]{natbib}
\usepackage{amsfonts}
\usepackage{jheppub}
\usepackage{longtable}
\usepackage{makecell}
\usepackage{amscd}
\usepackage{tabu}
\usepackage{bm}
\usepackage{enumerate}
\usepackage{pbox}
\usepackage{tikz}
\usetikzlibrary{decorations.pathmorphing}
\usepackage{pdflscape}
\usepackage{diagbox}
\usepackage{booktabs}
\usepackage{multirow}
\usepackage{pifont}
\usepackage{ifsym}
\usepackage{diagbox}
\usepackage{cleveref}
\crefname{table}{Table}{Tables}
\crefname{equation}{Eq.}{Eqs.}
\crefname{appendix}{App.}{Apps.}
\crefname{section}{Sec.}{Secs.}
\crefname{figure}{Fig.}{Figs.}

\makeatletter\g@addto@macro\bfseries{\boldmath}\makeatother

\newcommand{\dd}{\mathrm{d}}
\newcommand{\braket}[2]{\langle #1 \vert #2 \rangle}

\usetikzlibrary{calc,shapes,snakes}

\newcommand{\currsep}{0.2}
\newcommand{\bracketsep}{0.5}
\newcommand{\threepointsep}{0.5}
\newcommand{\currentmarker}[1]{
  \filldraw[fill=white,line width=1pt](#1)circle(.12);
  \draw[line width=.6pt] (#1) +(-135:.12) -- +(45:.12) +(-45:.12) -- +(135:.12);
}
\newcommand{\cut}{
  \hspace{5pt}\mathord{\begin{tikzpicture}[baseline=-0.65ex]
  \draw[dash pattern = on 5pt off 5pt] (0,1.2) -- (0,-1.2);
\end{tikzpicture}}\hspace{5pt}}

\newcommand{\angleone}{45}
\newcommand{\angletwo}{-135}
\newcommand{\anglethree}{-45}
\tikzset{iso0style/.style={circle,fill,inner sep=2pt}}
\tikzset{isohalfstyle/.style={circle,draw=black,fill=white,inner sep=2pt}}
\tikzset{iso1style/.style={rectangle,fill,inner sep=2pt}}
\tikzset{iso3halfstyle/.style={rectangle,draw=black,fill=white,inner sep=2pt}}
\tikzset{iso2style/.style={star,star points=5,fill,inner sep=2pt}}
\tikzset{labelstyle/.style={text=red,label distance=-5pt}}

\DeclareRobustCommand{\isozerokey}{\begin{tikzpicture}[baseline=-0.65ex] \node[iso0style] (A) at (0,0) {}; \end{tikzpicture}}
\DeclareRobustCommand{\isohalfkey}{\begin{tikzpicture}[baseline=-0.65ex] \node[isohalfstyle] (A) at (0,0) {}; \end{tikzpicture}}
\DeclareRobustCommand{\isoonekey}{\begin{tikzpicture}[baseline=-0.65ex] \node[iso1style] (A) at (0,0) {}; \end{tikzpicture}}
\DeclareRobustCommand{\isothreehalfkey}{\begin{tikzpicture}[baseline=-0.65ex] \node[iso3halfstyle] (A) at (0,0) {}; \end{tikzpicture}}
\DeclareRobustCommand{\isotwokey}{\begin{tikzpicture}[baseline=-0.65ex] \node[iso2style] (A) at (0,0) {}; \end{tikzpicture}}

\newcommand\encircle[1]{%
  \tikz[baseline=(X.base)] 
    \node (X) [draw, shape=circle, inner sep=0] {\strut #1};}

\usepackage{array}   
\newcolumntype{L}{>{$}l<{$}} 
\newcolumntype{C}{>{$}c<{$}} 

\definecolor{kinematic}{RGB}{46,139,87}
\definecolor{gauge}{RGB}{255,140,0}
\definecolor{flavour}{RGB}{139,0,139}

\newcommand{\phs}[1]{ {\color{kinematic} #1}}
\newcommand{\gge}[1]{ {\color{gauge} #1}}
\newcommand{\flv}[1]{ {\color{flavour} #1}}

\newcommand{\jvec}{\mathcal{J}^A}

\newcommand{\comment}[1]{}
\newcommand{\cutsub}{
  \hspace{5pt}\mathord{\begin{tikzpicture}[baseline=-0.65ex]
  \draw[dash pattern = on 5pt off 5pt] (0,1.2) -- (0,-1.2) node[right] {\scriptsize sub};
\end{tikzpicture}}\hspace{5pt}}

\newcommand{\ket}[1]{| #1 \rangle}

\newcommand{\st}{{\textit{s.t.}}~}

\newcommand{\ang}[2]{\langle #1 #2 \rangle} 
\newcommand{\sqr}[2]{\left[ #1 #2 \right]} 
\newcommand{\angbra}[1]{\langle #1 \rvert} 
\newcommand{\angket}[1]{\lvert #1 \rangle} 
\newcommand{\sqrbra}[1]{\left[ #1 \right|} 
\newcommand{\sqrket}[1]{\left| #1 \right]} 

\newcommand{\lag}{\ensuremath{\mathcal{L}}} 
\newcommand{\amp}{\ensuremath{\mathcal{A}}}
\newcommand{\op}{\mathcal{O}}

\newcommand{\Tr}{\mathrm{Tr}}

\newcommand{\nn}{\nonumber }

\newcommand{\beq}{\begin{equation}} 
\newcommand{\eeq}{\end{equation}} 
\newcommand{\ba}{\begin{array}}  
\newcommand{\ea}{\end{array}} 
\newcommand{\bea}{\begin{eqnarray}}  
\newcommand{\eea}{\end{eqnarray} }  
\newcommand{\be}{\begin{eqnarray}}  
\newcommand{\ee}{\end{eqnarray} }  
\newcommand{\bal}{\begin{align}}
\newcommand{\eal}{\end{align}}   
\newcommand{\ben}{\begin{enumerate}}  
\newcommand{\een}{\end{enumerate}}  
\newcommand{\bc}{\begin{center}}
\newcommand{\ec}{\end{center}} 
\newcommand{\bt}{\begin{table}}
\newcommand{\et}{\end{table}}  
\newcommand{\btb}{\begin{tabular}}
\newcommand{\etb}{\end{tabular}}

\newcommand{\cA}{{\mathcal A}}   

\def\ra{\rangle}
\def\la{\langle}  

\newcommand{\eps}{\epsilon}

\newcommand{\ct}{c_\frac{\theta}{2}}
\renewcommand{\st}{s_\frac{\theta}{2}}
\newcommand{\ifl}{\mathcal{I}}
\newcommand{\itfl}{\mathcal{I}_3}
\newcommand{\yfl}{\mathcal{Y}}

\newcommand{\scalarsym}{{\color{kinematic} 6 t}}
\newcommand{\scalarantisym}{{\color{kinematic} 2(s-u)}}
\newcommand{\scalarRH}{{\color{kinematic} 2 \ang41 \sqr13}}
\newcommand{\scalarLH}{{\color{kinematic} 2 \sqr41 \ang13}}
\newcommand{\LHLH}{{\color{kinematic} 2 \ang13 \sqr42}}
\newcommand{\LHantisym}{{\color{kinematic} 8 \ang13 \sqr42}}
\newcommand{\LHRH}{{\color{kinematic} 2 \ang14 \sqr32}}
\newcommand{\RHantisym}{{\color{kinematic} 8 \sqr13 \ang42}}
\newcommand{\RHRH}{{\color{kinematic} 2 \sqr13 \ang42}}

\newcommand{\sulsym}{{\color{gauge} \left(\delta\delta\right) }}
\newcommand{\sulantisym}{{\color{gauge} \left[\delta\delta\right]}}
\newcommand{\suldels}{{\color{gauge} \delta^{i}_{j} \delta^{k}_{l}}}
\newcommand{\sulsigmas}{{\color{gauge} [\sigma^I]^{i}_{j} [\sigma^I]^{k}_{l}}}
\newcommand{\suldel}{{\color{gauge} \delta^{i}_{j}}}
\newcommand{\sulepsu}{{\color{gauge} \epsilon^{ij}}}
\newcommand{\sulepsl}{{\color{gauge} \epsilon_{ij}}}

\newcommand{\colourdel}{{\color{gauge} \delta^{a}_{b}}}
\newcommand{\colourdels}{{\color{gauge} \delta^{a}_{b} \delta^{c}_{d}}}
\newcommand{\colourlambdas}{{\color{gauge} \frac14 [\lambda^A]^{a}_{b} [\lambda^A]^{c}_{d}}}
\newcommand{\coloursym}{{\color{gauge} \left(\delta\delta\right) }}
\newcommand{\colourantisym}{{\color{gauge} \left[\delta\delta\right] }}

\newcommand{\zerof}{{\color{flavour} c}}
\newcommand{\twof}{{\color{flavour} c^p_q}}
\newcommand{\fourf}{{\color{flavour} c^{pr}_{qs}  }}
\newcommand{\foursym}{{\color{flavour} \left( c \right) }}
\newcommand{\fourantisym}{{\color{flavour} \left[ c \right] }}
\newcommand{\minusf}{{\color{flavour} -}}

\author[a]{Camila S. Machado,}
\author[b]{Sophie Renner,}
\author[b]{and Dave Sutherland}

\affiliation[a]{Deutsches Elektronen-Synchrotron DESY, Notkestr.\ 85, 22607 Hamburg, Germany}
\affiliation[b]{School of Physics and Astronomy, University of Glasgow, Glasgow G12 8QQ, United Kingdom}

\emailAdd{camila.machado@desy.de}
\emailAdd{sophie.renner@glasgow.ac.uk}
\emailAdd{david.w.sutherland@glasgow.ac.uk}

\title{Building blocks of the flavourful SMEFT RG}

\abstract{
A powerful aspect of effective field theories is connecting scales through renormalisation group (RG) flow. 
The anomalous dimension matrix of the Standard Model Effective Field Theory (SMEFT) encodes clues to where to find relics of heavy new physics in data, but its unwieldy $2499 \times 2499$ size (at operator dimension 6) makes it difficult to draw general conclusions. 
In this paper, we study the flavour structure of the SMEFT one loop anomalous dimension matrix of dimension 6 current-current operators, a $1460 \times 1460$ submatrix.
We take an on-shell approach, laying bare simple patterns by factorising the entries of the matrix into their gauge, kinematic and flavour parts. We explore the properties of different diagram topologies, and make explicit the connection between the IR-finiteness of certain diagrams and their gauge and flavour structure.
Through a completely general flavour decomposition of the Wilson coefficient matrices, we uncover new flavour selection rules, from which small subsystems emerge which mix almost exclusively amongst themselves. We show that, for example, if we neglect all Yukawa couplings except for that of the top quark, the selection rules produce block diagonalisation within the current-current operators in which the largest block is a $61 \times 61$ matrix. 
 We provide all the ingredients of the calculations in comprehensive appendices, including SM and SMEFT helicity amplitudes, and explicit results for phase space integrals and gauge contractions. 
This deconstruction of the matrix, and its resulting block-diagonalisation, provides a first step to understanding the IR-relevant directions in the SMEFT parameter space, hence closing in on natural places for heavy new physics to make itself known.}

\preprint{DESY-22-161}

\begin{document}
\sloppy 

\makeatletter\renewcommand{\@fpheader}{\ }\makeatother

\maketitle

\section{Introduction}

While the reasons to expect new physics beyond the Standard Model (BSM) remain compelling, the fact that we haven't yet seen definitive evidence for new particles at the LHC implies a hierarchy between the electroweak scale and the heavier scale of new physics. In this case, the Standard Model effective field theory (SMEFT) provides a powerful framework to understand the phenomenology of heavy new physics in a model independent way, in terms of coefficients of higher dimensional operators.

The leading (lepton-number conserving) BSM effects begin at dimension 6 in the SMEFT. 
There are a total of 2499 independent baryon-number-conserving parameters in the dimension 6 Lagrangian, so a pressing issue is to identify the most important subsets of parameters. It is particularly necessary to identify significant flavour directions, since nearly all the parameters are elements of flavour matrices. Given the extremely strong constraints from measurements of flavour changing and/or CP violating processes, a new physics scale in the TeV or tens of TeV scale implies strong suppression in the Wilson coefficients of flavour changing or CP-odd operators, which can be achieved for example by imposing flavour and CP symmetries onto the theory. 

The large number of parameters in the SMEFT, and the relatively few very precisely measured observables, makes the inclusion of loop effects necessary for an understanding of the constraints on individual operators. The full dimension 6 one loop anomalous dimensions were first calculated in Refs.~\cite{Jenkins:2013zja,Jenkins:2013wua,Alonso:2013hga}, and have been implemented in public codes~\cite{Celis:2017hod,Fuentes-Martin:2020zaz,Aebischer:2018bkb,DiNoi:2022ejg}. Flavour is inherent in the SMEFT renormalisation group equations (RGEs), since the Higgs and top quark are kinematically accessible, meaning that significant Yukawa interactions enter into divergent loop diagrams. At one loop, flavourful observables are hence dependent on flavour-universal operators~(e.g.~\cite{Bobeth:2015zqa,Aebischer:2015fzz,Hurth:2019ula,Aoude:2020dwv}), and flavourless observables are dependent on flavour violating or non-universal operators (e.g.~\cite{deBlas:2015aea,Kumar:2021yod,Dawson:2022bxd}).

Knowledge of the anomalous dimensions therefore allows us to ask and answer questions relevant for phenomenology at the weak scale and below. Such questions which have received attention in the decade since the publication of the full SMEFT anomalous dimension matrices can often be phrased as ``Starting with particular operators at the high scale, which effects are generated by running?'' and ``Starting with a particular effect at the low scale, which operators at the high scale can generate it by running?''.
The quantitative answers to these questions are particularly easily obtained by public codes which solve the RGEs numerically, and can be used to put new constraints on operators, or gain insights into connected phenomenology across scales.
The inherent patterns and structure in the anomalous dimension matrix itself, however, remain comparatively unexplored, and are the focus of this paper. This structure represents important information about the SMEFT, since when diagonalised the anomalous dimension matrix tells us which directions in parameter space are enhanced or suppressed in the IR relative to their naive scaling dimension of six. Combined with a broad and model-independent prior on the important coefficients at the UV scale (for example based on the space of tree level weakly coupled UV completions~\cite{deBlas:2017xtg}, and/or on flavour and CP symmetries~\cite{Faroughy:2020ina,Greljo:2022cah}), an understanding of these directions can thus point to motivated search strategies for new physics in LHC and flavour data. 

Some substructure within the matrix has been found \cite{Alonso:2014rga}; in particular many zeroes in the matrix can be understood based on helicity arguments~\cite{Cheung:2015aba}. More recently, on-shell methods have been found to be powerful in calculating also its non-zero entries~\cite{Caron-Huot:2016cwu,Baratella:2020dvw,Baratella:2020lzz,Baratella:2022nog,DelleRose:2022ygn,Jiang:2020mhe,Bern:2020ikv,EliasMiro:2020tdv,AccettulliHuber:2021uoa,Shu:2021qlr}. In this paper, we explore the additional substructure that can be uncovered once gauge and flavour information is included. These aspects are often glossed over when studying EFTs on-shell, whose natural focus is on kinematic and helicity properties.\footnote{See however Refs.~\cite{Li:2020xlh,Li:2020gnx,AccettulliHuber:2021uoa,Li:2022tec} for amplitude EFT bases which take account of flavour indices.} But since they account for the vast majority of the SMEFT parameters, their effects in running cannot be ignored.

An advantage of working on-shell is that the gauge, flavour, and kinematic parts of amplitudes factorise simply. Moreover, in the case of indistinguishable particles each of these pieces inherits a definite transformation under crossing symmetry, which in turn maps neatly onto irreps of the gauge, flavour, and angular momentum groups \cite{Bellazzini:2014waa,Trott:2020ebl,Baratella:2020dvw}.
As a starting point, in this paper we focus on the operators with four fields and total helicity zero ($(n, \sum h)=(4,0)$), as nothing else mixes into these operators apart from by amounts proportional to small Yukawas. Moreover, they comprise the majority (1460) of the parameters\footnote{This number excludes the components of $Hud$ and $LedQ$ operators, for reasons outlined in \cref{sec:flavourStructure}.} in the dimension 6 SMEFT, and represent most of the operators that can be generated at tree level in weakly coupled UV completions~\cite{Einhorn:2013kja,deBlas:2017xtg,Craig:2019wmo}.
For the Wilson coefficients of these operators, we show a convenient and completely general decomposition based on irreducible representations (irreps) of the $SU(3)^5$ flavour group. In this basis, subsets of coefficients corresponding to various flavour symmetry assumptions can be easily isolated. 
 
We go on to show that we can understand each entry of the anomalous dimension matrix in terms of a gauge factor, a flavour factor, and a kinematic factor.  Each one of these factors is repeated in many places in the anomalous dimension matrix, generating patterns. We calculate each of these factors, and explain patterns and zeroes. We thereby provide the first check of many entries of the known SMEFT anomalous dimension matrix.

Finally, we derive new selection rules based on the quantum numbers of the Wilson coefficients under the flavour irrep decomposition. This allows us to block-diagonalise the anomalous dimension matrix (the size of the blocks depending on which small Yukawa couplings are neglected), where the phenomenology induced by the operators of each block can be broadly identified through their flavour quantum numbers. This analytic block diagonalisation is a significant step towards a fully diagonal matrix -- in the case where all light Yukawas (all except $y_t$) are neglected, the largest block allowed by these selection rules is $61\times 61$, a significant reduction on the $1460\times 1460$ block achieved by helicity selection rules alone.

The paper is organised as follows. In Section \ref{sec:SMEFTonshell}, we review the on-shell way of describing the dimension 6 SMEFT and the helicity non-renormalisation theorems, justifying our choice of studying the operators with four fields and vanishing total helicity. We introduce a compact notation to write the kinematic part of the relevant SM amplitudes as well as the amplitudes corresponding to the chosen set of dimension 6 operators. The symmetric and antisymmetric gauge structures are shown and we decompose the Wilson coefficients into their irreducible representations under the full flavour symmetry group of the SM, within a general spurion analysis. In Section \ref{sec:gammaUnitarity}, we review the on-shell techniques to compute anomalous dimensions, and show examples of how to compute the IR finite pieces as well as the soft and collinear divergent pieces. We also detail all possibilities for the gauge and flavour factors that can arise in the anomalous dimensions, making the connection with the corresponding diagram topology. In Section \ref{sec:anatomy}, we analyse the patterns and zeroes of the anomalous dimension matrix, beyond helicity arguments. We exhibit these patterns in submatrices of the full anomalous dimension matrix of the $(4,0)$ operators, separating the IR finite, soft and collinear pieces, and decomposing each entry by its kinematic, gauge and flavour factors.
Section \ref{sec:blockdiag} identifies flavour selection rules and studies how different approximations lead to different preserved quantum numbers. 
We also comment on how our analysis can be extended to operators beyond the $(4,0)$ operator block. In Section \ref{sec:pheno}, we discuss applications of our work, and give a concrete phenomenological example by focusing on lepton flavour non-universality in $B$ decays.

\section{SMEFT: the on-shell way}
\label{sec:SMEFTonshell}
The Lagrangian of the SMEFT at dimension 6 has a large number of parameters and many hidden redundancies, manifested as field redefinitions which relate different operator structures. $S$-matrix elements are unaffected by these redefinitions.  The $S$-matrix of the EFT is therefore a much simpler object than the Lagrangian, and the complexity induced by these redundancies can be avoided by studying scattering amplitudes ---
a basis of higher-dimensional effects can be constructed directly in term of on-shell massless \cite{Durieux:2019siw,Ma:2019gtx,Jiang:2021tqo} or massive \cite{Shadmi:2018xan,Aoude:2019tzn,Durieux:2019eor,Durieux:2020gip,Dong:2021yak} amplitudes. Properties of the SMEFT, obscured in the Lagrangian, become transparent in terms of these helicity amplitudes, for example powerful non-interference \cite{Azatov:2016sqh} and non-renormalisation \cite{Cheung:2015aba,Bern:2019wie} theorems. Moreover, it was recently shown how the non-zero entries of the anomalous dimension matrix can be calculated easily via generalised unitarity\footnote{For a review, see \cite{Britto:2010xq}.} methods \cite{Baratella:2022nog,Baratella:2021guc,Jiang:2020mhe,Baratella:2020dvw,Baratella:2020lzz,Caron-Huot:2016cwu,EliasMiro:2020tdv,Bern:2020ikv,Jin:2020pwh,Cao:2021cdt}. These methods have the advantage of allowing calculation of loops using on-shell tree-level amplitudes.

In this section, we describe the (massless) helicity amplitudes induced in the SMEFT and in the SM, and explain the helicity-based non-renormalisation arguments \cite{Cheung:2015aba} which justify our choice of operators to focus on as a starting point. For this set of operators, we discuss how the kinematic part of the amplitudes can be written in a compact form as a product of two currents that connects to its partial wave decomposition. We also use group theory to describe the possible gauge and flavour parts of the amplitudes. In particular, for the flavour structure, we decompose the Wilson coefficients into their irreducible representations under the full flavour symmetry of the SM, i.e.\ $SU(3)^5$, which allow us to perform a fully general spurion analysis. This naturally allows for a description in terms of new quantum numbers, similar to isospin and strangeness for the $SU(3)$ of light quark flavours. We further connect the quantum numbers to the range of phenomenology generated. This sets the groundwork to analyse the flavour structure of the anomalous dimension matrix in later sections.    

\subsection{Singling out the `$(4,0)$' operators} 
\label{sec:helicitynonrenormalization}
If we take all particles to be massless, we can define SM states of definite helicity as
\begin{equation}
    H, \psi^+, \psi^-, V^+, V^-,
\end{equation}
where $H$ represents the Higgs boson, $\psi^{\pm}$ represent any fermion and $V^{\pm}$ represent any vector, with helicities
\begin{equation}
    h=0,\frac{1}{2}, -\frac{1}{2}, 1, -1,
\end{equation}
respectively, defined for \emph{incoming} particles.\footnote{Taking particles to be incoming is slightly unusual, but perfectly self-consistent. See Appendix~\ref{app:conventions} for all our spinor conventions.} The various classes of SMEFT dimension 6 operators can then be labelled by coordinates $(n,\sum_i h_i)$, corresponding to the number of legs $n$ and total helicity $\sum_i h_i$ of the amplitudes they induce at tree level.\footnote{$(n,\sum_i h_i)$ are linear combinations of the holomorphic coordinates introduced in \cite{Cheung:2015aba}.} So, we refer to e.g.~the operators $\psi^2\bar{\psi}^2$\,,\, $\psi \bar{\psi}H^2D$ and $H^4D^2$ as the \emph{(4,0) operators}, since they induce 4-point amplitudes with zero total helicity.

\begin{figure}
\begin{center}
\includegraphics[width=0.4\textwidth]{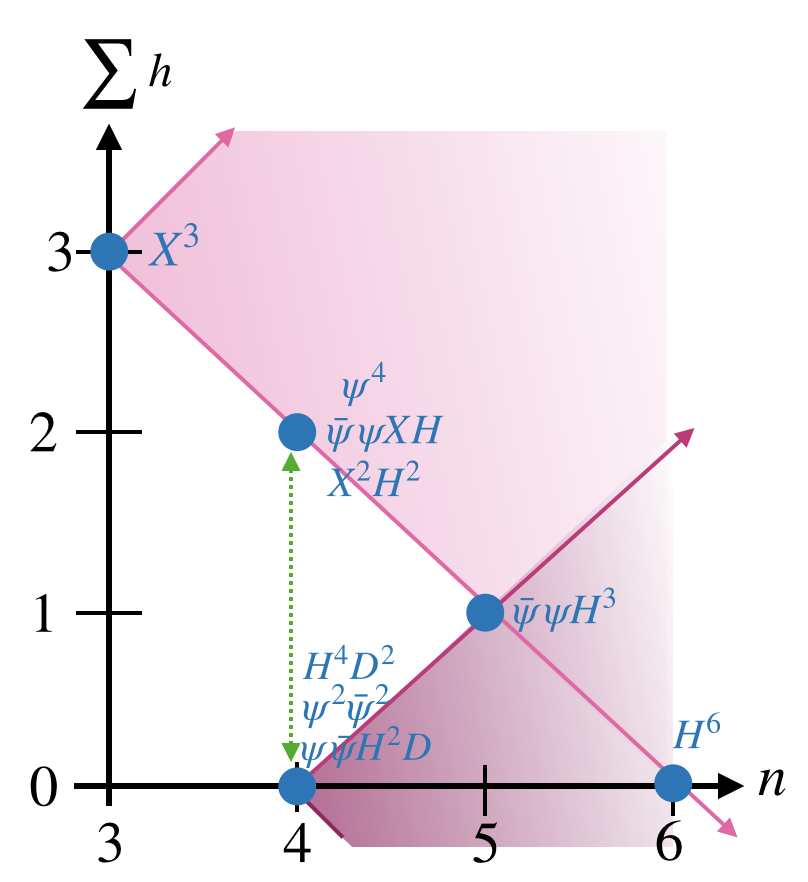}
\caption{\label{fig:nhdim6} The classes of dimension 6 SMEFT operators, plotted according to the number of legs $n$, and total helicity, $\Sigma h$, of their corresponding tree-level amplitudes. The diagram can be reflected about the $n$ axis by Hermitian conjugation to obtain the full set of operators. Pink cones delineate operators that can be renormalised by the operators at the apex of the cone. The green dotted line indicates exceptions to this rule due to SM amplitudes containing a product of two different Yukawa interactions.}
\end{center}
\end{figure}

This classification of operators in terms of their helicity structure leads to `non-interference'~\cite{Azatov:2016sqh} and `non-renormalisation'~\cite{Cheung:2015aba} theorems. These latter can be understood as they apply to the SMEFT at dimension 6, by firstly noticing that up to a few exceptions (discussed below), the $(n,\sum_i h_i)$ coordinates of tree-level SM amplitudes obey the rule
\begin{equation}
\label{eq:SMhelicity}
\left | \sum h_{SM} \right|\leq n_{SM}-4.
\end{equation}
Furthermore, by unitarity, if 2-cuts of a loop amplitude vanish, then the loop is finite, and cannot contribute to anomalous dimensions.\footnote{Caveat: if the loop has IR divergences, then the 2-cut may vanish in dimensional regularisation even though there may be a compensating UV divergence. But IR divergences can only arise in self-renormalisation diagrams, so this issue will not change the arguments of this section, which are focused on mixing between different classes of operators. We will come back to the issue of IR divergent diagrams in Sec.~\ref{sec:gammaUnitarity}.} This leads to the following relation between tree-level amplitudes and the amplitudes that they can renormalise \cite{Cheung:2015aba}:

\begin{equation}
\label{eq:nonrenormalizationhelicity}
\begin{tikzpicture}
\node[draw,circle,fill=gray,inner sep=8pt] (amp1) at (0,0) {$A$};
\node[draw,circle,fill=gray,inner sep=8pt] (amp2) at (3,0) {$B$};
\node[draw,circle,fill=gray,inner sep=8pt] (amp3) at (6,0) {$C$};
\node at (0,-1.6) {$\begin{pmatrix} n_A \\ \sum h_A \end{pmatrix}$};
\node at (1.5,-1.6) {$+ \!\! \begin{pmatrix} -4 \\ 0 \end{pmatrix} \!\! +$};
\node at (3,-1.6) {$\begin{pmatrix} n_B \\ \sum h_B \end{pmatrix}$};
\node at (6,-1.6) {$\begin{pmatrix} n_C \\ \sum h_C \end{pmatrix}$\,.};
\node at (4.5,-1.6) {$=$};
\node at (4.5,0) {$=$};
\foreach \ang in {100,140,...,260}
  \draw[thick] (amp1) -- (\ang:1);
\foreach \ang in {-80,-40,...,80}
  \draw[thick] (amp2) -- ++(\ang:1);
\foreach \ang in {0,36,...,324}
  \draw[thick] (amp3) -- ++(\ang:1);
\draw[thick] (amp1) to [out=45,in=180] (1,0.8) node[label=$\pm$] {};
\draw[thick] (amp1) to [out=-45,in=180] (1,-0.8) node[label=$\pm$] {};
\draw[thick] (amp2) to [out=135,in=0] (2,0.8) node[label=$\mp$] {};
\draw[thick] (amp2) to [out=-135,in=0] (2,-0.8) node[label=$\mp$] {};
\draw[dashed] (1.5,-1) -- (1.5,1);
\end{tikzpicture}
\end{equation}
The helicity of the legs on either side of the cut are equal and opposite, since the momenta of the legs of the tree-level amplitudes are all defined incoming. So if `$A$' and `$C$' are SMEFT amplitudes and `$B$' is a SM amplitude, when the relations in Eqs.~\eqref{eq:nonrenormalizationhelicity} and \eqref{eq:SMhelicity} are taken in combination, we find constraints on which `$C$' operators can be renormalised by which `$A$' operators, in terms of their respective helicities and number of legs. This is shown in Fig.~\ref{fig:nhdim6}: only operators whose $(n,\sum_i h_i)$ coordinates lie on or within any given pink cone can be renormalized by the operators at the apex of the cone, according to the restrictions on SM amplitudes~\eqref{eq:SMhelicity}. It can be seen that the $X^3$ (three field strength) operators with $(n,\sum_i h_i)=(3,3)$ are not renormalised by any other operator type, and can only run amongst themselves. The same is true of the $(4,0)$ operators $H^4 D^2$, $\psi^2 \bar \psi^2$, and $\psi\bar\psi H^2 D$.

However, we must now come back to the exceptional SM amplitudes, mentioned above Eq.~\eqref{eq:SMhelicity}. It turns out that there are SM amplitudes which break the relation~\eqref{eq:SMhelicity}, since they have 4 legs but $\Sigma_i h_i=\pm 2$, thus allowing renormalisation outside the pink cones. This is shown by the dotted green arrow in Fig.~\ref{fig:nhdim6}, allowing the $(4,0)$ operators to be renormalised by $(4,\pm 2)$ operators, and vice versa. However, these exceptional SM amplitudes can only be produced by Yukawa diagrams with one up-type ($Y_u$) and one down-type ($Y_d$ or $Y_e$) Yukawa interaction, and are therefore always suppressed by small Yukawa entries; the largest effect possible is suppressed by $y_b$~\cite{Alonso:2014rga,Cheung:2015aba}.

In summary, we see from these helicity arguments that nothing runs \emph{into} the $(4,0)$ operators apart from by amounts proportional to small Yukawas. 
On the other hand, the $(4,0)$ operators can contribute unsuppressed to the anomalous dimension matrix of $\bar \psi \psi H^3$ and $H^6$ operators, so they are not a closed system under renormalization. 
Studying the running only within the $(4,0)$ operators is nevertheless a good starting point for a general analysis of the structure of the anomalous dimension matrix. From a linear algebra point of view, the running \emph{out} of the $(4,0)$ block of the anomalous dimension matrix will not change the general conclusions of this restricted study, since the eigenvalues of the $(4,0)$ block will remain unchanged, while the eigenvectors will change only by the addition of admixtures of operators in the other classes.\footnote{
To see this, consider a general Wilson coefficient vector $\underline{c} = (\underline{c}_\text{in},\underline{c}_\text{out})$ which separates into parts that are respectively inside and outside the $(4,0)$ block. The anomalous dimension matrix --- modulo the small exceptional amplitudes considered above --- has the block triangular form
\newcommand{\mat}[1]{\underline{\underline{#1}}}
\begin{equation}
    \mat{\gamma} = \begin{pmatrix}
        \mat{\gamma}_{\text{in} \to \text{in}}  & 0 \\
        \mat{\gamma}_{\text{in} \to \text{out}}  & \mat{\gamma}_{\text{out} \to \text{out}}
    \end{pmatrix} \, .
\end{equation}
This means that the eigenvalue equation reads
\begin{equation}
    \begin{pmatrix}
        \mat{\gamma}_{\text{in} \to \text{in}}  & 0 \\
        \mat{\gamma}_{\text{in} \to \text{out}}  & \mat{\gamma}_{\text{out} \to \text{out}}
    \end{pmatrix} \begin{pmatrix}
        \underline{c}_\text{in} \\ \underline{c}_\text{out}
    \end{pmatrix} =
        \begin{pmatrix}
        \mat{\gamma}_{\text{in} \to \text{in}} \cdot \underline{c}_\text{in} \\
        \mat{\gamma}_{\text{in} \to \text{out}} \cdot \underline{c}_\text{in}  + \mat{\gamma}_{\text{out} \to \text{out}} \cdot  \underline{c}_\text{out}
    \end{pmatrix}
    = \lambda \begin{pmatrix}
        \underline{c}_\text{in} \\ \underline{c}_\text{out}
    \end{pmatrix} \, .
\end{equation}
for some eigenvalue $\lambda$. From the top line one can see that, unless $\underline{c}_\text{in}=\underline{0}$, both the eigenvalue $\lambda$ and the eigenvector's $\underline{c}_\text{in}$ composition is determined entirely by the $\mat{\gamma}_{\text{in} \to \text{in}}$ submatrix.}

Another powerful motivation for focusing on the $(4,0)$ operators is that they possess the majority of the parameters: 1460 real parameters, out of 2499 in total for the baryon- and lepton-number conserving dimension 6 SMEFT. They also represent most of the operators that can be generated at tree level in weakly coupled UV completions~\cite{Einhorn:2013kja,deBlas:2017xtg,Craig:2019wmo}, making them generally phenomenologically important.

\subsection{Formalism and factorisation for on-shell amplitudes}

2-cuts, as shown in Eq.~\eqref{eq:nonrenormalizationhelicity}, are not only useful for identifying zeros in the anomalous dimension matrix, but can also be used to calculate non-zero entries of the matrix~\cite{Jiang:2020mhe,Baratella:2020dvw,Baratella:2020lzz,Caron-Huot:2016cwu,EliasMiro:2020tdv,Bern:2020ikv,Jin:2020pwh,Cao:2021cdt}, as we will summarize in Sec.~\ref{sec:gammaUnitarity}. The ingredients needed are simply on-shell tree-level amplitudes, both at dimension 6 and dimension 4.

Each amplitude can be factorised into three parts: kinematic, gauge and flavour. The kinematic part carries only Lorentz indices, the gauge part carries only gauge indices (and can be further factorised into pieces carrying the indices of each SM gauge group), and the flavour part carries only flavour indices:
\begin{equation}
    \mathcal{A}=\text{(kinematics)}\times \text{(gauge)} \times \text{(flavour)}.
\end{equation}
We can therefore separately construct each of these parts, and decompose them in any way that simplifies the problem. The overall cut-constructible loop amplitude can then also be factorised into the tensor product of each of these separate parts.

For the kinematic parts of on-shell amplitudes it is convenient to use spinor-helicity variables, where the momenta for massless particles is written as
\begin{align}
    p_{\mu}\sigma^{\mu}_{\alpha \dot{\beta}} = |p\ra_{\alpha}[p|_{\dot{\beta}}, \qquad     p_{\mu}\bar{\sigma}^{\mu,\alpha \dot{\beta}} = |p\ra^{\alpha}[p|^{\dot{\beta}},
\end{align}
where $\sigma^\mu_{\alpha \dot \beta} = (1, \vec{\sigma})$, $\bar\sigma^{\mu,\dot\alpha \beta} = (1,-\vec{\sigma})$ and $\vec{\sigma}$ are the usual Pauli matrices. The bras and kets represent 2-component spinors, which can be contracted together via the fully antisymmetric tensor $\epsilon_{\alpha\beta}$. We have that
\begin{align}
    \la ij \ra[ji] = 2p_i \cdot p_j = s_{ij},
\end{align}
where $s_{ij}=(p_i+p_j)^2$ and we use the convention $s_{12}\equiv s$, $s_{13}\equiv t$ and $s_{14}\equiv u$. More details on the spinor-helicity formalism and the conventions used are given in App.~\ref{app:conventions}. Amplitudes can then be written as product of integer powers of angle and square brackets,
\begin{align}
    \cA(1^{h_1}2^{h_2}\cdots n^{h_n}) \propto \prod_{i<j} \la ij\ra^{a_{ij}}[ij]^{b_{ij}}.
\end{align}
The 3-point renormalisable interactions of the Standard Model can be written
\begin{align}
    \mathcal{A}_{\rm SM}(\psi^+\psi^+H^{(\dagger)})\propto Y \, [12], \qquad  \mathcal{A}_{\rm SM}(\psi^-\psi^-H^{(\dagger)})\propto Y^\dagger \,  \la 12\ra,
\end{align}
for Yukawa interactions (in our conventions, $Y$ is an element of the Yukawa matrices $Y_u,Y_d,Y_e$, and $Y^\dagger$ an element of their Hermitian conjugates) and
\begin{align}
    \cA_{\rm SM}({\psi^+\psi^-V^-}) \propto \frac{\la 23 \ra^2}{\la 12 \ra} ,~ \cA_{\rm SM}({\psi^+\psi^-V^+}) \propto \frac{[13]^2}{[12]}, ~ \cA_{\rm SM}(HH^{\dagger}V^-) \propto \frac{\la 13\ra \la 23\ra}{\la 12 \ra},
\end{align}
for gauge interactions. The gauge and flavour tensors then simply multiply these structures; the complete forms are given in App.~\ref{app:SM3point}.

A factorizable 4-point amplitude can be computed by `gluing' together two 3-point amplitudes and requiring correct factorization in all-channels.
Writing the amplitudes in this form allows us to easily unveil the helicity patterns of the SM amplitudes. The full list of 4-point SM amplitudes with zero total helicity, in this notation, is given in App.~\ref{app:SM4point}.

Moving to SMEFT interactions, we can set $n=4$ and ${\rm dim\{ spinor\}}=2$ (i.e. dimension 6 operators) and list all possible contact terms involving particles with helicity $|h_i| \le 1$:
\begin{align}
    &\cA_6(V^+V^+ H^\dagger H) \propto [12]^2 , \quad \cA_6(V^+ \psi^+ \psi^+ H) \propto [12][13], \quad \cA_6(\psi^+\psi^- H^\dagger H) \propto [132\ra,  \\
    &\cA_6(\psi^+\psi^+ \psi^+ \psi^+) \propto [12][34] ; [13][24],\quad \!\! \cA_6(\psi^+\psi^+ \psi^- \psi^-) \propto [12]\la 34\ra,\quad \!\! \cA_6(H^\dagger H H^\dagger H ) \propto s_{ij}, \!\nn
\end{align}
where the momentum insertion contracted with spinor brackets is represented by $[ijk\ra \equiv [i|p_j|k\ra$. We can see that there are not so many options for the spinor structure, which reveals the simplicity of the kinematic part of the dimension 6 SMEFT amplitudes in the unbroken phase \cite{Ma:2019gtx,Durieux:2019siw}. The proliferation of operators is instead due to the rich flavour and colour structure in the SM fields. Working directly with on-shell amplitudes will allow us to use the simplicity of the kinematic structure to further explore the complexity of the flavour patterns. The full list of $(4,0)$ SMEFT amplitudes, and the map to the operators in the Warsaw basis \cite{Grzadkowski:2010es} can be found in App.~\ref{app:basis}.
In particular, for operators with multiple copies of a field, we write a basis in terms of symmetric and antisymmetric objects which make explicit the exchange properties of the amplitudes, as per \cite{Ma:2019gtx}.

\subsection{$\mathcal{J}$-vector and partial waves decomposition}
\label{subsec:Jvector}
In order to calculate the anomalous dimensions within the $(4,0)$ operators via on-shell unitarity methods, we will need the tree-level amplitudes induced by those operators, as well as SM tree amplitudes which also have 4 legs and zero total helicity. All these amplitudes, both at dimension 4 and at dimension 6, can be written as a product of two currents, and so we can introduce a more compact form.

In spinor helicity notation, we can collect the currents into a vector $\jvec$ whose components correspond to right- or left-handed fermion currents (components $\mathcal{J}^1$ and $\mathcal{J}^3$ respectively) and a Higgs current (the $\mathcal{J}^2$ component):    
\begin{equation}
\label{eq:Jvectordef}
  \jvec(ij)^{\dot \alpha}_\beta \equiv \begin{pmatrix}
    \sqrket{i}^{\dot \alpha} \angket{j}_\beta \\
    \frac12 \left( \sqrket{i}^{\dot \alpha} \angket{i}_\beta -  \sqrket{j}^{\dot \alpha} \angket{j}_\beta \right) \\
    \sqrket{j}^{\dot \alpha} \angket{i}_\beta
  \end{pmatrix},
\end{equation}
which is normalized such that
\begin{equation}
\label{eq:normJ}
  \epsilon_{\dot\alpha\dot\beta} \epsilon^{\alpha\beta} \, \mathcal{J}^A(ij)^{\dot\alpha}_{\alpha} \mathcal{J}^B(ji)^{\dot\beta}_{\beta} = s_{ij} \begin{pmatrix}
    ~1~ & ~0~ & ~0~ \\ ~0~ & -\frac12~ & ~0~ \\ ~0~ & ~0~ & ~1~
  \end{pmatrix}.
\end{equation}
Note that $\mathcal{J}$ has the spinor indices of a Lorentz vector. Towards the end of this section we will show that $\mathcal{J}^A$ is a triplet of angular momentum in the two body centre-of-mass frame, but we begin by explaining how dim-4 and dim-6 amplitudes can be constructed explicitly from products of $\mathcal{J}$s.

We use the following conventions for the diagrammatic currents: $1,3$ always label particle momenta and $2,4$ always label antiparticle momenta (where the momenta are always incoming). The three currents are then depicted as
\begin{gather}
  \mathord{\begin{tikzpicture}[baseline=-0.65ex]
    \draw[thick,dashed] (-1,1) node[left] {$1$} -- (-\currsep,0) -- (-1,-1) node[left] {$2$};
    \currentmarker{-\currsep,0}
  \end{tikzpicture}}
  = \text{Higgs\,,~~~}
  \mathord{\begin{tikzpicture}[baseline=-0.65ex]
    \draw[thick] (-1,1) node[left] {$1,+$} -- (-\currsep,0) -- (-1,-1) node[left] {$2,-$};
    \currentmarker{-\currsep,0}
  \end{tikzpicture}}
  = \text{RH fermion\,,} 
  \mathord{\begin{tikzpicture}[baseline=-0.65ex]
    \draw[thick] (-1,1) node[left] {$1,-$} -- (-\currsep,0) -- (-1,-1) node[left] {$2,+$};
    \currentmarker{-\currsep,0}
  \end{tikzpicture}}
  = \text{LH fermion\,,}
\end{gather}
where RH (LH) denote right- (left-) handed. We further use dotted blue/red lines to depict a current which can be any of the above options, and solid blue/red lines to depict a current which can be either a left- or right-handed fermion current.
Then all the operator amplitudes in the $(4,0)$ block of the dimension 6 SMEFT can be drawn by putting two of these currents together.

In the case of \emph{distinguishable} currents, the kinematic parts of all $(4,0)$ operator amplitudes have the form
\begin{align}
\label{eq:JJdistinguishable}
\mathord{\begin{tikzpicture}[baseline=-0.65ex]
  \draw[thick,dotted,blue] (-1,1) node[left] {$1$} -- (-\currsep,0) -- (-1,-1) node[left] {$2$};
  \draw[thick,dotted,red] (1,1) node[right] {$3$} -- (\currsep,0) -- (1,-1) node[right] {$4$};
  \currentmarker{\currsep,0}
  \currentmarker{-\currsep,0}
\end{tikzpicture}}
&= -2 \mathcal{J}^{\color{blue}A}(12) \mathcal{J}^{\color{red}B}(34)
=   \begin{pmatrix}
    2 \sqr13 \ang42 & 2 \ang23 \sqr31 & 2 \sqr14 \ang32 \\
    2 \ang41 \sqr13 & t-u & 2 \sqr41 \ang13 \\
    2 \ang14 \sqr32 & 2 \sqr23 \ang31 & 2 \ang13 \sqr42 \\
  \end{pmatrix}, \end{align}
   where ${\color{blue}A}$ and ${\color{red} B}$ are the row and column indices of the matrix, and the contraction with $\epsilon_{\dot{\alpha}\dot{\beta}}\epsilon^{\alpha\beta}$ is implied whenever the Lorentz indices are omitted from the $\jvec$. We can easily identify (c.f.~the tables in App.~\ref{app:basis}) the amplitudes corresponding to operators of the form $\psi^2\bar{\psi}^2$ ($A,B \in \{1,3\}$, such as $\op_{LL},\op_{Le},\op_{ee}$), $\psi \bar{\psi}H^2$ ($A=2$ and $B \in \{1,3\}$ or vice-versa, such as $\op_{HL(1)},\op_{He}$). Note that it is not possible to have two distinguishable Higgs currents, so the entry for $A=B=2$ in Eq.~(\ref{eq:JJdistinguishable}) doesn't correspond to any SMEFT operator, so for four-Higgs operators we must instead consider the case of indistinguishable currents, which we do now.
   
In the case of \emph{indistinguishable} currents, the overall amplitude is (anti)symmetric under the exchange of the pair of bosons (fermions) or antibosons (antifermions). Necessarily $A=B$, and the kinematic parts of the operators for the symmetric and antisymmetric combinations, respectively, take the form
  \begin{align}
\mathord{\begin{tikzpicture}[baseline=-0.65ex]
  \draw[thick,dotted,blue] (-1,1) node[left] {$1$} -- (-\currsep,0) -- (-1,-1) node[left] {$2$};
  \draw[thick,dotted,blue] (1,1) node[right] {$3$} -- (\currsep,0) -- (1,-1) node[right] {$4$};
  \node at (-\bracketsep,0) {$($};
  \node at (\bracketsep,0) {$)$};
  \currentmarker{\currsep,0}
  \currentmarker{-\currsep,0}
\end{tikzpicture}}
&= -4 \mathcal{J}^{\color{blue}A}(12) \mathcal{J}^{\color{blue}A}(34)  -4 \mathcal{J}^{\color{blue}A}(32) \mathcal{J}^{\color{blue}A}(14)
= \begin{pmatrix}
  0 \\ 6 t \\ 0
\end{pmatrix}\,, \\
\mathord{\begin{tikzpicture}[baseline=-0.65ex]
  \draw[thick,dotted,blue] (-1,1) node[left] {$1$} -- (-\currsep,0) -- (-1,-1) node[left] {$2$};
  \draw[thick,dotted,blue] (1,1) node[right] {$3$} -- (\currsep,0) -- (1,-1) node[right] {$4$};
  \node at (-\bracketsep,0) {$[$};
  \node at (\bracketsep,0) {$]$};
  \currentmarker{\currsep,0}
  \currentmarker{-\currsep,0}
\end{tikzpicture}}
&= -4 \mathcal{J}^{\color{blue}A}(12) \mathcal{J}^{\color{blue}A}(34)  + 4 \mathcal{J}^{\color{blue}A}(32) \mathcal{J}^{\color{blue}A}(14)
= \begin{pmatrix}
  8 \sqr13 \ang42 \\ 2(s-u) \\ 8 \ang13 \sqr42
\end{pmatrix} \,,
\end{align}
where ${\color{blue}A}$ is the row index of the vector. Notice that for fermions only the antisymmetric combination is possible,\footnote{And gauge and flavour factors must enter in symmetric-symmetric or antisymmetric-antisymmetric combinations to make an overall exchange antisymmetric amplitude.} whereas for Higgs currents there are two distinct kinematic factors for the symmetric and antisymmetric cases.\footnote{These must be respectively multiplied by symmetric and antisymmetric $SU(2)$ factors to make overall exchange symmetric amplitudes.} We can then see, by referring to App.~\ref{app:basis}, that the (anti)symmetrised product of Higgs currents produces the kinematic parts of the four Higgs operator $\mathcal{O}_{HD+}$ ($\mathcal{O}_{HD-}$). And the antisymmetrised products of fermion currents produce the kinematic parts of the four fermion operators with identical fermions, e.g.~$\mathcal{O}_{dd}$, $\mathcal{O}_{LL}$, etc.

The kinematic factors associated with the dimension 4 (SM) amplitudes can also be written as products of two currents, with a propagator factor. The kinematic parts of the gauge amplitudes can all be written
\begin{align}
\label{eq:SMdiag}
  \mathord{\begin{tikzpicture}[baseline=-0.65ex]
    \draw[thick,dotted,blue] (-1,1) node[left] {$1$} -- (-\threepointsep,0) -- (-1,-1) node[left] {$2$};
    \draw[thick,dotted,red] (1,1) node[right] {$3$} -- (\threepointsep,0) -- (1,-1) node[right] {$4$};
    \draw[thick,decorate,decoration={snake}] (\threepointsep,0) -- (-\threepointsep,0);
  \end{tikzpicture}}
  &= \frac{1}{s_{12}} \times -2 \mathcal{J}^{\color{blue}A}(12) \mathcal{J}^{\color{red}B}(34)\,,
\end{align}
and the Yukawa amplitudes can be written (here, free indices refer to fermions, $\{A,B\}=\{1,3\}$, as indicated by the solid lines)
\begin{align}
  \mathord{\begin{tikzpicture}[baseline=-0.65ex]
    \draw[thick,blue] (-1,1) node[left] {$1$} -- (0,\threepointsep);
    \draw[thick,red] (0,\threepointsep) -- (1,-1) node[right] {$4$};
    \draw[thick,blue] (-1,-1) node[left] {$2$} -- (0,-\threepointsep);
    \draw[thick,red] (0,-\threepointsep) -- (1,1) node[right] {$3$};
    \draw[thick,dashed] (0,\threepointsep) -- (0,-\threepointsep);
  \end{tikzpicture}} &= \frac{1}{s_{14}} \times \mathcal{J}^{\color{blue}A}(12) \mathcal{J}^{\color{red}B}(34) \quad ({\color{blue}A} \neq {\color{red}B})\,,\\
  \mathord{\begin{tikzpicture}[baseline=-0.65ex]
    \draw[thick,blue] (-1,1) node[left] {$1$} -- (0,\threepointsep) -- (0,-\threepointsep) -- (-1,-1) node[left] {$2$};
    \draw[thick,dashed] (0,\threepointsep) -- (1,1) node[right] {$3$};
    \draw[thick,dashed] (0,-\threepointsep) -- (1,-1) node[right] {$4$};
  \end{tikzpicture}} &= \frac{1}{s_{13}} \times \mathcal{J}^{\color{blue}A}(12) \mathcal{J}^{2}(34)\,, \\
  \mathord{\begin{tikzpicture}[baseline=-0.65ex]
    \draw[thick,blue] (-1,1) node[left] {$1$} -- (0,\threepointsep) -- (0,-\threepointsep) -- (-1,-1) node[left] {$2$};
    \draw[thick,dashed] (0,\threepointsep) -- (1,-1) node[right] {$4$};
    \draw[thick,dashed] (0,-\threepointsep) -- (1,1) node[right] {$3$};
  \end{tikzpicture}} &= \frac{1}{s_{14}} \times \mathcal{J}^{\color{blue}A}(12) \mathcal{J}^{2}(34)\,.
\end{align}
So, all $(4,0)$ amplitudes at both dimension 6 and dimension 4 can be phrased as products of two $\jvec$ vectors.
It turns out that the entries of the vector can be projected into angular momentum eigenstates, and thus make the connection with the partial wave decomposition treatments of \cite{Baratella:2020dvw,Jiang:2021tqo,Li:2022abx}. For this, consider in the centre-of-mass frame, the $\theta,\phi$ dependence of a two particle state $\ket{J,M,h_1,h_2}$ with angular momentum quantum numbers $J$, $M$ (given that the particles have respective helicities $h_1$ and $h_2$). We can infer the two body state's dependence in $\theta,\phi$ by expanding the momenta $i$ and $j$ with respect to spinors built out of reference momenta pointing in the $z_+$ and $z_-$ directions (using the rotation matrices of \cite{Arkani-Hamed:2017jhn}):
\begin{align}
  \angket{i} &= c_{\theta/2} \angket{z_+} - s_{\theta/2} e^{i \phi} \angket{z_-}, \qquad ~~\,  \angket{j} = s_{\theta/2} e^{-i \phi} \angket{z_+} + c_{\theta/2} \angket{z_-}, \nonumber\\
  \sqrket{i} &= c_{\theta/2} \sqrket{z_+} - s_{\theta/2} e^{-i \phi} \sqrket{z_-},\qquad \sqrket{j} = s_{\theta/2} e^{i \phi} \sqrket{z_+} + c_{\theta/2} \sqrket{z_-}.
\end{align}
This allow us to compute $\braket{\theta,\phi}{\mathcal{J}^A(ij)}$ and identify the trigonometric functions that appear with the Wigner $d$-matrices. Now, given that \cite{Baratella:2020dvw}
\begin{equation}
  \braket{\theta,\phi}{J,M,h_1,h_2} = e^{i \phi (h_{12} - M)} \, \sqrt{2 J + 1} \, d^J_{M h_{12}}(\theta),
\end{equation}
where $h_{12} = h_1-h_2$, we can compute the $\mathcal{J}^A(ij)$ projection into different angular momentum eigenstates,
\begin{align}
  \braket{J,M,+\frac12,-\frac12}{\mathcal{J}^1(ij)} &=
\int \frac{\dd \Pi_2}{4 \pi}
  \braket{J,M, +\frac12,-\frac12}{\theta,\phi} \braket{\theta,\phi}{\mathcal{J}^1(ij)} \nonumber\\
  &= \frac{1}{\sqrt{2J+1}} \delta_{J,1} \delta_{M,1} \sqrket{z_+} \angket{z_-}.
\end{align}
where spinor indices are implicit and the integral measure over the 2-sphere is
\begin{equation}
\int \dd \Pi_2 \equiv \int_0^{2 \pi} \dd \phi \int_{-1}^1 \dd (\cos \theta) \, .
\label{eq:TwoBodyPS}
\end{equation}
Similarly,
\begin{align}
  \braket{J,M,0,0}{\mathcal{J}^2(ij)} &= \frac{1}{\sqrt{2J+1}} \delta_{J,1} \delta_{M,0} \frac12 (\sqrket{z_+} \angket{z_+}  - \sqrket{z_-} \angket{z_-}), \nonumber \\
  \braket{J,M,-\frac12,+\frac12}{\mathcal{J}^3(ij)} &=
  \frac{1}{\sqrt{2J+1}} \delta_{J,1} \delta_{M,-1} \angket{z_+} \sqrket{z_-}.
\end{align}
We see that the $\jvec$ vector simply collects the states corresponding to the triplet of $J=1$ partial waves. All operators within the $(4,0)$ block can be constructed from these $J=1$ currents.

For when it comes to cutting loop amplitudes in \cref{sec:gammaUnitarity}, we note therefore that two legs of the same current are always in a $J=1$ state, but two legs from different currents are not necessarily so \cite{Baratella:2020dvw}.

 \subsection{Gauge structure}
 
Amplitudes contain invariant tensors of $SU(3)_c$ and $SU(2)_L$, built out of the external states' gauge indices. For the $(4,0)$ operators, we need to consider two cases. First, with one upper (fundamental) and one lower (anti-fundamental) index of $SU(N)$; in this case the only invariant tensor is the Kronecker delta $ \delta^a_b$. Second, with two upper (fundamental) and two lower (anti-fundamental) indices of $SU(N)$;  in this case there are two possible invariants.
 For distinguishable currents, similar to the Warsaw basis,\footnote{In this paper, as in the Warsaw basis \cite{Grzadkowski:2010es}, there is an extra factor of $\frac14$ for the $SU(3)$ invariant $[\lambda^A]^a_b [\lambda^A]^c_d$ in `$(8)$' operators.} we choose the invariants to be
 \begin{align}
   \delta \, \delta \equiv \delta^a_b \delta^c_d \, , \qquad
   \lambda \, \lambda \equiv [\lambda^A]^a_b [\lambda^A]^c_d \, ,
 \end{align}
 where the lambda matrices satisfy $\Tr [ \lambda^A \lambda^B ] = 2 \delta^{AB}$ and reduce to the Pauli (Gell-Mann) matrices when $N=2(3)$. For indistinguishable currents, it is more convenient to decompose in terms of combinations with manifest exchange symmetry, $\delta \, \delta = (\delta \, \delta) + [\delta \, \delta] \,$, where
 \begin{align}
    (\delta\,\delta) \equiv \delta^{(a}_{(b} \delta^{c)}_{d)} \equiv  \frac12 \left( \delta^{a}_{b} \delta^{c}_{d} + \delta^{a}_{d} \delta^{c}_{b} \right) \, , \qquad
   [\delta\,\delta] \equiv  \delta^{[a}_{[b} \delta^{c]}_{d]} \equiv \frac12 \left( \delta^{a}_{b} \delta^{c}_{d} - \delta^{a}_{d} \delta^{c}_{b} \right) \, .
  \end{align}
 The above pairs of invariants satisfy the Fierz relations 
 \begin{gather}
 \label{eq:symmFierz}
   (\delta \, \delta) = \left( \frac{N+1}{2 N} \right) \delta \, \delta + \frac14 \lambda \, \lambda \, ; \qquad
   [\delta \, \delta] = \left( \frac{N-1}{2 N} \right) \delta \, \delta - \frac14 \lambda \, \lambda \, .
 \end{gather}

 \subsection{Flavour structure\label{sec:flavourStructure}}
 
To understand the flavour structure of the amplitudes, it will be advantageous to decompose the Wilson coefficients into their irreducible representations (irreps) under
 \begin{equation}
   SU(3)^5 = SU(3)_Q \times SU(3)_u \times SU(3)_d \times SU(3)_L \times SU(3)_e \, .
 \end{equation}
 This is the flavour symmetry group of the SM gauge and kinetic terms, with the five global $U(1)$s of the full SM Lagrangian factored out (these are trivially conserved by the dimension 6 SMEFT RG). This $SU(3)^5$ is preserved by the gauge interactions, and hierarchically broken into smaller subgroups by the Yukawa matrices, which we define via the Lagrangian\footnote{N.B.~Our Yukawa matrices are the Hermitian conjugates of those in~\cite{Jenkins:2013wua,Jenkins:2013zja,Alonso:2013hga}.}
\begin{equation}
\label{eq:YukawaLag}
\mathcal{L}_{Yuk}=-Y_u \,\bar Q u \tilde{H}-Y_d \,\bar Q d H -  Y_e \,\bar L e H.
\end{equation}
 Thus, we can get a better handle on the approximate flavour-space subsystems in the RG in an appropriate basis of $SU(3)^5$ irreps. By charging the Wilson coefficients under this flavour group, we will implicitly perform a \emph{fully general spurion analysis}, in which the full SMEFT Lagrangian up to dimension 6 is invariant under $SU(3)^5$.
 
 Note, however, that we are not restricting the form of the Wilson coefficients, nor reducing the number of parameters, but for two exceptions. In the $(4,0)$ block $\op_{Hud}$ and $\op_{LedQ}$ do not take the form of the product of two currents of given particle species, and so we drop them from our treatment. The mixing of $\op_{Hud}$ with the rest of the $(4,0)$ block is suppressed by a factor of $Y_u \times Y_d$, the same parametric suppression that isolates $(4,0)$ operators from the $(4,2)$ operators \cite{Cheung:2015aba}. The mixing of $\op_{LedQ}$ with the rest of the $(4,0)$ block is even more suppressed, by a factor of $Y_e \times Y_d$. Dropping $\op_{Hud}$ and $\op_{LedQ}$ in this manner means we need only consider three classes of Hermitian Wilson coefficients: $c$, $c^p_q$ and $c^{pr}_{qs}$.\footnote{These are Hermitian in the sense that $(c)^* = c$, $(c^p_q)^* = c^q_p$ and $(c^{pr}_{qs})^* = c^{qs}_{pr}$.}

The $H^4 D^2$ operators, with real Wilson coefficient $c$, are trivially singlets of $SU(3)^5$. Of the two-fermion operators, $\psi \bar \psi H^2 D$, all their Wilson coefficients $c^p_q$ transform as a
 \begin{equation}
   \mathbf{3}_F \otimes \overline{\mathbf{3}}_F = \mathbf{1}_F \oplus \mathbf{8}_F \, ,
   \label{eq:TwoF}
 \end{equation}
 under some $SU(3)_F$, $F \in \{Q,u,d,L,e\}$, with the exception of $\op_{Hud}$, whose Wilson coefficient is an irreducible $\mathbf{3}_d \otimes \overline{\mathbf{3}}_u$. The convention for the indices here are that upper indices on the Wilson coefficients label the flavour of a fermion in the operator (i.e. fields that transform as a $\mathbf{3}_F$), while lower indices label the flavour of an antifermion (i.e. fields that transform as a $\bar{\mathbf{3}}_F$). On the fields themselves, fermions carry lower indices and antifermions carry upper indices. 
 
 Consider now the four-fermion current-current operators, $\psi^2 \bar \psi^2$. In the case of distinguishable currents $F_1,F_2 \in \{Q,u,d,L,e\}$ and $F_1 \neq F_2$, the Wilson coefficients $c^{pr}_{qs}$ transform as a direct product of two current decompositions
 \begin{equation}
   \mathbf{3}_{F_1} \otimes \overline{\mathbf{3}}_{F_1} \otimes \mathbf{3}_{F_2} \otimes \overline{\mathbf{3}}_{F_2} = \left(\mathbf{1}_{F_1} \otimes \mathbf{1}_{F_2} \right) \oplus \left(\mathbf{1}_{F_1} \otimes \mathbf{8}_{F_2} \right) \oplus \left(\mathbf{8}_{F_1} \otimes \mathbf{1}_{F_2} \right) \oplus \left(\mathbf{8}_{F_1} \otimes \mathbf{8}_{F_2} \right)\, .
 \end{equation}

 In the case of indistinguishable currents, as with the kinematic and gauge structures above, it is convenient to decompose the Wilson coefficients in flavour space in terms of components that are entirely symmetric, or entirely antisymmetric, in the identical particles:
 \begin{equation}
   c^{pr}_{qs}=c^{(pr)}_{(qs)}+c^{[pr]}_{[qs]}\,,
 \end{equation}
 where
 \begin{subequations}
 \begin{align}
 (c) &\equiv c^{(pr)}_{(qs)} \equiv \frac14 \left(c^{pr}_{qs}+c^{rp}_{qs} +c^{pr}_{sq} +c^{rp}_{sq} \right) , \\
 [c] &\equiv c^{[pr]}_{[qs]} \equiv \frac14 \left(c^{pr}_{qs}-c^{rp}_{qs} -c^{pr}_{sq} +c^{rp}_{sq} \right) .
 \end{align}
 \end{subequations}
 The (anti)symmetrised components in turn map neatly onto $SU(3)_F$ irreps. The symmetric coefficient $c^{(pr)}_{(qs)}$ decomposes as
 \begin{equation}
 \left(\mathbf{3}_{F} \otimes \mathbf{3}_{F}\right)_\text{sym} \otimes \left(\overline{\mathbf{3}}_{F} \otimes \overline{\mathbf{3}}_{F} \right)_\text{sym} =  \mathbf{6}_F \otimes \overline{\mathbf{6}}_F = \mathbf{1}_F \oplus \mathbf{8}_F \oplus \mathbf{27}_F \, ,
 \label{eq:FourSymF}
 \end{equation}
 whereas the antisymmetric coefficient $c^{[pr]}_{[qs]}$ decomposes as
 \begin{equation}
   \left(\mathbf{3}_{F} \otimes \mathbf{3}_{F}\right)_\text{antisym} \otimes \left(\overline{\mathbf{3}}_{F} \otimes \overline{\mathbf{3}}_{F} \right)_\text{antisym} = \overline{\mathbf{3}}_F \otimes \mathbf{3}_F  = \mathbf{1}_F \oplus \mathbf{8}_F \, .
   \label{eq:FourAntisymF}
 \end{equation}
 Note that, due to the absence of an antisymmetric gauge tensor, the $\mathcal{O}_{ee}$ operator does not support an antisymmetric flavour tensor $c^{[pq]}_{[rs]}$ \cite{Alonso:2013hga}, as its combination with the necessarily antisymmetric kinematic structure would result in an overall exchange symmetric amplitude, in violation of spin statistics.

 To formulate the explicit decomposition into irreps, we use the conventions of \cite{deSwart:1963pdg,Kaeding:1995vq} developed for the $SU(3)$ of light flavours $u,d,s$, and their familiar quantum numbers of isospin and strangeness. Each component of a Wilson coefficient's decomposition is indexed by
 \begin{itemize}
   \item $d_{\rm irrep}$, the irrep dimension;
   \item $\ifl$, total \emph{``lightspin''} (the generational analogue of isospin);
   \item $\itfl$, third component of \emph{``lightspin''}, and
   \item $\yfl$, \emph{``thirdness''} (the generational analogue of strangeness, a.k.a.\ hypercharge).
 \end{itemize}
  Acting on the fundamental $\mathbf{3}_F$, these three quantum numbers $\{\ifl,\itfl,\yfl\}$ are realised by matrices proportional to
 \begin{equation}
   \lambda_1^2+\lambda_2^2+\lambda_3^2,~ \lambda_3,~\lambda_8\,,
 \end{equation}
respectively, in terms of the Gell-Mann matrices.\footnote{We work in a convention where these are defined $\lambda_1=\tiny{\begin{pmatrix}
   \cdot & 1 & \cdot \\
   1& \cdot  & \cdot \\
   \cdot & \cdot & \cdot
 \end{pmatrix}}$, $\lambda_2=\tiny{\begin{pmatrix}
   \cdot & -i & \cdot \\
   i& \cdot  & \cdot \\
   \cdot & \cdot & \cdot
 \end{pmatrix}}$, $\lambda_3=\tiny{\begin{pmatrix}
  1 & \cdot  & \cdot \\
   \cdot & -1  & \cdot \\
   \cdot & \cdot & \cdot
 \end{pmatrix}}$, $\lambda_8=\frac{1}{\sqrt{3}}\tiny{\begin{pmatrix}
  1 & \cdot  & \cdot \\
   \cdot & 1  & \cdot \\
   \cdot & \cdot & -2
 \end{pmatrix}}$.} The $\{\ifl,\itfl,\yfl\}$ values of components of the fundamental and antifundamental irreps are shown in \cref{fig:fundCharges}. It can be seen here that only the first two generations have non-zero $\itfl$ (hence the suggestion of the name ``lightspin'') while the third generation has only non-zero $\yfl$ (hence ``thirdness''). 
 
We pause here to define what we mean by the generation indices. As far as possible, it is most convenient to align these with the mass basis, which is straightforward and implies no loss of generality for all fields except for $Q$ (so, for example the right-handed up-type fields $u_R,c_R,t_R$ can be simply labelled $u_R^1, u_R^2, u_R^3$). Due to the CKM misalignment between up-type and down-type left handed quarks, for the $Q$ generations we can choose either to work in the basis where $Y_u$ is diagonalised (`up-basis') or where $Y_d$ is diagonalised (`down-basis'). These are related by a rotation in quark flavour, i.e.~by an $SU(3)_Q$ rotation, under which components within an $SU(3)_Q$ irrep  rotate amongst themselves, but the irreps themselves remain unchanged. In the rest of this paper, we choose to work in the `up-basis', meaning that the flavour index on the quark doublet is identified with that of the up-type quarks: $Q^p\equiv (u^p_L, V_q^p d^q_L)^T$. This is a convenient basis in which to understand the anomalous dimension matrix, since the large top Yukawa in this basis is aligned entirely to the third generation, which makes its effects more transparent. 
 
 \begin{figure}
  \centering
  \begin{tikzpicture}[scale=1.5]
    \draw[->] (-1,0) -- (1,0) node[right] {$\itfl$};
    \draw[->] (0,-1) -- (0,1) node[above] {$\yfl$};
    \draw (0.5,0) -- (0.5,-0.1) node[below] {\small$\frac12$};
    \draw (0,-0.66) -- (-0.1,-0.66) node[left] {\small$-\frac23$};
    \node at (0.5,0.33) [isohalfstyle,label={[labelstyle]\angleone:$c_{3,1}$}]{};
    \node at (-0.5,0.33) [isohalfstyle,label={[labelstyle]\angleone:$c_{3,2}$}]{};
    \node at (0,-0.66) [iso0style,label={[labelstyle]\angleone:$c_{3,3}$}]{};
  \end{tikzpicture}
  \begin{tikzpicture}[scale=1.5]
    \draw[->] (-1,0) -- (1,0) node[right] {$\itfl$};
    \draw[->] (0,-1) -- (0,1) node[above] {$\yfl$};
    \draw (0.5,0) -- (0.5,0.1) node[above] {\small$\frac12$};
    \draw (0,0.66) -- (-0.1,0.66) node[left] {\small$\frac23$};
    \node at (-0.5,-0.33) [isohalfstyle,label={[labelstyle]\angleone:$c_{\bar{3},1}$}]{};
    \node at (0.5,-0.33) [isohalfstyle,label={[labelstyle]\angleone:$c_{\bar{3},2}$}]{};
    \node at (0,0.66) [iso0style,label={[labelstyle]\angleone:$c_{\bar{3},3}$}]{};
  \end{tikzpicture}
  \caption{Components of the fundamental irreps $\mathbf{3}$ and $\overline{\mathbf{3}}$. Total lightspin ($\ifl$) key: \isozerokey$\,=0$, \isohalfkey$\,=\frac12$.
  \label{fig:fundCharges}}
 \end{figure}

We return now to the flavour decomposition of Wilson coefficients. If the dimension $d_{\rm irrep}$ of its irrep is larger than 1, the coefficient is flavour \emph{non-universal}. Meanwhile, the `magnetic' quantum numbers of the generational $SU(3)$, $\itfl$ and $\yfl$, parameterise the degree of \emph{off-diagonality}. In terms of the number $n_i$ of $i$th generation particle (i.e. upstairs) indices in the Wilson coefficient (and the number $\overline{n}_i$ of $i$th generation antiparticle (i.e. downstairs) indices)
 \begin{align}
   \itfl &= \frac12 \left( n_1 - n_2 \right) - \frac12 \left( \overline{n}_1 - \overline{n}_2 \right) \, ,\label{eq:I3ns} \\
   \yfl &= \frac13 \left( n_1 + n_2 - 2 n_3 \right) - \frac13 \left( \overline{n}_1 + \overline{n}_2 - 2 \overline{n}_3 \right) \, \label{eq:Yns}.
 \end{align}
 For example, the coefficient $c^{11}_{23}$ has $n_1=2,\, \overline{n}_2=1,\, \overline{n}_3=1$ and hence $(\itfl,\yfl)=(\frac32,1)$. 
 In Fig.~\ref{fig:irrepCharges} we show the components of the singlet, octet and 27-plet irreps on the $\{\ifl,\itfl,\yfl\}$ plane.
 
  \begin{figure}
   \centering
   \begin{tikzpicture}[scale=1.5]
    \draw[->] (-0.6,0) -- (0.6,0) node[right] {$\itfl$};
    \draw[->] (0,-0.6) -- (0,0.6) node[above] {$\yfl$};
    \node at (0,0) [iso0style,label={[labelstyle]\angleone:$c_{1,1}$}]{};
  \begin{scope}[shift = {(2.5,0)}]
    \coordinate (e1) at (0.5,1);
    \coordinate (e2) at (-0.5,1);
    \coordinate (e3) at (1,0);
    \coordinate (e4) at (0,0);
    \coordinate (e5) at (-1,0);
    \coordinate (e6) at (0,0);
    \coordinate (e7) at (0.5,-1);
    \coordinate (e8) at (-0.5,-1);
    \draw[gray,dashed] (e1) -- (e2) -- (e5) -- (e8) -- (e7) -- (e3) -- cycle;
    \draw[->] (-1.2,0) -- (1.2,0) node[right] {$\itfl$};
    \draw[->] (0,-1.2) -- (0,1.2) node[above] {$\yfl$};
    \draw (1,0) -- (1,-0.1) node[below] {\small$1$};
    \draw (0,1) -- (0.1,1) node[right] {\small$1$};
    \node at (e1) [isohalfstyle,label={[labelstyle]\angleone:$c_{8,1}$}]{};
    \node at (e2) [isohalfstyle,label={[labelstyle]\angleone:$c_{8,2}$}]{};
    \node at ($(e3) + (\angleone:0.1)$) [iso1style,label={[labelstyle]\angleone:$c_{8,3}$}]{};
    \node at ($(e4) + (\angleone:0.1)$) [iso1style,label={[labelstyle]\angleone:$c_{8,4}$}]{};
    \node at ($(e5) + (\angleone:0.1)$) [iso1style,label={[labelstyle]\angleone:$c_{8,5}$}]{};
    \node at ($(e6) + (\angletwo:0.1)$) [iso0style,label={[labelstyle]\angletwo:$c_{8,6}$}]{};
    \node at (e7) [isohalfstyle,label={[labelstyle]\angleone:$c_{8,7}$}]{};
    \node at (e8) [isohalfstyle,label={[labelstyle]\angleone:$c_{8,8}$}]{};
      \end{scope}
  \begin{scope}[shift = {(6.5,0)}]
    \draw[->] (-2.2,0) -- (2.2,0) node[right] {$\itfl$};
    \draw[->] (0,-2.2) -- (0,2.2) node[above] {$\yfl$};
    \draw (2,0) -- (2,-0.1) node[below] {\small$2$};
    \draw (0,2) -- (-0.1,2) node[left] {\small$2$};
    \draw[gray,dash dot] (1,2) -- (-1,2) -- (-2,0) -- (-1,-2) -- (1,-2) -- (2,0) -- cycle;
    \draw[gray,dashed] (0.5,1) -- (-0.5,1) -- (-1,0) -- (-0.5,-1) -- (0.5,-1) -- (1,0) -- cycle;
    \node at (1,2) [iso1style,label={[labelstyle]\angleone:$c_{27,1}$}]{};
    \node at (0,2) [iso1style,label={[labelstyle]\angleone:$c_{27,2}$}]{};
    \node at (-1,2) [iso1style,label={[labelstyle]\angleone:$c_{27,3}$}]{};

    \node at ($(1.5,1) + (\angleone:0.1)$) [iso3halfstyle,label={[labelstyle]\angleone:$c_{27,4}$}]{};
    \node at ($(0.5,1) + (\angleone:0.1)$) [iso3halfstyle,label={[labelstyle]\angleone:$c_{27,5}$}]{};
    \node at ($(-0.5,1) + (\angleone:0.1)$) [iso3halfstyle,label={[labelstyle]\angleone:$c_{27,6}$}]{};
    \node at ($(-1.5,1) + (\angleone:0.1)$) [iso3halfstyle,label={[labelstyle]\angleone:$c_{27,7}$}]{};
    \node at ($(0.5,1) + (\angletwo:0.1)$) [isohalfstyle,label={[labelstyle]\anglethree:$c_{27,8}$}]{};
    \node at ($(-0.5,1) + (\angletwo:0.1)$) [isohalfstyle,label={[labelstyle]\angletwo:$c_{27,9}$}]{};

    \node at ($(2,0) + (\angleone:0.1)$) [iso2style,label={[labelstyle]\angleone:$c_{27,10}$}]{};
    \node at ($(1,0) + (\angleone:0.1)$) [iso2style,label={[labelstyle]\angleone:$c_{27,11}$}]{};
    \node at ($(0,0) + (\angleone:0.1)$) [iso2style,label={[labelstyle]\angleone:$c_{27,12}$}]{};
    \node at ($(-1,0) + (\angleone:0.1)$) [iso2style,label={[labelstyle]\angleone:$c_{27,13}$}]{};
    \node at ($(-2,0) + (\angleone:0.1)$) [iso2style,label={[labelstyle]\angleone:$c_{27,14}$}]{};

    \node at ($(1,0) + (\angletwo:0.1)$) [iso1style,label={[labelstyle]\anglethree:$c_{27,15}$}]{};
    \node at ($(0,0) + (\angletwo:0.1)$) [iso1style,label={[labelstyle]\angletwo:$c_{27,16}$}]{};
    \node at ($(-1,0) + (\angletwo:0.1)$) [iso1style,label={[labelstyle]\angletwo:$c_{27,17}$}]{};
    \node at ($(0,0) + (\anglethree:0.1)$) [iso0style,label={[labelstyle]\anglethree:$c_{27,18}$}]{};

    \node at ($(1.5,-1) + (\angleone:0.1)$) [iso3halfstyle,label={[labelstyle]\angleone:$c_{27,19}$}]{};
    \node at ($(0.5,-1) + (\angleone:0.1)$) [iso3halfstyle,label={[labelstyle]\angleone:$c_{27,20}$}]{};
    \node at ($(-0.5,-1) + (\angleone:0.1)$) [iso3halfstyle,label={[labelstyle]\angleone:$c_{27,21}$}]{};
    \node at ($(-1.5,-1) + (\angleone:0.1)$) [iso3halfstyle,label={[labelstyle]\angleone:$c_{27,22}$}]{};
    \node at ($(0.5,-1) + (\angletwo:0.1)$) [isohalfstyle,label={[labelstyle]\anglethree:$c_{27,23}$}]{};
    \node at ($(-0.5,-1) + (\angletwo:0.1)$) [isohalfstyle,label={[labelstyle]\angletwo:$c_{27,24}$}]{};

    \node at (1,-2) [iso1style,label={[labelstyle]\angleone:$c_{27,25}$}]{};
    \node at (0,-2) [iso1style,label={[labelstyle]\angleone:$c_{27,26}$}]{};
    \node at (-1,-2) [iso1style,label={[labelstyle]\angleone:$c_{27,27}$}]{};
          \end{scope}
  \end{tikzpicture}
  \caption{Components of the irreps $\mathbf{1}$, $\mathbf{8}$ and $\mathbf{27}$ as a function of their flavour charges. Total lightspin ($\ifl$) key: \isozerokey$\,=0$, \isohalfkey$\,=\frac12$, \isoonekey$\,=1$, \isothreehalfkey$\,=\frac32$, \isotwokey$\,=2$.
  \label{fig:irrepCharges}}
\end{figure}

\subsubsection{Phenomenological interpretation of flavour quantum numbers}
Explicitly, for the two-fermion operators transforming as $\mathbf{3}_F \otimes \overline{\mathbf{3}}_F$, the components of the singlet  ($c_{1,1}$) and the octet ($c_{8,i}$ where $i=\{1,\cdots,8\}$) are related to the components of the Warsaw basis coefficients $c^p_q$ by\footnote{Recall $(c^p_q)^* = c^q_p$, $(c^{pr}_{qs})^* = c^{qs}_{pr}$.}
\begin{align}
c_{1,1}&=\frac{c_1^1}{\sqrt{3}}+\frac{c_2^2}{\sqrt{3}}+\frac{c_3^3}{\sqrt{3}},~~
c_{8,1}=c_3^1, ~~ c_{8,2}=c_3^2, ~~ c_{8,3}=c_2^1,~~ c_{8,4}=\frac{c_2^2}{\sqrt{2}}-\frac{c_1^1}{\sqrt{2}},\nonumber \\ c_{8,5}&=-c_1^2,~~c_{8,6}=-\frac{c_1^1}{\sqrt{6}}-\frac{c_2^2}{\sqrt{6}}+\sqrt{\frac{2}{3}} c_3^3,~~ c_{8,7}=c_2^3,~~ c_{8,8}=-c_1^3.
\end{align}
With these relations we can see more deeply the physical meaning of the $SU(3)_F$ quantum numbers $\{\ifl,\itfl,\yfl\}$ shown in Fig.~\ref{fig:irrepCharges}. The singlet $c_{1,1}$ is the diagonal and flavour-universal part of the Wilson coefficient, while the two components of the octet which also sit at the origin in the $\{\itfl,\yfl\}$ plane ($c_{8,4}$ and $c_{8,6}$) are diagonal but flavour non-universal. All components away from the origin are flavour off-diagonal; furthermore if they have zero thirdness ($\yfl=0$) then they involve only the first two generations ($c_{8,3}$ and $c_{8,5}$), while components with non-zero thirdness are off-diagonal involving the third generation ($c_{8,1}$, $c_{8,2}$, $c_{8,7}$, $c_{8,8}$), as expected from Eq.~\eqref{eq:Yns}.

For the 27-plet irrep of the 4-fermion operators, this pattern continues. Specifically, the three irrep components at the origin of the $\{\itfl,\yfl\}$ plane (shown on the right in Fig.~\ref{fig:irrepCharges}) are diagonal but flavour non-universal, as seen in their relation to the Warsaw basis coefficients:
\begin{align}
c_{27,12}&=\frac{1}{\sqrt{6}}\left( c_{11}^{11}+c_{22}^{22}-4c_{(12)}^{(12)} \right),~~c_{27,16}=\frac{1}{\sqrt{10}}\left(c_{11}^{11}-c_{22}^{22}-4c_{(13)}^{(13)}+4c_{(23)}^{(23)}\right),\nonumber\\
c_{27,18}&=\frac{1}{\sqrt{30}}\left( c_{11}^{11}+c_{22}^{22}+c_{33}^{33}+2c_{(12)}^{(12)}-6  c_{(13)}^{(13)}-6 c_{(23)}^{(23)}\right).
\end{align}
Any component away from the origin has some degree of off-diagonality. For components lying on the inner dashed hexagon of the 27-plet (i.e.\ that are at the same position on the plane as an octet component), this off-diagonality is not maximal, meaning that one of the fermion flavours is unchanged in the interaction. For example, the two components at the point $(1, 0)$ can be written in terms of the Warsaw basis coefficients as
\begin{align}
c_{27,11}&=c_{22}^{(12)} - c_{(12)}^{11},~~
c_{27,15}=\frac{1}{\sqrt{5}}\left(4c_{(23)}^{(13)}-c_{(12)}^{11}-c_{22}^{(12)}\right),
\end{align}
where it can be seen that there is one flavour change between the first two generations, and none in the third generation. This is in keeping with what we might expect for something with zero thirdness $\yfl$ but non-zero lightspin $\ifl$, based on what we have seen for the octet case. Conversely, the components on the inner dashed hexagon with $\yfl \neq 0$ are each off-diagonal in the third generation but flavour-conserving in one of the first two generations. 

Finally, the components on the outer dot-dashed hexagon of the 27-plet are maximally flavour violating, meaning that no fermion involved in the interaction retains its flavour. Again, their $\yfl$ and $\itfl$ values indicate which generations are involved. For the components with zero thirdness, $c_{27,14}$ and $c_{27,10}$, only the first two generations are implicated in the interaction; in terms of the Warsaw basis coefficients these are simply $c_{27,10}=c_{22}^{11}$ and $c_{27,14}=c_{11}^{22}$. Physically, if these describe a four-quark operator with down-type quarks, they are the coefficients mediating $K^0-\overline{K}^0$ mixing (or for a four-lepton operator, muonium-antimuonium oscillations). The components on the other vertices of the outer hexagon are each maximally off-diagonal between the third generation and one of the first two generations (and hence would mediate $B_s$ or $B_d$ mixing in a down-quark operator).  The components at the midpoints of the sides of the outer hexagon are instead off-diagonal in all three generations. We provide relations between all the Wilson coefficients in the mass basis to the irrep basis, and vice versa, in Appendix \ref{app:CGdecomp}.

To expand on these arguments, in Tab.~\ref{tab:irrepPheno} we give general examples of the types of phenomenology that can be generated by operators whose coefficients have various charges under the quark and lepton flavour groups. The \encircle{C} row and \encircle{3} column refer to components lying at the vertices of the hexagon in the corresponding octet (see Fig.~\ref{fig:irrepCharges}), as well as components with the same $\{\ifl, \itfl, \yfl\}$ charges in the 27-plet. The last row and column, \encircle{4} and \encircle{D}, refer to components lying on the larger hexagon in the corresponding 27-plet. In this case, the operator must be a singlet under all other flavour groups, since at dimension 6 the 27-plet irrep can only exist as part of a 4-fermion operator with identical currents. By `MFV FCNCs' we refer to flavour changing neutral currents between left-handed down-type quarks $d_{pL} \to d_{qL}$, suppressed by CKM factors $V_{tp}^* V_{tq}$ as in the SM. These can arise from operators involving only the third generation of quark doublets (N.B.~we are working in the `up-basis' where the quark doublet is defined $Q^p\equiv(u_L^p, V_q^p d_L^q)^T$). By `non-MFV FCNCs' we refer to all other types of FCNCs, for example between up-type quarks, or right handed down-type quarks, and which do not come with CKM suppressions. The `MFV FCNC' phenomenology is the only part of Tab.~\ref{tab:irrepPheno} which depends on our choice of `up-basis': in the equivalent `down-basis', Wilson coefficients with these charges would induce flavour changing neutral currents between left-handed up-type quarks $u_{pL} \to u_{qL}$ instead, suppressed by CKM factors $V_{qb}^* V_{pb}$.

Since this framework is couched in flavour groups, it is also straightforward to identify the subsets of coefficients that may arise from flavour symmetric BSM physics. If we assume the new physics respects an exact $U(3)_F$ flavour symmetry, then the only non-zero Wilson coefficients are the ones for which $d_F=1$ (i.e.\ singlets). If instead the new physics couples in a way which respects an exact $U(2)_F$ symmetry, then any dimensionality is allowed, but all the other quantum numbers $\{\ifl, \itfl, \yfl \}_F=0$, so only one component of each of the octet and 27-plet survives. 

In summary, by writing all the Wilson coefficients of the $(4,0)$ block in this fully general `flavour irrep basis', we can categorise their flavour effects in terms of the quantum numbers $\{d_{\text{irrep}}, \ifl ,\itfl, \yfl\}_F$.  Another major advantage of this basis is that it renders the anomalous dimension matrix automatically block-diagonal, as we will show in Section \ref{sec:blockdiag}.

\begin{table}[]
    \centering
   \resizebox{\textwidth}{!}{%
    \begin{tabular}{|c||c|c|c|c|}
    \hline
     & \encircle{1} & \encircle{2} &\encircle{3} & \encircle{4}\\
     \diagbox[width=55mm]{Lepton}{Quark} & $d_{\{Q,u,d\}}=1$ & $d_{\{Q,u\}}>1$,  $\{\itfl,\yfl\}_{\{Q,u\}}=0$ & $(\ifl^2+\frac34 \yfl^2 )_{\{Q,u,d\}}=1$ & $(\ifl^2+\frac34 \yfl^2 )_{\{Q,u,d\}}>1$\\
    \hline
    \hline
    \encircle{A} & & & & \\
    $d_{\{L,e\}}=1$ & Higgs, EW, ... & top, MFV FCNCs & non-MFV FCNCs & e.g.\ meson mixing \\
    \hline 
     \encircle{B} & LFUV (quark flavour conserved) & LFUV in MFV FCNCs & LFUV in non-MFV FCNCs & - \\
     $d_{\{L,e\}}>1$, $\{\itfl,\yfl\}_{\{L,e\}}=0$ & e.g.\ LFUV in $Z$ decays & & e.g.\ $R_K$ &  \\
        \hline 
         \encircle{C}& LFV (quark flavour conserved) & LFV in MFV FCNCs & LFV in non-MFV FCNCs & - \\ 
  $(\ifl^2+\frac34 \yfl^2 )_{\{L,e\}}=1$ & e.g.\ $\mu \to 3 e$, $H\to \tau \mu$ & & e.g. $B\to K \mu e$ &  \\
    \hline
     \encircle{D} & e.g.\ muonium oscillations, & - & - & -\\
   $(\ifl^2+\frac34 \yfl^2 )_{\{L,e\}}>1$ & $\tau^+ \to \mu^- e^+ e^+$& & & \\
    \hline
    \end{tabular}}
    \caption{Coarse overview of phenomenology generated by operators whose Wilson coefficients have different flavour charges. The rows (labelled by circled letters) give four different scenarios for leptonic flavour charges. We have grouped together $L$ and $e$ here, but for each case we assume that only one carries the flavour charges given. Similar logic applies to the columns (labelled by circled numbers), which list charges for $Q$, $u$ and $d$ (note that the scenarios listed are not exhaustive). Within the table, we give examples of phenomenology that can be generated by operators carrying the specified charges, including lepton flavour violation (LFV), lepton flavour universality violation (LFUV) and flavour changing neutral currents (FCNCs).} 
    \label{tab:irrepPheno}
\end{table}

\section{Anomalous dimension matrix from generalized unitarity}
\label{sec:gammaUnitarity}

In this section, we recap the approach of \cite{Jiang:2020mhe,Baratella:2020dvw,Baratella:2020lzz} to calculating anomalous dimensions in theories such as the SMEFT. We explain how IR finite and IR divergent kinetic pieces of the anomalous dimensions of the (4,0) dimension 6 operators are calculated in this approach, using the $\mathcal{J}$ vector notation of Sec.~\ref{subsec:Jvector}. We then explain the calculation of the gauge and flavour pieces of the anomalous dimension matrices, and advertise some of their features and implications, to be discussed in more detail in the next section.

The 1-loop amplitudes can be expressed using the Passarino-Veltman (PV) \cite{Passarino:1978jh} decomposition such that
\begin{align}
\mathcal{A}_{\rm 1-loop}= \sum_i d_i \, I_{4,i}^{\rm box} + \sum_i c_i \, I_{3,i}^{\rm triangle} + \sum_i b_i I_{2,i}^{\rm bubble} + \sum_i a_i I_{1,i}^{\rm tadpole} + R,
\end{align}
where $I_{4,i}^{\rm box}$, $I_{3,i}^{\rm triangle}$, $I_{2,i}^{\rm bubble}$ and $I_{1,i}^{\rm tadpole}$ are $D$-dimensional scalar integrals with four, three, two, and one propagator(s), respectively, and $i$ indexes the different arrangements of the external leg momenta at their vertices. The coefficients $d_i$, $c_i$, $b_i$, $a_i$, and $R$ are rational functions of kinematic invariants (and also include gauge and flavour factors). The practicalities of calculating the anomalous dimensions from cut loop amplitudes depends on the integrals appearing in their Passarino-Veltman decomposition: we can usefully classify the amplitudes as a) IR finite, b) containing soft IR divergences, and c) containing collinear IR divergences.

In practice, we only need consider three scalar integrals that appear in $2 \to 2$ amplitudes \cite{Baratella:2020lzz}:
\begin{align}
    I_2^{\rm bub,massive} &=     \mathord{\begin{tikzpicture}[baseline=-0.65ex]
    \coordinate (n1) at (0,0);
    \coordinate (n2) at (1,0);
  \draw[ultra thick] (n1) -- ($(n1)+(135:0.6)$) node[left] {$1$};
  \draw[ultra thick] (n1) -- ($(n1)+(-135:0.6)$) node[left] {$2$};
    \draw[ultra thick] (n2) -- ($(n2)+(45:0.6)$) node[right] {$3$};
  \draw[ultra thick] (n2) -- ($(n2)+(-45:0.6)$) node[right] {$4$};
  \draw[ultra thick] (n1) to [out=45,in=135] (n2);
  \draw[ultra thick] (n1) to [out=-45,in=-135] (n2);
\end{tikzpicture}} = \int \frac{\dd^4 k}{(2\pi)^4 i} \frac{1}{k^2 (k+p_3+p_4)^2}
\, , \\
    I_2^{\rm bub,massless} &=      \mathord{\begin{tikzpicture}[baseline=-0.65ex]
    \coordinate (n1) at (0,0);
    \coordinate (n2) at (1,0);
  \draw[ultra thick] (n1) -- ($(n1)+(135:0.6)$);
    \draw[ultra thick] (n1) -- ($(n1)+(180:0.6)$);
  \draw[ultra thick] (n1) -- ($(n1)+(-135:0.6)$);
    \draw[ultra thick] (n2) -- ($(n2)+(0:0.6)$) node[right] {$i$};
  \draw[ultra thick] (n1) to [out=45,in=135] (n2);
  \draw[ultra thick] (n1) to [out=-45,in=-135] (n2);
\end{tikzpicture}} = \int \frac{\dd^4 k}{(2\pi)^4 i} \frac{1}{k^2 (k+p_i)^2}
\, , \\
  I_3^{\rm triangle} &=   
    \mathord{\begin{tikzpicture}[baseline=-0.65ex]
    \coordinate (n1) at (0,0);
    \coordinate (n2) at (0.6,0.4);
    \coordinate (n3) at (0.6,-0.4);
  \draw[ultra thick] (n1) -- ($(n1)+(135:0.6)$) node[left] {$1$};
  \draw[ultra thick] (n1) -- ($(n1)+(-135:0.6)$) node[left] {$2$};
    \draw[ultra thick] (n2) -- ($(n2)+(0:0.6)$) node[right] {$3$};
  \draw[ultra thick] (n3) -- ($(n3)+(0:0.6)$) node[right] {$4$};
  \draw[ultra thick] (n1) -- (n2) -- (n3) -- (n1);
\end{tikzpicture}} =  \int \frac{\dd^4 k}{(2\pi)^4 i} \frac{-1}{k^2 (k+p_3)^2 (k+p_3+p_4)^2} \, .
\end{align}

It will be convenient to factorise the product of the dimension 6 and SM amplitudes coming from $\text{Cut} \left[ \mathcal{A}_{\rm 1-loop} \right]$ into a product of their respective kinematic, gauge, and flavour factors:
\begin{align}
    \mathcal{A}_6 \times \mathcal{A}_{\rm SM} = ({\rm kinematics}) \times ({\rm gauge})\times ({\rm flavour})\,.
\end{align}
In the remainder of this section, we derive each of these pieces in turn for the anomalous dimensions of the $(4,0)$ operators. We begin with the kinematic factors in the next subsection. The gauge factors are discussed in Sec. \ref{sec:gauge} and the flavour factors in Sec. \ref{sec:flavour}. 

\subsection{Kinematic factors}
\label{sec:kinematicFactors}

We calculate the kinematic pieces of the anomalous dimension matrices using the PV decomposition of the loop diagrams, classifying them by their IR properties. As we will show, just a few phase space integrals appear. The magic behind this simplicity is angular momentum conservation, which was explored in \cite{Baratella:2020dvw,Jiang:2021tqo}. The summary of our results using the diagrammatic representation presented in Sec.~\ref{subsec:Jvector} is shown in Appendix \ref{app:summaryPhaseSpace} and we detail a few examples in the following. 

\subsubsection{IR finite pieces}
\label{sec:IRfinitephase}

Many of the 4-point loop amplitudes that appear in the calculation of the anomalous dimensions are IR finite. This includes all amplitudes responsible for mixing between operators with different species of particles. Their PV decomposition comprises a single scalar bubble\footnote{Generally, one might worry about tadpole diagrams, but they do not contribute to the anomalous dimension of the operator at one loop, nor do they contribute to the discontinuities in the amplitude (as can be seen in dimensional regularisation, where they are trivially scaleless and zero in the case of massless particles).}
\begin{equation}
\mathcal{A}_{\rm 1-loop}(1,2,3,4) = b \; 
    \mathord{\begin{tikzpicture}[baseline=-0.65ex]
    \coordinate (n1) at (0,0);
    \coordinate (n2) at (1,0);
  \draw[ultra thick] (n1) -- ($(n1)+(135:0.6)$) node[left] {$1$};
  \draw[ultra thick] (n1) -- ($(n1)+(-135:0.6)$) node[left] {$2$};
    \draw[ultra thick] (n2) -- ($(n2)+(45:0.6)$) node[right] {$3$};
  \draw[ultra thick] (n2) -- ($(n2)+(-45:0.6)$) node[right] {$4$};
  \draw[ultra thick] (n1) to [out=45,in=135] (n2);
  \draw[ultra thick] (n1) to [out=-45,in=-135] (n2);
\end{tikzpicture}}
\, ,
\end{equation}
where we have written $I_2^{\rm bub,massive}$ diagrammatically. If we define a `Cut' operation that places two scalar propagators on-shell,\footnote{E.g., via the substitution $\frac{i}{k^2} \to 2 \pi \theta(k^0) \delta(k^2)$ acting on two scalar propagators \cite[\S 2.1]{Abreu:2014cla}} then the RHS is trivialised
\begin{equation}
\text{Cut} \left[bI_2^{\rm bub,massive} \right] 
\propto b \left( \int \dd \Pi_2 \right) \, ,
\end{equation}
where $\dd \Pi_2$ is the two-body phase space measure in the center-of-mass frame, \cref{eq:TwoBodyPS}. For the loop amplitudes relevant for the anomalous dimension matrix, the LHS evaluates to a product of a tree amplitude $\mathcal{A}_6$, containing a dimension 6 operator, and a Standard Model tree amplitude, $\mathcal{A}_\text{SM}$.
\begin{equation}
    \text{Cut} \left[ \mathcal{A}_{\rm 1-loop}(1,2,3,4) \right] \propto  \int \dd \Pi_2 \,\, \mathcal{A}_6(1,2,3^\prime,4^\prime) \times \mathcal{A}_\text{SM}(-4^\prime,-3^\prime,3,4)\,,
\end{equation}
where the primed momenta are a $(\theta,\phi)$ rotation of their unprimed counterparts in the centre-of-mass frame.

Evaluating the above, we extract directly the coefficient $b_i$, and therefore infer the UV divergence of the 1-loop amplitude. Normalising the above expressions appropriately leads to the master formula of \cite{Baratella:2020lzz} in the IR finite case,
\begin{gather}
    \mathord{\begin{tikzpicture}[baseline=-0.65ex]
  \draw[thick] (0,0) -- (45:1) node[right] {$3^\prime$};
  \draw[thick] (0,0) -- (135:1) node[left] {$1$};
    \draw[thick] (0,0) -- (-45:1) node[right] {$4^\prime$};
  \draw[thick] (0,0) -- (-135:1) node[left] {$2$};
  \node[draw,thick,circle,fill=white] at (0,0) {$\amp_{6,j}$};
\end{tikzpicture}}
\cut 
\mathord{\begin{tikzpicture}[baseline=-0.65ex]
    \draw[thick] (0,0) -- (45:1) node[right] {$3$};
  \draw[thick] (0,0) -- (135:1) node[left] {$-3^\prime$};
    \draw[thick] (0,0) -- (-45:1) node[right] {$4$};
  \draw[thick] (0,0) -- (-135:1) node[left] {$-4^\prime$};
  \node[draw,thick,circle,fill=white] at (0,0) {$\amp_{\rm SM}$};
\end{tikzpicture}}
=~\gamma_{ij}
 \mathord{\begin{tikzpicture}[baseline=-0.65ex]
  \draw[thick] (0,0) -- (45:1) node[right] {$3$};
  \draw[thick] (0,0) -- (135:1) node[left] {$1$};
    \draw[thick] (0,0) -- (-45:1) node[right] {$4$};
  \draw[thick] (0,0) -- (-135:1) node[left] {$2$};
  \node[draw,thick,circle,fill=white] at (0,0) {$\amp_{6,i}$};
\end{tikzpicture}} \nn \\
 - 2 (i)^{n_{\psi}} \int \frac{\dd \Pi_2}{2^\rho \cdot 4\pi}  \sum_\text{ext.} \, \cA_{6,j}(1,2,3',4') \times \cA_{\rm SM}(-4',-3',3,4) = \gamma_{ij}\,\cA_{ 6, i}(1,2,3,4) \, ,
 \label{eq:IRfiniteMasterFormula}
\end{gather}
where $\int d\Pi_2$ is the 2-sphere measure (Eq.~(\ref{eq:TwoBodyPS})), $n_{\psi}$ is the number of fermion lines cut and $\rho=0(1)$ if the particles being cut are (in)distinguishable. $\sum_\text{ext}$ denotes a sum over possible ways of arranging pairs of the external legs on different sides of the cut. The anomalous dimension matrix, $\gamma_{ij}$, is defined at the level of Wilson coefficients as
 \begin{equation}
     \frac{\dd C_{\mathcal{O}_i}}{ \dd \ln \mu} = \sum_j \frac{1}{16 \pi^2} \gamma_{ij}C_{\mathcal{O}_j}\,.
 \end{equation}

The SM gauge amplitudes with \emph{distinguishable} currents consist only of the $s$-channel diagram, so the product of the possible dimension 6 and SM amplitudes is given by
\begin{align}
\label{eq:gaugemain}
\mathord{\begin{tikzpicture}[baseline=-0.65ex]
  \draw[thick,dotted,blue] (-1,1) -- (-\currsep,0) -- (-1,-1);
  \draw[thick,dotted] (1,1) -- (\currsep,0) -- (1,-1);
  \currentmarker{\currsep,0}
  \currentmarker{-\currsep,0}
\end{tikzpicture}}
\cut
\mathord{\begin{tikzpicture}[baseline=-0.65ex]
  \draw[thick,dotted] (-1,1) -- (-\threepointsep,0) -- (-1,-1);
  \draw[thick,dotted,red] (1,-1) -- (\threepointsep,0) -- (1,1);
  \draw[thick,decorate,decoration={snake}] (\threepointsep,0) -- (-\threepointsep,0);
\end{tikzpicture}}
= \gamma_{\rm loop}^{\rm dist}
\mathord{\begin{tikzpicture}[baseline=-0.65ex]
  \draw[thick,dotted,blue] (-1,1) -- (-\currsep,0) -- (-1,-1);
  \draw[thick,dotted,red] (1,1) -- (\currsep,0) -- (1,-1);
  \currentmarker{\currsep,0}
  \currentmarker{-\currsep,0}
\end{tikzpicture}} 
 \,,
\end{align}
where the dotted black line (i.e. the particle in the loop) can be either a fermion $\psi$ or a Higgs $H$, which leads to the phase space factors $\gamma_{\psi}^{\rm dist}$ and $\gamma_{H}^{\rm dist}$, respectively.
Writing the amplitudes in terms of $\mathcal{J}$ vector as shown in Sec.~\ref{subsec:Jvector}, we have that the LHS of Eq.~(\ref{eq:gaugemain}) is then given by\footnote{Note that the particle and antiparticle of a single current are always distinguishable.}
\begin{align}
 - 2 (i)^{n_{\psi}} \int \frac{\dd \Pi_2}{4\pi}
 (-2\mathcal{J}^{\color{blue}A}(12)\mathcal{J}^B(3'4'))\frac{(-2\mathcal{J}^B(4'3')\mathcal{J}^{\color{red}C}(34))}{s_{34}}\,.
\end{align}
We use the trick of \cite{Caron-Huot:2016cwu} to rotate the spinors associated with the internal loop momenta such that:
\begin{align}
\label{eq:rotation}
\begin{pmatrix}
\angket{3^\prime} \\ \angket{4^\prime} \end{pmatrix} &= \begin{pmatrix}
\ct  &- \st e^{i\phi}\\
\st e^{-i\phi} & \ct
\end{pmatrix} \begin{pmatrix}
\angket{3} \\ \angket{4} \end{pmatrix},  \qquad
\begin{pmatrix}
\sqrket{3^\prime} \\ \sqrket{4^\prime} \end{pmatrix} &= \begin{pmatrix}
\ct &- \st e^{-i\phi}\\
\st e^{i\phi} & \ct
\end{pmatrix} \begin{pmatrix}
\sqrket{3} \\ \sqrket{4} \end{pmatrix}\,.
\end{align}
 Notice that the little group in the LHS and RHS of Eq.~\eqref{eq:IRfiniteMasterFormula} 
 has to be the same, so the spinors associated with the external particles cancel out.
After rotating the internal spinors, we can write
\begin{align}
\mathcal{J}^B(3'4') = M_L^{BX}\mathcal{J}^{X}(34), \quad \mathcal{J}^B(4'3') = M_R^{BX}\mathcal{J}^{X}(43),
\end{align}
where we define the phase-space matrix, which carries all angular dependence, as
\begin{equation}
\label{eq:ML}
   M_L^{BX}  = \begin{pmatrix}
    \ct^2 & 2 \st \ct e^{-i\phi} & -\st^2 e^{-2i\phi} \\
    - \st \ct e^{i\phi} & \ct^2-\st^2 & -\st \ct e^{-i\phi} \\
    -\st^2 e^{2i\phi} & 2 \st \ct e^{i\phi} & \ct^2
  \end{pmatrix}\,,
\end{equation}
and the phase space matrix $M_R^{BX}$ 
can be obtained switching $3 \leftrightarrow 4$, $e^{i\phi} \leftrightarrow e^{-i\phi}$ and $\st \leftrightarrow -\st$. 
This allows us to have all angular information encoded in the product of the matrices $M_L^{BX} M_R^{BY}$ (note that the index $B$ is not summed over). 
Fixing $B=1,2,3$ we can get the phase-space factors when cutting a right-handed fermion, Higgs or left-handed fermion, respectively. The phase-space factor defined in Eq.~(\ref{eq:gaugemain}) is then given by 
\begin{align}
    \gamma_{\rm loop}^{\rm dist} =  - 2 (i)^{n_{\psi}} \int \frac{\dd \Pi_2}{4\pi}
    M_L^{BX}M_R^{BY} \frac{(-2\mathcal{J}^Y(43)\mathcal{J}^C(34))}{s_{34}}\,.
\end{align}
Next, using Eqs.~(\ref{eq:normJ}) and (\ref{eq:ML}), we find that the phase space factors depend only on the degrees of freedom in the loop, i.e.
\begin{align}
     \gamma_{{\rm loop}, \psi}^{\rm dist}= 
     \frac{4}{3} \quad (\text{for}~B=1,3) ~~ \text{and}~~
    \gamma_{{\rm loop}, H}^{\rm dist}= 
    \frac{2}{3} \quad (\text{for}~B=2)\,.
\end{align}

We now turn to loop diagrams involving SM Yukawas. In the case where a single current is cut, the SM amplitudes on the right of the cut have poles only in the $t$ and/or $u$-channels. For any allowable rearrangement of the helicities, the $t$-channel diagram gives
\begin{align}
\label{eq:Yukawatmain}
  \mathord{\begin{tikzpicture}[baseline=-0.65ex]
    \draw[thick,dotted,blue] (-1,1) -- (-\currsep,0) -- (-1,-1);
    \draw[thick,dotted,black] (1,1) -- (\currsep,0) -- (1,-1);
    \currentmarker{\currsep,0}
    \currentmarker{-\currsep,0}
  \end{tikzpicture}}
  \cut
  \mathord{\begin{tikzpicture}[baseline=-0.65ex]
  \draw[thick,dotted,black] (-1,1) -- (0,\threepointsep);
  \draw[thick,dotted,black] (0,-\threepointsep) -- (-1,-1);
  \draw[thick,dotted,red] (0,\threepointsep) -- (1,1);
  \draw[thick] (0,\threepointsep) -- (0,-\threepointsep);
  \draw[thick,dotted,red] (0,-\threepointsep) -- (1,-1);
  \end{tikzpicture}}
  =\gamma_{{\rm \color{red} ext}}^{\rm dist}
\mathord{\begin{tikzpicture}[baseline=-0.65ex]
  \draw[thick,dotted,blue] (-1,1) -- (-\currsep,0) -- (-1,-1);
  \draw[thick,dotted,red] (1,1) -- (\currsep,0) -- (1,-1);
  \currentmarker{\currsep,0}
  \currentmarker{-\currsep,0}
\end{tikzpicture}} 
  \,,
\end{align}
where $\gamma_{{\rm \color{red} ext}}^{\rm dist}$ indicates that the phase space factor in this case depends on the {\color{red} external} particles on the right side of the cut. In terms of the $\mathcal{J}$-vectors we have that the LHS of Eq.~(\ref{eq:Yukawatmain}) is given by
\begin{align}
\label{eq:extYuk}
 - 2 (i)^{n_{\psi}} \int \frac{\dd \Pi_2}{4\pi} (-2\mathcal{J}^{\color{blue}A}(12)\mathcal{J}^B(3'4'))\frac{\mathcal{J}^B(4'3')\mathcal{J}^{\color{red}C}(34)}{s_{3'3}}.
\end{align}
The only difference with the gauge $s$-channel calculation is that we have to include an extra angular factor that appears due to the pole in the SM amplitude, i.e.  $s_{3^\prime 3} = \st^2 s_{34}$.
 \begin{align}
    \gamma_{{\rm \color{red} ext}}^{\rm dist} =  - 2 (i)^{n_{\psi}} \int \frac{\dd \Pi_2}{4\pi}  M_L^{BX}M_R^{BY} \frac{\mathcal{J}^Y(43)\mathcal{J}^C(34)}{\st^2 s_{34}}\,.
\end{align}
Notice that, for example for $B=3$ (and similarly for $B=1$), the phase space integral gives
 \begin{align}
     - 2 (i)^{n_{\psi}} \int \frac{\dd \Pi_2}{4\pi} \frac{M_L^{3X}M_R^{3Y}}{\st^2} = 
\begin{pmatrix}
  ~1 & ~0 & ~0~ \\
  ~0 & - 4 & ~0~ \\
  ~0 & ~0 & ~ ~
\end{pmatrix},
 \end{align}
 where the final entry in the matrix is blank because it corresponds to a topology that doesn't exist: a Yukawa diagram cannot be constructed out of four left-handed fermion fields.
 We then find that, in contrast to the gauge case, the phase space integral is independent of the particle in the loop and changes if the external current is fermionic or scalar. The calculation for Yukawa $u$-channel SM diagram (which can have a fermion or a Higgs propagator) is analogous and leads to exactly the same result, i.e.
  \begin{align}
    \gamma_{{\rm \color{red}ext},\psi}^{\rm dist} = 1 \, ,
    \qquad  \gamma_{{\rm \color{red}ext},H}^{\rm dist} =2
    , \quad ({\rm for~any}~B).
    \end{align}

    Another possibility is  the case where two currents are cut. In contrast to the previous case, here the phase space factors differ depending on the SM diagram topology and the external particles. In the case of only external fermions we have that
    \begin{align}
\mathord{\begin{tikzpicture}[baseline=-0.65ex]
  \draw[thick] (-1,1) -- (0,\currsep) -- (1,1);
  \draw[thick] (-1,-1) -- (0,-\currsep) -- (1,-1);
  \currentmarker{0,\currsep}
  \currentmarker{0,-\currsep}
\end{tikzpicture}}
\cut
\mathord{\begin{tikzpicture}[baseline=-0.65ex]
  \draw[thick] (-1,1) -- (0,\threepointsep) -- (1,-1);
  \draw[thick] (-1,-1) -- (0,-\threepointsep) -- (1,1);
  \draw[thick,dashed] (0,\threepointsep) -- (0,-\threepointsep);
\end{tikzpicture}}
 &= \gamma_{\psi^4,\,u}^{\rm dist}
\mathord{\begin{tikzpicture}[baseline=-0.65ex]
  \draw[thick] (-1,1) -- (0,\currsep) -- (1,1);
  \draw[thick] (-1,-1) -- (0,-\currsep) -- (1,-1);
  \currentmarker{0,\currsep}
  \currentmarker{0,-\currsep}
\end{tikzpicture}},\\
\mathord{\begin{tikzpicture}[baseline=-0.65ex]
  \draw[thick] (-1,1) -- (0,\currsep) -- (1,1);
  \draw[thick] (-1,-1) -- (0,-\currsep) -- (1,-1);
  \currentmarker{0,\currsep}
  \currentmarker{0,-\currsep}
\end{tikzpicture}}
\cut
\mathord{\begin{tikzpicture}[baseline=-0.65ex]
  \draw[thick] (-1,1) -- (-\threepointsep,0) -- (-1,-1);
  \draw[thick] (1,-1) -- (\threepointsep,0) -- (1,1);
  \draw[thick,dashed] (\threepointsep,0) -- (-\threepointsep,0);
\end{tikzpicture}}
 &= \gamma_{\psi^4,\,s}^{\rm dist} 
\mathord{\begin{tikzpicture}[baseline=-0.65ex]
  \draw[thick] (-1,1) -- (0,\currsep) -- (1,1);
  \draw[thick] (-1,-1) -- (0,-\currsep) -- (1,-1);
  \currentmarker{0,\currsep}
  \currentmarker{0,-\currsep}
\end{tikzpicture}},
\end{align}
and in the case in which there is a scalar current: 
\begin{align}
\mathord{\begin{tikzpicture}[baseline=-0.65ex]
  \draw[thick,dashed] (-1,1) -- (0,\currsep) -- (1,1);
  \draw[thick] (-1,-1) -- (0,-\currsep) -- (1,-1);
  \currentmarker{0,\currsep}
  \currentmarker{0,-\currsep}
\end{tikzpicture}}
\cut
\mathord{\begin{tikzpicture}[baseline=-0.65ex]
  \draw[thick,dashed] (-1,1) -- (0,\threepointsep);
  \draw[thick,dashed] (1,1) -- (0,-\threepointsep);
  \draw[thick] (-1,-1) -- (0,-\threepointsep) -- (0,\threepointsep) -- (1,-1);
\end{tikzpicture}}
 &= \gamma_{ \psi^2H^2, \,u}^{\rm dist}
\mathord{\begin{tikzpicture}[baseline=-0.65ex]
  \draw[thick,dashed] (-1,1) -- (0,\currsep) -- (1,1);
  \draw[thick] (-1,-1) -- (0,-\currsep) -- (1,-1);
  \currentmarker{0,\currsep}
  \currentmarker{0,-\currsep}
\end{tikzpicture}}, \\
\mathord{\begin{tikzpicture}[baseline=-0.65ex]
  \draw[thick,dashed] (-1,1) -- (0,\currsep) -- (1,1);
  \draw[thick] (-1,-1) -- (0,-\currsep) -- (1,-1);
  \currentmarker{0,\currsep}
  \currentmarker{0,-\currsep}
\end{tikzpicture}}
\cut
\mathord{\begin{tikzpicture}[baseline=-0.65ex]
  \draw[thick] (-1,-1) -- (-\threepointsep,0) -- (\threepointsep,0) -- (1,-1);
  \draw[thick,dashed] (-1,1) -- (-\threepointsep,0);
  \draw[thick,dashed] (\threepointsep,0) -- (1,1);
\end{tikzpicture}}
 &= \gamma_{\psi^2H^2,\,s}^{\rm dist}
\mathord{\begin{tikzpicture}[baseline=-0.65ex]
  \draw[thick,dashed] (-1,1) -- (0,\currsep) -- (1,1);
  \draw[thick] (-1,-1) -- (0,-\currsep) -- (1,-1);
  \currentmarker{0,\currsep}
  \currentmarker{0,-\currsep}
\end{tikzpicture}},
\end{align}
where the result is independent of the helicity configuration allowed. 
The cases with the $u$-channel SM diagram can be calculated using  
\begin{align}
- 2 (i)^{n_{\psi}} \int \frac{\dd \Pi_2}{4\pi} (-2\mathcal{J}^{\color{blue}A}(13')\mathcal{J}^B(24'))\frac{\mathcal{J}^B(4'3')\mathcal{J}^{\color{red}C}(34)}{s_{4'3}}\,,
\end{align}
with the appropriate $A,C$ for each case. For the $s$-channel diagram we just need to replace $s_{4'4} \rightarrow s_{34}$. The result for the phase space factors is
\begin{align}
   & \gamma_{\psi^4,\,u}^{\rm dist}=-1
   , ~\qquad \gamma_{\psi^4,\,s}^{\rm dist} = -2
   \,, \nn\\
   & \gamma_{ \psi^2H^2, \,u}^{\rm dist}=2
   , \qquad \gamma_{\psi^2H^2,\,s}^{\rm dist}=1
   \,.
\end{align}

In the case of \emph{indistinguishable} 
currents, there is just the $s$-channel gauge contribution. The possible cases are
\begin{align}
\mathord{\begin{tikzpicture}[baseline=-0.65ex]
  \draw[thick,dashed] (-1,1) -- (-\currsep,0) -- (-1,-1);
  \draw[thick,dashed] (1,1) -- (\currsep,0) -- (1,-1);
  \currentmarker{\currsep,0}
  \currentmarker{-\currsep,0}
  \node at (-\bracketsep,0) {$($};
  \node at (\bracketsep,0) {$)$};
\end{tikzpicture}}
\cut
\mathord{\begin{tikzpicture}[baseline=-0.65ex]
  \draw[thick,dashed] (-1,1) -- (-\threepointsep,0) -- (-1,-1);
  \draw[thick,dotted,red] (1,-1) -- (\threepointsep,0) -- (1,1);
  \draw[thick,decorate,decoration={snake}] (\threepointsep,0) -- (-\threepointsep,0);
\end{tikzpicture}}
 &= \gamma^{\rm indist,S}_{H}
\mathord{\begin{tikzpicture}[baseline=-0.65ex]
  \draw[thick,dashed] (-1,1) -- (-\currsep,0) -- (-1,-1);
  \draw[thick,dotted,red] (1,1) -- (\currsep,0) -- (1,-1);
  \currentmarker{\currsep,0}
  \currentmarker{-\currsep,0}
\end{tikzpicture} \,,} \\
\mathord{\begin{tikzpicture}[baseline=-0.65ex]
  \draw[thick,dashed] (-1,1) -- (-\currsep,0) -- (-1,-1);
  \draw[thick,dashed] (1,1) -- (\currsep,0) -- (1,-1);
  \currentmarker{\currsep,0}
  \currentmarker{-\currsep,0}
  \node at (-\bracketsep,0) {$[$};
  \node at (\bracketsep,0) {$]$};
\end{tikzpicture}}
\cut
\mathord{\begin{tikzpicture}[baseline=-0.65ex]
  \draw[thick,dashed] (-1,1) -- (-\threepointsep,0) -- (-1,-1);
  \draw[thick,dotted,red] (1,-1) -- (\threepointsep,0) -- (1,1);
  \draw[thick,decorate,decoration={snake}] (\threepointsep,0) -- (-\threepointsep,0);
\end{tikzpicture} }
 &= \gamma^{\rm indist,AS}_{H}
\mathord{\begin{tikzpicture}[baseline=-0.65ex]
  \draw[thick,dashed] (-1,1) -- (-\currsep,0) -- (-1,-1);
  \draw[thick,dotted,red] (1,1) -- (\currsep,0) -- (1,-1);
  \currentmarker{\currsep,0}
  \currentmarker{-\currsep,0}
\end{tikzpicture} \,,} \\
\mathord{\begin{tikzpicture}[baseline=-0.65ex]
  \draw[thick] (-1,1) -- (-\currsep,0) -- (-1,-1);
  \draw[thick] (1,1) -- (\currsep,0) -- (1,-1);
  \currentmarker{\currsep,0}
  \currentmarker{-\currsep,0}
  \node at (-\bracketsep,0) {$[$};
  \node at (\bracketsep,0) {$]$};
\end{tikzpicture}}
\cut
\mathord{\begin{tikzpicture}[baseline=-0.65ex]
  \draw[thick] (-1,1) -- (-\threepointsep,0) -- (-1,-1);
  \draw[thick,dotted,red] (1,-1) -- (\threepointsep,0) -- (1,1);
  \draw[thick,decorate,decoration={snake}] (\threepointsep,0) -- (-\threepointsep,0);
\end{tikzpicture}}
 &= \gamma^{\rm indist,AS}_{\psi}
\mathord{\begin{tikzpicture}[baseline=-0.65ex]
  \draw[thick] (-1,1) -- (-\currsep,0) -- (-1,-1);
  \draw[thick,dotted,red] (1,1) -- (\currsep,0) -- (1,-1);
  \currentmarker{\currsep,0}
  \currentmarker{-\currsep,0}
\end{tikzpicture} \,,}
\end{align}
where $S$ stands for the symmetric case and $AS$ for the antisymmetric. One can show that when cutting a fermion current that belongs to a four fermion operator with identical particles, we can just multiply the phase space factor in the distinguishable case by a factor $4$. For the cases with Higgs lines we need to compute
\begin{align}
- 2 (i)^{n_{\psi}} \int \frac{\dd \Pi_2}{4\pi} (-4J^2(12) J^{2}(3'4'))_{\rm S/AS}(-2 J^2(4'3')J^2(34))\frac{1}{ s_{34}} \,.
\end{align}
The final result is then
\begin{align}
    \gamma^{\rm indist,S}_{H} = 
    2, \quad \gamma^{\rm indist,AS}_{H}= 
    \frac23, \quad \gamma^{\rm indist,AS}_{\psi} = 
    \frac{16}{3}\,.
\end{align}

\subsubsection{IR divergent (soft)}
\label{sec:softphase}
Soft IR divergences can complicate the calculation of the anomalous dimensions, by introducing additional discontinuities which mean that the triangle diagrams in the PV decomposition, despite being UV finite, can still contribute to the 2-particle cut. These pieces therefore need to be subtracted off. There are only a limited set of circumstances where these soft divergences can appear, namely, in self-renormalisation diagrams involving SM gauge interactions.\footnote{`Self-renormalisation' here includes the mixing of different invariant colour structures in operators built from the same species of particles.}
The PV decomposition of these diagrams is:
\begin{align}
\mathcal{A}_{\rm 1-loop} &=    b \; 
    \mathord{\begin{tikzpicture}[baseline=-0.65ex]
    \coordinate (n1) at (0,0);
    \coordinate (n2) at (1,0);
  \draw[ultra thick] (n1) -- ($(n1)+(135:0.6)$) node[left] {$1$};
  \draw[ultra thick] (n1) -- ($(n1)+(-135:0.6)$) node[left] {$2$};
    \draw[ultra thick] (n2) -- ($(n2)+(45:0.6)$) node[right] {$3$};
  \draw[ultra thick] (n2) -- ($(n2)+(-45:0.6)$) node[right] {$4$};
  \draw[ultra thick] (n1) to [out=45,in=135] (n2);
  \draw[ultra thick] (n1) to [out=-45,in=-135] (n2);
\end{tikzpicture}}
+
 c \; 
    \mathord{\begin{tikzpicture}[baseline=-0.65ex]
    \coordinate (n1) at (0,0);
    \coordinate (n2) at (0.6,0.4);
    \coordinate (n3) at (0.6,-0.4);
  \draw[ultra thick] (n1) -- ($(n1)+(135:0.6)$) node[left] {$1$};
  \draw[ultra thick] (n1) -- ($(n1)+(-135:0.6)$) node[left] {$2$};
    \draw[ultra thick] (n2) -- ($(n2)+(0:0.6)$) node[right] {$3$};
  \draw[ultra thick] (n3) -- ($(n3)+(0:0.6)$) node[right] {$4$};
  \draw[ultra thick] (n1) -- (n2) -- (n3) -- (n1);
\end{tikzpicture}}
\, ,
\end{align}
where the triangle contains a soft IR divergence. Applying the same Cut operation to both sides
\begin{equation}
\text{Cut} \left[ \mathcal{A}_{\rm 1-loop} \right] \propto b \left( \int \dd \Pi_2 \right) - c \left( \frac{1}{s_{34}} \int \dd \Pi_2 \frac{1}{\sin^2 \frac{\theta}{2}} \right) \, , 
\end{equation}
we see the uncut scalar propagator of the triangle graph diverges as $\theta \to 0$. Both $\text{Cut} \left[ \mathcal{A}_{\rm 1-loop} \right] $ and the contribution of the cut triangle graph are therefore infinite. Their combination must be finite however, and allows us to extract the coefficient of the bubble graph.
\begin{equation}
\text{Cut}_\text{sub} \left[ \mathcal{A}_{\rm 1-loop} \right]  = \text{Cut} \left[ \mathcal{A}_{\rm 1-loop} \right] - k \left( \frac{1}{s_{34}} \int \dd \Pi_2 \frac{1}{\sin^2 \frac{\theta}{2}} \right) \propto b \left( \int \dd \Pi_2 \right) \, ,
\end{equation}
where the coefficient $k$ is determined to be whatever is necessary to make the integral finite. The dimensionful $k$ can always be expressed in terms of a soft factor $F_{\text{soft}}$ and the tree-level amplitude $\cA_{ 6, i}(1,2,3,4)$~\cite{Baratella:2020dvw}.\footnote{The soft factor $F_{\text{soft}}$ can be calculated as proportional to a product of the appropriate gauge generators~\cite{Baratella:2020dvw,Catani:1998bh}, but is also completely determined by demanding cancellation of the divergence, as shown in the example below.}
This leads to the master formula
\begin{align}
   \centering
    \mathord{\begin{tikzpicture}[baseline=-0.65ex]
  \draw[thick] (0,0) -- (45:1) node[right] {$3^\prime$};
  \draw[thick] (0,0) -- (135:1) node[left] {$1$};
    \draw[thick] (0,0) -- (-45:1) node[right] {$4^\prime$};
  \draw[thick] (0,0) -- (-135:1) node[left] {$2$};
  \node[draw,thick,circle,fill=white] at (0,0) {$\amp_{6,j}$};
\end{tikzpicture}}
\cutsub 
\mathord{\begin{tikzpicture}[baseline=-0.65ex]
    \draw[thick] (0,0) -- (45:1) node[right] {$3$};
  \draw[thick] (0,0) -- (135:1) node[left] {$-3^\prime$};
    \draw[thick] (0,0) -- (-45:1) node[right] {$4$};
  \draw[thick] (0,0) -- (-135:1) node[left] {$-4^\prime$};
  \node[draw,thick,circle,fill=white] at (0,0) {$\amp_{\rm SM}$};
\end{tikzpicture}}
&=~\gamma_{ij}
 \mathord{\begin{tikzpicture}[baseline=-0.65ex]
  \draw[thick] (0,0) -- (45:1) node[right] {$3$};
  \draw[thick] (0,0) -- (135:1) node[left] {$1$};
    \draw[thick] (0,0) -- (-45:1) node[right] {$4$};
  \draw[thick] (0,0) -- (-135:1) node[left] {$2$};
  \node[draw,thick,circle,fill=white] at (0,0) {$\amp_{6,i}$};
\end{tikzpicture}} \nonumber \\
 - 2 (i)^{n_{\psi}} \int \frac{\dd \Pi_2}{2^\rho \cdot 4\pi}  \Bigg( \sum_\text{ext.} \cA_{6,j}(1,2,3',4') \times \cA_{\rm SM}(-4',-3',3,4) \nonumber \\
 - F_{\text{soft}} \frac{1}{\sin^2 \frac{\theta}{2}} \cA_{ 6, i}(1,2,3,4)\Bigg) &= \gamma_{ij}\,\cA_{ 6, i}(1,2,3,4) \, .
 \label{eq:SoftMasterFormula}
\end{align}
where $F_{\text{soft}}$ is chosen to make the LHS finite.

For example, in the diagram
\begin{align}
\mathord{\begin{tikzpicture}[baseline=-0.65ex]
  \draw[thick,dotted,blue] (-1,1) -- (-\currsep,0) -- (-1,-1);
  \draw[thick] (1,1) -- (\currsep,0) -- (1,-1);
  \currentmarker{\currsep,0}
  \currentmarker{-\currsep,0}
\end{tikzpicture}}
\cutsub
\mathord{\begin{tikzpicture}[baseline=-0.65ex]
  \draw[thick] (-1,1) -- (0,\threepointsep) -- (1,1);
  \draw[thick] (-1,-1) -- (0,-\threepointsep) -- (1,-1);
  \draw[thick,decorate,decoration={snake}] (0,\threepointsep) -- (0,-\threepointsep);
\end{tikzpicture}}
 &= \gamma 
\mathord{\begin{tikzpicture}[baseline=-0.65ex]
  \draw[thick,dotted,blue] (-1,1) -- (-\currsep,0) -- (-1,-1);
  \draw[thick] (1,1) -- (\currsep,0) -- (1,-1);
  \currentmarker{\currsep,0}
  \currentmarker{-\currsep,0}
\end{tikzpicture} \,,} 
\end{align}
per \cref{eq:SoftMasterFormula}, we can write the (purely kinematic part of the) LHS as 
\begin{equation}
 - 2 (i)^{n_{\psi}} \int \frac{\dd \Pi_2}{4\pi}  \left[ (-2\mathcal{J}^{\color{blue}A}(12)\mathcal{J}^B(3'4'))\frac{(-2\mathcal{J}^B(4'3')\mathcal{J}^{C}(34))}{s_{3'3}} - F_{\text{soft}} \frac{1}{\st^2} (-2\mathcal{J}^{\color{blue}A}(12)\mathcal{J}^{C}(34)) \right]\,,
\end{equation}
where $B=C=1,3$, $s_{3'3}=\st^2s_{34}$. As before, we can rotate the internal momenta using Eqn.~\eqref{eq:rotation}. Notice that even though the diagram is distinct, the IR divergent part has the same form as the contribution from the Yukawa diagram shown in Eq.~(\ref{eq:extYuk}) (apart from a overall $-2$ factor coming from the SM amplitude). This means that before the $\theta$-integration we have for $B=3$:
\begin{align}
\!\!\!- 2 (i)^{n_{\psi}}\! \int \frac{\dd \Pi_2}{4\pi}  M_L^{3X}M_R^{3Y}\frac{(-2\mathcal{J}^3(43)\mathcal{J}^{3}(34))}{\st^2 s_{34}} = 
4 \int_{-1}^1 \frac{\dd (\cos \theta)}{2}
\begin{pmatrix}
  -\st^2 & ~0 & ~0~ \\
  ~0 & -2\ct^2 & ~0~ \\
  ~0 & ~0 & - \frac{\ct^4}{\st^2} ~
\end{pmatrix}.
 \end{align}
So we need to set $F_{\text{soft}}=1$ in order to make the $\theta$-integration finite for the $C=3$ component:\footnote{N.B.~the gauge factors have been factorised out of the amplitudes, since in this subsection we focus on the kinematic part of $\gamma$. To compare with the definition of the soft factors in Ref.~\cite{Baratella:2020dvw} as products of gauge generators, the overall gauge factors (described below in Sec.~\ref{sec:gauge}) should be multiplied back into the calculation.}
\begin{align}
   \gamma = 4 \int_{-1}^1 \frac{\dd (\cos \theta)}{2} \frac{1}{\st^2} \left(-\ct^4 + 1 \right)= 6 \,.
\end{align}
We collect all the soft IR divergent phase space factors in Appendix \ref{app:IRdiv}.

\subsubsection{IR divergent (collinear)\label{sec:collinear}}
 There are also pieces of the anomalous dimension matrix which cannot be calculated by 2-cuts. This is because they arise from graphs which reduce to a massless bubble. Diagrammatically,
\begin{equation}
    \mathcal{A}_{\rm 1-loop}= b \; 
    \mathord{\begin{tikzpicture}[baseline=-0.65ex]
    \coordinate (n1) at (0,0);
    \coordinate (n2) at (1,0);
  \draw[ultra thick] (n1) -- ($(n1)+(135:0.6)$);
    \draw[ultra thick] (n1) -- ($(n1)+(180:0.6)$);
  \draw[ultra thick] (n1) -- ($(n1)+(-135:0.6)$);
    \draw[ultra thick] (n2) -- ($(n2)+(0:0.6)$);
  \draw[ultra thick] (n1) to [out=45,in=135] (n2);
  \draw[ultra thick] (n1) to [out=-45,in=-135] (n2);
\end{tikzpicture}}
.
\end{equation}
Massless bubble scalar integrals vanish under all cuts (and indeed are simply zero in dimensional regularisation). Nevertheless, they contain UV divergences that we must account for, and which are masked by equal and opposite collinear IR divergences. On-shell, the IR (and therefore also UV) divergence may be calculated from the real part of $2 \to 3$ diagrams \cite{Jiang:2020mhe} or via the IR singularities of the stress-energy tensor \cite{Caron-Huot:2016cwu}; off-shell, the UV divergences are accounted for by wavefunction renormalisation factors.
These pieces therefore only affect self-renormalisation diagrams, although (unlike the soft IR divergences described above) they can change flavour indices since they can involve Yukawa interactions. 
They can be calculated as
\begin{equation}
\gamma^{\rm coll} \mathcal{A}_{\text{tree}}= \sum_{l} \gamma(l) \cdot \mathcal{A}_{\text{tree}}\,,
\end{equation}
where gauge and flavour indices on $\mathcal{A}_{\text{tree}}$ and $\gamma(l)$ have been suppressed, and the sum is over all legs of the tree amplitude, labelled as $l$.
The full set of the collinear factors $\gamma(f)$ was calculated in \cite{Jiang:2020mhe,AccettulliHuber:2021uoa}. In our conventions they are collected in Appendix \ref{app:collinear}.

For example, in the case of the 
$ed \rightarrow ed$ anomalous dimension matrix, we have
\begin{align}
\gamma_{\rm coll} \cdot c = [\gamma(d)]^r_v c^{pv}_{qs} + [\gamma(d)]^v_s c^{pr}_{qv} + [\gamma(e)]^p_v c^{vr}_{qs} + [\gamma(e)]^v_q c^{pr}_{vs}
\,,
\end{align}
where
\begin{align}
[\gamma(e)]^p_q&=\left(\frac{N_L}{2}[Y_e^\dagger Y_e]^p_q-3y_e^2g_1^2\delta^p_q\right) \, , \\
[\gamma(d)]^p_q&=\left(\frac{N_L}{2}[Y_d^\dagger Y_d]^p_q-3g_3^2 \, , C_2(N_c)\delta^p_q-3y_d^2g_1^2\delta^p_q\right)\,,
\end{align}
and $C_2(3)=4/3$. Considering only the $g_3,g_1$ contributions we have
\begin{align}
\gamma_{\rm coll} = - 8g_3^2 - 6y_d^2g_1^2 - 6y_e^2g_1^2 \,,
\end{align}
$y_d$ and $y_e$ being the respective hypercharges of $d$ and $e$.

\subsection{Gauge factors}
\label{sec:gauge}
The gauge structure of the 1-loop amplitudes can be understood as the contraction of the gauge structure of the tree amplitudes on either side of the cut. Since both the SMEFT and the SM tree amplitudes that we consider can be constructed as products of currents (see Sec.~\ref{subsec:Jvector}), the gauge factors of the amplitudes only involve (products of) structures with two indices, i.e.~$\delta$ and the $SU(N)$ generators $\lambda$.  
The SM 4-point amplitudes only admit a few gauge structures (see App.~\ref{app:SM4point}). These are, for the gauge amplitudes:
\begin{align}
  U(1)_Y:~  \mathord{\begin{tikzpicture}[baseline=-0.65ex]
   \draw[thick,dotted] (-1,1) -- (-\currsep,0) -- (-1,-1);
  \currentmarker{-\currsep,0}
  \draw[thick,dotted] (1,1) -- (\currsep,0) -- (1,-1);
  \currentmarker{\currsep,0}
  \node at (-0.6,0) {$\delta^a_b$};
  \node at (0.6,0) {$\delta^c_d$};
\end{tikzpicture}},~~~~~~~~~~
  SU(2)_L \text{~or~} SU(3)_c:~  \mathord{\begin{tikzpicture}[baseline=-0.65ex]
   \draw[thick,dotted] (-1,1) -- (-\currsep,0) -- (-1,-1);
  \currentmarker{-\currsep,0}
  \draw[thick,dotted] (1,1) -- (\currsep,0) -- (1,-1);
  \currentmarker{\currsep,0}
  \node at (-0.85,0) {$[\lambda^A]^a_b$};
  \node at (0.85,0) {$[\lambda^A]^c_d$};
\end{tikzpicture}},
\end{align}
where $\lambda$ refer to the Pauli (Gell-Mann) matrices for the case of $SU(2)_L$ ($SU(3)_c$). For the Yukawa amplitudes, things are a little more complicated. Amplitudes with only two legs charged under each gauge group are contracted trivially as:
\begin{gather}
  \mathcal{A}_{SM}(e^+e^-L^{i-}L_j^+):~  \mathord{\begin{tikzpicture}[baseline=-0.65ex]
   \draw[thick,blue] (-1,1)-- (-\currsep,0) -- (-1,-1);
  \currentmarker{-\currsep,0}
  \draw[thick,red] (1,1) -- (\currsep,0) -- (1,-1);
  \currentmarker{\currsep,0}
  \node at (0.6,0) {$\delta^i_j$};
\end{tikzpicture}}, ~~~~~
  \mathcal{A}_{SM}(e^+e^-H^iH^\dagger_j):~  \mathord{\begin{tikzpicture}[baseline=-0.65ex]
   \draw[thick,blue] (-1,1)-- (-\currsep,0) -- (-1,-1);
  \currentmarker{-\currsep,0}
  \draw[thick,dashed] (1,1) -- (\currsep,0) -- (1,-1);
  \currentmarker{\currsep,0}
  \node at (0.6,0) {$\delta^i_j$};
\end{tikzpicture}},\\
  \mathcal{A}_{SM}(u^{+a}u^{-b}H^iH^\dagger_j):~  \mathord{\begin{tikzpicture}[baseline=-0.65ex]
   \draw[thick,blue] (-1,1)-- (-\currsep,0) -- (-1,-1);
  \currentmarker{-\currsep,0}
  \draw[thick,dashed] (1,1) -- (\currsep,0) -- (1,-1);
  \currentmarker{\currsep,0}
  \node at (-0.6,0) {$\delta^a_b$};
  \node at (0.6,0) {$\delta^i_j$};
\end{tikzpicture}},
\end{gather}
where we now explicitly label $SU(2)_L$ indices with $i,j$ and $SU(3)_c$ indices with $a,b$, and show currents of LH fermions ($Q,L$) in red and currents of RH fermions ($u,d,e$) in blue. The amplitude with RH down-type quarks and Higgses, $\mathcal{A}_{SM}(d^{+a}d^{-b}H^iH^\dagger_j)$, has the same form as $\mathcal{A}_{SM}(u^{+a}u^{-b}H^iH^\dagger_j)$. 
In Yukawa amplitudes in which all four legs are charged under the same gauge group, we end up with more than one type of gauge structure between the currents:
\begin{align}
\label{eq:gaugeuuQQSM}
  \mathcal{A}_{SM}(u^{+a}u^{-}_bQ^{ci-}Q_{dj}^+):& \mathord{\begin{tikzpicture}[baseline=-0.65ex]
   \draw[thick,blue] (-1,1)-- (-\currsep,0) -- (-1,-1);
  \currentmarker{-\currsep,0}
  \draw[thick,red] (1,1) -- (\currsep,0) -- (1,-1);
  \currentmarker{\currsep,0}
  \node at (0,0.45) {$\delta^a_d$};
  \node at (0,-0.45) {$\delta^c_b$};
  \node at (0.6,0) {$\delta^i_j$};
\end{tikzpicture}}\, = \frac{1}{N_c}
\mathord{\begin{tikzpicture}[baseline=-0.65ex]
   \draw[thick,blue] (-1,1)-- (-\currsep,0) -- (-1,-1);
  \currentmarker{-\currsep,0}
  \draw[thick,red] (1,1) -- (\currsep,0) -- (1,-1);
  \currentmarker{\currsep,0}
  \node at (-0.6,0) {$\delta^a_b$};
  \node at (0.8,0) {$\delta^c_d\delta^i_j$};
\end{tikzpicture}}+\frac{1}{2}
\mathord{\begin{tikzpicture}[baseline=-0.65ex]
   \draw[thick,blue] (-1,1)-- (-\currsep,0) -- (-1,-1);
  \currentmarker{-\currsep,0}
  \draw[thick,red] (1,1) -- (\currsep,0) -- (1,-1);
  \currentmarker{\currsep,0}
  \node at (-0.8,0) {$[\lambda^A]^a_b$};
  \node at (1.0,0) {$[\lambda^A]^c_d\delta^i_j$};
\end{tikzpicture}},\\
\label{eq:gaugeLLHHSM}
  \mathcal{A}_{SM}(L^{-i} L^{+}_{j} H^k H^\dagger_{l}):& \mathord{\begin{tikzpicture}[baseline=-0.65ex]
   \draw[thick,red] (-1,1)-- (-\currsep,0) -- (-1,-1);
  \currentmarker{-\currsep,0}
  \draw[thick,dashed] (1,1) -- (\currsep,0) -- (1,-1);
  \currentmarker{\currsep,0}
  \node at (0,0.45) {$\delta^k_j$};
  \node at (0,-0.45) {$\delta^i_l$};
\end{tikzpicture}}\, = \frac{1}{2}
\mathord{\begin{tikzpicture}[baseline=-0.65ex]
   \draw[thick,red] (-1,1)-- (-\currsep,0) -- (-1,-1);
  \currentmarker{-\currsep,0}
  \draw[thick,dashed] (1,1) -- (\currsep,0) -- (1,-1);
  \currentmarker{\currsep,0}
  \node at (-0.6,0) {$\delta^i_j$};
  \node at (0.6,0) {$\delta^k_l$};
\end{tikzpicture}}+\frac{1}{2}
\mathord{\begin{tikzpicture}[baseline=-0.65ex]
   \draw[thick,red] (-1,1)-- (-\currsep,0) -- (-1,-1);
  \currentmarker{-\currsep,0}
  \draw[thick,dashed] (1,1) -- (\currsep,0) -- (1,-1);
  \currentmarker{\currsep,0}
  \node at (-0.8,0) {$[\sigma^I]^i_j$};
  \node at (0.8,0) {$[\sigma^I]^k_l$};
\end{tikzpicture}},
\end{align}
where $N_c=3$. For the $\mathcal{A}_{SM}(Q^{-ia}Q^+_{jb} H^k H^\dagger_{l})$ amplitude, there are two different $SU(2)_L$ structures. The piece proportional to $Y_dY_d^\dagger$ has the gauge structure $\delta_j^k \delta_l^i \delta^a_b$, which is identical to that in Eq.~\eqref{eq:gaugeLLHHSM}, with the quark colour indices trivially contracted. The piece proportional to $Y_uY_u^\dagger$ instead has the gauge structure $\eps_{jl} \eps^{ik} \delta_a^b$. The $\eps$ matrices can be related back to the $\delta$s and $\sigma$s via the Fierz relation
\begin{equation}
\eps_{jl} \eps^{ik} = \frac12 \delta^i_j \delta^k_l - \frac12 [\sigma^I]^i_j [\sigma^I]^k_l,
\end{equation}
so overall we can write:
\begin{align}
\label{eq:gaugeQQHHSM}
  \mathcal{A}_{SM}(Q^{-ia}Q^+_{jb} H^k H^\dagger_{l}):
  \begin{cases}
 \frac{1}{2}
\mathord{\begin{tikzpicture}[baseline=-0.65ex]
   \draw[thick,red] (-1,1)-- (-\currsep,0) -- (-1,-1);
  \currentmarker{-\currsep,0}
  \draw[thick,dashed] (1,1) -- (\currsep,0) -- (1,-1);
  \currentmarker{\currsep,0}
  \node at (-0.8,0) {$\delta^a_b \delta^i_j$};
  \node at (0.6,0) {$\delta^k_l$};
\end{tikzpicture}}-\frac{1}{2}
\mathord{\begin{tikzpicture}[baseline=-0.65ex]
   \draw[thick,red] (-1,1)-- (-\currsep,0) -- (-1,-1);
  \currentmarker{-\currsep,0}
  \draw[thick,dashed] (1,1) -- (\currsep,0) -- (1,-1);
  \currentmarker{\currsep,0}
  \node at (-1.0,0) {$\delta^a_b [\sigma^I]^i_j$};
  \node at (0.8,0) {$[\sigma^I]^k_l$};
\end{tikzpicture}} & Y_u Y^\dagger_u \text{ piece}, \\[30pt]
 \frac{1}{2}
\mathord{\begin{tikzpicture}[baseline=-0.65ex]
   \draw[thick,red] (-1,1)-- (-\currsep,0) -- (-1,-1);
  \currentmarker{-\currsep,0}
  \draw[thick,dashed] (1,1) -- (\currsep,0) -- (1,-1);
  \currentmarker{\currsep,0}
  \node at (-0.8,0) {$\delta^a_b \delta^i_j$};
  \node at (0.6,0) {$\delta^k_l$};
\end{tikzpicture}}+\frac{1}{2}
\mathord{\begin{tikzpicture}[baseline=-0.65ex]
   \draw[thick,red] (-1,1)-- (-\currsep,0) -- (-1,-1);
  \currentmarker{-\currsep,0}
  \draw[thick,dashed] (1,1) -- (\currsep,0) -- (1,-1);
  \currentmarker{\currsep,0}
  \node at (-1.0,0) {$\delta^a_b [\sigma^I]^i_j$};
  \node at (0.8,0) {$[\sigma^I]^k_l$};
\end{tikzpicture}} & Y_d Y^\dagger_d \text{ piece}.
\end{cases}
\end{align}

The full range of contractions needed in the anomalous dimensions is then given in Tab.~\ref{tab:colorfactors}. In the first column, the structures on the left of `$\times$' come from the dimension 6 amplitude, while structures on the right come from the SM amplitude. Of course in a general amplitude there can be more than one type of gauge index, and/or some legs may be uncharged under a gauge group, in which case these relations can be straightforwardly generalised. For SMEFT operators with identical currents, it is convenient to (anti)symmetrise the gauge indices (see Sec.~\ref{sec:gauge}). We provide equivalent gauge contractions in terms of these (anti)symmetrised structures in App.~\ref{app:gaugeantisymm}.

We note that zeroes will arise in the anomalous dimension matrix whenever a $SU(N)$ generator $\lambda$ is traced over. This means that if one of the tree amplitudes in the cut diagram has a $\lambda$ matrix connecting the cut legs, the other tree amplitude must also connect the cut legs with a $\lambda$ matrix, or else the loop amplitude is zero:
\begin{align}
\label{eq:zerolambdatrace}
& \mathord{\begin{tikzpicture}[baseline=-0.65ex]
  \draw[thick,dotted] (1,1) -- (\currsep,0) -- (1,-1);
  \currentmarker{\currsep,0}
  \node at (-0.5,0) {$[\lambda^{A}]^a_b$};
\end{tikzpicture}}
\cut
\mathord{\begin{tikzpicture}[baseline=-0.65ex]
  \draw[thick,dotted] (-1,1) -- (-\currsep,0) -- (-1,-1);
  \currentmarker{-\currsep,0}
  \node at (0.3,0) {$\delta_a^b$};
\end{tikzpicture}}
=0\,.
\end{align}
However, if one (or both) amplitude contains four particles all charged under the same gauge group, there may be different options for the cut orientations. Even if one amplitude is constructed with a $\lambda\lambda$ structure and the other has a $\delta\delta$ structure, in some cut directions the $\lambda$ matrix will not be traced over (or equivalently, will be contracted with another $\lambda$ via the Fierz relations~\eqref{eq:symmFierz}):
{
\allowdisplaybreaks
\begin{align}
\mathord{\begin{tikzpicture}[baseline=-0.65ex]
  \draw[thick,dotted] (1,1) -- (\currsep,0) -- (1,-1);
  \currentmarker{\currsep,0}
  \node at (-0.5,0) {$[\lambda^{A}]^a_b$};
\end{tikzpicture}}
\cut
  \mathord{\begin{tikzpicture}[baseline=-0.65ex]
    \draw[thick,dotted] (1,-1)  -- (0,-\currsep) -- (-1,-1);
    \draw[thick,dotted] (-1,1) -- (0,\currsep) -- (1,1);
    \currentmarker{0,\currsep}
    \currentmarker{0,-\currsep}
    \node at (0,0.6) {$\delta_a^c$};
    \node at (0,-0.6) {$\delta_d^b$};
  \end{tikzpicture}}
    &= \mathord{\begin{tikzpicture}[baseline=-0.65ex]
  \draw[thick,dotted] (1,1) -- (\currsep,0) -- (1,-1);
  \currentmarker{\currsep,0}
  \node at (-0.5,0) {$[\lambda^{A}]^c_d$};
\end{tikzpicture}}
\times \text{kinematics} \times \text{flavour\,,}\,\\
\label{eq:deltagaugeIRdiv}
\mathord{\begin{tikzpicture}[baseline=-0.65ex]
  \draw[thick,dotted] (1,1) -- (\currsep,0) -- (1,-1);
  \currentmarker{\currsep,0}
  \node at (-0.3,0) {$\delta^a_b$};
\end{tikzpicture}}
\cut
  \mathord{\begin{tikzpicture}[baseline=-0.65ex]
    \draw[thick,dotted] (1,-1)  -- (0,-\currsep) -- (-1,-1);
    \draw[thick,dotted] (-1,1) -- (0,\currsep) -- (1,1);
    \currentmarker{0,\currsep}
    \currentmarker{0,-\currsep}
    \node at (0,0.8) {$[\lambda^A]_a^c$};
    \node at (0,-0.8) {$[\lambda^A]_d^b$};
  \end{tikzpicture}}
    &= 2\frac{N^2-1}{N}\, \mathord{\begin{tikzpicture}[baseline=-0.65ex]
  \draw[thick,dotted] (1,1) -- (\currsep,0) -- (1,-1);
  \currentmarker{\currsep,0}
  \node at (-0.77,0) {$\delta^c_d$};
\end{tikzpicture}}
\times \text{kinematics} \times \text{flavour \,,}\,\\
\label{eq:lambdagaugeIRdiv}
\mathord{\begin{tikzpicture}[baseline=-0.65ex]
  \draw[thick,dotted] (1,1) -- (\currsep,0) -- (1,-1);
  \currentmarker{\currsep,0}
  \node at (-0.5,0) {$[\lambda^{A}]^a_b$};
\end{tikzpicture}}
\cut
  \mathord{\begin{tikzpicture}[baseline=-0.65ex]
    \draw[thick,dotted] (1,-1)  -- (0,-\currsep) -- (-1,-1);
    \draw[thick,dotted] (-1,1) -- (0,\currsep) -- (1,1);
    \currentmarker{0,\currsep}
    \currentmarker{0,-\currsep}
    \node at (0,0.8) {$[\lambda^A]_a^c$};
    \node at (0,-0.8) {$[\lambda^A]_d^b$};
  \end{tikzpicture}}
   &= -\frac{1}{2N}\mathord{\begin{tikzpicture}[baseline=-0.65ex]
  \draw[thick,dotted] (1,1) -- (\currsep,0) -- (1,-1);
  \currentmarker{\currsep,0}
  \node at (-0.6,0) {$\,[\lambda^{A}]^c_d$};
\end{tikzpicture}}\times \text{kinematics} \times \text{flavour \,,}\,
\end{align}}\noindent
In the cases above, we nevertheless note that a $\lambda\lambda$ operator can still only be renormalised by another $\lambda\lambda$ operator, as seen by the fact that the gauge structure on the right-hand side is always the same as the structure in the leftmost current.

In fact, the only situation in which a $\lambda\lambda$ operator can mix into a $\delta\delta$ operator is if the cut involves two currents in \emph{both} the amplitudes on either side of the cut, and both are constructed from currents with $\lambda$ structures, i.e.:

\begin{align}
  \mathord{\begin{tikzpicture}[baseline=-0.65ex]
    \draw[thick,dotted] (1,-1)  -- (0,-\currsep) -- (-1,-1);
    \draw[thick,dotted] (-1,1) -- (0,\currsep) -- (1,1);
    \currentmarker{0,\currsep}
    \currentmarker{0,-\currsep}
    \node at (0,0.8) {$[\lambda^A]_b^a$};
    \node at (0,-0.8) {$[\lambda^A]_d^c$};
  \end{tikzpicture}}
\cut
  \mathord{\begin{tikzpicture}[baseline=-0.65ex]
    \draw[thick,dotted] (1,-1)  -- (0,-\currsep) -- (-1,-1);
    \draw[thick,dotted] (-1,1) -- (0,\currsep) -- (1,1);
    \currentmarker{0,\currsep}
    \currentmarker{0,-\currsep}
    \node at (0,0.8) {$[\lambda^B]_e^b$};
    \node at (0,-0.8) {$[\lambda^B]_d^f$};
  \end{tikzpicture}}
   &=  4\frac{N^2-1}{N^2} \,  \mathord{\begin{tikzpicture}[baseline=-0.65ex]
    \draw[thick,dotted] (1,-1)  -- (0,-\currsep) -- (-1,-1);
    \draw[thick,dotted] (-1,1) -- (0,\currsep) -- (1,1);
    \currentmarker{0,\currsep}
    \currentmarker{0,-\currsep}
    \node at (0,0.8) {$[\lambda^C]_e^a$};
    \node at (0,-0.8) {$[\lambda^C]_c^f$};
  \end{tikzpicture}}\times \text{kinematics} \times \text{flavour \,,}\,\nn \\
  &+ 2\frac{N^2-2}{N} \,  \mathord{\begin{tikzpicture}[baseline=-0.65ex]
    \draw[thick,dotted] (1,-1)  -- (0,-\currsep) -- (-1,-1);
    \draw[thick,dotted] (-1,1) -- (0,\currsep) -- (1,1);
    \currentmarker{0,\currsep}
    \currentmarker{0,-\currsep}
    \node at (0,0.6) {$\delta_e^a$};
    \node at (0,-0.6) {$\delta_c^f$};
  \end{tikzpicture}}\times \text{kinematics} \times \text{flavour}\,. \label{eq:doublecutIRdivgauge}
\end{align}
As we will see in the next section, the rarity of this setup amongst the SMEFT amplitudes leads to many zeroes in the anomalous dimension matrix between operators with different gauge structures. 

 \begin{table}
 \begin{center}
 \scalebox{0.8}{
 \begin{tabular}{| L | L | L | L |}
 \hline
 \mathcal{A}_6 \times \mathcal{A}_{\rm SM}~~{\rm (gauge ~part)}& {\rm General~} SU(N) & N=2 & N=3 \\
 \hline\hline
 \delta^a_b \delta^e_f \times \delta^f_e \delta^c_d & = N \delta^a_b \delta^c_d & = 2 \delta^a_b \delta^c_d & =3 \delta^a_b \delta^c_d\\
 \hline
 [\lambda^A]^a_b [\lambda^A]^e_f \times \delta^f_e \delta^c_d &= 0 &=0 &=0\\
 \hline
 \delta^a_b \delta^e_f \times \frac14 [\lambda^A]^f_e [\lambda^A]^c_d &  =0 & =0 & =0\\
 \hline
    [\lambda^A]^a_b [\lambda^A]^e_f \times \frac14 [\lambda^B]^f_e [\lambda^B]^c_d &= \frac12 [\lambda^C]^a_b [\lambda^C]^c_d & =\frac12 [\lambda^C]^a_b [\lambda^C]^c_d & =\frac12 [\lambda^C]^a_b [\lambda^C]^c_d  \\
   \hline  \hline
  \delta^a_b \delta^e_f \times \delta^f_d \delta^c_e & = \delta^a_b \delta^c_d  & =  \delta^a_b \delta^c_d & = \delta^a_b \delta^c_d\\
     \hline
     \delta^a_b \delta^e_f \times \frac{1}{4} [\lambda^A]^f_d [\lambda^A]^c_e & = \frac{N^2-1}{2N} \delta^a_b \delta^c_d  & =  \frac{3}{4}\delta^a_b \delta^c_d & = \frac{4}{3}\delta^a_b \delta^c_d\\
 \hline
 [\lambda^A]^a_b [\lambda^A]^e_f \times \delta^f_d \delta^c_e &= [\lambda^B]^a_b [\lambda^B]^c_d  & =[\lambda^B]^a_b [\lambda^B]^c_d & =[\lambda^B]^a_b [\lambda^B]^c_d\\
  \hline
   [\lambda^A]^a_b [\lambda^A]^e_f \times \frac{1}{4} [\lambda^B]^f_d [\lambda^B]^c_e &= -\frac{1}{2N}[\lambda^C]^a_b [\lambda^C]^c_d  & =-\frac{1}{4}[\lambda^C]^a_b [\lambda^C]^c_d  & =-\frac{1}{6} [\lambda^C]^a_b [\lambda^C]^c_d  \\
   \hline \hline
  [\lambda^A]^e_b [\lambda^A]^a_f \times  \frac{1}{4} [\lambda^B]^f_d [\lambda^B]^c_e &= \frac{N^2-1}{N^2}\delta^a_b \delta^c_d +\frac{N^2-2}{2N}[\lambda^C]^a_b [\lambda^C]^c_d & = \frac{3}{4}\delta^a_b \delta^c_d +\frac{1}{2}[\lambda^C]^a_b [\lambda^C]^c_d &  = \frac{8}{9}\delta^a_b \delta^c_d +\frac{7}{6}[\lambda^C]^a_b [\lambda^C]^c_d\\
  \hline
    [\lambda^A]^e_b [\lambda^A]^f_d \times  \frac{1}{4} [\lambda^B]^c_f [\lambda^B]^a_e &= \frac{N^2-1}{N^2}\delta^a_b \delta^c_d -\frac{1}{N}[\lambda^C]^a_b [\lambda^C]^c_d & = \frac{3}{4}\delta^a_b \delta^c_d -\frac{1}{2}[\lambda^C]^a_b [\lambda^C]^c_d &  = \frac{8}{9}\delta^a_b \delta^c_d -\frac{1}{3}[\lambda^C]^a_b [\lambda^C]^c_d\\
    \hline
 \end{tabular}}
 \caption{\label{tab:colorfactors}Possible gauge contractions appearing in $\mathcal{A}_6 \times \mathcal{A}_{\rm SM}$. The identities are valid for $SU(N)$ generators $\lambda^A$ normalised such that $\Tr[\lambda^A \lambda^B] = 2 \delta^{AB}$, and which reduce to the Pauli (Gell-Mann) matrices when $N=2\,(3)$.}
 	\end{center}
 \end{table}

 \subsection{Flavour factors}
\label{sec:flavour}
The different topologies of loop diagrams introduce different flavour factors into the anomalous dimension matrix.

The SM gauge amplitudes are very simple in their flavour structure; any flavour indices are simply contracted with a $\delta$ within each current:
\begin{align}
\mathord{\begin{tikzpicture}[baseline=-0.65ex]
   \draw[thick] (-1,1) -- (-\currsep,0) -- (-1,-1);
  \currentmarker{-\currsep,0}
  \draw[thick] (1,1) -- (\currsep,0) -- (1,-1);
  \currentmarker{\currsep,0}
  \node at (-0.6,0) {$\delta^p_q$};
  \node at (0.6,0) {$\delta^r_s$};
\end{tikzpicture} \,.}
\end{align}For the Yukawa amplitudes, the flavour structure depends on the external fields. The different amplitudes have the form (where currents of LH fermions ($Q,L$) are shown in red and currents of RH fermions ($u,d,e$) are shown in blue):
\begin{align}
\label{eq:Yukfactors}
\mathord{\begin{tikzpicture}[baseline=-0.65ex]
   \draw[thick,blue] (-1,1) node[left]{$p$}  node[right]{$+$} -- (-\currsep,0) -- (-1,-1) node[left]{$q$} node[right]{$-$};
  \currentmarker{-\currsep,0}
  \draw[thick,dashed] (1,1) -- (\currsep,0) -- (1,-1);
  \currentmarker{\currsep,0}
  \node at (-1,0) {$[Y^\dagger Y]^p_q$};
\end{tikzpicture}},~~~~~~ \mathord{\begin{tikzpicture}[baseline=-0.65ex]
   \draw[thick,red] (-1,1) node[left]{$p$} node[right]{$-$} -- (-\currsep,0) -- (-1,-1) node[left]{$q$} node[right]{$+$};
  \currentmarker{-\currsep,0}
  \draw[thick,dashed] (1,1) -- (\currsep,0) -- (1,-1);
  \currentmarker{\currsep,0}
  \node at (-1,0) {$[YY^\dagger]^p_q$};
\end{tikzpicture}},~~~~~~
  \mathord{\begin{tikzpicture}[baseline=-0.65ex]
   \draw[thick,blue] (-1,1) node[left]{$p$} node[right]{$+$} -- (-\currsep,0) -- (-1,-1) node[left]{$q$} node[right]{$-$};
   \currentmarker{-\currsep,0}
   \draw[thick,red] (1,1) node[right]{$r$} node[left]{$-$} -- (\currsep,0) -- (1,-1) node[right]{$s$} node[left]{$+$};
   \currentmarker{\currsep,0}
  \node at (0,0.5) {$Y^p_s$};
  \node at (0,-0.5) {$[Y^\dagger]^r_q$};
 \end{tikzpicture} \,.}
\end{align}
The different ways that the SM flavour tensors can be contracted with the SMEFT flavour tensors then depends on the arrangement of the diagram. The only diagram which produces true self-renormalisation (i.e. the flavour structure of the renormalised operator is identical to the renormalising operator) is the gauge diagram
\begin{align}
  \mathord{\begin{tikzpicture}[baseline=-0.65ex]
    \draw[thick,dotted] (1,-1)  -- (0,-\currsep) -- (-1,-1);
    \draw[thick,dotted] (-1,1) -- (0,\currsep) -- (1,1);
    \currentmarker{0,\currsep}
    \currentmarker{0,-\currsep}
    \node at (-0.9,0.0) {$(c_{ws}^{pv})\, c_{w}^p$};
  \end{tikzpicture}}
\cut
  \mathord{\begin{tikzpicture}[baseline=-0.65ex]
    \draw[thick,dotted] (1,-1)  -- (0,-\currsep) -- (-1,-1);
    \draw[thick,dotted] (-1,1) -- (0,\currsep) -- (1,1);
    \currentmarker{0,\currsep}
    \currentmarker{0,-\currsep}
    \node at (0,0.7) {$(\delta_v^r)$};
    \node at (0,-0.7) {$\delta_q^w$};
  \end{tikzpicture}}
   &=  \,  \mathord{\begin{tikzpicture}[baseline=-0.65ex]
    \draw[thick,dotted] (1,-1)  -- (0,-\currsep) -- (-1,-1);
    \draw[thick,dotted] (-1,1) -- (0,\currsep) -- (1,1);
    \currentmarker{0,\currsep}
    \currentmarker{0,-\currsep}
    \node at (-0.9,0) {$(c_{qs}^{pr})\, c_{q}^p$};
  \end{tikzpicture}}\times \text{kinematics} \times \text{gauge}\,, \label{eq:doublecutIRdivflavour}
\end{align}
where the Wilson coefficient is in the form $(c_{qs}^{pr})~ c_{q}^p$ for (four) two fermion operators respectively. 

All other types of cut diagram will give rise to an operator with a different flavour structure. For mixing between operators with equal numbers of fermions ($4f\to 4f$ and $2f\to 2f$), the other (flavour-changing) options are
{\allowdisplaybreaks\begin{align}
\label{eq:schannelgaugeflavour} \mathord{\begin{tikzpicture}[baseline=-0.65ex]
  \draw[thick] (1,1) -- (\currsep,0) -- (1,-1);
  \currentmarker{\currsep,0}
   \node at (-0.6,0) {$(c^{vr}_{ws})\, c_w^v$};
\end{tikzpicture}}
\cut
\mathord{\begin{tikzpicture}[baseline=-0.65ex]
   \draw[thick] (-1,1) -- (-\currsep,0) -- (-1,-1);
  \currentmarker{-\currsep,0}
  \draw[thick] (1,1) -- (\currsep,0) -- (1,-1);
  \currentmarker{\currsep,0}
  \node at (-0.6,0) {$\delta^w_v$};
  \node at (0.6,0) {$\delta^p_q$};
\end{tikzpicture}}
&= \mathord{\begin{tikzpicture}[baseline=-0.65ex]
  \draw[thick] (1,1) -- (\currsep,0) -- (1,-1);
  \currentmarker{\currsep,0}
   \node at (-1.0,0) {$(c_{vq}^{vp}\delta^p_q)\, c_{v}^v\delta^p_q$};
\end{tikzpicture}}\times \text{kinematics} \times \text{gauge}\,,\\
\label{eq:schannelYukflavour} \mathord{\begin{tikzpicture}[baseline=-0.65ex]
  \draw[thick, blue] (1,1) -- (\currsep,0) -- (1,-1);
  \currentmarker{\currsep,0}
   \node at (-0.6,0) {$(c^{vr}_{ws})\, c_w^v$};
\end{tikzpicture}}
\cut
  \mathord{\begin{tikzpicture}[baseline=-0.65ex]
   \draw[thick,blue] (-1,1) -- (-\currsep,0) -- (-1,-1);
   \currentmarker{-\currsep,0}
   \draw[thick,red] (1,1) -- (\currsep,0) -- (1,-1);
   \currentmarker{\currsep,0}
  \node at (0,0.5) {$Y^{w}_q$};
  \node at (0,-0.5) {$[Y^{\dagger}]^p_v$};
 \end{tikzpicture}}
&= \mathord{\begin{tikzpicture}[baseline=-0.65ex]
  \draw[thick,red] (1,1) -- (\currsep,0) -- (1,-1);
  \currentmarker{\currsep,0}
   \node at (-2.2,0) {$\left( [Y^{\dagger}]^p_v c^{vr}_{ws} Y^{w}_q \right)\, [Y^{\dagger}]^p_v c^v_w Y^{w}_q$};
\end{tikzpicture}}\times \text{kinematics} \times \text{gauge}\,,\\
\label{eq:schannelYukflavour2} \mathord{\begin{tikzpicture}[baseline=-0.65ex]
  \draw[thick, red] (1,1) -- (\currsep,0) -- (1,-1);
  \currentmarker{\currsep,0}
   \node at (-0.6,0) {$(c^{vr}_{ws})\, c_w^v$};
\end{tikzpicture}}
\cut
  \mathord{\begin{tikzpicture}[baseline=-0.65ex]
   \draw[thick,red] (-1,1) -- (-\currsep,0) -- (-1,-1);
   \currentmarker{-\currsep,0}
   \draw[thick,blue] (1,1) -- (\currsep,0) -- (1,-1);
   \currentmarker{\currsep,0}
  \node at (0,0.5) {$Y^p_v$};
  \node at (0,-0.5) {$[Y^{\dagger}]^w_q$};
 \end{tikzpicture}}
&= \mathord{\begin{tikzpicture}[baseline=-0.65ex]
  \draw[thick,blue] (1,1) -- (\currsep,0) -- (1,-1);
  \currentmarker{\currsep,0}
   \node at (-2.2,0) {$\left( Y^p_v c^{vr}_{ws} [Y^{\dagger}]^w_q \right)\, Y^p_v c^v_w [Y^{\dagger}]^w_q $};
\end{tikzpicture}}\times \text{kinematics} \times \text{gauge}\,,\\
  \mathord{\begin{tikzpicture}[baseline=-0.65ex]
    \draw[thick,red] (1,-1)  -- (0,-\currsep) -- (-1,-1);
    \draw[thick,blue] (-1,1) -- (0,\currsep) -- (1,1);
    \currentmarker{0,\currsep}
    \currentmarker{0,-\currsep}
    \node at (-0.5,0.0) {$c_{ws}^{pv}$};
  \end{tikzpicture}}
\cut
  \mathord{\begin{tikzpicture}[baseline=-0.65ex]
    \draw[thick,red] (1,-1)  -- (0,-\currsep) -- (-1,-1);
    \draw[thick,blue] (-1,1) -- (0,\currsep) -- (1,1);
    \currentmarker{0,\currsep}
    \currentmarker{0,-\currsep}
  \node at (0.5,0.0) {$Y^r_q$};
  \node at (-0.7,0.0) {$[Y^\dagger]^w_v$};
  \end{tikzpicture}}
   &=  \,  \mathord{\begin{tikzpicture}[baseline=-0.65ex]
    \draw[thick,red] (1,-1)  -- (0,-\currsep) -- (-1,-1);
    \draw[thick,blue] (-1,1) -- (0,\currsep) -- (1,1);
    \currentmarker{0,\currsep}
    \currentmarker{0,-\currsep}
    \node at (-1.2,0) {$[Y^\dagger]^w_v c_{ws}^{pv} Y^r_q$};
  \end{tikzpicture}}\times \text{kinematics} \times \text{gauge}\,, \label{eq:doublecutIRdivYukflavour}\\
    \mathord{\begin{tikzpicture}[baseline=-0.65ex]
    \draw[thick] (1,-1)  -- (0,-\currsep) -- (-1,-1);
    \draw[thick,dashed] (-1,1) -- (0,\currsep) -- (1,1);
    \currentmarker{0,\currsep}
    \currentmarker{0,-\currsep}
    \node at (0.0,-0.8) {$c_{q}^{v}$};
  \end{tikzpicture}}
\cut
  \mathord{\begin{tikzpicture}[baseline=-0.65ex]
    \draw[thick] (1,-1)  -- (0,-\currsep) -- (-1,-1);
    \draw[thick,dashed] (-1,1) -- (0,\currsep) -- (1,1);
    \currentmarker{0,\currsep}
    \currentmarker{0,-\currsep}
  \node at (0.0,-0.8) {$M^p_v$};
  \end{tikzpicture}}
   &=  \,  \mathord{\begin{tikzpicture}[baseline=-0.65ex]
    \draw[thick] (1,-1)  -- (0,-\currsep) -- (-1,-1);
    \draw[thick,dashed] (-1,1) -- (0,\currsep) -- (1,1);
    \currentmarker{0,\currsep}
    \currentmarker{0,-\currsep}
    \node at (-1.1,0) {$M^p_v c_{q}^{v}$};
  \end{tikzpicture}}\times \text{kinematics} \times \text{gauge}\,, \label{eq:doublecuthiggsfermYukflavour}
\end{align}}\noindent
where $M\equiv Y^\dagger Y$ for operators with $SU(2)_L$ singlet fermions, or $M\equiv YY^\dagger$ for operators with $SU(2)_L$ doublet fermions. Wavefunction renormalisation diagrams will result in flavour factors similar to those in Eq.~\eqref{eq:doublecutIRdivflavour} and Eq.~\eqref{eq:doublecuthiggsfermYukflavour}, for gauge and Yukawa loops respectively.

Finally, mixing between operators with different numbers of fermions can occur via cut diagrams involving a SM amplitude with one Higgs and one fermion current:
\begin{align}
   \label{eq:changingfermionnumbersflavour} \mathord{\begin{tikzpicture}[baseline=-0.65ex]
  \draw[thick] (1,1) -- (\currsep,0) -- (1,-1);
  \currentmarker{\currsep,0}
\end{tikzpicture}}
\cut
\mathord{\begin{tikzpicture}[baseline=-0.65ex]
   \draw[thick] (-1,1) -- (-\currsep,0) -- (-1,-1);
  \currentmarker{-\currsep,0}
  \draw[thick,dashed] (1,1) -- (\currsep,0) -- (1,-1);
  \currentmarker{\currsep,0}
\end{tikzpicture}}, ~~~~~~~~ \mathord{\begin{tikzpicture}[baseline=-0.65ex]
  \draw[thick,dashed] (1,1) -- (\currsep,0) -- (1,-1);
  \currentmarker{\currsep,0}
\end{tikzpicture}}
\cut
\mathord{\begin{tikzpicture}[baseline=-0.65ex]
   \draw[thick,dashed] (-1,1) -- (-\currsep,0) -- (-1,-1);
  \currentmarker{-\currsep,0}
  \draw[thick] (1,1) -- (\currsep,0) -- (1,-1);
  \currentmarker{\currsep,0}
\end{tikzpicture}}.
\end{align}
In the first case, two indices on the Wilson coefficient are contracted with $\delta$ or $M$, for a SM gauge or Yukawa amplitude respectively. In the second case, $\delta$ or $M$ is introduced as a new factor with free indices. The various flavour factors are collected in Tab.~\ref{tab:flavourfactors}.

To understand how Wilson coefficient components mix, when phrased in the irrep basis (Sec.~\ref{sec:flavourStructure}), it is useful to decompose the gauge and Yukawa factors in the same way. Since the weights of representations add in their tensor product, we can then deduce the weights of the renormalized operator coefficients from those of the two amplitudes that make up the loop diagram.

Gauge interactions only involve currents which are flavour singlets, so in spurionic language, the gauge couplings are also singlets. By contrast, the Yukawa matrices must famously be charged under appropriate flavour groups in order for the SM Lagrangian to be formally invariant under the flavour $SU(3)^5$ group.
Specifically, with the Yukawas defined via the Lagrangian of Eqn.~\eqref{eq:YukawaLag},
then $Y_u$ is charged as a $3_Q \times \bar 3_u$, $Y_d$ as a $3_Q \times \bar 3_d$, and $Y_e$ as a $3_L \times \bar 3_e$. If we are in a basis in which a given Yukawa matrix is diagonal, $Y_f=\text{diag}(y_1,y_2,y_3)$, then the $\{\ifl, \itfl, \yfl\}_{l,r}$ charges of each of its three components are as follows, where $SU(3)_l$ is the flavour group of the corresponding left handed field ($l=Q,L$), and $SU(3)_r$ is the flavour group of the corresponding right handed field ($r=u,d,e$): \\
\begin{equation}
\label{eq:YukawaWeights}
\begin{tabular}{|c || c| c|  c| c|}\hline
  & $\ifl_{3,l}$  & $\yfl_l$  & $\ifl_{3,r}$  & $\yfl_r$ \\
 \hline \hline
 $y_1$  &$\frac12$ &  $\frac13$  & $-\frac12$ &  $-\frac13$ \\
  $y_2$ &$-\frac12$ &  $\frac13$  & $\frac12$ &  $-\frac13$ \\
  $y_3$  & $0$ &  $\frac23$ &  $0$ &  $-\frac23$ \\ \hline
\end{tabular}
\end{equation}
The charges of the components of the diagonal daggered Yukawa,  $Y_f^\dagger$, are respectively equal and opposite to these.

The SM Yukawa amplitudes relevant for running within the $(4,0)$ current-current operators always contain a product of a Yukawa and its Hermitian conjugate,\footnote{Given that we neglect $O_{Hud}$ and $O_{LedQ}$ in the analysis.} as seen in Eq.~\eqref{eq:Yukfactors}. These may or may not be contracted together, and the overall flavour factor of the loop diagram is some tensor product of $c$, $Y^\dagger$ and $Y$, where $c$ is the flavour factor of the SMEFT amplitude involved in the loop. From \eqref{eq:YukawaWeights}, it's clear that any entry of a diagonal Yukawa has charges such that $\ifl_{3,l}=-\ifl_{3,r}$ and $\yfl_l=-\yfl_r$. This means that running involving products of diagonal Yukawas will always preserve the sums of the left and right handed charges, $\ifl_{3,l+r}\equiv \ifl_{3,l}+ \ifl_{3,r}$ and $\yfl_{l+r}\equiv \yfl_l+ \yfl_r$.

If we work under the approximation that the third generation Yukawa $y_3$ is the only non-negligible component of the Yukawa matrix, then any product of a diagonal Yukawa and its Hermitian conjugate will have zero overall charge, since in the product the charges of $y_3$ will always be added to the equal and opposite charges of $y_3^\dagger$. Therefore, under this approximation, running will always preserve the left and right handed charges individually. 

In the case where we cannot work in a basis in which all the non-negligible Yukawa matrices are diagonal, then in general the running will not preserve any charges. This is the case for running with operators involving quark doublets if we do not neglect the CKM or down-type Yukawas. Therefore in general, the SMEFT running will not preserve charges under $SU(3)_Q$. We will summarize all these flavour selection rules, under which approximations they hold,  and how this can lead to block-diagonalisation of the anomalous dimension matrix, in Sec.~\ref{sec:flavourselectionrules}.

\begin{table}
    \centering
    \begin{tabular}{|c||c|c|}
    \hline
    Diagram type & Diagram & Flavour factor \\
    \hline
    \hline
    $0f \to 2f$ & \eqref{eq:changingfermionnumbersflavour} & $\delta_q^p$ or $M_q^p$ \\
    \hline 
    $2f \to 0f$ & \eqref{eq:changingfermionnumbersflavour} & $c_w^v \delta^w_v$ or $c_w^v M^w_v$\\
    \hline
$2f\to 2f$   & \eqref{eq:doublecutIRdivflavour} and collinear gauge & $c_q^p$\\
      $2f\to 2f$   & \eqref{eq:schannelgaugeflavour} & $c_w^v \delta^w_v \delta^q_p$ \\
      $2f\to 2f$   & \eqref{eq:schannelYukflavour} and \eqref{eq:schannelYukflavour2} & $[Y^{\dagger}]^p_v c^v_w Y^{w}_q$ or $Y^p_v c^v_w [Y^{\dagger}]^w_q$ \\
         $2f\to 2f$   & \eqref{eq:doublecuthiggsfermYukflavour} and collinear Yukawa & $M_q^w c^p_w+ c^v_q M_v^p$ \\
         \hline
    $2f \to 4f$ & \eqref{eq:changingfermionnumbersflavour} &  $c_s^r \delta_q^p$ or $c_s^r M_q^p$\\
      \hline
$4f\to 4f$   & \eqref{eq:doublecutIRdivflavour} and collinear gauge & $c^{pr}_{qs}$ \\
 $4f \to 4f$& \eqref{eq:schannelgaugeflavour} & $c^{vr}_{ws}\delta_v^w \delta^p_q$ \\
          $4f \to 4f$ & \eqref{eq:schannelYukflavour} and  \eqref{eq:schannelYukflavour2} 
          &  $[Y^{\dagger}]^p_v c^{vr}_{ws} Y^{w}_q$ or $Y^p_v c^{vr}_{ws} [Y^{\dagger}]^w_q$
          \\
    $4f\to 4f$ & collinear Yukawa & $M_q^w c^{pr}_{ws} + c^{vr}_{qs} M_v^p+M_s^w c^{pr}_{qw} + c^{pv}_{qs} M_v^r$ \\
    \hline
    $4f \to 2f$ & \eqref{eq:changingfermionnumbersflavour} &  $c^{vr}_{ws} \delta_v^w$ or $c^{vr}_{ws} M_v^w$ \\
\hline
    \end{tabular}
    \caption{Flavour factors that appear in contributions to the anomalous dimension matrix. $M_p^q\equiv [Y^\dagger Y]_p^q$ or $[YY^\dagger]_p^q$ (depending on the operator). $c$ is the Wilson coefficient of the dimension 6 operator in the loop diagram. Here we have not specified whether 4-fermion operators are distinguishable, symmetric or antisymmetric: the corresponding Wilson coefficients can be (anti)-symmetrised in an obvious way.}
    \label{tab:flavourfactors}
\end{table}

\section{Anatomy of the anomalous dimension matrix}
\label{sec:anatomy}

\begin{table}[]
    \centering
    \begin{tabular}{|c|c|c|c|c|}
    \hline
    $\gamma$ contribution & Table & Cut topology & Gauge action & Flavour action \\
    \hline
    \hline
     IR-finite gauge  & \ref{tab:colourlessSMEFTIRfinitegauge} &  \begin{tikzpicture}[baseline=-0.65ex]
  \draw[thick,dotted] (0.4,0.4) -- (-0.4,-0.4);
  \draw[thick,dotted] (-0.4,0.4) -- (0.4,-0.4);
  \draw[thick,dashed] (0.5,0.5)--(0.5,-0.5);
   \draw[thick,dotted] (0.6,0.4) -- (0.9,0) -- (0.6,-0.4);
  \draw[thick,decorate,decoration={snake}] (0.9,0) -- (1.5,0);
  \draw[thick,dotted] (1.8,0.4) -- (1.5,0) -- (1.8,-0.4);
\end{tikzpicture} & $\delta\delta \leftrightarrow \delta\delta$, $\lambda\lambda \leftrightarrow \lambda\lambda$ & singlets $\leftrightarrow$ singlets \\
\hline
     IR-finite Yukawa & \ref{tab:colourlessSMEFTIRfiniteYuk} &  \begin{tikzpicture}[baseline=-0.65ex]
  \draw[thick,dotted] (0.4,0.4) -- (-0.4,-0.4);
  \draw[thick,dotted] (-0.4,0.4) -- (0.4,-0.4);
  \draw[thick,dashed] (0.5,0.5)--(0.5,-0.5);
   \draw[thick,dotted] (0.6,0.4) -- (1.1,0.2) -- (1.7,0.4);
  \draw[thick,dotted] (1.1,0.2) -- (1.1,-0.2);
\draw[thick,dotted] (0.6,-0.4) -- (1.1,-0.2) -- (1.7,-0.4);
\end{tikzpicture} &  $\delta\delta \leftrightarrow \delta\delta$, $\lambda\lambda \leftrightarrow \lambda\lambda$ & mixes irreps \\
\hline
     IR-divergent gauge &\ref{tab:colourlessSMEFTIRdivergent} & \begin{tikzpicture}[baseline=-0.65ex]
  \draw[thick,dotted] (0.4,0.4) -- (-0.4,-0.4);
  \draw[thick,dotted] (-0.4,0.4) -- (0.4,-0.4);
  \draw[thick,dashed] (0.5,0.5)--(0.5,-0.5);
   \draw[thick,dotted] (0.6,0.4) -- (1.1,0.2) -- (1.7,0.4);
  \draw[thick,decorate,decoration={snake}] (1.1,0.2) -- (1.1,-0.2);
\draw[thick,dotted] (0.6,-0.4) -- (1.1,-0.2) -- (1.7,-0.4);
\end{tikzpicture} & $\lambda\lambda \leftrightarrow \delta\delta$ &   blind \\
& & and collinear & & \\
\hline
     IR-divergent Yukawa  & \ref{tab:colourlessSMEFTIRdivergentYuk} & 
   collinear
     & blind & mixes irreps \\
     \hline
    \end{tabular}
    \caption{Different types of contributions to the anomalous dimensions, and how they act. $\delta\delta$ and $\lambda\lambda$ refer to the gauge structure of the operator: $\delta\delta$ means that the operator is constructed out of two currents which are singlets under a gauge group, while $\lambda\lambda$ means that the operator is constructed out of two currents both of which are either $SU(2)_L$ triplets or $SU(3)_c$ octets}
    \label{tab:topologyrules}
\end{table}

In the previous section, we collected all the ingredients for calculating the one-loop anomalous dimensions of the $(4,0)$ operators of the SMEFT. We can now take stock of these ingredients, and the patterns that they impose on the anomalous dimension matrix, determined by the helicity, gauge and flavour properties of the SM and SMEFT amplitudes. There are a very limited set of helicity amplitudes at both dimension 4 and dimension 6, demonstrated by the fact that they can all be written as products of the 3-component $\mathcal{J}$ vector~\eqref{eq:Jvectordef}.
Furthermore, gauge and flavour combines with helicity in ways prescribed by the fields of the SM. For these reasons, we expect to see many repeated patterns throughout the anomalous dimension matrix.

\subsection{Patterns and zeroes}
\label{sec:patternsandzeroes}

We lay out the different contributions to the anomalous dimension matrix in Tabs.~\ref{tab:colourlessSMEFTIRfinitegauge}, \ref{tab:colourlessSMEFTIRfiniteYuk}, \ref{tab:colourlessSMEFTIRdivergent}, and \ref{tab:colourlessSMEFTIRdivergentYuk} where for simplicity here we only include in the tables operators without quarks. All entries in this table agree with \cite{Jenkins:2013wua,Jenkins:2013zja,Alonso:2013hga}, appropriately transformed, and provide the first check (of which we are aware) of the RGEs of this subset of quarkless two- and four-lepton operators. The tables are separated into different types of contributions: IR-finite gauge and Yukawa pieces (which are the only contributions to running between different operators) and IR-divergent gauge and Yukawa pieces (which contribute to self-renormalisation). How each of these pieces behaves schematically in terms of the gauge and flavour properties of the operators is laid out in Tab.~\ref{tab:topologyrules}. The full anomalous dimension is the sum of these pieces.

Within the tables, the phase space factors (in {\color{kinematic} green}), gauge factors (in {\color{gauge} orange}) and flavour factors (in\flv{purple}) have been separated out. The Higgs quartic terms, whose flavour effects are trivial, are shown in Tab.~\ref{tab:higgsquartic}. Zeroes are represented in the tables either by `0' or `\ding{53}'.

For all the current-current operators, the relevant phase space factors can be found in App.~\ref{app:summaryPhaseSpace}. The gauge factors can be found in Tabs.~\ref{tab:colorfactors}, \ref{tab:colorfactorsnonsymmtosymm} and \ref{tab:colorfactorssymmtononsymm}. The flavour factors can be found in Tab.~\ref{tab:flavourfactors}. 
\subsubsection{`\ding{53}' entry}
A `\ding{53}' entry signifies that there is no corresponding diagram that can be drawn at one-loop. Reasons for this are as follows:
\begin{itemize}
    \item One loop diagrams can never involve more than two legs of a SMEFT amplitude. So, operators differing by more than two legs can never renormalise each other.
    E.g.,~four Higgs operators never renormalise four fermion operators, or vice versa.
    \item IR finite Yukawa diagrams leading to a non-zero entry in Tab.~\ref{tab:colourlessSMEFTIRfiniteYuk} must always be drawable via a cut diagram with a SM Yukawa amplitude on one side of the cut (i.e.~diagrams in Eq.~\eqref{eq:Yukfactors}). These SM amplitudes determine which one-loop diagrams can be drawn, for example the fact that there is no SM Yukawa amplitude with four external Higgs legs means that four-Higgs operators cannot self-renormalise in this way. The same is true of operators with four left-handed (or four right-handed) fermions.
    \item On the other hand, collinear Yukawa factors simply act as wavefunction renormalisation on the fields of an operator. So in this case, as seen in Tab.~\ref{tab:colourlessSMEFTIRdivergentYuk}, self-renormalisation of any operator is possible. However this can never change the external field species, or the gauge structure of the operator, so all but the diagonal entries of Tab.~\ref{tab:colourlessSMEFTIRdivergentYuk} are marked with a `\ding{53}'.
    \item Finally, soft and collinear gauge pieces can only cause renormalisation between operators with the same external states (but can change the gauge structure of the operators). This explains the crossed entries in Tab.~\ref{tab:colourlessSMEFTIRdivergent}.
\end{itemize}
\subsubsection{`0' entry\label{sec:zeroentry}}
Zeroes represented by `0' instead arise when a cut diagram can be drawn, but evaluates to zero for other reasons. In the tables, these zeroes are always due to the gauge factors, as explained in Sec.~\ref{sec:gauge}. In Tabs.~\ref{tab:colourlessSMEFTIRfinitegauge} and~\ref{tab:colourlessSMEFTIRfiniteYuk}, all `0' entries are ultimately due to the tracelessness of the Pauli $\sigma$ matrices. For the same reason, many of the non-zero entries are proportional only to $g_1^2$ and do not have a $g_2^2$ part.
In Tab.~\ref{tab:colourlessSMEFTIRdivergent}, the `0' entries on the diagonal can be understood as cancellations between the soft and collinear diagrams, or equivalently as the non-renormalisation of number currents. The off-diagonal `0' entries are again due to the tracelessness of the Pauli matrices, or the exchange symmetry of the (anti)symmetric gauge factors in the case of indistinguishable currents.

When dealing with indistinguishable currents, the power of treating operators in a factorised basis of (anti)symmetric gauge and flavour tensors becomes apparent. In the Warsaw basis, the flavour and gauge parts of the $\op_{QQ}^{(1)}$ and $\op_{QQ}^{(3)}$ operators appear to mix under (naively flavour-blind) soft gauge diagrams; for example, $C^{QQ(1)}_{1123}$ can be generated by $C^{QQ(3)}_{1321}$. However, the running is entirely diagonal in terms of the (anti)symmetrised $\op_{QQ++}$, $\op_{QQ+-}$, $\op_{QQ-+}$, and $\op_{QQ--}$ operators defined here. This is because the unperturbed (anti)symmetric flavour tensor will always preserve the corresponding symmetry of the gauge piece.

\subsubsection{Non-zero entry}
As for the non-zero entries, it is clear that there are many repeated factors in the tables. This is due to the same gauge and kinematic factors arising in many different diagrams. We see in particular in Tab.~\ref{tab:colourlessSMEFTIRfinitegauge} that entries proportional to $g_2^2$ are rather sparse, and only arise between operators containing structures of the `$\sigma \sigma$' form. There is further structure to be found in the flavour factors, which leads not only to zeros but even to block diagonalisation, to be discussed in the next section.

\begin{table}[]
 \centering
    \resizebox{\columnwidth}{!}{%
  \begin{tabular}{| p{12mm} || p{20mm}| p{20mm} || p{25mm} | p{30mm} | p{20mm} || p{35mm} | p{35mm} | p{25mm} | p{30mm} | }
  \hline
  & $HD+$ & $HD-$ & $HL(1)$ & $HL(3)$ & $He$ & $LL+$ & $LL-$ & $Le$ & $ee$ \\ \hline \hline
$HD+$
  & $\phs{2}\cdot\gge{\frac32}\cdot\flv{g_{1HH}^2}$ + $\phs{2}\cdot\gge{\frac18}\cdot\flv{g_2^2}$ & $\phs{\frac23}\cdot\gge{\frac12}\cdot\flv{g_{1HH}^2}$ + $\phs{\frac23}\cdot\gge{-\frac18}\cdot\flv{g_2^2}$
  & $\phs{\frac43}\cdot\gge{2}\cdot\flv{g_{1HL}^2\Tr[c]}$ & $\phs{\frac43}\cdot\gge{\frac12}\cdot\flv{g_2^2\Tr[c]}$ & $\phs{\frac43}\cdot\gge{1}\cdot\flv{g_{1He}^2\Tr[c]}$
  & \ding{53} & \ding{53} & \ding{53} & \ding{53} \\ \hline
$HD-$
  & $\phs{2}\cdot\gge{\frac32}\cdot\flv{g_{1HH}^2}$ + $\phs{2}\cdot\gge{-\frac38}\cdot\flv{g_2^2}$ & $\phs{\frac23}\cdot\gge{\frac12}\cdot\flv{g_{1HH}^2}$ + $\phs{\frac23}\cdot\gge{\frac38}\cdot\flv{g_2^2}$
  & $\phs{\frac43}\cdot\gge{2}\cdot\flv{g_{1HL}^2\Tr[c]}$ & $\phs{\frac43}\cdot\gge{-\frac32}\cdot\flv{g_2^2\Tr[c]}$ & $\phs{\frac43}\cdot\gge{1}\cdot\flv{g_{1He}^2\Tr[c]}$
  & \ding{53} & \ding{53} & \ding{53} & \ding{53} \\ \hline \hline
$HL(1)$
  & $\phs{2}\cdot\gge{\frac32}\cdot\flv{g_{1HL}^2\delta^p_q}$ & $\phs{\frac23}\cdot\gge{\frac12}\cdot\flv{g_{1HL}^2\delta^p_q}$
  & $\phs{\frac43}\cdot\gge{2}\cdot\flv{g_{1LL}^2 \Tr[c] \delta^p_q}$ + $\phs{\frac23}\cdot\gge{2}\cdot\flv{g_{1HH}^2 c^p_q}$ & 0 & $\phs{\frac43}\cdot\gge{1}\cdot\flv{g_{1Le}^2 \Tr[c] \delta^p_q}$
  & $\phs{\frac{16}{3}}\cdot\gge{\frac32}\cdot\flv{g_{1HL}^2 c^{(pv)}_{(qv)}}$ & $\phs{\frac{16}{3}}\cdot\gge{\frac12}\cdot\flv{g_{1HL}^2 c^{[pv]}_{[qv]}}$ & $\phs{\frac43}\cdot\gge{1}\cdot \flv{g_{1He}^2 c^{pv}_{qv}}$ & \ding{53} \\ \hline
$HL(3)$
  & $\phs{2}\cdot\gge{\frac18}\cdot\flv{g_2^2\delta^p_q}$ & $\phs{\frac23}\cdot\gge{-\frac18}\cdot\flv{g_2^2\delta^p_q}$
  & 0 & $\phs{\frac43}\cdot\gge{\frac12}\cdot\flv{g_2^2 \Tr[c] \delta^p_q}$+ $\phs{\frac23}\cdot\gge{\frac12}\cdot\flv{g_2^2 c^p_q}$ & 0
  & $\phs{\frac{16}{3}}\cdot\gge{\frac18}\cdot\flv{g_2^2 c^{(pv)}_{(qv)}}$ & $\phs{\frac{16}{3}}\cdot\gge{-\frac18}\cdot\flv{g_2^2 c^{[pv]}_{[qv]}}$ & 0 & \ding{53} \\ \hline
$He$
  & $\phs{2}\cdot\gge{\frac32}\cdot\flv{g_{1He}^2\delta^p_q}$ & $\phs{\frac23}\cdot\gge{\frac12}\cdot\flv{g_{1He}^2\delta^p_q}$
  & $\phs{\frac43}\cdot\gge{2}\cdot\flv{g_{1Le}^2\Tr[c] \delta^p_q}$ & 0 & $\phs{\frac43}\cdot\gge{1}\cdot\flv{g_{1ee}^2\Tr[c] \delta^p_q}$ + $\phs{\frac23}\cdot\gge{2}\cdot\flv{g_{1HH}^2 c^p_q}$
  & \ding{53} & \ding{53} & $\phs{\frac43}\cdot\gge{2}\cdot \flv{g_{1HL}^2 c^{vp}_{vq}}$ & $\phs{\frac{16}{3}}\cdot\gge{1}\cdot\flv{g_{1He}^2 c^{(pv)}_{(qv)}}$ \\ \hline \hline
$LL+$
  & \ding{53} & \ding{53}
  & $\phs{\frac23}\cdot\gge{2}\cdot\flv{g_{1HL}^2(\delta c)}$ & $\phs{\frac23}\cdot\gge{\frac12}\cdot\flv{g_2^2(\delta c)}$ & \ding{53}
  & $\phs{\frac{16}{3}}\cdot\gge{\frac32}\cdot\flv{g_{1LL}^2 c^{(v(p)}_{(v(q)} \delta^{r)}_{s)}}$ + $\phs{\frac{16}{3}}\cdot\gge{\frac18}\cdot\flv{g_2^2 c^{(v(p)}_{(v(q)} \delta^{r)}_{s)}}$ & $\phs{\frac{16}{3}}\cdot\gge{\frac12}\cdot\flv{g_{1LL}^2 c^{[v(p]}_{[v(q]} \delta^{r)}_{s)}}$ + $\phs{\frac{16}{3}}\cdot\gge{-\frac18}\cdot\flv{g_2^2 c^{[v(p]}_{[v(q]} \delta^{r)}_{s)}}$ & $\phs{\frac43}\cdot\gge{1}\cdot\flv{g^2_{1Le} \delta^{(r}_{(s} c^{p)v}_{q)v}} $ & \ding{53} \\ \hline
$LL-$
  & \ding{53} & \ding{53}
  & $\phs{\frac23}\cdot\gge{2}\cdot\flv{g_{1HL}^2[\delta c]}$ & $\phs{\frac23}\cdot\gge{-\frac32}\cdot\flv{g_2^2[\delta c]}$ & \ding{53}
  & $\phs{\frac{16}{3}}\cdot\gge{\frac32}\cdot\flv{g_{1LL}^2 c^{(v[p)}_{(v[q)} \delta^{r]}_{s]}}$ + $\phs{\frac{16}{3}}\cdot\gge{-\frac38}\cdot\flv{g_2^2 c^{(v[p)}_{(v[q)} \delta^{r]}_{s]}}$ & $\phs{\frac{16}{3}}\cdot\gge{\frac12}\cdot\flv{g_{1LL}^2 c^{[v[p]}_{[v[q]} \delta^{r]}_{s]}}$ + $\phs{\frac{16}{3}}\cdot\gge{\frac38}\cdot\flv{g_2^2 c^{[v[p]}_{[v[q]} \delta^{r]}_{s]}}$ & $\phs{\frac43}\cdot\gge{1}\cdot\flv{g^2_{1Le} \delta^{[r}_{[s} c^{p]v}_{q]v}} $ & \ding{53} \\ \hline
$Le$
  & \ding{53} & \ding{53}
  & $ \phs{\frac23}\cdot\gge{2}\cdot\flv{g_{1He}^2 \delta c}$ & 0 & $ \phs{\frac23}\cdot\gge{2}\cdot\flv{g_{1HL}^2 \delta c}$
  & $\phs{\frac{16}{3}}\cdot\gge{\frac32}\cdot\flv{g_{1Le}^2 c^{(vp)}_{(vq)} \delta^{r}_{s}}$ & $\phs{\frac{16}{3}}\cdot\gge{\frac12}\cdot\flv{g_{1Le}^2 c^{[vp]}_{[vq]} \delta^{r}_{s}}$ & $\phs{\frac43}\cdot\gge{2}\cdot\flv{g^2_{1LL} c^{vr}_{vs} \delta^p_q}$ + $\phs{\frac43}\cdot\gge{1}\cdot\flv{g^2_{1ee} c^{pv}_{qv} \delta^r_s}$ & $\phs{\frac{16}{3}}\cdot\gge{1}\cdot\flv{g_{1Le}^2 c^{(vr)}_{(vs)} \delta^{p}_{q}}$ \\ \hline
$ee$
  & \ding{53} & \ding{53}
  & \ding{53} & \ding{53} & $\phs{\frac23}\cdot\gge{2}\cdot\flv{g_{1He}^2(\delta c)}$
  & \ding{53} & \ding{53} & $\phs{\frac43}\cdot\gge{2}\cdot\flv{g^2_{1Le} c^{v(p}_{v(q} \delta^{r)}_{s)}}$ & $\phs{\frac{16}{3}}\cdot\gge{1}\cdot\flv{g_{1ee}^2 c^{(v(p)}_{(v(q)} \delta^{r)}_{s)}}$ \\ \hline
  \end{tabular}}
  \caption{IR finite gauge pieces of the $(4,0)$ block of the colourless SMEFT anomalous dimension matrix. $g_{1XY}^2 = g_1^2 y_X y_Y$ includes the product of the hypercharges of the two species. Operators are defined in \cref{tab:basis}. Phase space factors,  in\phs{green}, can be found in App.~\ref{app:IRfinitePhase} and are explained in Sec.~\ref{sec:IRfinitephase}. Gauge factors, in\gge{orange}, can be found in Tabs.~\ref{tab:colorfactors}, \ref{tab:colorfactorsnonsymmtosymm} and \ref{tab:colorfactorssymmtononsymm}, and are explained in Sec.~\ref{sec:gauge}. The flavour factors, in\flv{purple}, labelled by $p,q,r,s$ indices, can be found in Tab.~\ref{tab:flavourfactors}, and are explained in Sec.~\ref{sec:flavour}.  \label{tab:colourlessSMEFTIRfinitegauge}}
\end{table}

\begin{table}[]
 \centering
    \resizebox{\columnwidth}{!}{%
\begin{tabular}{|p{10mm} || p{10mm} |p{10mm} || p{23mm} | p{23mm} | p{23mm} || p{30mm} | p{30mm} | p{33mm}|  p{28mm} |}
\hline
& $HD+$ & $HD-$ & $HL(1)$ & $HL(3)$ & $He$ & $LL+$ & $LL-$ & $Le$ & $ee$ \\ \hline \hline
$HD+$
& \ding{53} & \ding{53} & $\phs{2}\cdot\gge{1}\cdot \flv{-\Tr[cM_l]}$ & $\phs{2}\cdot\gge{1}\cdot \flv{-\Tr[cM_l]}$ & $\phs{2}\cdot\gge{1}\cdot\flv{\Tr[c M_e]}$ & \ding{53} & \ding{53} & \ding{53} & \ding{53} \\ \hline
$HD-$
& \ding{53} & \ding{53} & $\phs{2}\cdot\gge{1}\cdot \flv{-\Tr[cM_l]}$ & $\phs{2}\cdot\gge{-3}\cdot \flv{-\Tr[cM_l]}$ & $\phs{2}\cdot\gge{1}\cdot\flv{\Tr[cM_e]}$ & \ding{53} & \ding{53} & \ding{53} & \ding{53} \\ \hline \hline
$HL(1)$
& $\phs{3}\cdot\gge{\frac34}\cdot \flv{- M_l}$ & $\phs{1}\cdot\gge{\frac14}\cdot \flv{- M_l}$
& $(\phs{1}\cdot\gge{\frac12}+\phs{2}\cdot\gge{\frac12})\cdot(\flv{cM_l}+\flv{M_lc})$ & $(\phs{1}\cdot\gge{\frac32}+\phs{2}\cdot\gge{\frac32})\cdot(\flv{cM_l}+\flv{M_lc})$ & $\phs{1}\cdot\gge{1}\cdot\flv{-Y_e c Y_e^\dagger}$
& $\phs{8}\cdot\gge{\frac34}\cdot \flv{- c^{(pv)}_{(qw)} [M_l]^w_v}$ & $\phs{8}\cdot\gge{\frac14}\cdot \flv{- c^{[pv]}_{[qw]} [M_l]^w_v}$ & $\phs{2}\cdot\gge{1}\cdot \flv{ c^{pv}_{qw} [M_e]^w_v}$ & \ding{53} \\ \hline
$HL(3)$
& $\phs{3}\cdot\gge{\frac14}\cdot \flv{- M_l}$ & $\phs{1}\cdot\gge{-\frac14}\cdot \flv{-M_l}$
& $(\phs{1}\cdot\gge{\frac12}+\phs{2}\cdot\gge{\frac12})\cdot(\flv{cM_l}+\flv{M_lc})$ & $(\phs{1}\cdot\gge{\frac32}+\phs{2}\cdot\gge{-\frac12})\cdot(\flv{cM_l}+\flv{M_lc})$ & 0
& $\phs{8}\cdot\gge{\frac14}\cdot \flv{- c^{(pv)}_{(qw)} [M_l]^w_v}$ & $\phs{8}\cdot\gge{-\frac14}\cdot \flv{- c^{[pv]}_{[qw]} [M_l]^w_v}$  & 0 & \ding{53} \\ \hline
$He$
& $\phs{3}\cdot\gge{\frac32}\cdot \flv{M_e}$ & $\phs{1}\cdot\gge{\frac12}\cdot \flv{ M_e}$
& $\phs{1}\cdot\gge{2}\cdot\flv{- Y_e^\dagger c Y_e}$ & 0 & $(\phs{1}\cdot\gge{1}+\phs{2}\cdot\gge{1})\cdot (\flv{c M_e} + \flv{M_e c})$
& \ding{53} & \ding{53} & $\phs{2}\cdot\gge{1}\cdot \flv{- c^{vp}_{wq} [M_l]^w_v}$ & $\phs{8}\cdot\gge{1}\cdot \flv{c^{(pv)}_{(qw)} [M_e]^w_v}$ \\ \hline \hline
$LL+$
& \ding{53} & \ding{53} & $\phs{1}\cdot\gge{1}\cdot \flv{-(c M_l)}$ & $\phs{1}\cdot\gge{1}\cdot \flv{-(c M_l)}$ & 0 & \ding{53} & \ding{53} & $\phs{1}\cdot\gge{1}\cdot \flv{- [Y_e^\dagger]^w_{(s} [Y_e]^{(r}_v c^{p)v}_{q)w}} $ & \ding{53} \\ \hline
$LL-$
& \ding{53} & \ding{53} & $\phs{1}\cdot\gge{1}\cdot \flv{-[c M_l]}$ & $\phs{1}\cdot\gge{-3}\cdot \flv{-[c M_l]}$ & 0 & \ding{53} & \ding{53} & $\phs{1}\cdot\gge{1}\cdot \flv{-[Y_e^\dagger]^w_{[s} [Y_e]^{[r}_v c^{p]v}_{q]w}}$ & \ding{53} \\ \hline
$Le$
& \ding{53} & \ding{53} & $\phs{1}\cdot\gge{2}\cdot \flv{c M_e}$ & 0 & $\phs{1}\cdot\gge{1}\cdot \flv{- c M_l}$
& $\phs{4}\cdot\gge{\frac32}\cdot \flv{ - c^{(pv)}_{(qw)} [Y_e^\dagger]^r_v [Y_e]^w_s}$ & $\phs{4}\cdot\gge{\frac12}\cdot \flv{- c^{[pv]}_{[qw]} [Y_e^\dagger]^r_v [Y_e]^w_s}$ & $\phs{-2}\cdot\gge{1}\cdot\flv{-c^{vr}_{qw} [Y_e^\dagger]^w_v [Y_e]^p_s}\phs{-2}\cdot\gge{1}\cdot\flv{-c^{pw}_{vs} [Y_e^\dagger]^r_q [Y_e]^v_w} \phs{-1}\cdot\gge{1}\cdot\flv{-c^{vw}_{qs} [Y_e^\dagger]^r_v [Y_e]^p_w} \phs{-1}\cdot\gge{1}\cdot\flv{-c^{pr}_{vw} [Y_e^\dagger]^w_q [Y_e]^v_s}$
 & $\phs{4}\cdot\gge{1}\cdot \flv{- c^{(rv)}_{(sw)} [Y_e^\dagger]^w_q [Y_e]^p_v}$ \\ \hline
$ee$
& \ding{53} & \ding{53} & \ding{53} & \ding{53} & $\phs{1}\cdot\gge{2}\cdot \flv{(c M_e)}$ & \ding{53} & \ding{53} & $\phs{1}\cdot\gge{2}\cdot \flv{- c^{v(p}_{w(q}  [Y_e]^w_{s)} [Y_e^\dagger]^{r)}_v}$ & \ding{53} \\\hline
\end{tabular}}
\caption{IR finite Yukawa pieces of the $(4,0)$ block of the colourless SMEFT anomalous dimension matrix. We define $M_l = Y_e Y_e^\dagger,M_e= Y_e^\dagger Y_e,$. Operators are defined in \cref{tab:basis}. Phase space factors, in\phs{green}, can be found in App.~\ref{app:IRfinitePhase} and are explained in Sec.~\ref{sec:IRfinitephase}. Gauge factors, in\gge{orange}, can be found in Tabs.~\ref{tab:colorfactors}, \ref{tab:colorfactorsnonsymmtosymm} and \ref{tab:colorfactorssymmtononsymm}, and are explained in Sec.~\ref{sec:gauge}. The flavour factors, in\flv{purple}, labelled by $p,q,r,s$ indices, can be found in Tab.~\ref{tab:flavourfactors}, and are explained in Sec.~\ref{sec:flavour}. \label{tab:colourlessSMEFTIRfiniteYuk}}
\end{table}

\begin{table}[]
 \centering
    \resizebox{\columnwidth}{!}{%
\begin{tabular}{|p{12mm} || p{33mm}| p{33mm} || p{10mm} | p{23mm}| p{5mm} || p{30mm}| p{30mm}| p{28mm}| p{33mm} |}
\hline
   &$HD+$ & $HD-$ & $HL(1)$ & $HL(3)$ & $He$ & $LL+$ & $LL-$ & $Le$ & $ee$ \\ \hline \hline
$HD+$
 & $(\phs{26}+\mathbf{2}\cdot\phs{-2}+\mathbf{4}\cdot\phs{-4})\cdot \flv{g_{1HH}^2 c}$ & $\phs{-6}\cdot\gge{-\frac14}\cdot \flv{g_2^2 c}$ & \ding{53} & \ding{53} & \ding{53} & \ding{53} & \ding{53} & \ding{53} & \ding{53} \\ \hline
$HD-$
 & $\phs{-18}\cdot\gge{-\frac34}\cdot\flv{g_2^2 c}$ & $(\phs{14}+\mathbf{2}\cdot\phs{-8}+\mathbf{4}\cdot\phs{-4})\cdot\flv{g_{1HH}^2 c}$ & \ding{53} & \ding{53} & \ding{53} & \ding{53} & \ding{53} & \ding{53} & \ding{53}  \\ \hline \hline
$HL(1)$
 & \ding{53} & \ding{53} & 0 & 0 & \ding{53} & \ding{53} & \ding{53} & \ding{53} & \ding{53}  \\ \hline
$HL(3)$
 & \ding{53} & \ding{53} & 0 & $(\phs{8}\cdot\gge{-\frac14}+\phs{6}\cdot\gge{-\frac14}+\mathbf{2}\cdot\phs{-4}\cdot\gge{-\frac12}+\mathbf{2}\cdot\phs{4}\cdot\gge{\frac12}+\phs{-8}\cdot\gge{\frac34}+\phs{-6}\cdot\gge{\frac34})\cdot\flv{g_2^2 c^p_q}$ & \ding{53} & \ding{53} & \ding{53} & \ding{53} & \ding{53}  \\ \hline
$He$
 & \ding{53} & \ding{53} & \ding{53} & \ding{53} & 0 & \ding{53} & \ding{53} & \ding{53} & \ding{53}  \\ \hline \hline
$LL+$
 & \ding{53} & \ding{53} & \ding{53} & \ding{53} & \ding{53} & $(\phs{24}+\mathbf{4}\cdot\phs{-3})\cdot\flv{g^2_{1LL} c^{pr}_{qs}}+(\phs{24}\cdot\gge{\frac12}+\mathbf{4}\cdot\phs{-3}\cdot\gge{\frac34})\cdot\flv{g_2^2 c^{pr}_{qs}}$ & 0 & \ding{53} & \ding{53}  \\ \hline
$LL-$
 & \ding{53} & \ding{53} & \ding{53} & \ding{53} & \ding{53} & 0 & $(\phs{24}+\mathbf{4}\cdot\phs{-3})\cdot\flv{g^2_{1LL} c^{pr}_{qs}}+\mathbf{4}\cdot\phs{-3}\cdot\gge{\frac34}\cdot \flv{g_2^2 c^{pr}_{qs}}$ & \ding{53} & \ding{53}  \\ \hline
$Le$
 & \ding{53} & \ding{53} & \ding{53} & \ding{53} & \ding{53} & \ding{53} & \ding{53} & $\mathbf{2}\cdot\phs{-6}\cdot\flv{g^2_{1Le} c^{pr}_{qs}}$ & \ding{53} \\ \hline
$ee$
 & \ding{53} & \ding{53} & \ding{53} & \ding{53} & \ding{53} & \ding{53} & \ding{53} & \ding{53} & $(\phs{24}+\mathbf{4}\cdot\phs{-3})\cdot\flv{g^2_{1ee} c^{pr}_{qs}}$ \\ \hline
  \end{tabular}}
  \caption{Soft plus collinear gauge pieces. Operators are defined in \cref{tab:basis}. Phase space factors, in\phs{green}, can be found in Apps.~\ref{app:IRdiv} and~\ref{app:collinear} and are explained in Secs.~\ref{sec:softphase} and~\ref{sec:collinear}. Gauge factors, in\gge{orange}, can be found in Tabs.~\ref{tab:colorfactors}, \ref{tab:colorfactorsnonsymmtosymm} and \ref{tab:colorfactorssymmtononsymm}, and are explained in Sec.~\ref{sec:gauge}. The flavour factors, in\flv{purple}, labelled by $p,q,r,s$ indices, can be found in Tab.~\ref{tab:flavourfactors}, and are explained in Sec.~\ref{sec:flavour}. {\bf Bold} numerical factors are combinatorial: they correspond to several equivalent diagrams for the same process.
  \label{tab:colourlessSMEFTIRdivergent}}
\end{table}

\begin{table}[]
 \centering
    \resizebox{\columnwidth}{!}{%
\begin{tabular}{|p{12mm} || p{25mm}| p{25mm} || p{30mm} | p{30mm} | p{30mm} || p{40mm} | p{40mm}| p{40mm}| p{40mm} |}
\hline
   &$HD+$ & $HD-$ & $HL(1)$ & $HL(3)$ & $He$ & $LL+$ & $LL-$ & $Le$ & $ee$ \\ \hline \hline
$HD+$
 & ${\mathbf 4}\cdot\phs{1}\cdot\gge{N_c}\cdot \flv{\text{Tr} [M_u+M_d] c}+{\mathbf 4}\cdot\phs{1}\cdot\gge{1}\cdot \flv{\text{Tr} [M_e] c}$ & \ding{53} & \ding{53} & \ding{53} & \ding{53} & \ding{53} & \ding{53} & \ding{53} & \ding{53} \\ \hline
$HD-$
 & \ding{53} & ${\mathbf 4}\cdot\phs{1}\cdot\gge{N_c}\cdot \flv{\text{Tr} [M_u+M_d] c}+{\mathbf 4}\cdot\phs{1}\cdot\gge{1}\cdot \flv{\text{Tr} [M_e] c}$ & \ding{53} & \ding{53} & \ding{53} & \ding{53} & \ding{53} & \ding{53} & \ding{53}  \\ \hline \hline
$HL(1)$
 & \ding{53} & \ding{53} & ${\mathbf 2}\cdot\phs{1}\cdot\gge{N_c}\cdot \flv{\text{Tr} [M_u+M_d] c^p_q}+{\mathbf 2}\cdot\phs{1}\cdot\gge{1}\cdot \flv{\text{Tr} [M_e] c^p_q}+\phs{\frac12} \cdot\gge{1}\cdot \flv{ (M_l c + c M_l)}$ & \ding{53} & \ding{53} & \ding{53} & \ding{53} & \ding{53} & \ding{53}  \\ \hline
$HL(3)$
 & \ding{53} & \ding{53} & \ding{53} & ${\mathbf 2}\cdot\phs{1}\cdot\gge{N_c}\cdot \flv{\text{Tr} [M_u+M_d] c^p_q}+{\mathbf 2}\cdot\phs{1}\cdot\gge{1}\cdot \flv{\text{Tr} [M_e] c^p_q}+\phs{\frac12} \cdot\gge{1}\cdot \flv{ (M_l c + c M_l)}$ & \ding{53} & \ding{53} & \ding{53} & \ding{53} & \ding{53}  \\ \hline
$He$
 & \ding{53} & \ding{53} & \ding{53} & \ding{53} & ${\mathbf 2}\cdot\phs{1}\cdot\gge{N_c}\cdot \flv{\text{Tr} [M_u+M_d] c^p_q}+\cdot\phs{1}\cdot\gge{1}\cdot \flv{\text{Tr} [M_e] c^p_q}+\phs{\frac12} \cdot\gge{2}\cdot \flv{ (M_e c + c M_e)}$ & \ding{53} & \ding{53} & \ding{53} & \ding{53}  \\ \hline \hline
$LL+$
 & \ding{53} & \ding{53} & \ding{53} & \ding{53} & \ding{53} & ${\mathbf 2}\cdot\phs{\frac12} \cdot\gge{2}\cdot \flv{ ( c^{pr}_{v(s} [M_l]^v_{q)} +  c^{w(r}_{qs} [M_l]^{p)}_w )}$ & \ding{53} & \ding{53} & \ding{53}  \\ \hline
$LL-$
 & \ding{53} & \ding{53} & \ding{53} & \ding{53} & \ding{53} & \ding{53} & ${\mathbf 2}\cdot\phs{\frac12} \cdot\gge{2}\cdot \flv{ ( c^{pr}_{v[s} [M_l]^v_{q]} +  c^{w[r}_{qs} [M_l]^{p]}_w )}$ & \ding{53} & \ding{53}  \\ \hline
$Le$
 & \ding{53} & \ding{53} & \ding{53} & \ding{53} & \ding{53} & \ding{53} & \ding{53} & $\phs{\frac12} \cdot\gge{1}\cdot \flv{ ([M_l]^v_q c^{pr}_{vs} + [M_l]^p_w c^{wr}_{qs})}+\phs{\frac12} \cdot\gge{2}\cdot \flv{ ([M_e]^v_s c^{pr}_{qv} + [M_e]^r_w c^{pw}_{qs})}$ & \ding{53} \\ \hline
$ee$
 & \ding{53} & \ding{53} & \ding{53} & \ding{53} & \ding{53} & \ding{53} & \ding{53} & \ding{53} & ${\mathbf 2}\cdot\phs{\frac12} \cdot\gge{2}\cdot \flv{ ( c^{pr}_{v(s} [M_e]^v_{q)} \!+\!  c^{w(r}_{qs} [M_e]^{p)}_w )}$ \\ \hline
  \end{tabular}}
  \caption{Collinear Yukawa pieces. $M_{r=u/d/e} = Y^\dagger_r Y_r, M_l = Y_e Y^\dagger_e$. Operators are defined in \cref{tab:basis}. Phase space factors, given in\phs{green}, can be found in App.~\ref{app:collinear} and are explained in Secs.~\ref{sec:collinear}. Gauge factors, given in\gge{orange}, can be found in Tabs.~\ref{tab:colorfactors}, \ref{tab:colorfactorsnonsymmtosymm} and \ref{tab:colorfactorssymmtononsymm}, and are explained in Sec.~\ref{sec:gauge}. The flavour factors, in\flv{purple}, labelled by $p,q,r,s$ indices, can be found in Tab.~\ref{tab:flavourfactors}, and are explained in Sec.~\ref{sec:flavour}. {\bf Bold} numerical factors are combinatorial: they correspond to several equivalent diagrams, e.g.~two or four external legs which can be renormalised in the same way by wavefunction renormalisation.
  \label{tab:colourlessSMEFTIRdivergentYuk}}
\end{table}

\newcommand{\xmark}{\text{\ding{53}}}

\subsection{The very flavourful anomalous dimension matrix}
\label{sec:sparseg3}
The strong coupling $g_3$ is the largest parameter in the anomalous dimension matrix. This might lead one to guess that focussing only on $g_3$-induced running would give a good approximation. However, $g_3$ appears very sparsely in the anomalous dimension matrix, a fact which can be easily understood via the gauge factors laid out in Sec.~\ref{sec:gauge}.

The SM amplitude that gives $g_3$ factors in the anomalous dimensions of the $(4,0)$ operators is of the form
\begin{align}
\mathord{\begin{tikzpicture}[baseline=-0.65ex]
   \draw[thick] (-1,1) -- (-\currsep,0) -- (-1,-1);
  \currentmarker{-\currsep,0}
  \draw[thick] (1,1) -- (\currsep,0) -- (1,-1);
  \currentmarker{\currsep,0}
  \node at (-1,0) {$\delta^p_q [\lambda^A]^a_b$};
  \node at (1,0) {$\delta^r_s [\lambda^A]^c_d$};
\end{tikzpicture}},
\end{align}
where $a,b$ are colour indices and $p,q$ are flavour indices.
In order to produce a $g_3$ contribution to operator mixing, \emph{both} operators involved in the mixing must be constructed from colour-octet currents (i.e.~have a $\lambda\lambda$ structure), and at least one of the currents must furthermore be a flavour singlet. Otherwise, in either colour or flavour space we get a zero by Eq.~\eqref{eq:zerolambdatrace}. The only operators that can mix in this way are the appropriate flavour components of $\{O_{Qu}^{(8)}, O_{Qd}^{(8)}, O_{ud}^{(8)}, O_{uu}, O_{dd},O_{QQ}^{(1,3)}\}$. Importantly, these gauge mediated diagrams are flavour\emph{ful}, as they treat flavour universal and non-universal pieces differently.

This gauge diagram can also mediate self-renormalisation of particular operators, which is truly flavourless. This occurs when both currents of the SM amplitude are involved in the cut. By looking at Eq.~\eqref{eq:deltagaugeIRdiv}, it would seem that colour-singlet currents (from $\delta\delta$-type operators) could be renormalised in this way. However, since this is renormalisation of a number current, this diagram in fact cancels against the wavefunction renormalisation of the two cut legs, and the overall result is zero. Again, we find that only $\lambda\lambda$-type operators can be self-renormalised by strong interactions, via the diagram in Eq.~\eqref{eq:lambdagaugeIRdiv}.

Finally, there is the situation shown in Eq.~\eqref{eq:doublecutIRdivgauge}, which can allow not only self-renormalisation of a 4-quark $\lambda\lambda$-type operator, but also allows it to mix into the equivalent $\delta\delta$-type operator, e.g. $O_{ud}^{(8)}$ into $O_{ud}^{(1)}$. The flavour structure of the gauge diagram ensures that the flavour indices are completely unchanged between the two operators. 

Overall, it is clear that $g_3$ can only appear in the anomalous dimensions of 4-quark operators, and even then only in specific flavour directions. Operators with distinguishable quark currents and a $\delta\delta$ colour structure can only be run \emph{into}, by operators with identical flavour structures but with $\lambda\lambda$ colour structure. Operators with only one quark current are not renormalized by strong interactions.

Similar arguments follow for $g_2$; we can see in Tab.~\ref{tab:colourlessSMEFTIRfinitegauge} that it only appears in the anomalous dimensions of operators with a $\sigma\sigma$ structure, or operators with identical $SU(2)_L$-charged currents which contain some $\sigma\sigma$ parts in their (anti)symmetrised forms.

By contrast, Yukawa interactions have a less restrictive structure, in terms of both gauge and flavour, allowing them to enter into the anomalous dimensions of a wider range of operators. Many of the entries of the Yukawa matrices are small, of course, but the top Yukawa $y_t$ is of the same order of magnitude as $g_3$, as seen by the comparisons:
\begin{align}
    &\alpha_t(m_Z) \approx 0.08, ~\qquad\alpha_s(m_Z) \approx 0.12, \nn\\
    &\alpha_t(10\, \text{TeV}) \approx 0.05, \quad \alpha_s(10\, \text{TeV}) \approx 0.07,
\end{align}
where $\alpha_t=y_t^2/(4\pi)$. The top Yukawa can enter into the anomalous dimension of any operator containing $Q_3$, $u_3$ and/or $H$. The operator does not have to be flavour-conserving. Moreover, the fact that two different types of gauge structures appear in each of Eqs.~\eqref{eq:gaugeLLHHSM}, \eqref{eq:gaugeQQHHSM} and \eqref{eq:gaugeuuQQSM} means that there are fewer possibilities of zeroes arising. 

In summary, we see that in the running of the SMEFT, the top Yukawa plays a much more widespread role than the strong (or weak) interactions. This makes the anomalous dimension matrix an inherently flavourful object, that cannot be understood without tackling its flavour-breaking nature. In the next section, we will show how the flavour structure of the SM interactions can be used to vastly simplify the problem, and to block-diagonalise the matrix.

\section{Flavour selection rules and block-diagonalisation}
\label{sec:blockdiag}

As discussed in Sec.~\ref{sec:helicitynonrenormalization}, some zeroes in the SMEFT anomalous dimension matrix can be derived simply from the helicity structure of the amplitudes~\cite{Cheung:2015aba}. These helicity arguments, by definition, cannot predict any structure within the large $(4,0)$ operator block that we focus on. But having set up and understood the gauge and flavour parts of the amplitudes and the anomalous dimensions, we are now in a position to ask if there are any additional selection rules or preserved quantum numbers that can be extracted from these pieces. 

On the gauge side, the tracelessness of the Gell-Mann and Pauli matrices ensures that zeroes arise in the mixing of operators with different gauge structures, i.e.~between $\delta\delta$-type operators and $\lambda\lambda$-type operators, as discussed in Sec.~\ref{sec:patternsandzeroes}. For example, of the $\psi \bar \psi \phi^2 D$ type operators, the triplet current operators $\mathcal{O}_{HL}^{(3)}$ and $\mathcal{O}_{HQ}^{(3)}$ are renormalised by only the triplet combination of the four-Higgs operators $(H^\dagger i \overleftrightarrow{D}^\mu \sigma^I H)^2\propto O_{H\Box}$, while the singlet-current operators $\mathcal{O}_{He}, \mathcal{O}_{Hd},\mathcal{O}_{Hu}, \mathcal{O}_{HL}^{(1)}, \mathcal{O}_{HQ}^{(1)}$ are renormalised by the singlet-current combination $(H^\dagger i \overleftrightarrow{D}^\mu H)^2$. This was previously pointed out in Ref.~\cite{Jiang:2020mhe}. However, some operators with different gauge structures but with the same field content \emph{can} mix, via the diagram in Eq.~\eqref{eq:doublecutIRdivgauge} (i.e.~the four-Higgs operators can mix into each other), so this $\delta\delta$ vs $\lambda\lambda$ gauge structure property is not preserved overall by the SMEFT RG. It therefore cannot be used to block-diagonalise the anomalous dimension matrix.

On the other hand, the full flavour $SU(3)^5$ symmetry group and all its associated quantum numbers (see Sec.~\ref{sec:flavourStructure}) are preserved by gauge running, but not by Yukawa running. However, as we can gather from the arguments of Sec.~\ref{sec:flavour}, not all the flavour quantum numbers are broken, and if we neglect small entries of the Yukawa matrices, many are conserved. This allows us to identify and categorise blocks of the full anomalous dimension matrix which --- to a high degree of approximation --- do not renormalise each other. 

\subsection{Flavour non-renormalisation theorems}
\label{sec:flavourselectionrules}

Let us consider the flavour quantum numbers of \cref{sec:flavourStructure} that are preserved by renormalisation when we progressively turn on the interactions of the Standard Model residing in the anomalous dimension matrix.
In each case the quantum numbers which are preserved under running are determined by the flavour symmetry of the Standard Model interactions under the simplifying assumptions. Sometimes additional quantum numbers are accidentally conserved within the $(4,0)$ block, as explained in Sec.~\ref{sec:flavour}. Note that \emph{we do not impose any restrictions or symmetry assumptions on the form of the SMEFT Wilson coefficients themselves.}
The summary of how these conserved quantum numbers allow the anomalous dimension matrix to be decomposed into blocks which are closed under RG is given in Tab.~\ref{tab:blockDiagonal}. We consider a number of different approximations on the SM interactions, listed in the row headings, ranging from the assumption that only the gauge couplings are important, to the case where no interactions are neglected and the full Yukawa matrices are used. It can be seen at a glance, from the block sizes in the fourth column, that depending on the approximations made, the total 1460 parameters of the $(4,0)$ operators are immediately diagonalised into much smaller blocks. The resulting matrices are shown pictorially in Fig.~\ref{fig:diagpic}. If light fermion Yukawas are neglected (an excellent approximation), a large fraction of the parameters end up in $1\times 1$ or $2 \times 2$ blocks, trivially diagonalising much of the otherwise unwieldy anomalous dimension matrix. 
In the following, we explain how this block diagonalisation comes about.

\begin{figure}
    \centering
    \includegraphics[width=\textwidth]{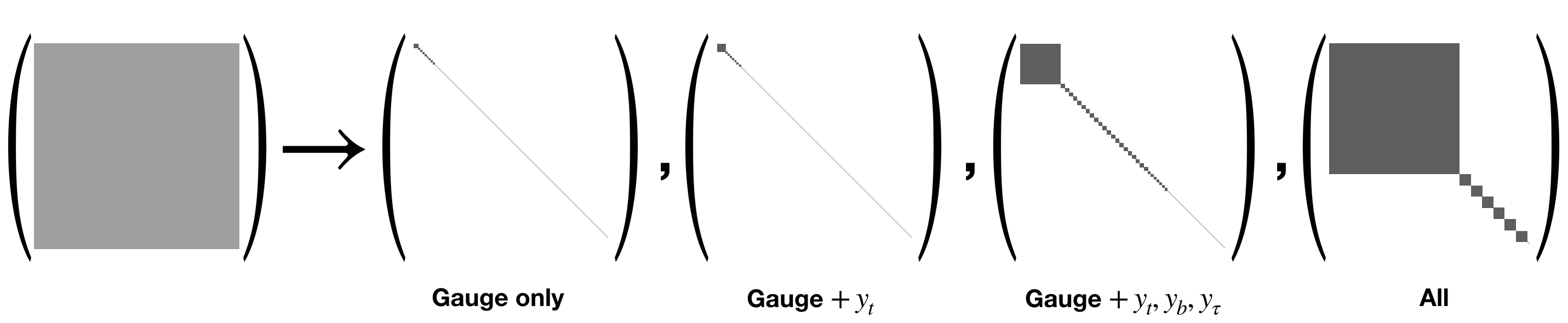}
    \caption{Scale representation of the block diagonalisation of the anomalous dimension matrix of the $(4,0)$ operators that occurs under various approximations of the SM interactions, as listed in Tab.~\ref{tab:blockDiagonal}. (Here, the ``Gauge $+y_t, y_b, y_\tau$'' scenario is assuming the full (non-unit) CKM, i.e.~it corresponds to the penultimate row in Tab.~\ref{tab:blockDiagonal}.)}
    \label{fig:diagpic}
\end{figure}

\subsubsection{Explanation of the conserved quantum numbers}

Here we explain in more detail how the running, under various approximations of the SM interactions, preserves different flavour quantum numbers, as given in the second column of Tab.~\ref{tab:blockDiagonal}. Recall that the magnetic quantum numbers $\itfl$ and $\yfl$, measure the degree of off-diagonality of a Wilson coefficient.

\begin{itemize}
    \item \textbf{Gauge only} If we only turn on the gauge interactions, $SU(3)^5$ flavour symmetry is exactly conserved by the SM. This implies that all the quantum numbers,
   \begin{equation}
      \{d_\text{irrep},\ifl,\itfl,\yfl\}_F, \forall F \in \{Q,u,d,L,e\} \, ,%
\end{equation}
are conserved.
\item \textbf{Gauge $+y_t$} Now if we also turn on only the top Yukawa, which is by far the largest parameter in the Yukawa matrices, then the symmetry of the up-type Yukawa interaction in the SM,
\begin{equation}
    \mathcal{L}_{Y} \supset \bar{Q}^p [Y_u]_p^q u_q \tilde{H}
\end{equation}
is now broken to $SU(2)_{Q}\times SU(2)_u \times U(1)_{Q+u}$. Overall any interaction involving this Yukawa matrix must therefore preserve $\ifl_Q$, $\ifl_u$, $\ifl_{3,Q}$, $\ifl_{3,u}$, $\yfl_{Q}+\yfl_{u}$. However, for running within the $(4,0)$ block, the flavour factors always involve either two or zero factors of $Y_u$, never just one, and so by the arguments of Sec.~\ref{sec:flavour}, the running also preserves $\yfl_Q$ and $\yfl_u$ independently.\footnote{In general, running to/from $C_{Hud}$ is an exception to this, since it involves a factor of $Y_u Y_d^\dagger$. Here we exclude $C_{Hud}$ from our analysis, but in any case this drops out if $Y_d$ is approximated to zero.} 
Since the flavour groups for the other fermion species are unbroken, this approximation also conserves $\{d_\text{irrep},\ifl,\itfl,\yfl\}_F, \forall F \in \{d,L,e\}$.
\item \textbf{Gauge $+y_t, y_b, y_{\tau}$ $(V=1)$} If the CKM is approximated to the unit matrix, then the case of taking all three third generation Yukawas non-zero simply follows the arguments above, but now for all fermion species. So while the dimensionality $d_\text{irrep}$ is broken for all species, each of $\{\ifl,\itfl,\yfl\}_F, \forall F \in \{Q,u,d,L,e\}$ are separately conserved.
\item \textbf{Gauge $+y_t, y_b, y_{\tau}$ (full $V$)} With the full CKM, all $Q$ charges are broken, but the $d$, $u$, $L$ and $e$ conserved charges are as before, i.e., $\{\ifl,\itfl,\yfl\}_F, \forall F \in \{u,d,L,e\}$. This can be seen by the fact that the running into, out of, or among operators with $Q$ charges can be proportional both to products of $Y_u^{(\dagger)}$ and to products of $Y_d^{(\dagger)}$, which are not simultaneously diagonalisable, and hence can ultimately break all $Q$ charges. On the other hand, operators with $u$ charges are connected to each other and to other operators only by products of $Y_u^{(\dagger)}$, so the situation for these is exactly as in the $V=1$ case above. The same is true for operators with $d$, $L$ and/or $e$ charges.
\item \textbf{All} If we do not neglect any of the SM Yukawa couplings at all, then all flavour charges are broken, except two combinations of the lepton charges,
$\yfl_{L+e}\equiv \yfl_L+\yfl_e$ and $\ifl_{3,L+e}\equiv \ifl_{3,L}+\ifl_{3,e}$. This follows from the fact that the sum of left- and right-handed charges are always conserved by diagonal Yukawas, as argued in Sec.~\ref{sec:flavour}.
Another way of seeing that these charges must be conserved is noticing that they are linear combinations of the individual lepton numbers $N_e$, $N_\mu$ and $N_\tau$, which are accidentally conserved by the SM (remember that overall lepton number has been factored out at the beginning of our analysis, so only two charges remain). Explicitly, from Eq.~\eqref{eq:Yns},
\begin{align}
   \ifl_{3,L+e} = \frac12 \left(N_e - N_\mu \right), \,  \quad \yfl_{L+e} = \frac13 \left( N_e + N_\mu - 2 N_\tau \right).
 \end{align}
 \end{itemize}
 
\subsubsection{Explanation of the blocks, their sizes and implications}

Having identified the flavour charges preserved by running under different approximations to the SM interactions, we now turn to understanding which coefficients can renormalise each other. The sizes and multiplicities of the blocks found in this way are given in columns 4 and 5 of Tab.~\ref{tab:blockDiagonal}. These numbers can be derived by simply counting parameters with the appropriate conserved charges, and we have checked them against the anomalous dimension matrix using DSixTools~\cite{Fuentes-Martin:2020zaz,Celis:2017hod}. By applying our flavour decomposition, we check both that the given charges are conserved under running, and also that all other charges in our flavour decomposition are broken.

\begin{itemize}
    \item \textbf{Gauge only} 
The largest block of the anomalous dimension matrix of the $(4,0)$ operators is that which contains all the flavour singlets: a $34 \times 34$ block. Then there are many smaller blocks which contain each of the eight components of an octet irrep.  For example, there are eight $6\times 6$ blocks: each block contains an ${\bf 8}_e$ octet component of each of the six coefficients $\{C_{He}, C_{ed},C_{eu},C_{Qe},C_{Le}, C_{ee}\}$ (where in each case, the coefficient is a singlet under all other flavour groups). Similar arguments hold for the octet components of the other flavour groups, with the differing block sizes determined by the number of Wilson coefficients which contain such an octet. Finally, we have hundreds of $2 \times 2$ and $1\times 1$ blocks, containing coefficients with other charges. If there are two coefficients with the same flavour charges but different gauge structures, for example a component of both $C_{ud}^{(1)}$ and $C_{ud}^{(8)}$, then those which do not fall into any of the previous blocks will sit in $2\times 2$ blocks. These mix via IR-divergent gauge diagrams (see Tab.~\ref{tab:topologyrules}). Finally, flavour components for which there is only one possible gauge structure (and also for $QQ$ operators, see \cref{sec:zeroentry}), and which have different charges than those of the larger blocks, will sit in $1\times 1$ blocks, and can only renormalise themselves. 
\item \textbf{Gauge $+y_t$} Here, the largest block now contains also all coefficients with $\ifl_{\{Q,u\}}=0$ and which are singlets under all the other flavour groups ($\{d,L,e\}$). This expands its size to $61\times 61$ coefficients. The multiplicity of two of the smaller blocks is reduced from 8 to 7, since the $\ifl = 0$ ($c_{8,6}$) components of the ${\bf 8}_Q$ and ${\bf 8}_u$ octets are now contained within the largest block (if a singlet under other groups). Additionally, the sizes of the blocks are larger than in the gauge-only case due to the relaxed requirements on the $Q$, $u$ charges. Finally, there are still a large number of $2\times 2$ and $1\times 1$ blocks, which have charges not contained in any of the larger blocks.
\item \textbf{Gauge $+y_t,y_b, y_\tau$ ($V=1$)} The largest block now contains all coefficients with $\ifl_{\{Q,u,d,L,e\}}=0$. This expands its size to $100\times 100$ coefficients. The multiplicity of five of the smaller blocks are reduced from 8 to 7, since the $c_{8,6}$ component of any octet ${\bf 8}_F$ is now contained within the largest block. Again, there are a large number of $2\times 2$ and $1\times 1$ blocks, which have charges not contained in any of the larger blocks.
\item \textbf{Gauge $+y_t,y_b, y_\tau$ (full $V$)} The largest block now contains almost 300 coefficients since it contains all coefficients with $I_{\{u,d,L,e\}}=0$ and with any $Q$ charges. There are now no blocks that can be defined by their $Q$ charges. The number of $2\times2$ blocks hence has decreased significantly compared to the cases with more approximations, since many coefficients with more than one gauge structure also have $Q$ charges (e.g.~$C_{QQ}^{(1,3)}$, $C_{Qu}^{(1,8)}$), which are now included in larger groups.
\item \textbf{All} If we do not neglect any of the Yukawa or gauge couplings, then there are far fewer blocks than in the other cases, and in particular there are no longer any $1\times1$ blocks. However, the matrix still does block diagonalise, into 19 blocks in total. By far the largest block is the one for which both $\ifl_{3,L+e}$ ($\equiv \ifl_{3,L}+\ifl_{3,e}$) and $\yfl_{L+e}$ ($\equiv \yfl_{L}+\yfl_e$) are zero. The next largest blocks, of size $81\times 81$ each, consist of coefficients whose total $\ifl_{3,L+e}$ and $\yfl_{L+e}$ values lie on the vertices of the hexagon in the middle diagram of Fig.~\ref{fig:irrepCharges} (or equivalently the smaller hexagon on the rightmost diagram of the same figure). Next we have six $4\times 4$ blocks, these are coefficients whose $\ifl_{3,L+e}$ and $\yfl_{L+e}$ values correspond to the six components at the midpoints of the larger hexagon in the rightmost diagram of Fig.~\ref{fig:irrepCharges}. The $4\times 4$ size of each block can be counted as: one component of the ${\bf 27}_e$ multiplet of $C_{ee}$, one component of the ${\bf 27}_L$ multiplet of $C_{LL}$, and two components of the ${\bf 8}_e \times {\bf 8}_L$ multiplet of $C_{Le}$ (there are two different ways to add the individual $\{\itfl,\yfl\}_{L,e}$ charges to get each $(\ifl_{3,L+e},\yfl_{L+e})$ coordinate). Finally, there are six $3\times 3$ blocks, these are coefficients whose $\ifl_{3,L+e}$ and $\yfl_{L+e}$ values correspond to the six components at the vertices of the larger hexagon in the rightmost diagram of Fig.~\ref{fig:irrepCharges}. The $3\times 3$ size of each block can be counted as: one component of the ${\bf 27}_e$ multiplet of $C_{ee}$, one component of the ${\bf 27}_L$ multiplet of $C_{LL}$, and one component of the ${\bf 8}_e \times {\bf 8}_L$ multiplet of $C_{Le}$ (for each of these $(\ifl_{3,L+e},\yfl_{L+e})$ coordinates, there is only one way to add the individual $\{\itfl,\yfl\}_{L,e}$ charges).
\end{itemize}
In the final column of Tab.~\ref{tab:blockDiagonal}, the letters and numbers refer to the example phenomenology as outlined in Tab.~\ref{tab:irrepPheno}. This is to give an indication of the types of phenomenology that can be linked by RG. For example, if the Yukawa couplings of the first two generations and the off-diagonals of the CKM are neglected, then the A4-type coefficients are disconnected under RG, lying in either $1\times 1$ or $2\times 2$ blocks. This means that coefficients responsible for meson mixing are individually closed under renormalisation, at least to zeroth order in light Yukawas or small CKM elements.

An example of the innards of a block is given in \cref{tab:exampleQuarklessFlavourBlock}, which shows the part of the largest $100\times 100$ block in the ``gauge $+y_t,y_b, y_\tau$ ($V=1$)'' approximation which mixes operators without quarks. We observe that there are many more $y_\tau$ entries than gauge entries, behaviour which is mirrored by $y_t$ in the parts of the block (not shown) that mix quark operators, as justified in Sec.~\ref{sec:sparseg3}. Moreover, the block is a dense matrix: most of the operators mix. This demonstrates the power of the flavour decomposition in removing most of the zeroes from the full anomalous dimension matrix.

\begin{table}[]
    \centering
    \scalebox{0.67}{
    \begin{tabular}{|c||l|c|c|c|c|}
    \hline
\textbf{SM}  & Conserved quantum & Block quantum numbers &  \textbf{Block} & \textbf{\# of} & Example \\
\textbf{approx.} & numbers & \small{(unspecified numbers distinguish same size blocks)} & \textbf{size} & \textbf{blocks} & phenomenology\\
\hline
\hline 
\multirow{5}{4em}{Gauge only} &\multirow{5}{4em}{$\{d,\ifl,\itfl,\yfl\}_{\{Q,u,d,L,e\}}$}  &$d_{\{Q,u,d,L,e\}}=1$ & $34\times 34$ & 1 & A1 \\
& & $d_Q=8$, $d_{\{u,d,L,e\}}=1$ & $13 \times 13$ & 8 & A2, A3\\
& & $d_u=8$, $d_{\{Q,d,L,e\}}=1$ & $9\times 9$ & 8 & A2, A3\\
& & $d_d=8$, $d_{\{Q,u,L,e\}}=1$ & $9\times 9$ & 8 & A3\\
& & $d_L=8$, $d_{\{Q,u,d,e\}}=1$ & $9\times 9$ & 8 & B1, C1 \\
& & $d_e=8$, $d_{\{Q,u,d,L\}}=1$ & $6\times 6$ & 8 & B1, C1\\
& & other & $2\times 2$ & 256 & A4, B2, C2, B3, C3, D1\\
& & other & $1\times 1$ & 546 & A4, B2, C2, B3, C3, D1\\
\hline
\multirow{5}{4em}{Gauge $+y_t$} &\multirow{5}{4em}{$\{\ifl,\itfl,\yfl\}_{\{Q,u\}}$, $\{d,\ifl,\itfl,\yfl\}_{\{d,L,e\}}$ } & $\ifl_{\{Q,u\}}=0$, $d_{\{d,L,e\}}=1$ & $61\times 61$ & 1 & A1, A2 \\
& & $\ifl_{Q}\in\{\frac{1}{2},1\}$, $\{\itfl,\yfl \}_Q\leq 1$, $\ifl_{u}=0$, $d_{\{d,L,e\}}=1$ & $17\times 17$ & 7 & A3\\
& & $\ifl_{\{Q,u\}}=0$, $d_d=8$, $d_{\{L,e\}}=1$ & $13\times 13$ & 8 & A3\\
& & $\ifl_{u}\in\{\frac{1}{2},1\}$, $\{\itfl,\yfl \}_u\leq 1$, $\ifl_{Q}=0$, $d_{\{d,L,e\}}=1$ & $12\times 12$ & 7 & A3\\
& & $\ifl_{\{Q,u\}}=0$, $d_L=8$, $d_{\{d,e\}}=1$ & $12\times 12$ & 8 & B1, B2, C1, C2\\
& & $\ifl_{\{Q,u\}}=0$, $d_e=8$, $d_{\{d,L\}}=1$ & $8\times 8$ & 8 & B1, B2, C1, C2\\
& & other & $2\times 2$ & 217 & B3, C2, C3, A4, D1\\
& & other & $1\times 1$ & 498 & B3, C2, C3, A4, D1\\
\hline
\multirow{5}{4em}{Gauge $+y_t,y_b,y_\tau$ ($V=1$)} & \multirow{5}{4em}{$\{\ifl,\itfl,\yfl\}_{\{Q,u,d,L,e\}}$} & $\ifl_{\{Q,u,d,L,e\}}=0$ & $100\times 100$ & 1 & A1, A2, B1, B2\\
& & $\ifl_{Q}\in\{\frac{1}{2},1\}$, $\{\itfl,\yfl \}_Q\leq 1$, $\ifl_{\{u,d,L,e\}}=0$ & $22\times 22$ & 7 & A3, B3\\
& & $\ifl_{u}\in\{\frac{1}{2},1\}$, $\{\itfl,\yfl \}_u\leq 1$, $\ifl_{\{Q,d,L,e\}}=0$ & $16\times 16$ & 7 & A3, B3 \\
& & $\ifl_{d}\in\{\frac{1}{2},1\}$, $\{\itfl,\yfl \}_d\leq 1$, $\ifl_{\{Q,u,L,e\}}=0$ & $16\times 16$ & 7 & A3, B3 \\
& & $\ifl_{L}\in\{\frac{1}{2},1\}$, $\{\itfl,\yfl \}_L\leq 1$, $\ifl_{\{Q,u,d,e\}}=0$ & $15\times 15$ & 7 & C1, C2 \\
& & $\ifl_{e}\in\{\frac{1}{2},1\}$, $\{\itfl,\yfl \}_e\leq 1$, $\ifl_{\{Q,u,d,L\}}=0$ & $11\times 11$ & 7 & C1, C2 \\
& & other & $2\times 2$ & 196 & C3, A4, D1 \\
& & other & $1\times 1$& 408 & C3, A4, D1 \\
\hline
\multirow{5}{4em}{Gauge $+y_t,y_b,y_\tau$ (full $V$)} & \multirow{5}{4em}{$\{\ifl,\itfl,\yfl\}_{\{u,d,L,e\}}$} & $\ifl_{\{u,d,L,e\}}=0$ & $292\times 292$ & 1 & A1--A4, B1--B3\\
& & $\ifl_{u}\in\{\frac{1}{2},1\}$, $\{\itfl,\yfl \}_u\leq 1$, $\ifl_{\{d,L,e\}}=0$ & $30\times 30$ & 7 & A2, A3, B2, B3\\
& & $\ifl_{d}\in\{\frac{1}{2},1\}$, $\{\itfl,\yfl \}_d\leq 1$, $\ifl_{\{u,L,e\}}=0$ & $30\times 30$ & 7 & A2, A3, B2, B3\\
& & $\ifl_{L}\in\{\frac{1}{2},1\}$, $\{\itfl,\yfl \}_L\leq 1$, $\ifl_{\{u,d,e\}}=0$ & $29\times 29$ & 7 & B1--B3, C1--C3 \\
& & $\ifl_{e}\in\{\frac{1}{2},1\}$, $\{\itfl,\yfl \}_e\leq 1$, $\ifl_{\{u,d,L\}}=0$ & $18\times 18$ & 7 & B1--B3, C1--C3\\
& & other & $2\times 2$ & 49 & A4, D1\\
& & other & $1\times 1$ & 321 & A4, D1\\
\hline
\multirow{4}{4em}{All} & \multirow{4}{4em}{$\ifl_{3,L+e},~\yfl_{L+e}$} & $\ifl_{3,L+e}=0$ ,$\yfl_{L+e}=0$ & $932\times 932$ & 1 & A1--A4, B1--B4\\
& & $|\ifl_{3,L+e}|\in\{\frac{1}{2},1\}$, $|\yfl_{L+e}|\leq 1$ & $81\times 81$ & 6 & C1, C2, C3\\
& & $\ifl_{3,L+e}^2+\frac34 \yfl_{L+e}^2= 3$ & $4\times 4$ & 6 & D1\\
& & $\ifl_{3,L+e}^2+\frac34 \yfl_{L+e}^2= 4$  & $3\times 3$ & 6 & D1 \\
\hline
    \end{tabular}}
    \caption{The flavour quantum numbers of the blocks of the anomalous dimension matrix under different approximations of the SM interactions. The $\ifl, \itfl, \yfl$ quantum numbers are given in the basis where the $Y_u$ and $Y_e$ Yukawa matrices are diagonal. We recall that a block multiplicity of $7$ implies there is one block with the charges of each component of an octet apart from the isospin singlet piece.}
    \label{tab:blockDiagonal}
\end{table}

\begin{table}
\resizebox{\columnwidth}{!}{%
\begin{tabular}{|C|CC|CCCCCC|CCCCCCCCCCCC|} \hline
g_1^2 & HD+ & HD- 
& HL1_{(1,1)} & HL1_{(8,6)} & HL3_{(1,1)} & HL3_{(8,6)} & He_{(1,1)} & He_{(8,6)}
& LL+_{(1,1)} & LL+_{(8,6)} & LL+_{(27,18)}
& LL-_{(1,1)} & LL-_{(8,6)} 
& Le_{(1,1,1,1)} & Le_{(8,6,1,1)} & Le_{(1,1,8,6)} & Le_{(8,6,8,6)}
& ee_{(1,1)} & ee_{(8,6)} & ee_{(27,18)} \\ \hline
 HD+ &\frac{9}{4} & \frac{1}{12} & -\frac{2}{\sqrt{3}} & 0 & 0 & 0 & -\frac{2}{\sqrt{3}} & 0 & \xmark & \xmark & \xmark & \xmark & \xmark & \xmark & \xmark & \xmark & \xmark & \xmark & \xmark & \xmark \\
 HD- &\frac{3}{4} & -\frac{53}{12} & -\frac{2}{\sqrt{3}} & 0 & 0 & 0 & -\frac{2}{\sqrt{3}} & 0 & \xmark & \xmark & \xmark & \xmark & \xmark & \xmark & \xmark & \xmark & \xmark & \xmark & \xmark & \xmark \\ \hline
 HL1_{(1,1)} &-\frac{3 \sqrt{3}}{4}  & -\frac{1}{4 \sqrt{3}} & \frac{7}{3} & 0 & 0 & 0 & 2 & 0 & -2 \sqrt{2} & 0 & 0 & \frac{2}{3} & 0 & -\frac{2}{\sqrt{3}} & 0 & 0 & 0 & \xmark & \xmark & \xmark \\
 HL1_{(8,6)} &0 & 0 & 0 & \frac{1}{3} & 0 & 0 & 0 & 0 & 0 & -\sqrt{5} & 0 & 0 & -\frac{1}{3} & 0 & -\frac{2}{\sqrt{3}} & 0 & 0 & \xmark & \xmark & \xmark \\
 HL3_{(1,1)} &0 & 0 & 0 & 0 & 0 & 0 & 0 & 0 & 0 & 0 & 0 & 0 & 0 & 0 & 0 & 0 & 0 & \xmark & \xmark & \xmark \\
 HL3_{(8,6)} &0 & 0 & 0 & 0 & 0 & 0 & 0 & 0 & 0 & 0 & 0 & 0 & 0 & 0 & 0 & 0 & 0 & \xmark & \xmark & \xmark \\
 He_{(1,1)} &-\frac{3 \sqrt{3}}{2}  & -\frac{1}{2 \sqrt{3}} & 4 & 0 & 0 & 0 & \frac{13}{3} & 0 & \xmark & \xmark & \xmark & \xmark & \xmark & -\frac{2}{\sqrt{3}} & 0 & 0 & 0 & -\frac{8 \sqrt{2}}{3}  & 0 & 0 \\
 He_{(8,6)} &0 & 0 & 0 & 0 & 0 & 0 & 0 & \frac{1}{3} & \xmark & \xmark & \xmark & \xmark & \xmark & 0 & 0 & -\frac{2}{\sqrt{3}} & 0 & 0 & -\frac{4 \sqrt{5}}{3}  & 0 \\ \hline
 LL+_{(1,1)} & \xmark & \xmark & -\frac{\sqrt{2}}{3} & 0 & 0 & 0 & \xmark & \xmark & 7 & 0 & 0 & -\frac{2 \sqrt{2}}{3}  & 0 & 2 \sqrt{\frac{2}{3}} & 0 & 0 & 0 & \xmark & \xmark & \xmark \\
  LL+_{(8,6)} & \xmark & \xmark & 0 & -\frac{\sqrt{5}}{6} & 0 & 0 & \xmark & \xmark & 0 & \frac{11}{2} & 0 & 0 & \frac{\sqrt{5}}{6} & 0 & \sqrt{\frac{5}{3}} & 0 & 0 & \xmark & \xmark & \xmark \\
 LL+_{(27,18)} & \xmark & \xmark & 0 & 0 & 0 & 0 & \xmark & \xmark & 0 & 0 & 3 & 0 & 0 & 0 & 0 & 0 & 0 & \xmark & \xmark & \xmark \\
 LL-_{(1,1)} & \xmark & \xmark & \frac{1}{3} & 0 & 0 & 0 & \xmark & \xmark & -2 \sqrt{2} & 0 & 0 & \frac{11}{3} & 0 & -\frac{2}{\sqrt{3}} & 0 & 0 & 0 & \xmark & \xmark & \xmark \\
 LL-_{(8,6)} & \xmark & \xmark & 0 & -\frac{1}{6} & 0 & 0 & \xmark & \xmark & 0 & \frac{\sqrt{5}}{2} & 0 & 0 & \frac{19}{6} & 0 & \frac{1}{\sqrt{3}} & 0 & 0 & \xmark & \xmark & \xmark \\
 Le_{(1,1,1,1)} & \xmark & \xmark & -\frac{2}{\sqrt{3}} & 0 & 0 & 0 & -\frac{1}{\sqrt{3}} & 0 & 4 \sqrt{6} & 0 & 0 & -\frac{4}{\sqrt{3}} & 0 & 0 & 0 & 0 & 0 & 8 \sqrt{\frac{2}{3}} & 0 & 0 \\
 Le_{(8,6,1,1)} & \xmark & \xmark & 0 & -\frac{2}{\sqrt{3}} & 0 & 0 & 0 & 0 & 0 & 2 \sqrt{15} & 0 & 0 & \frac{2}{\sqrt{3}} & 0 & -2 & 0 & 0 & 0 & 0 & 0 \\
 Le_{(1,1,8,6)} & \xmark & \xmark & 0 & 0 & 0 & 0 & 0 & -\frac{1}{\sqrt{3}} & 0 & 0 & 0 & 0 & 0 & 0 & 0 & -4 & 0 & 0 & 4 \sqrt{\frac{5}{3}} & 0 \\
 Le_{(8,6,8,6)}
& \xmark & \xmark & 0 & 0 & 0 & 0 & 0 & 0 & 0 & 0 & 0 & 0 & 0 & 0 & 0 & 0 & -6 & 0 & 0 & 0 \\
 ee_{(1,1)} & \xmark & \xmark & \xmark & \xmark & \xmark & \xmark & -\frac{2 \sqrt{2}}{3}  & 0 & \xmark & \xmark & \xmark & \xmark & \xmark & 4 \sqrt{\frac{2}{3}} & 0 & 0 & 0 & \frac{68}{3} & 0 & 0 \\
 ee_{(8,6)} & \xmark & \xmark & \xmark & \xmark & \xmark & \xmark & 0 & -\frac{\sqrt{5}}{3} & \xmark & \xmark & \xmark & \xmark & \xmark & 0 & 0 & 2 \sqrt{\frac{5}{3}} & 0 & 0 & \frac{56}{3} & 0 \\
 ee_{(27,18)} & \xmark & \xmark & \xmark & \xmark & \xmark & \xmark & 0 & 0 & \xmark & \xmark & \xmark & \xmark & \xmark & 0 & 0 & 0 & 0 & 0 & 0 & 12 \\ \hline
\end{tabular}}

\vspace*{2ex}

\resizebox{\columnwidth}{!}{%
\begin{tabular}{|C|CC|CCCCCC|CCCCCCCCCCCC|} \hline
g_2^2 & HD+ & HD- 
& HL1_{(1,1)} & HL1_{(8,6)} & HL3_{(1,1)} & HL3_{(8,6)} & He_{(1,1)} & He_{(8,6)}
& LL+_{(1,1)} & LL+_{(8,6)} & LL+_{(27,18)}
& LL-_{(1,1)} & LL-_{(8,6)} 
& Le_{(1,1,1,1)} & Le_{(8,6,1,1)} & Le_{(1,1,8,6)} & Le_{(8,6,8,6)}
& ee_{(1,1)} & ee_{(8,6)} & ee_{(27,18)} \\ \hline
 HD+ & \frac{1}{4} & \frac{17}{12} & 0 & 0 & \frac{2}{\sqrt{3}} & 0 & 0 & 0 & \xmark & \xmark & \xmark & \xmark & \xmark & \xmark & \xmark & \xmark & \xmark & \xmark & \xmark & \xmark \\
 HD- & \frac{51}{4} & \frac{1}{4} & 0 & 0 & -2 \sqrt{3} & 0 & 0 & 0 & \xmark & \xmark & \xmark & \xmark & \xmark & \xmark & \xmark & \xmark & \xmark & \xmark & \xmark & \xmark \\ \hline
 HL1_{(1,1)} & 0 & 0 & 0 & 0 & 0 & 0 & \xmark & \xmark & 0 & 0 & 0 & 0 & 0 & 0 & 0 & 0 & 0 & \xmark & \xmark & \xmark \\
 HL1_{(8,6)} & 0 & 0 & 0 & 0 & 0 & 0 & \xmark & \xmark & 0 & 0 & 0 & 0 & 0 & 0 & 0 & 0 & 0 & \xmark & \xmark & \xmark \\
 HL3_{(1,1)} & \frac{\sqrt{3}}{4} & -\frac{1}{4 \sqrt{3}} & 0 & 0 & -\frac{11}{3} & 0 & \xmark & \xmark & \frac{2 \sqrt{2}}{3} & 0 & 0 & \frac{2}{3} & 0 & 0 & 0 & 0 & 0 & \xmark & \xmark & \xmark \\
 HL3_{(8,6)} & 0 & 0 & 0 & 0 & 0 & -\frac{17}{3} & \xmark & \xmark & 0 & \frac{\sqrt{5}}{3} & 0 & 0 & -\frac{1}{3} & 0 & 0 & 0 & 0 & \xmark & \xmark & \xmark \\
 He_{(1,1)} & 0 & 0 & 0 & 0 & 0 & 0 & 0 & 0 & \xmark & \xmark & \xmark & \xmark & \xmark & 0 & 0 & 0 & 0 & 0 & 0 & 0 \\
 He_{(8,6)} & 0 & 0 & 0 & 0 & 0 & 0 & 0 & 0 & \xmark & \xmark & \xmark & \xmark & \xmark & 0 & 0 & 0 & 0 & 0 & 0 & 0 \\ \hline
 LL+_{(1,1)} & \xmark & \xmark & 0 & 0 & \frac{\sqrt{2}}{3} & 0 & 0 & 0 & \frac{13}{3} & 0 & 0 & \frac{2 \sqrt{2}}{3} & 0 & 0 & 0 & 0 & 0 & \xmark & \xmark & \xmark \\
 LL+_{(8,6)} & \xmark & \xmark & 0 & 0 & 0 & \frac{\sqrt{5}}{6} & 0 & 0 & 0 & \frac{23}{6} & 0 & 0 & -\frac{\sqrt{5}}{6} & 0 & 0 & 0 & 0 & \xmark & \xmark & \xmark \\
 LL+_{(27,18)} & \xmark & \xmark & 0 & 0 & 0 & 0 & 0 & 0 & 0 & 0 & 3 & 0 & 0 & 0 & 0 & 0 & 0 & \xmark & \xmark & \xmark \\
 LL-_{(1,1)} & \xmark & \xmark & 0 & 0 & 1 & 0 & 0 & 0 & 2 \sqrt{2} & 0 & 0 & -7 & 0 & 0 & 0 & 0 & 0 & \xmark & \xmark & \xmark \\
 LL-_{(8,6)} & \xmark & \xmark & 0 & 0 & 0 & -\frac{1}{2} & 0 & 0 & 0 & -\frac{\sqrt{5}}{2} & 0 & 0 & -\frac{17}{2} & 0 & 0 & 0 & 0 & \xmark & \xmark & \xmark \\
 Le_{(1,1,1,1)} & \xmark & \xmark & 0 & 0 & 0 & 0 & 0 & 0 & 0 & 0 & 0 & 0 & 0 & 0 & 0 & 0 & 0 & 0 & 0 & 0 \\
 Le_{(8,6,1,1)} & \xmark & \xmark & 0 & 0 & 0 & 0 & 0 & 0 & 0 & 0 & 0 & 0 & 0 & 0 & 0 & 0 & 0 & 0 & 0 & 0 \\
 Le_{(1,1,8,6)} & \xmark & \xmark & 0 & 0 & 0 & 0 & 0 & 0 & 0 & 0 & 0 & 0 & 0 & 0 & 0 & 0 & 0 & 0 & 0 & 0 \\
 Le_{(8,6,8,6)} & \xmark & \xmark & 0 & 0 & 0 & 0 & 0 & 0 & 0 & 0 & 0 & 0 & 0 & 0 & 0 & 0 & 0 & 0 & 0 & 0 \\
 ee_{(1,1)} & \xmark & \xmark & \xmark & \xmark & \xmark & \xmark & 0 & 0 & \xmark & \xmark & \xmark & \xmark & \xmark & 0 & 0 & 0 & 0 & 0 & 0 & 0 \\
 ee_{(8,6)} & \xmark & \xmark & \xmark & \xmark & \xmark & \xmark & 0 & 0 & \xmark & \xmark & \xmark & \xmark & \xmark & 0 & 0 & 0 & 0 & 0 & 0 & 0 \\
 ee_{(27,18)} & \xmark & \xmark & \xmark & \xmark & \xmark & \xmark & 0 & 0 & \xmark & \xmark & \xmark & \xmark & \xmark & 0 & 0 & 0 & 0 & 0 & 0 & 0 \\ \hline
\end{tabular}}

  \vspace*{2ex}
  
\resizebox{\columnwidth}{!}{%
\begin{tabular}{|C|CC|CCCCCC|CCCCCCCCCCCC|} \hline
y_\tau^2 & HD+ & HD- 
& HL1_{(1,1)} & HL1_{(8,6)} & HL3_{(1,1)} & HL3_{(8,6)} & He_{(1,1)} & He_{(8,6)}
& LL+_{(1,1)} & LL+_{(8,6)} & LL+_{(27,18)}
& LL-_{(1,1)} & LL-_{(8,6)} 
& Le_{(1,1,1,1)} & Le_{(8,6,1,1)} & Le_{(1,1,8,6)} & Le_{(8,6,8,6)}
& ee_{(1,1)} & ee_{(8,6)} & ee_{(27,18)} \\ \hline
HD+ & 4 & 0 & -\frac{2}{\sqrt{3}} & -2 \sqrt{\frac{2}{3}} & -\frac{2}{\sqrt{3}} & -2 \sqrt{\frac{2}{3}} & \frac{2}{\sqrt{3}} & 2 \sqrt{\frac{2}{3}} & \xmark & \xmark & \xmark & \xmark & \xmark & \xmark & \xmark & \xmark & \xmark & \xmark & \xmark & \xmark \\
 HD- & 0 & 4 & -\frac{2}{\sqrt{3}} & -2 \sqrt{\frac{2}{3}} & 2 \sqrt{3} & 2 \sqrt{6} & \frac{2}{\sqrt{3}} & 2 \sqrt{\frac{2}{3}} & \xmark & \xmark & \xmark & \xmark & \xmark & \xmark & \xmark & \xmark & \xmark & \xmark & \xmark & \xmark \\ \hline
 HL1_{(1,1)} &  -\frac{3 \sqrt{3}}{4}  & -\frac{1}{4 \sqrt{3}} & \frac{10}{3} & \frac{4 \sqrt{2}}{3} & 3 & 3 \sqrt{2} & -\frac{1}{3} & -\frac{\sqrt{2}}{3} & -2 \sqrt{2} & -\sqrt{10} & 0 & \frac{2}{3} & -\frac{\sqrt{2}}{3} & \frac{2}{\sqrt{3}} & 0 & 2 \sqrt{\frac{2}{3}} & 0 & \xmark & \xmark & \xmark \\
 HL1_{(8,6)} &-\frac{3}{2} \sqrt{\frac{3}{2}} & -\frac{1}{2 \sqrt{6}} & \frac{4 \sqrt{2}}{3} & \frac{14}{3} & 3 \sqrt{2} & 6 & -\frac{\sqrt{2}}{3} & -\frac{2}{3} & -1 & -\frac{7}{\sqrt{5}} & -\frac{9}{\sqrt{5}} & -\frac{\sqrt{2}}{3} & \frac{1}{3} & 0 & \frac{2}{\sqrt{3}} & 0 & 2 \sqrt{\frac{2}{3}} & \xmark & \xmark & \xmark \\
 HL3_{(1,1)} &-\frac{\sqrt{3}}{4} & \frac{1}{4 \sqrt{3}} & 1 & \sqrt{2} & \frac{8}{3} & \frac{2 \sqrt{2}}{3} & 0 & 0 & -\frac{2 \sqrt{2}}{3}  & -\frac{\sqrt{10}}{3} & 0 & -\frac{2}{3} & \frac{\sqrt{2}}{3} & 0 & 0 & 0 & 0 & \xmark & \xmark & \xmark \\
 HL3_{(8,6)} &-\frac{\sqrt{\frac{3}{2}}}{2} & \frac{1}{2 \sqrt{6}} & \sqrt{2} & 2 & \frac{2 \sqrt{2}}{3} & \frac{10}{3} & 0 & 0 & -\frac{1}{3} & -\frac{7}{3 \sqrt{5}} & -\frac{3}{\sqrt{5}} & \frac{\sqrt{2}}{3} & -\frac{1}{3} & 0 & 0 & 0 & 0 & \xmark & \xmark & \xmark \\
 He_{(1,1)} & \frac{3 \sqrt{3}}{2} & \frac{1}{2 \sqrt{3}} & -\frac{2}{3} & -\frac{2 \sqrt{2}}{3}  & 0 & 0 & \frac{14}{3} & \frac{8 \sqrt{2}}{3} & \xmark & \xmark & \xmark & \xmark & \xmark & -\frac{2}{\sqrt{3}} & -2 \sqrt{\frac{2}{3}} & 0 & 0 & \frac{8 \sqrt{2}}{3} & \frac{4 \sqrt{10}}{3} & 0 \\
 He_{(8,6)} & 3 \sqrt{\frac{3}{2}} & \frac{1}{\sqrt{6}} & -\frac{2 \sqrt{2}}{3}  & -\frac{4}{3} & 0 & 0 & \frac{8 \sqrt{2}}{3} & \frac{22}{3} & \xmark & \xmark & \xmark & \xmark & \xmark & 0 & 0 & -\frac{2}{\sqrt{3}} & -2 \sqrt{\frac{2}{3}} & \frac{4}{3} & \frac{28}{3 \sqrt{5}} & \frac{12}{\sqrt{5}} \\ \hline
 LL+_{(1,1)} & \xmark & \xmark & -\frac{\sqrt{2}}{3} & -\frac{1}{6} & -\frac{\sqrt{2}}{3} & -\frac{1}{6} & \xmark & \xmark & \frac{2}{3} & \frac{\sqrt{5}}{3} & 0 & 0 & 0 & -\frac{\sqrt{\frac{2}{3}}}{3} & -\frac{1}{6 \sqrt{3}} & -\frac{2}{3 \sqrt{3}} & -\frac{1}{3 \sqrt{6}} & \xmark & \xmark & \xmark \\
 LL+_{(8,6)} & \xmark & \xmark & -\frac{\sqrt{\frac{5}{2}}}{3} & -\frac{7}{6 \sqrt{5}} & -\frac{\sqrt{\frac{5}{2}}}{3} & -\frac{7}{6 \sqrt{5}} & \xmark & \xmark & \frac{\sqrt{5}}{3} & \frac{17}{15} & \frac{3}{5} & 0 & 0 & -\frac{\sqrt{\frac{5}{6}}}{3} & -\frac{7}{6 \sqrt{15}} & -\frac{\sqrt{\frac{5}{3}}}{3} & -\frac{7}{3 \sqrt{30}} & \xmark & \xmark & \xmark \\
 LL+_{(27,18)} & \xmark & \xmark & 0 & -\frac{3}{2 \sqrt{5}} & 0 & -\frac{3}{2 \sqrt{5}} & \xmark & \xmark & 0 & \frac{3}{5} & \frac{6}{5} & 0 & 0 & 0 & -\frac{\sqrt{\frac{3}{5}}}{2} & 0 & -\sqrt{\frac{3}{10}} & \xmark & \xmark & \xmark \\
 LL-_{(1,1)} & \xmark & \xmark & \frac{1}{3} & -\frac{1}{3 \sqrt{2}} & -1 & \frac{1}{\sqrt{2}} & \xmark & \xmark & 0 & 0 & 0 & \frac{2}{3} & -\frac{\sqrt{2}}{3} & \frac{1}{3 \sqrt{3}} & -\frac{1}{3 \sqrt{6}} & \frac{\sqrt{\frac{2}{3}}}{3} & -\frac{1}{3 \sqrt{3}} & \xmark & \xmark & \xmark \\
 LL-_{(8,6)} & \xmark & \xmark & -\frac{1}{3 \sqrt{2}} & \frac{1}{6} & \frac{1}{\sqrt{2}} & -\frac{1}{2} & \xmark & \xmark & 0 & 0 & 0 & -\frac{\sqrt{2}}{3} & \frac{1}{3} & -\frac{1}{3 \sqrt{6}} & \frac{1}{6 \sqrt{3}} & -\frac{1}{3 \sqrt{3}} & \frac{1}{3 \sqrt{6}} & \xmark & \xmark & \xmark \\
 Le_{(1,1,1,1)} & \xmark & \xmark & \frac{2}{\sqrt{3}} & 0 & 0 & 0 & -\frac{1}{\sqrt{3}} & 0 & -2 \sqrt{\frac{2}{3}} & -\sqrt{\frac{10}{3}} & 0 & \frac{2}{3 \sqrt{3}} & -\frac{\sqrt{\frac{2}{3}}}{3} & \frac{5}{3} & \sqrt{2} & \frac{4 \sqrt{2}}{3} & \frac{4}{3} & -\frac{4}{3} \sqrt{\frac{2}{3}} & -\frac{2}{3} \sqrt{\frac{10}{3}} & 0 \\
 Le_{(8,6,1,1)} & \xmark & \xmark & 0 & \frac{2}{\sqrt{3}} & 0 & 0 & -\sqrt{\frac{2}{3}} & 0 & -\frac{1}{\sqrt{3}} & -\frac{7}{\sqrt{15}} & -3 \sqrt{\frac{3}{5}} & -\frac{\sqrt{\frac{2}{3}}}{3} & \frac{1}{3 \sqrt{3}} & \sqrt{2} & \frac{8}{3} & \frac{4}{3} & 2 \sqrt{2} & -\frac{8}{3 \sqrt{3}} & -\frac{4}{3} \sqrt{\frac{5}{3}}  & 0 \\
 Le_{(1,1,8,6)} & \xmark & \xmark & 2 \sqrt{\frac{2}{3}} & 0 & 0 & 0 & 0 & -\frac{1}{\sqrt{3}} & -\frac{4}{\sqrt{3}} & -2 \sqrt{\frac{5}{3}} & 0 & \frac{2 \sqrt{\frac{2}{3}}}{3} & -\frac{2}{3 \sqrt{3}} & \frac{4 \sqrt{2}}{3} & \frac{4}{3} & 3 & \frac{5 \sqrt{2}}{3} & -\frac{2}{3 \sqrt{3}} & -\frac{14}{3 \sqrt{15}} & -2 \sqrt{\frac{3}{5}} \\
 Le_{(8,6,8,6)} & \xmark & \xmark & 0 & 2 \sqrt{\frac{2}{3}} & 0 & 0 & 0 & -\sqrt{\frac{2}{3}} & -\sqrt{\frac{2}{3}} & -7 \sqrt{\frac{2}{15}} & -3 \sqrt{\frac{6}{5}} & -\frac{2}{3 \sqrt{3}} & \frac{\sqrt{\frac{2}{3}}}{3} & \frac{4}{3} & 2 \sqrt{2} & \frac{5 \sqrt{2}}{3} & \frac{14}{3} & -\frac{2}{3} \sqrt{\frac{2}{3}}  & -\frac{14}{3} \sqrt{\frac{2}{15}}  & -2 \sqrt{\frac{6}{5}} \\
 ee_{(1,1)} & \xmark & \xmark & \xmark & \xmark & \xmark & \xmark & \frac{2 \sqrt{2}}{3} & \frac{1}{3} & \xmark & \xmark & \xmark & \xmark & \xmark & -\frac{2}{3} \sqrt{\frac{2}{3}}  & -\frac{4}{3 \sqrt{3}} & -\frac{1}{3 \sqrt{3}} & -\frac{\sqrt{\frac{2}{3}}}{3} & \frac{4}{3} & \frac{2 \sqrt{5}}{3} & 0 \\
 ee_{(8,6)} & \xmark & \xmark & \xmark & \xmark & \xmark & \xmark & \frac{\sqrt{10}}{3} & \frac{7}{3 \sqrt{5}} & \xmark & \xmark & \xmark & \xmark & \xmark & -\frac{\sqrt{\frac{10}{3}}}{3} & -\frac{2}{3} \sqrt{\frac{5}{3}} & -\frac{7}{3 \sqrt{15}} & -\frac{7}{3} \sqrt{\frac{2}{15}}  & \frac{2 \sqrt{5}}{3} & \frac{34}{15} & \frac{6}{5} \\
 ee_{(27,18)} & \xmark & \xmark & \xmark & \xmark & \xmark & \xmark & 0 & \frac{3}{\sqrt{5}} & \xmark & \xmark & \xmark & \xmark & \xmark & 0 & 0 & -\sqrt{\frac{3}{5}} & -\sqrt{\frac{6}{5}} & 0 & \frac{6}{5} & \frac{12}{5} \\ \hline
\end{tabular}}
  \caption{The gauge and Yukawa pieces of the running among the quarkless current-current operators, when electron and muon Yukawas are neglected. Shown are the 20 operators with flavour charges $\yfl_L=\yfl_e=\ifl_{3,L}=\ifl_{3,e}=\ifl_{L}=\ifl_{e}=0$, which are conserved by the running in the no-light-Yukawa approximation, as discussed in \cref{sec:flavourselectionrules}. Top: the entries proportional to $g_1^2$ (with $g_1^2$ factored out), which result from the flavour decomposition of the sum of the IR finite and divergent pieces in \cref{tab:colourlessSMEFTIRfinitegauge,tab:colourlessSMEFTIRdivergent}. Middle: similarly for $g_2^2$. Bottom: the entries proportional to tau Yukawa $y_\tau^2$ (with $y_\tau^2$ factored out), which result from the flavour decomposition of the sum of the IR finite and divergent pieces in \cref{tab:colourlessSMEFTIRfiniteYuk,tab:colourlessSMEFTIRdivergentYuk}. `$\xmark$' indicates the absence of a diagram for a process.
  \label{tab:exampleQuarklessFlavourBlock}}
\end{table}

\subsection{Beyond the $(4,0)$ operators}
We now comment on to what extent this block-diagonalisation remains true when the other operators of the SMEFT, ignored so far, are included in the analysis. As shown in Refs.~\cite{Alonso:2014rga,Cheung:2015aba}, and summarised in Sec.~\ref{sec:helicitynonrenormalization}, there exist zeroes in the entries of the anomalous dimension matrix between the $(4,0)$ operators that we study, and operators with a different helicity structure (such as $H^6$, dipole operators). Crucially, at one loop there is no running \emph{into} the $(4,0)$ class from operators of a different total helicity, with the exception of entries proportional to products of two different Yukawas. For example, $C_{ud}^{(1)}$ can be renormalised by $C_{QuQd}^{(1)}$, but only by an amount proportional to $Y_u Y_d$.  All of these contributions are zero in the case that all Yukawas except $y_t$ are neglected. 

But even under a $y_t$-only approximation of the Yukawa interactions, there is still some running \emph{out} of the (4,0) coefficients.  This will mean that the extra coefficients that they run into, and those that these in turn are connected to by running, will ultimately need to be included in the blocks. However, the fact that there is no feeding back into the $(4,0)$ coefficients means that the blocks we have identified in Table~\ref{tab:blockDiagonal} cannot merge, and their $(4,0)$ contents cannot change, they can only gain new members from outside the $(4,0)$ category. This justifies our initial assertion that focusing on the $(4,0)$ operators is a significant first step towards the full (block-)diagonalisation of the anomalous dimension matrix. 

If we do not neglect $y_b$ (and/or $y_\tau$), things become more complicated, and a fuller flavour decomposition analysis encompassing also the $\bar \psi \psi X H$, $(\bar L R)(\bar L R)$ and $(\bar L R)(\bar R L)$ operators must be performed, as well as the $\op_{Hud}$ and $\op_{LedQ}$ operators that we excluded from the analysis.\footnote{It still applies that operators with more than 4 fields, such as $\bar \psi \psi H^3$, can only be run \emph{into}.} We leave this for future work.

\section{Phenomenology and applications}
\label{sec:pheno}

As indicated schematically in the final column of Tab.~\ref{tab:blockDiagonal}, an obvious application of block-diagonalisation of the anomalous dimension matrix is the ability to see at-a-glance which types of SMEFT phenomenology can or cannot be linked by RG, and how strongly (i.e.~dependent on which SM parameters). An illustration of this, for the well-studied case of lepton flavour non-universality in $B$ decays, is given in Sec.~\ref{sec:Banomalies} below. 

Since the blocks are labelled in terms of Clebsch-Gordan coefficients of the SM flavour group, it is also easy to identify the appropriate subsystems when specific symmetry-based flavour structures \cite{Faroughy:2020ina,Greljo:2022cah} are imposed on the SMEFT from the UV. This could be used as a starting point for a more in-depth study of the behaviour of flavour assumptions under RG flow. 

Building these blocks is of course also a significant step towards complete diagonalisation of the SMEFT anomalous dimension matrix. In a basis where the matrix is diagonal, it is straightforward to identify IR-enhanced directions which could be theoretically-motivated targets for searches. In particular, due to the large size of the anomalous dimension matrix, it is likely that the magnitude of a few of its eigenvalues will be very large as compared to the average value of its entries (in a non-diagonal basis)\footnote{This intuition that the numerical spread of eigenvalues should increase with the size of the matrix can be justified in simple cases through Wigner's semicircle law~\cite{10.2307/1970079} or the Circular law (\cite{10.1214/aop/1024404298} and references therein).}, and the corresponding eigenoperators can therefore provide a natural starting point for further investigation.

Moreover, by understanding the anomalous dimension matrix in on-shell language, we can more directly make connections with other on-shell features of SMEFT operators at dimension 6, $e.g.$~dispersion relations and sum rules~\cite{Remmen:2020uze,Gu:2020thj,Azatov:2021ygj,Remmen:2022orj}. This could lead to a clearer overview of the interplay of RG flow with these theoretical constraints.\footnote{See \cite{Bellazzini:2020cot,Chala:2021wpj} for studies of the scale dependence of constraints from dispersion relations in general and at dimension eight in SMEFT.}

\subsection{Case study: lepton flavour non-universality in $B$ decays}
\label{sec:Banomalies}
Measurements of semileptonic and rare $B$ decays at $B$ factories and LHCb, especially the ratios $R_{D^{(*)}}$ and $R_{K^{(*)}}$~\cite{LHCb:2021trn,LHCb:2017avl,HFLAV:2019otj,HFLAV:2022pwe,Belle:2019rba,LHCb:2017smo,Belle:2016dyj,LHCb:2015gmp,Belle:2015qfa,BaBar:2012obs}, seem to hint at lepton flavour non-universal new physics.

It is possible to fit both sets of anomalies by a contribution to a single linear combination of operators, in a basis which has some small misalignment with the mass basis of both the quarks and the leptons~\cite{Glashow:2014iga,Feruglio:2016gvd}:
\begin{align}
\label{eq:pureMixing}
    \mathcal{L}_{\text{NP}}&= \frac{C}{\Lambda^2}\left( (\bar Q^\prime_3\gamma^\mu Q^\prime_3)(\bar L^\prime_3\gamma_\mu L^\prime_3)+(\bar Q^\prime_3\gamma^\mu \sigma^I Q^\prime_3)(\bar L^\prime_3\gamma_\mu \sigma^I L^\prime_3) \right).
\end{align}
The primed fields are transformed to the mass basis by unitary transformations:
\begin{align}
    u_L = V_u u_L^\prime, ~~d_L = V_d d_L^\prime, ~~ e_L = V_e e_L^\prime
\end{align}
where $V_u^\dagger V_d=V_{\text{CKM}}$. Then the experimental values of $R_K^{(*)}$ and $R_D^{(*)}$ can be reproduced by appropriate values of the seven parameters $\frac{C}{\Lambda^2}$, $\lambda^f_{33}$, $\lambda^f_{32}$ and $\lambda^f_{22}$, where $\lambda^f_{ij} \equiv V_{f3i}^* V_{f3j}$ is a Hermitian matrix, and $\lambda^f_{33} \gg \lambda^f_{32} \gg \lambda^f_{22}$~\cite{Feruglio:2016gvd}.

However, this simple model is in trouble with bounds on lepton flavour universality in $\tau$ and $Z$ decays~\cite{Feruglio:2016gvd,Feruglio:2017rjo}. The model must be amended by adding some amount of an explicitly quark flavour violating Wilson coefficient connecting the second and third generations~\cite{Buttazzo:2017ixm}, which can also contribute to $R_K^{(*)}$ and $R_D^{(*)}$, but does not create (significant) radiative effects in these other LFUV observables.

We can now get a simple understanding of this situation, by expanding the Wilson coefficients into flavour irreps, as in Sec.~\ref{sec:flavour}. The `pure mixing' scenario in Eq.~\eqref{eq:pureMixing}, in the primed basis, is decomposed as
\begin{align}
    C= c_{33}^{33} = \frac{1}{3} \left( c_{\underbrace{1,1}_{L^\prime},\underbrace{1,1}_{Q^\prime}}+ \sqrt{2} c_{\underbrace{8,6}_{L^\prime},\underbrace{1,1}_{Q^\prime}} + \sqrt{2} c_{\underbrace{1,1}_{L^\prime},\underbrace{8,6}_{Q^\prime}} +2 c_{\underbrace{8,6}_{L^\prime},\underbrace{8,6}_{Q^\prime}} \right) \, .
\end{align}
Of course, we actually need the decomposition in the mass basis of the unprimed fields, and specifically, for comparison with Tab.~\ref{tab:blockDiagonal}, we need to be in the up-basis. Since the relations between the primed and unprimed fields are just unitary rotations, they preserve the $SU(3)_f$ irrep structure, and in any basis the overall octet piece and the total singlet piece should remain the same. Moreover, if we make the approximation of only keeping $O(1)$ pieces $\propto \lambda^f_{33}$, then the third generation of the primed and unprimed bases are the same, so we have simply
\begin{align}
\label{eq:CUpbasis}
    C= c_{33}^{33} = \frac{1}{3} \lambda^d_{33}\lambda^e_{33} \left( c_{\underbrace{1,1}_{L},\underbrace{1,1}_{Q}}+ \sqrt{2} c_{\underbrace{8,6}_{L},\underbrace{1,1}_{Q}} + \sqrt{2} c_{\underbrace{1,1}_{L},\underbrace{8,6}_{Q}} +2 c_{\underbrace{8,6}_{L},\underbrace{8,6}_{Q}} \right) \, .
\end{align}

 Now we can look at how this coefficient runs by looking at Tab.~\ref{tab:blockDiagonal}. 
If we neglect all SM interactions except for the gauge interactions and $y_t$, then the piece of the irrep decomposition with $\ifl_{\{Q,u\}}=0$ and in an octet component of lepton flavour falls within a $12\times 12$ block. In Eq.~\eqref{eq:CUpbasis}, this is the piece
\begin{equation}
    C \supset  \frac{\sqrt{2}}{3} \lambda^d_{33}\lambda^e_{33} \left(  c_{\underbrace{8,6}_{L},\underbrace{1,1}_{Q}} + \sqrt{2} c_{\underbrace{8,6}_{L},\underbrace{8,6}_{Q}} \right) \, .
\end{equation}
The 12 coefficients in the block that can mix under RG are the $(c_{8,6})_L$ lepton octet component of each of\footnote{The reason that there are two parameters for each of $C_{LQ}^{(1)}$ , $C_{LQ}^{(3)}$ and $C_{Lu}$ are that for each choice of lepton charges, there are two quark components with $I_{\{Q,u\}}=0$: $c_{1,1}$ and $c_{8,6}$ (c.f.~Fig.~\ref{fig:irrepCharges}). Meanwhile, the two parameters for $C_{LL}$ correspond to the symmetric and antisymmetric combinations.}
\begin{align}
    C_{LQ}^{(1)}\, (\times 2),~C_{LQ}^{(3)}(\times 2),~C_{Lu}\,(\times 2),~C_{Ld},~C_{LL}\, (\times 2), ~C_{Le}, ~C_{HL}^{(1)},~C_{HL}^{(3)}.
\end{align}
Identifying this block hence tells us immediately which observables may have important radiative effects from the operator in Eq.~\eqref{eq:pureMixing}. The coefficients in this block induce LFUV in $Z$ couplings ($C_{HL}^{(1)},~C_{HL}^{(3)}$), LFUV in $W$ couplings ($C_{HL}^{(3)}$) and leptonic $\tau$ decays ($C_{LL},~C_{Le}$). These observables have been found to put important constraints on the model~\cite{Feruglio:2016gvd,Feruglio:2017rjo}.

On the other hand, instead of relying on the quark flavour changing effects induced by basis- and CKM-rotation from Eq.~\ref{eq:pureMixing}, both sets of anomalies can also be fit by NP with explicit quark flavour violation $\propto c^{?3}_{?2}$. This Wilson coefficient has $\ifl_Q \neq 0$ and does not contribute to the $12\times 12$ block discussed above (or at most only by an amount proportional to small off-diagonal CKM elements). But there is a large contribution to coefficients within octet components of both quarks and leptons $c_{\underbrace{8,6}_{L},\underbrace{8,7}_{Q}}$, which fall within a $2 \times 2$ block in the gauge $+y_t$ scenario.
Here, the only sizeable running is between the $SU(2)_L$-triplet ($C_{LQ}^{(3)}$) and $SU(2)_L$-singlet ($C_{LQ}^{(1)}$) coefficients with the same flavour structure, hence constraints from electroweak and leptonic processes can be evaded. The $C_{LQ}^{(3)}-C_{LQ}^{(1)}$ mixing is a consequence of IR divergent gauge diagrams, and since these are flavour-blind, it cannot be avoided under any flavour assumption. This effect is nevertheless important, because it means that the $C_{LQ}^{(1)}=C_{LQ}^{(3)}$ condition, as seen in the Lagrangian~\ref{eq:pureMixing}, and which is chosen to eliminate effects in $b\to s \nu \nu$, is not radiatively stable, and running must be included in order to derive robust constraints and predictions for $b\to s \nu \nu$ processes. 

One final note is that \emph{any} lepton flavour non-universal operator coefficients will decompose into singlet pieces ($c_{1,1}$) and octet pieces ($c_{8,6}$ and $c_{8,4}$) under the lepton flavour group. These pieces each fall into different blocks, and the running of the singlet piece will depend on many more operator coefficients than that of the octet pieces. So the degree of lepton flavour non-universality will change with scale.

These facts about the running of the important operators for the flavour anomalies are already well known in the literature. But the flavour irrep basis and the resulting block-diagonalisation makes it very straightforward to identify the limited number of other operator coefficients that can be produced by running from the different possible flavour structures.

\section{Conclusions}

The story of the LHC's future will be written in EFTs. Understanding the ways in which heavy new physics can show up at low energies, and the general connections between phenomenology at different scales, is a crucial goal to make the most of its unprecedented volumes of data. In this context, the anomalous dimension matrix of the SMEFT is a powerful object, which is worth decoding fully. Its rich flavour structure cannot be ignored, making this effort both interesting and challenging.

In this article, we have calculated the anomalous dimension matrix of the operators of the dimension six SMEFT with four fields and zero total helicity (a large class of operators comprising $\psi^2\bar{\psi}^2$\,,\, $\psi \bar{\psi}\phi^2D$ and $\phi^4D^2$) using generalised unitarity. The amplitude-based approach has allowed us to understand the origin of each entry of the anomalous dimension matrix in terms of its gauge, flavour and kinematic factors individually, as well as in terms of the IR properties of the diagrams responsible. We have thus identified sources of many repeated patterns and zeroes in the matrix as due to the gauge and flavour tensors present in SM four-point amplitudes. A notable aspect of the SMEFT anomalous dimension matrix is the scarcity of entries dependent on the strong coupling $g_3$. This contributes to making the SMEFT RG flow particularly flavourful, since it is left heavily dependent on $y_t$. In our analysis the reasons behind this are clarified, in terms of the gauge structure of SM and SMEFT tree amplitudes.

We have shown that by performing a fully general Clebsch-Gordan decomposition on the flavour space of the Wilson coefficients, the anomalous dimension matrix is block diagonalised. The extent of the block diagonalisation depends on which, if any, of the SM couplings are neglected. Under the common and well justified approximation of neglecting all but the gauge interactions and the top Yukawa, this block diagonalisation reduces the largest block in the (4,0) operators to $61\times 61$ entries, as well as diagonalising a large fraction of the matrix into $1\times 1$ or $2\times 2$ blocks. 

This block diagonalisation not only enables at-a-glance understanding of the loop level phenomenology of individual operators, but is also a step towards full diagonalisation of the anomalous dimension matrix. With a diagonal anomalous dimension matrix, the most IR-relevant operator directions, natural targets for BSM searches, can be automatically identified. Furthermore, the flavour decomposition we employ, based as it is on the flavour groups of the SM, should generalise easily in the effective field theory below the electroweak scale (known as WET or LEFT)~\cite{Jenkins:2017jig,Jenkins:2017dyc}. Here, the remnant flavour symmetry is exactly conserved by the RG. Future work within this effective theory will therefore allow extension of the block diagonalisation down to mesonic scales.

Although this article has focussed only on the zero helicity, four field operators of the SMEFT, this is a substantial starting point for two reasons. Firstly, the coefficients of these operators comprise the majority (1460) of the parameters at dimension six, and an even greater majority of those generated at tree level by weakly-coupled UV completions. And second, the corresponding block of the anomalous dimension matrix is somewhat disconnected from the other operators: nothing runs \emph{into} it, apart from by amounts proportional to small Yukawas. This means in particular that the block diagonalisation we find will still hold in the full matrix, up to small admixtures of other operators.
Nevertheless, future work to incorporate also the rest of the operator classes into the framework will be important, especially in the case of the dipole ($\bar\psi \psi X H$) and Yukawa-like ($\bar \psi \psi H^3$) operators, which can impact flavour phenomenology.

\section*{Acknowledgments}

We are grateful to Timothy Cohen, Nathaniel Craig, and Christoph Englert for useful comments on the manuscript. SR and DS are grateful for the hospitality of the CERN theory group and the DESY theory group while parts of this work was done.
This work was performed in part at Aspen Center for Physics, which is supported by National Science Foundation grant PHY-1607611. This work was partially supported by a grant from the Simons Foundation. SR is supported by UKRI Stephen Hawking Fellowship EP/W005433/1. CSM is supported
by the Deutsche Forschungsgemeinschaft under Germany’s Excellence Strategy EXC 2121
“Quantum Universe” - 390833306.


\appendix

\allowdisplaybreaks

\section{Spinor conventions}
\label{app:conventions}

We use the spinor conventions of \cite{Dreiner:2008tw} in mostly minus metric $\eta_{\mu\nu}=\mathrm{diag}(+1,-1,-1,-1)$ and $\epsilon^{12}=-\epsilon^{21}=\epsilon_{21}=-\epsilon_{12}=1$. All momenta are \emph{ingoing} and \emph{massless}. Undotted and dotted indices are contracted as
\begin{equation}
  \ang{p}{q} \equiv \angbra{p}^\alpha \angket{q}_\alpha, \qquad  \sqr{p}{q} \equiv \sqrbra{p}_{\dot\alpha} \sqrket{q}^{\dot\alpha}.
\end{equation}
In terms of the Pauli sigma matrices and identity, we have $\sigma^\mu_{\alpha \dot \beta} = (1, \vec{\sigma})$ and $\bar\sigma^{\mu,\dot\alpha \beta} = (1,-\vec{\sigma})$. For a general four momentum $p^{\mu}=(p^0,p^1,p^2,p^3)$, the spinors can be written as
\begin{gather}
    \angket{p}_{\alpha} = \frac{1}{\sqrt{p^0+p^3}} \begin{pmatrix}
      -p^1 + i p^2 \\ p^0 + p^3
    \end{pmatrix}, \quad \sqrket{p}^{\dot\alpha} =  \frac{1}{\sqrt{p^0+p^3}} \begin{pmatrix}
      p^0 + p^3 \\ p^1 + i p^2
    \end{pmatrix}, \nn \\
    \sqrbra{p}_{\dot\alpha} = (\angket{p}_{\alpha})^\dagger = \frac{1}{\sqrt{p^0+p^3}} \begin{pmatrix}
      -p^1 - i p^2 & p^0 + p^3
    \end{pmatrix},  \nn \\
    \angbra{p}^{\alpha} = (\sqrket{p}^{\dot\alpha})^\dagger = \frac{1}{\sqrt{p^0+p^3}} \begin{pmatrix}
      p^0 + p^3 & p^1 - i p^2
    \end{pmatrix}.
\end{gather}
We also use the short-hand notation for the Lorentz invariant spinor contractions:
\begin{equation}
\langle i j \rangle \equiv \epsilon_{\alpha \beta} \lambda_i^\alpha \lambda_j^\beta,
~~~~ [ i j ] \equiv \epsilon_{\alpha \beta} \tilde{\lambda}_i^\alpha \tilde{\lambda}_j^\beta,
\end{equation}
which satisfy the spinor relations
\begin{align}
    \ang{i}{j} = - \ang{j}{i}, \quad \ang{i}{j} = - \ang{j}{i}, \quad \ang{i}{j} \sqr{j}{i} = 2 p_i \cdot p_j = s_{ij}, \quad \angbra{i} \sigma^\mu \sqrket{j} = \sqrbra{j} \bar\sigma^\mu \angket{i}.
\end{align}
The Schouten identities are  given by
\begin{align}
    \ang12 \ang34 + \ang13 \ang42 + \ang14 \ang23 = 0, \qquad  \sqr12 \sqr34 + \sqr13 \sqr42 + \sqr14 \sqr23 = 0 ,
    \end{align}
    and the Fierz relations are given by
    \begin{align}
  & \angbra1 \sigma^\mu \sqrket2 \angbra3 \sigma_\mu \sqrket4 = 2 \ang13 \sqr42, \qquad
    \,\sqrbra1 \bar\sigma^\mu \angket2 \sqrbra3 \bar\sigma_\mu \angket4 = 2 \sqr13 \ang42, \nn\\
  & \sqrbra1 \bar\sigma^\mu \angket2 \sqrbra3 \bar\sigma_\mu \angket4 = 2 \sqr13 \ang42, \qquad  \angbra1 \sigma^\mu \sqrket2 \sqrbra3 \bar\sigma_\mu \angket4 = -2 \ang14 \sqr23.
\end{align}

When using formulae for amplitude cuts, and all amplitudes are defined with ingoing momenta, it's necessary to use spinors with negative four-momenta $-p$ on one side of the cut, which are related to their positive counterparts by:
\begin{align}
     \angket{-p}_\alpha = i \angket{p}_\alpha , \quad
    \sqrket{-p}^{\dot\alpha} = i \sqrket{p}^{\dot\alpha}, \quad
    \angbra{-p}^\alpha = i \angbra{p}^\alpha, \quad
    \sqrbra{-p}_{\dot\alpha} = i \sqrbra{p}_{\dot\alpha}.
\end{align}
Hence we have
\begin{align}
  \angket{p} \sqrbra{-p} = i p \cdot \sigma, \qquad   \sqrket{p} \angbra{-p} = i p \cdot \bar\sigma, \qquad
  \sum_{\lambda=\pm} \epsilon_\lambda^\mu(p) \epsilon_{-\lambda}^\nu(-p) = g^{\mu\nu}.
\end{align}
In order to get the correct propagator in the uncut diagram, we need include a factor $(i)^{n_{\psi}}$ (where $n_{\psi}$ is the number of fermion in the cut) in the master formula showed in Eq.~\eqref{eq:IRfiniteMasterFormula}. This is consistent with the convention that the particles with negative momentum are in the right side of the cut.

\section{Standard Model amplitudes}
\label{app:SMamplitudes}
\subsection{3-point}
\label{app:SM3point}
The kinematic parts of the 3-point amplitudes are fixed only by the little group and locality. In the case of Yukawa interactions, we have that $\mathcal{A}(\psi^+\psi^+H^{(\dagger)})=[12]$ or $\mathcal{A}(\psi^-\psi^-H^{(\dagger)})=\la 12\ra$. Dressing the amplitudes with their flavour and gauge indices leads to
\begin{align}
&\mathcal{A}_{SM}(Q_{p i a}^+ u^{+q b} H_j^\dagger) = [Y_u]_{p}^q \,[12] \,\epsilon_{ij}\delta_a^b, \quad ~\mathcal{A}_{SM}(Q_{p i a}^+ d^{+q b} H^j) = [Y_d]_{p}^q\, [12] \,\delta_i^j\delta_a^b,  \\
&\mathcal{A}_{SM}(Q^{-p i a} u_{q b}^- H^j) = [Y_u^\dagger]_q^p\, \langle 12\rangle\, \epsilon^{ij}\delta^a_b, \quad\mathcal{A}_{SM}(Q^{-p i a} u_{q b}^- H_j^\dagger) = [Y_d^\dagger]_q^p\, \langle 12\rangle\, \delta^i_j \delta^a_b,
\end{align}
and
\begin{align}
\mathcal{A}_{SM}(L_{p i}^+ e^{+q} H^j) = [Y_e]_{p}^q\, [12] \,\delta_i^j, \quad\mathcal{A}_{SM}(L^{-p i} e_{q}^- H_j^\dagger) = [Y_e^\dagger]_q^p\, \langle 12\rangle\, \delta^i_j,
\end{align}
for quarks and leptons, where $p,q$ are flavour indices, $i,j$ are $SU(2)_L$ indices, and $a,b$ are $SU(3)_c$ indices. For gauge interactions, the relevant amplitudes are $\cA({\psi^+\psi^-V^-}) =\la 23 \ra^2/\la 12 \ra$, $\cA({\psi^+\psi^-V^+}) =[13]^2/[12]$ and $\cA(HH^{\dagger}V^-)=\la 13\ra \la 23\ra/\la 12 \ra$. Adding the colour/flavor indices leads to (where $y_x$ is the hypercharge of field $x$)
\begin{align}
 & \mathcal{A}_{\rm SM}(e^{+p} e^-_q B^+) =  \sqrt{2} g_1 y_e \frac{{\sqr13}^2}{\sqr12} \delta^p_q, \qquad~~~~~
  \mathcal{A}_{\rm SM}(e^{+p} e^-_q B^-) =  -\sqrt{2} g_1 y_e \frac{{\ang23}^2}{\ang12} \delta^p_q ,\\
 & \mathcal{A}_{\rm SM}(L^{-pi} L^+_{qj} B^+) =  -\sqrt{2} g_1 y_L \frac{{\sqr23}^2}{\sqr12} \delta^i_j\delta^p_q, \quad
  \mathcal{A}_{\rm SM}(L^{-pi} L^+_{qj} B^-) =  \sqrt{2} g_1 y_L \frac{{\ang13}^2}{\ang12} \delta^i_j\delta^p_q,  \\
  &\mathcal{A}_{\rm SM}(H^{i} H^\dagger_{j} B^+) =  -\sqrt{2} g_1 y_H \frac{\sqr13 \sqr23}{\sqr12} \delta^i_j, \quad~
  \mathcal{A}_{\rm SM}(H^{i} H^\dagger_{j} B^-) =  -\sqrt{2} g_1 y_H \frac{\ang13 \ang23}{\ang12} \delta^i_j,
\end{align} and mutatis mutandis for the $W$ and $G$ interactions.

\subsection{4-point}
\label{app:SM4point}

The 4-point amplitudes can be obtained by gluing 3-point amplitudes together and requiring locality and unitarity of the full amplitude. Parity conjugate amplitudes can be obtained replacing angle $\leftrightarrow$ square-brackets. Including the gauge and flavour indices we can write
\begin{align}
\mathcal{A}_{\rm SM}(u^{+pa} u^-_{qb} H^i H_j^\dagger) &= [Y_u^\dagger Y_u]^p_q \delta_j^i\delta_b^a \frac{[ 31 ]}{ [32 ]} \nn \\
&=\phs{ \frac{J^1(12)J^2(34)}{u} } \cdot \gge{\delta^i_j \delta^a_b} \cdot \flv{-[Y_u^\dagger Y_u]^p_q} , \\
\mathcal{A}_{\rm SM}(d^{+pa} d^-_{qb} H^i H_j^\dagger) &= -[Y_d^\dagger Y_d]^p_q \delta^i_j\delta_a^b \frac{\langle 23 \rangle}{\langle 13 \rangle}  \nn \\
&= \phs{ \frac{J^1(12)J^2(34)}{t} } \cdot \gge{\delta^i_j \delta^a_b} \cdot \flv{[Y_d^\dagger Y_d]^p_q} \, \\
\mathcal{A}_{\rm SM}(u^{+pa} d^-_{qb}  H_i^\dagger H_j^\dagger) &= [Y_d^\dagger Y_u]^p_q \epsilon_{ij}\delta_a^b\left( \frac{ [31] }{ [32] }+ \frac{ \ang23 }{ \ang13 }\right) \nn \\
&= \phs{J^1(12)J^2(34) \left( \frac{1}{t}+\frac{1}{u}\right)} \cdot \gge{\epsilon_{ij}\delta^a_b} \cdot \flv{-[Y_d^\dagger Y_u]^p_q} \,, \\
\mathcal{A}_{\rm SM}(e^{+p} e^-_{q} H^i H_j^\dagger) &= -[Y_e^\dagger Y_e]^p_q \delta^i_j \frac{\langle 23 \rangle}{\langle 13 \rangle} \nn\\
&= \phs{ \frac{J^1(12)J^2(34)}{t} } \cdot \gge{\delta^i_j} \cdot \flv{[Y_e^\dagger Y_e]^p_q} \,, \\
\mathcal{A}_{SM}(L^{-p i}L_{q j}^+  H^k H_l^\dagger)&= [Y_e Y_e^\dagger]_q^p \delta_j^k \delta_l^i \frac{ \ang 31 }{ \ang 32 }\nn\\
&= \phs{ \frac{J^3(12)J^2(34)}{u} } \cdot \gge{\delta^k_j \delta^i_l} \cdot \flv{- [Y_e Y_e^\dagger]^p_q} \,, \\
\mathcal{A}_{SM}( Q^{-p i a} Q_{q j b}^+ H^k H_l^\dagger)
  &= [Y_d Y_d^\dagger]^p_q \delta_j^k \delta_l^i \delta_b^a\frac{ \ang 31 }{ \ang 32 }-[Y_u Y_u^\dagger]^p_q \epsilon_{jl} \epsilon^{ik}\delta_b^a \frac{ [23] }{ [13] } \, \\
  &= \phs{ \frac{J^3(12)J^2(34)}{u} } \cdot \gge{\delta^i_l \delta^k_j \delta^a_b} \cdot \flv{- [Y_d Y_d^\dagger]^p_q}  + \phs{ \frac{J^3(12)J^2(34)}{t} } \cdot \gge{\epsilon^{ik} \epsilon_{jl} \delta^a_b} \cdot \flv{[Y_u Y_u^\dagger]^p_q} \, \nn,
\end{align}
and
\begin{align}
\mathcal{A}_{\rm SM}(L^{- p i} L^+_{q j} e^{+r} e^-_{s}) &= [Y_e]^r_q [Y_e^\dagger]_s^p \delta^i_j \frac{[32]}{[41]} \,\nn \\
&= \phs{\frac{J^3(12)J^1(34)}{ u}} \cdot \gge{\delta^i_j} \cdot \flv{-[Y_e]^r_q [Y_e^\dagger]_s^p} \,, \\
\mathcal{A}_{\rm SM}(Q^{- p i a} Q^+_{q j b} u^{+r c} u^-_{s d})&= [Y_u]^r_q [Y_u^\dagger]_s^p \delta^i_j \delta^a_d \delta^c_b \frac{[32]}{[14]} \, \nn \\
&= \phs{\frac{J^3(12)J^1(34)}{ u}} \cdot \gge{\delta^i_j \delta^a_b \delta^c_d} \cdot \flv{- [Y_u]^j_i [Y_u^\dagger]_l^k} \, , \\
\mathcal{A}_{\rm SM}(Q^{- pia } Q^+_{qjb} d^{+rc} d^-_{sd})&= [Y_d]^r_q [Y_d^\dagger]_s^p \delta^i_j \delta^a_d \delta^c_b \frac{[32]}{[14]} \, \nn \\
&= \phs{\frac{J^3(12)J^1(34)}{ u}} \gge{\delta^i_j \delta^a_b \delta^c_d} \cdot \flv{-[Y_d]^j_i [Y_d^\dagger]_l^k} \,  .
\end{align}
The $\eps$ matrices of $SU(2)_L$ can be related back to $\delta$ and $\lambda$ by the Fierz relation
\begin{equation}
    \eps_{ij} \eps^{kl} = \delta^k_i \delta^l_j - \delta^k_j \delta^l_i = \frac12 \delta^k_i \delta^l_j - \frac12 [\sigma^I]^k_i [\sigma^I]^l_j.
\end{equation}
For the gauge amplitudes we proceed analogously. We have, for distinct fermion species $\psi$ and $\chi$, schematically
\begin{align}
  \amp_\text{\rm SM}\left( \psi^+ \bar \psi^- \chi^+ \bar \chi^- \right) &= \frac{2 \sqr13 \ang42}{s}  \sum_\text{gauge groups} g^2 T \otimes T\,,  \\
  \amp_\text{\rm SM}\left( \psi^+ \bar \psi^- \chi^- \bar \chi^+ \right) &=\frac{2 \sqr14 \ang32}{s}  \sum_\text{gauge groups} g^2 T \otimes T \,,\\
  \amp_\text{\rm SM}\left( \psi^- \bar \psi^+ \chi^- \bar \chi^+ \right) &=\frac{2 \ang13 \sqr42}{s}  \sum_\text{gauge groups} g^2 T \otimes T\,.  
\end{align}
with respective examples (where $y$ below is a hypercharge generator),
\begin{align}
 & \amp_\text{\rm SM}\left( d^{+pa} d^{-}_{qb} u^{+rc} u^{-}_{sd} \right) = \frac{2 \sqr13 \ang42}{s} \delta^p_q \delta^r_s \left( g_1^2 y_u y_d \delta^a_b \delta^c_d + g_3^2 \frac12 (\lambda^A)^a_b \frac12 (\lambda^A)^c_d \right), \\
 & \amp_\text{\rm SM}\left( d^{+pa} d^{-}_{qb} L^{-ri} L^{+}_{sj} \right)  = \frac{2 \sqr14 \ang32}{s} \delta^p_q \delta^r_s \left( g_1^2 y_d y_L  \delta^a_b \delta^i_j \right) , \\
 & \amp_\text{\rm SM}\left( Q^{-pia} Q^{+}_{qjb} L^{-rk} L^{+}_{sl} \right) =  \frac{2 \ang13 \sqr42}{s} \delta^p_q \delta^r_s \delta^a_b \left( g_1^2 y_Q y_L \delta^i_j \delta^k_l + g_2^2 \frac12 (\sigma^I)^i_j \frac12 (\sigma^I)^k_l \right)    .
\end{align}
For identical fermion currents we need to add the crossed diagram, e.g.,
\begin{align}
&  \amp_\text{\rm SM}\left( Q^{-pia} Q^{+}_{qjb} Q^{-rkc} Q^{+}_{sld} \right) = \\
  &=\frac{2 \ang13 \sqr42}{s} \delta^p_q \delta^r_s \left( g_1^2 y_Q y_Q \delta^a_b \delta^c_d  \delta^i_j \delta^k_l + g_2^2 \delta^a_b \delta^c_d  \frac12 (\sigma^I)^i_j \frac12 (\sigma^I)^k_l + g_3^2 \frac12 (\lambda^A)^a_b \frac12 (\lambda^A)^c_d  \delta^i_j \delta^k_l \right)
                                        \nn \\
                                        &- \text{crossing } \{1,a,i,p\} \leftrightarrow \{3, c,k,r\}\,. \nn
\end{align}
For Higgs-fermion gauge pieces we have schematically (for right handed and left handed fermion particles respectively)
\begin{align}
  \amp_\text{\rm SM}\left(H H^\dagger  \psi^+ \bar \psi^- \right) &= \frac{2 \ang41 \sqr13}{s}  \sum_\text{gauge groups} g^2 T \otimes T\,,  \\
  \amp_\text{\rm SM}\left(H H^\dagger \psi^- \bar \psi^+ \right) &=\frac{2 \sqr41 \ang13}{s}  \sum_\text{gauge groups} g^2 T \otimes T\,,
\end{align}
e.g.\
\begin{align}
  \amp_\text{\rm SM}\left( H^i H^\dagger_j e^{+p} e^{-}_q \right) = \frac{2 \ang41 \sqr13}{s} \left( g_1^2 y_H y_e \delta^i_j \delta^p_q \right) .
\end{align}
For the four Higgs gauge amplitude
\begin{align}
  \amp_\text{\rm SM}\left( H^i H^\dagger_j H^k H^\dagger_l \right) =& \frac{t-u}{s} \left( g_1^2 y_H^2 \delta^i_j \delta^k_l + g_2^2 \frac14 (\sigma^I)^i_j (\sigma^I)^k_l \right) \nn \\
  +& \frac{t-s}{u} \left( g_1^2 y_H^2 \delta^i_l \delta^k_j + g_2^2 \frac14 (\sigma^I)^i_l (\sigma^I)^k_j \right). 
\end{align}
For the four Higgs quartic amplitude
\begin{align}
\amp_\text{\rm SM}\left( H^i H^\dagger_j H^k H^\dagger_l \right)=-2\lambda \cdot \delta^{(i}_{(j} \delta^{k)}_{l)},
\end{align}
where $\delta^{(i}_{(j} \delta^{k)}_{l)} \equiv \frac12 \left( \delta^{i}_{j} \delta^{k}_{l} + \delta^{k}_{j} \delta^{i}_{l} \right)$.

\section{Dimension-6 SMEFT amplitudes }
\label{app:basis}

The SMEFT Lagrangian is defined schematically as\footnote{Note that fields with downstairs indices annihilate incoming particles with upstairs indices, e.g., $H)j \ket{H^i} \propto \delta^i_j$.}
\begin{align}
  \lag &= -\frac14 \sum_{F=B,W,G} FF + \sum_{f=Q,u,d,L,e} i \bar f \slashed{D} f + |DH|^2 -V(H)  \\
        & + \left( -[Y_u]^q_p \epsilon_{ij} \bar Q^{pi} H^{\dagger j} u_q  - [Y_d]^q_p \delta^j_i \bar Q^{pi} H_j d_q - [Y_e]^q_p \delta^j_i \bar L^{pi} H_j e_q + \text{h.c.} \right) + \sum c_i \mathcal{O}_i \nn ,
\end{align}
where $D_\mu Q = \partial_\mu Q - i g_3 G^A_\mu T^A Q - i g_2 W^I_\mu \tau^I Q - i g_1 B_\mu \frac16 Q$ and  $T^A = \frac12 \lambda^A, \tau^I= \frac12 \sigma^I$ are the $SU(3)$ and $SU(2)$ generators respectively. We bypass the Lagrangian and operators and write directly the amplitudes at order $1/\Lambda^2$. For convenience, we display the relation between the on-shell amplitudes and for the (4,0) set of operators $\mathcal{O}_i$ in the \emph{Warsaw basis}, i.e. four fields and total helicity zero, in Table \ref{tab:basis}.
We define symmetric and antisymmetric combinations of gauge tensors:
  \begin{align}
    (\delta\delta) \equiv \delta^{(a}_{(b} \delta^{c)}_{d)} \equiv  \frac12 \left( \delta^{a}_{b} \delta^{c}_{d} + \delta^{a}_{d} \delta^{c}_{b} \right), \qquad
   [\delta\delta] \equiv  \delta^{[a}_{[b}  \delta^{c]}_{d]} \equiv \frac12 \left( \delta^{a}_{b} \delta^{c}_{d} - \delta^{a}_{d} \delta^{c}_{b} \right),
  \end{align}
  and similarly for the coefficients carrying the flavour indices:
\begin{align}
(c) \equiv c_{(qs)}^{(pr)} = \frac 14 \left(c_{qs}^{pr} +c_{sq}^{pr} +c_{qs}^{rp} +   c_{sq}^{rp} \right) , \qquad [c] \equiv c_{[qs]}^{[pr]} = \frac14 \left( c_{qs}^{pr} -c_{sq}^{pr} -c_{qs}^{rp} +   c_{sq}^{rp} \right). 
\end{align}
\begin{table}[]
    \centering
    \scalebox{0.790}{
    \begin{tabular}{|L|L|}
\hline
\text{Operator} & \text{Amplitude} \\
\hline\hline
\phi^4D^2 &  \\
\hline \hline
\zerof \, \op_{HD+} = 3 \zerof \, (\op_{HD} + \frac12 \op_{H\Box})	&	\amp_{HD+}\left( H^{i} \, H^\dagger_{j} \,  H^{k}  \, H^\dagger_{l} \right) = \scalarsym \cdot \sulsym \cdot \zerof \\
\hline
\zerof \, \op_{HD-} =  \zerof \, (\op_{HD} - \frac12 \op_{H\Box})	&	\amp_{HD-}\left( H^{i} \,  H^\dagger_{j}  \, H^{k}  \, H^\dagger_{l} \right) = \scalarantisym \cdot \sulantisym \cdot \zerof \\
\hline
  \hline
\psi \bar{\psi}\phi^2D &  \\
\hline\hline
\twof \, [\op_{HL(1)}]_p^q & \amp_{HL(1)}\left(H^{i} \, H^{\dagger}_{j} \, L^{-pk} \, L^{+}_{ql}\right) = \scalarLH \cdot \suldels \cdot \twof \\
\hline
\twof \, [\op_{HL(3)}]_p^q & \amp_{HL(3)}\left(H^{i} \, H^{\dagger}_{j} \, L^{-pk} \, L^{+}_{ql}\right) = \scalarLH \cdot \sulsigmas \cdot \twof \\
\hline
\twof \, [\op_{HQ(1)}]_p^q & \amp_{HQ(1)}\left(H^{i} \, H^{\dagger}_{j} \, Q^{-pak} \, Q^{+}_{qbl}\right) = \scalarLH \cdot \colourdel \, \suldels \cdot \twof \\
\hline
\twof \, [\op_{HQ(3)}]_p^q & \amp_{HQ(3)}\left(H^{i} \, H^{\dagger}_{j} \, Q^{-pak} \, Q^{+}_{qbl}\right) = \scalarLH \cdot \colourdel \, \sulsigmas \cdot \twof \\
\hline
  \twof \, [\op_{He}]_p^q &	\amp_{He}\left(H^{i} \, H^{\dagger}_{j} \, e^{+p} \, e^{-}_{q}\right) = \scalarRH \cdot \suldel \cdot \twof \\
  \hline
  \twof \, [\op_{Hu}]_p^q &	\amp_{Hu}\left(H^{i} \, H^{\dagger}_{j} \, u^{+pa}  \, u^{-}_{qb} \right) = \scalarRH \cdot \colourdel \, \suldel \cdot \twof \\
  \hline
  \twof \, [\op_{Hd}]_p^q &	\amp_{Hd}\left(H^{i} \, H^{\dagger}_{j} \, d^{+pa}  \, d^{-}_{qb} \right) = \scalarRH \cdot \colourdel \, \suldel \cdot \twof \\
  \hline
  \twof \, [\op_{Hud}]_p^q	&	\amp_{Hud}\left( H^{i}  \, H^{j} \,  d^{+pa}  \, u^{-}_{qb}\right) = \scalarRH \cdot \colourdel \, \sulepsu \cdot \twof \\
  \hline
  \twof \, [\op_{Hud}^\dagger]_p^q	&	\amp_{Hud\dagger}\left( H^\dagger_{i}  \, H^\dagger_{j}  \, u^{+pa} \,  d^{-}_{qb}\right) = \scalarRH \cdot \colourdel \,  \sulepsl \cdot \minusf \twof \\ \hline
  \hline
\psi^2\bar{\psi}^2 &  \\
\hline\hline
\foursym \, [\op_{LL}]_{(pr)}^{(qs)} 	&	\amp_{LL+}\left(L^{-pi} \,  L^+_{qj} \,  L^{-rk} \, L^+_{sl}\right) = \LHantisym \cdot \sulsym \cdot \foursym \\ \hline
\fourantisym \, [\op_{LL}]_{[pr]}^{[qs]}	&	\amp_{LL-}\left(L^{-pi} \,  L^+_{qj}  \, L^{-rk} \, L^+_{sl}\right) = \LHantisym \cdot \sulantisym \cdot \fourantisym \\ \hline
\foursym \, \left( [\op_{QQ+}] = \frac34 [\op_{QQ(1)}] + \frac14 [\op_{QQ(3)}] \right)_{(pr)}^{(qs)}	&	\amp_{QQ++}\left(Q^{-pi a}  \, Q^+_{qj b} \,  Q^{-rk c} \,  Q^+_{sl d}\right) = \LHantisym \cdot \coloursym \, \sulsym \cdot \foursym \\ \hline
\fourantisym \, \left( [\op_{QQ+}] = \frac34 [\op_{QQ(1)}] + \frac14 [\op_{QQ(3)}] \right)_{[pr]}^{[qs]} &	\amp_{QQ+-}\left(Q^{-pi a} \,  Q^+_{qj b} \,  Q^{-rk c} \,  Q^+_{sl d}\right) = \LHantisym \cdot \colourantisym \, \sulsym \cdot \fourantisym \\ \hline
\foursym \, \left( [\op_{QQ-}] = \frac14 [\op_{QQ(1)}] - \frac14 [\op_{QQ(3)}] \right)_{(pr)}^{(qs)}	&	\amp_{QQ-+}\left(Q^{-pi a} \,  Q^+_{qj b} \,  Q^{-rk c}  \, Q^+_{sl d}\right) = \LHantisym \cdot \colourantisym \, \sulantisym \cdot \foursym \\ \hline
\fourantisym \, \left( [\op_{QQ-}] = \frac14 [\op_{QQ(1)}] - \frac14 [\op_{QQ(3)}] \right)_{[pr]}^{[qs]} &	\amp_{QQ--}\left(Q^{-pi a}  \, Q^+_{qj b} \,  Q^{-rk c} \,  Q^+_{sl d}\right) = \LHantisym \cdot \coloursym \, \sulantisym \cdot \fourantisym \\
\hline
\fourf \, [\op_{LQ(1)}]_{pr}^{qs} & \amp_{LQ(1)}\left( L^{-pi}  \, L^+_{qj}  \, Q^{-rk a} \,  Q^+_{s l b} \right) = \LHLH \cdot \colourdel \suldels \cdot \fourf \\ \hline
\fourf \, [\op_{LQ(3)}]_{pr}^{qs} & \amp_{LQ(3)}\left( L^{-pi}  \, L^+_{qj}  \, Q^{-rk a} \,  Q^+_{s l b} \right) = \LHLH \cdot \colourdel \sulsigmas \cdot \fourf \\ 
\hline\hline
\fourf \, [\op_{Le}]_{pr}^{qs}	&	\amp_{Le}\left(L^{-pi} \,  L^+_{qj} \, e^{+r}  \, e^-_{s}\right) = \LHRH \cdot \suldel \cdot \fourf \\ \hline
\fourf \, [\op_{Lu}]_{pr}^{qs} &	\amp_{Lu}\left(L^{-pi}  \, L^+_{qj}  \, u^{+r a} \, u^-_{s b}\right) = \LHRH \cdot \colourdel \, \suldel \cdot \fourf \\ \hline
\fourf \, [\op_{Ld}]_{pr}^{qs} &	\amp_{Ld}\left(L^{-pi}  \, L^+_{qj}  \, d^{+r a} \, d^-_{s b}\right) = \LHRH \cdot \colourdel \, \suldel \cdot \fourf \\ \hline
\fourf \, [\op_{Qe}]_{pr}^{qs}	&	\amp_{Qe}\left(Q^{-pi a}  \, Q^+_{qj b} \, e^{+r} \,  e^-_{s}\right) = \LHRH \cdot \colourdel \, \suldel \cdot \fourf \\ \hline
\fourf \, [\op_{Qu(1)}]_{pr}^{qs}	&	\amp_{Qu(1)}\left(Q^{-pi a} \,  Q^+_{qj b}  \, u^{+r c}  \, u^-_{s d}\right) = \LHRH \cdot \colourdels \, \suldel \cdot \fourf \\ \hline
\fourf \, [\op_{Qu(8)}]_{pr}^{qs}	&	\amp_{Qu(8)}\left(Q^{-pi a}  \, Q^+_{qj b} \,  u^{+r c}  \, u^-_{s d}\right) = \LHRH \cdot \colourlambdas \, \suldel \cdot \fourf \\ \hline
\fourf \, [\op_{Qd(1)}]_{pr}^{qs}	&	\amp_{Qd(1)}\left(Q^{-pi a}  \, Q^+_{qj b}  \, d^{+r c}  \, d^-_{s d}\right) = \LHRH \cdot \colourdels \, \suldel \cdot \fourf \\ \hline
\fourf \, [\op_{Qd(8)}]_{pr}^{qs}	&	\amp_{Qd(8)}\left(Q^{-pi a}  \, Q^+_{qj b} \,  d^{+r c} \,  d^-_{s d}\right) = \LHRH \cdot \colourlambdas \, \suldel \cdot \fourf \\ \hline
  \hline
\foursym \, [\op_{ee}]_{(pr)}^{(qs)} &	\amp_{ee}\left(e^{+p} \, e^-_{q} \, e^{+r} \, e^-_{s}\right) = \RHantisym \cdot \foursym \\ \hline
\foursym \, [\op_{uu}]_{(pr)}^{(qs)}&	\amp_{uu+}\left(u^{+p a} \, u^-_{q b} \, u^{+r c} \, u^-_{s d}\right) = \RHantisym \cdot \coloursym \cdot \foursym \\ \hline
\fourantisym \, [\op_{uu}]_{[pr]}^{[qs]}	&	\amp_{uu-}\left(u^{+p a} \, u^-_{q b} \, u^{+r c} \, u^-_{s d}\right) = \RHantisym \cdot \colourantisym \cdot \fourantisym \\ \hline
 \foursym \, [\op_{dd}]_{(pr)}^{(qs)} &	\amp_{dd+}\left(d^{+p a} \, d^-_{q b} \, d^{+r c} \, d^-_{s d}\right) = \RHantisym \cdot \coloursym \cdot \foursym \\ \hline
 \fourantisym \, [\op_{dd}]_{[pr]}^{[qs]} &	\amp_{dd-}\left(d^{+p a} \, d^-_{q b} \, d^{+r c} \, d^-_{s d}\right) = \RHantisym \cdot \colourantisym \cdot \fourantisym \\ \hline
\fourf \, [\op_{eu}]_{pr}^{qs} &	\amp_{eu}\left(e^{+p} \, e^-_{q} \, u^{+r a} \, u^-_{s b}\right) = \RHRH \cdot \colourdel \cdot \fourf \\ \hline
\fourf \, [\op_{ed}]_{pr}^{qs} &	\amp_{ed}\left(e^{+p} \, e^-_{q} \, d^{+r a} \, d^-_{s b}\right) = \RHRH \cdot \colourdel \cdot \fourf \\ \hline
\fourf \, [\op_{ud(1)}]_{pr}^{qs} &	\amp_{ud(1)}\left(u^{+p a} \, u^-_{q b} \, d^{+r c} \, d^-_{s d}\right) = \RHRH \cdot \colourdels \cdot \fourf \\ \hline
\fourf \, [\op_{ud(8)}]_{pr}^{qs} &	\amp_{ud(8)}\left(u^{+p a} \, u^-_{q b} \, d^{+r c} \, d^-_{s d}\right) = \RHRH \cdot \colourlambdas \cdot \fourf \\ \hline
    \end{tabular}}
    \caption{Correspondence between $(4,0)$ dimension 6 operators in the Warsaw basis and on-shell amplitudes. We represent in\phs{green} the kinematic factors, in\gge{orange} the gauge factors and in\flv{purple} the flavour factors. The symbols $(\delta \delta)$, $(c)$ and $[\delta \delta]$, $[c]$ stand for symmetric and antisymmetric indices combinations, respectively.}
    \label{tab:basis}
\end{table}

\section{Collinear anomalous dimensions}
\label{app:collinear}

We collect here collinear anomalous dimensions for the SM fields~\cite{Jenkins:2013zja,Jenkins:2013wua,Alonso:2013hga,Jiang:2020mhe,AccettulliHuber:2021uoa}:
\begin{align}
&\gamma(H)=\left(\phs{1}\cdot\gge{N_c}\cdot \flv{\text{Tr} [Y_u Y_u^\dagger+Y_d Y_d^\dagger]}+\phs{1}\cdot\gge{1}\cdot \flv{\text{Tr} [Y_e Y_e^\dagger]}\phs{-4}\cdot\gge{1}\cdot\flv{g_{1HH}^2}\phs{-4}\cdot\gge{C_2(2)}\cdot\flv{g_2^2}\right),\\
&[\gamma(L)]^p_q=\left(\phs{\frac12} \cdot\gge{1}\cdot \flv{ [Y_e Y_e^\dagger]^p_q} \phs{-3}\cdot\gge{C_2(2)}\cdot\flv{g_2^2 \delta^p_q}  \phs{-3}\cdot\gge{1}\cdot\flv{g_{1LL}^2 \delta^p_q} \right),\\
&[\gamma(e)]^p_q=\left(\phs{\frac12} \cdot\gge{N_L}\cdot \flv{[Y_e^\dagger Y_e]^p_q} \phs{-3}\cdot\gge{1}\cdot\flv{g_{1ee}^2 \delta^p_q}\right),\\
&[\gamma(Q)]^p_q= \!\!\left(\phs{\frac12} \cdot\gge{1}\cdot \flv{[Y_u Y_u^\dagger+Y_d Y_d^\dagger]^p_q}\phs{-3}\cdot\gge{C_2(N_c)}\cdot\flv{g_3^2 \delta^p_q}  \phs{-3}\cdot\gge{C_2(2)}\cdot\flv{g_2^2 \delta^p_q}  \phs{-3}\cdot\gge{1}\cdot\flv{g_{1QQ}^2 \delta^p_q}\right)\!,\!\\
&[\gamma(d)]^p_q=\left(\phs{\frac12} \cdot\gge{N_L}\cdot \flv{[Y_d^\dagger Y_d]^p_q}\phs{-3}\cdot\gge{C_2(N_c)}\cdot\flv{g_3^2 \delta^p_q}  \phs{-3}\cdot\gge{1}\cdot\flv{g_{1dd}^2 \delta^p_q}\right),\\
&[\gamma(u)]^p_q=\left(\phs{\frac12} \cdot\gge{N_L}\cdot \flv{[Y_u^\dagger Y_u]^p_q}\phs{-3}\cdot\gge{C_2(N_c)}\cdot\flv{g_3^2 \delta^p_q}  \phs{-3}\cdot\gge{1}\cdot\flv{g_{1uu}^2 \delta^p_q}\right),
\end{align}
where $N_L=2$, $N_c=3$, and $C_2(N)=\frac{N^2-1}{2N}$.\footnote{Note that the terms (2.15) of \cite{Jiang:2020mhe} are the coefficients of $\frac{1}{\epsilon}$ poles in the collinear pieces, which need to be multiplied by $-2$ to get the anomalous dimension.}

\section{Phase space factors}
\label{app:summaryPhaseSpace}

Individual currents are denoted as follows:
\begin{align}
  \mathord{\begin{tikzpicture}[baseline=-0.65ex]
    \draw[thick,dotted,blue] (-1,1) -- (-\currsep,0) -- (-1,-1);
    \currentmarker{-\currsep,0}
  \end{tikzpicture}}
  \, , \,
  \mathord{\begin{tikzpicture}[baseline=-0.65ex]
    \draw[thick,dotted,red] (-1,1) -- (-\currsep,0) -- (-1,-1);
    \currentmarker{-\currsep,0}
  \end{tikzpicture}} = \text{Any current} 
  \mathord{\begin{tikzpicture}[baseline=-0.65ex]
    \draw[thick,dashed] (-1,1) node[left] {$1$} -- (-\currsep,0) -- (-1,-1) node[left] {$2$};
    \currentmarker{-\currsep,0}
  \end{tikzpicture}}
  = \text{Higgs current} 
  \mathord{\begin{tikzpicture}[baseline=-0.65ex]
    \draw[thick] (-1,1) node[left] {$1$} -- (-\currsep,0) -- (-1,-1) node[left] {$2$};
    \currentmarker{-\currsep,0}
  \end{tikzpicture}}
  = \text{Fermion current}
\end{align}
 where $1,3$ always label particles and $2,4$ always label antiparticles and we use the convention of all incoming particles. The cut is an integral over two body phase space defined as:
\begin{equation}
  \cut = - 2 (i)^{n_{\psi}} \int \frac{\dd \Pi_2}{2^\rho \cdot 4\pi}  \mathcal{A}_6 \times \mathcal{A}_{\rm SM}\,,
\end{equation}
$n_{\psi}$ is the number of fermion lines cut and $\rho=0(1)$ if the particles being cut are (in)distinguishable. The subtracted cut removes the contribution of the divergent triangle diagram:
\begin{equation}
  \cutsub =  - 2 (i)^{n_{\psi}} \int \frac{\dd \Pi_2}{2^\rho \cdot 4\pi}  \left[ \mathcal{A}_6 \times \mathcal{A}_{\rm SM} - k \frac{1}{s_{34}} \frac{1}{\st^2} \right]\,,
\end{equation}
where the coefficient $k$ is such to make the overall integral finite.

\subsection{IR finite}
\label{app:IRfinitePhase}

$\bullet$ $s$-channel gauge:
\begin{align}
\mathord{\begin{tikzpicture}[baseline=-0.65ex]
  \draw[thick,dotted,blue] (-1,1) -- (-\currsep,0) -- (-1,-1);
  \draw[thick] (1,1) -- (\currsep,0) -- (1,-1);
  \currentmarker{\currsep,0}
  \currentmarker{-\currsep,0}
\end{tikzpicture}}
\cut
\mathord{\begin{tikzpicture}[baseline=-0.65ex]
  \draw[thick] (-1,1) -- (-\threepointsep,0) -- (-1,-1);
  \draw[thick,dotted,red] (1,-1) -- (\threepointsep,0) -- (1,1);
  \draw[thick,decorate,decoration={snake}] (\threepointsep,0) -- (-\threepointsep,0);
\end{tikzpicture}}
 &= \frac43
\mathord{\begin{tikzpicture}[baseline=-0.65ex]
  \draw[thick,dotted,blue] (-1,1) -- (-\currsep,0) -- (-1,-1);
  \draw[thick,dotted,red] (1,1) -- (\currsep,0) -- (1,-1);
  \currentmarker{\currsep,0}
  \currentmarker{-\currsep,0}
\end{tikzpicture}} \\
\mathord{\begin{tikzpicture}[baseline=-0.65ex]
  \draw[thick,dotted,blue] (-1,1) -- (-\currsep,0) -- (-1,-1);
  \draw[thick,dashed] (1,1) -- (\currsep,0) -- (1,-1);
  \currentmarker{\currsep,0}
  \currentmarker{-\currsep,0}
\end{tikzpicture}}
\cut
\mathord{\begin{tikzpicture}[baseline=-0.65ex]
  \draw[thick,dashed] (-1,1) -- (-\threepointsep,0) -- (-1,-1);
  \draw[thick,dotted,red] (1,-1) -- (\threepointsep,0) -- (1,1);
  \draw[thick,decorate,decoration={snake}] (\threepointsep,0) -- (-\threepointsep,0);
\end{tikzpicture}}
 &= \frac23
\mathord{\begin{tikzpicture}[baseline=-0.65ex]
  \draw[thick,dotted,blue] (-1,1) -- (-\currsep,0) -- (-1,-1);
  \draw[thick,dotted,red] (1,1) -- (\currsep,0) -- (1,-1);
  \currentmarker{\currsep,0}
  \currentmarker{-\currsep,0}
\end{tikzpicture}} 
\end{align}
\begin{align}
\mathord{\begin{tikzpicture}[baseline=-0.65ex]
  \draw[thick,dashed] (-1,1) -- (-\currsep,0) -- (-1,-1);
  \draw[thick,dashed] (1,1) -- (\currsep,0) -- (1,-1);
  \currentmarker{\currsep,0}
  \currentmarker{-\currsep,0}
  \node at (-\bracketsep,0) {$($};
  \node at (\bracketsep,0) {$)$};
\end{tikzpicture}}
\cut
\mathord{\begin{tikzpicture}[baseline=-0.65ex]
  \draw[thick,dashed] (-1,1) -- (-\threepointsep,0) -- (-1,-1);
  \draw[thick,dotted,red] (1,-1) -- (\threepointsep,0) -- (1,1);
  \draw[thick,decorate,decoration={snake}] (\threepointsep,0) -- (-\threepointsep,0);
\end{tikzpicture}}
 &=  2
\mathord{\begin{tikzpicture}[baseline=-0.65ex]
  \draw[thick,dashed] (-1,1) -- (-\currsep,0) -- (-1,-1);
  \draw[thick,dotted,red] (1,1) -- (\currsep,0) -- (1,-1);
  \currentmarker{\currsep,0}
  \currentmarker{-\currsep,0}
\end{tikzpicture}} \\
\mathord{\begin{tikzpicture}[baseline=-0.65ex]
  \draw[thick,dashed] (-1,1) -- (-\currsep,0) -- (-1,-1);
  \draw[thick,dashed] (1,1) -- (\currsep,0) -- (1,-1);
  \currentmarker{\currsep,0}
  \currentmarker{-\currsep,0}
  \node at (-\bracketsep,0) {$[$};
  \node at (\bracketsep,0) {$]$};
\end{tikzpicture}}
\cut
\mathord{\begin{tikzpicture}[baseline=-0.65ex]
  \draw[thick,dashed] (-1,1) -- (-\threepointsep,0) -- (-1,-1);
  \draw[thick,dotted,red] (1,-1) -- (\threepointsep,0) -- (1,1);
  \draw[thick,decorate,decoration={snake}] (\threepointsep,0) -- (-\threepointsep,0);
\end{tikzpicture}}
 &= \frac23
\mathord{\begin{tikzpicture}[baseline=-0.65ex]
  \draw[thick,dashed] (-1,1) -- (-\currsep,0) -- (-1,-1);
  \draw[thick,dotted,red] (1,1) -- (\currsep,0) -- (1,-1);
  \currentmarker{\currsep,0}
  \currentmarker{-\currsep,0}
\end{tikzpicture}} \\
\mathord{\begin{tikzpicture}[baseline=-0.65ex]
  \draw[thick] (-1,1) -- (-\currsep,0) -- (-1,-1);
  \draw[thick] (1,1) -- (\currsep,0) -- (1,-1);
  \currentmarker{\currsep,0}
  \currentmarker{-\currsep,0}
  \node at (-\bracketsep,0) {$[$};
  \node at (\bracketsep,0) {$]$};
\end{tikzpicture}}
\cut
\mathord{\begin{tikzpicture}[baseline=-0.65ex]
  \draw[thick] (-1,1) -- (-\threepointsep,0) -- (-1,-1);
  \draw[thick,dotted,red] (1,-1) -- (\threepointsep,0) -- (1,1);
  \draw[thick,decorate,decoration={snake}] (\threepointsep,0) -- (-\threepointsep,0);
\end{tikzpicture}}
 &=  \frac{16}{3}
\mathord{\begin{tikzpicture}[baseline=-0.65ex]
  \draw[thick] (-1,1) -- (-\currsep,0) -- (-1,-1);
  \draw[thick,dotted,red] (1,1) -- (\currsep,0) -- (1,-1);
  \currentmarker{\currsep,0}
  \currentmarker{-\currsep,0}
\end{tikzpicture}}
\end{align}
where the coloured dotted lines can be a Higgs or a fermion current, and the brackets denote (anti)symmetrisation in the case of indistinguishable currents.
\\

\noindent $\bullet$ $t/u$-channel Yukawa (single current cut):

\begin{align}
  \mathord{\begin{tikzpicture}[baseline=-0.65ex]
    \draw[thick,black] (-1,1) -- (-\currsep,0) -- (-1,-1);
    \draw[thick,dashed,black] (1,1) -- (\currsep,0) -- (1,-1);
    \currentmarker{\currsep,0}
    \currentmarker{-\currsep,0}
  \end{tikzpicture}}
  \cut
  \mathord{\begin{tikzpicture}[baseline=-0.65ex]
  \draw[thick,dashed,black] (-1,1) -- (0,\threepointsep);
  \draw[thick,dashed,black] (0,-\threepointsep) -- (-1,-1);
  \draw[thick,black] (0,\threepointsep) -- (1,1);
  \draw[thick] (0,\threepointsep) -- (0,-\threepointsep);
  \draw[thick,black] (0,-\threepointsep) -- (1,-1);
  \end{tikzpicture}}
  \quad \text{or} \quad \mathord{\begin{tikzpicture}[baseline=-0.65ex]
  \draw[thick,dashed,black] (-1,1) -- (0,\threepointsep);
  \draw[thick,dashed] (0,-\threepointsep) -- (-1,-1);
  \draw[thick] (0,\threepointsep) -- (1,-1);
  \draw[thick] (0,\threepointsep) -- (0,-\threepointsep);
  \draw[thick
  ] (0,-\threepointsep) -- (1,1);
  \end{tikzpicture}} 
  &= 1
  \times
  \mathord{\begin{tikzpicture}[baseline=-0.65ex]
   \draw[thick,black] (-1,1) -- (-\currsep,0) -- (-1,-1);
   \draw[thick,black] (1,1) -- (\currsep,0) -- (1,-1);
   \currentmarker{\currsep,0}
   \currentmarker{-\currsep,0}
  \end{tikzpicture}} \\
   \mathord{\begin{tikzpicture}[baseline=-0.65ex]
    \draw[thick,dashed,black] (-1,1) -- (-\currsep,0) -- (-1,-1);
    \draw[thick,black] (1,1) -- (\currsep,0) -- (1,-1);
    \currentmarker{\currsep,0}
    \currentmarker{-\currsep,0}
  \end{tikzpicture}}
  \cut
  \mathord{\begin{tikzpicture}[baseline=-0.65ex]
  \draw[thick,black] (-1,1) -- (0,\threepointsep);
  \draw[thick,black] (0,-\threepointsep) -- (-1,-1);
  \draw[thick,dashed, black] (0,\threepointsep) -- (1,1);
  \draw[thick] (0,\threepointsep) -- (0,-\threepointsep);
  \draw[thick,dashed] (0,-\threepointsep) -- (1,-1);
  \end{tikzpicture}}
  \quad \text{or} \quad \mathord{\begin{tikzpicture}[baseline=-0.65ex]
  \draw[thick,black] (-1,1) -- (0,\threepointsep);
  \draw[thick] (0,-\threepointsep) -- (-1,-1);
  \draw[thick,dashed] (0,\threepointsep) -- (1,-1);
  \draw[thick] (0,\threepointsep) -- (0,-\threepointsep);
  \draw[thick,dashed
  ] (0,-\threepointsep) -- (1,1);
  \end{tikzpicture}} 
  &= 2
  \times
  \mathord{\begin{tikzpicture}[baseline=-0.65ex]
   \draw[thick,dashed,black] (-1,1) -- (-\currsep,0) -- (-1,-1);
   \draw[thick,dashed,black] (1,1) -- (\currsep,0) -- (1,-1);
   \currentmarker{\currsep,0}
   \currentmarker{-\currsep,0}
  \end{tikzpicture}} \\
   \mathord{\begin{tikzpicture}[baseline=-0.65ex]
    \draw[thick,dashed,black] (-1,1) -- (-\currsep,0) -- (-1,-1);
    \draw[thick,dashed,black] (1,1) -- (\currsep,0) -- (1,-1);
    \currentmarker{\currsep,0}
    \currentmarker{-\currsep,0}
      \node at (-\bracketsep,0) {$($};
  \node at (\bracketsep,0) {$)$};
  \end{tikzpicture}}
  \cut
  \mathord{\begin{tikzpicture}[baseline=-0.65ex]
  \draw[thick,dashed,black] (-1,1) -- (0,\threepointsep);
  \draw[thick,dashed,black] (0,-\threepointsep) -- (-1,-1);
  \draw[thick,black] (0,\threepointsep) -- (1,1);
  \draw[thick] (0,\threepointsep) -- (0,-\threepointsep);
  \draw[thick,black] (0,-\threepointsep) -- (1,-1);
  \end{tikzpicture}}
  \quad \text{or} \quad \mathord{\begin{tikzpicture}[baseline=-0.65ex]
  \draw[thick,dashed,black] (-1,1) -- (0,\threepointsep);
  \draw[thick,dashed] (0,-\threepointsep) -- (-1,-1) ;
  \draw[thick] (0,\threepointsep) -- (1,-1);
  \draw[thick] (0,\threepointsep) -- (0,-\threepointsep);
  \draw[thick
  ] (0,-\threepointsep) -- (1,1);
  \end{tikzpicture}} 
  &= 3
  \times
  \mathord{\begin{tikzpicture}[baseline=-0.65ex]
   \draw[thick,dashed,black] (-1,1) -- (-\currsep,0) -- (-1,-1);
   \draw[thick,black] (1,1) -- (\currsep,0) -- (1,-1);
   \currentmarker{\currsep,0}
   \currentmarker{-\currsep,0}
  \end{tikzpicture}} \\
   \mathord{\begin{tikzpicture}[baseline=-0.65ex]
    \draw[thick,dashed,black] (-1,1) -- (-\currsep,0) -- (-1,-1);
    \draw[thick,dashed,black] (1,1) -- (\currsep,0) -- (1,-1);
    \currentmarker{\currsep,0}
    \currentmarker{-\currsep,0}
      \node at (-\bracketsep,0) {$[$};
  \node at (\bracketsep,0) {$]$};
  \end{tikzpicture}}
  \cut
  \mathord{\begin{tikzpicture}[baseline=-0.65ex]
  \draw[thick,dashed,black] (-1,1) -- (0,\threepointsep);
  \draw[thick,dashed,black] (0,-\threepointsep) -- (-1,-1);
  \draw[thick,black] (0,\threepointsep) -- (1,1);
  \draw[thick] (0,\threepointsep) -- (0,-\threepointsep);
  \draw[thick,black] (0,-\threepointsep) -- (1,-1);
  \end{tikzpicture}}
  \quad \text{or} \quad \mathord{\begin{tikzpicture}[baseline=-0.65ex]
  \draw[thick,dashed,black] (-1,1) -- (0,\threepointsep);
  \draw[thick,dashed] (0,-\threepointsep) -- (-1,-1) ;
  \draw[thick] (0,\threepointsep) -- (1,-1);
  \draw[thick] (0,\threepointsep) -- (0,-\threepointsep);
  \draw[thick
  ] (0,-\threepointsep) -- (1,1);
  \end{tikzpicture}} 
  &= 1
  \times
  \mathord{\begin{tikzpicture}[baseline=-0.65ex]
   \draw[thick,dashed,black] (-1,1) -- (-\currsep,0) -- (-1,-1);
   \draw[thick,black] (1,1) -- (\currsep,0) -- (1,-1);
   \currentmarker{\currsep,0}
   \currentmarker{-\currsep,0}
  \end{tikzpicture}} \\
    \mathord{\begin{tikzpicture}[baseline=-0.65ex]
    \draw[thick,black] (-1,1) -- (-\currsep,0) -- (-1,-1);
    \draw[thick,black] (1,1) -- (\currsep,0) -- (1,-1);
    \currentmarker{\currsep,0}
    \currentmarker{-\currsep,0}
    \node at (-\bracketsep,0) {$[$};
  \node at (\bracketsep,0) {$]$};
  \end{tikzpicture}}
  \cut
  \mathord{\begin{tikzpicture}[baseline=-0.65ex]
  \draw[thick,black] (-1,1) -- (0,\threepointsep);
  \draw[thick,black] (0,-\threepointsep) -- (-1,-1);
  \draw[thick,dashed, black] (0,\threepointsep) -- (1,1);
  \draw[thick] (0,\threepointsep) -- (0,-\threepointsep);
  \draw[thick,dashed] (0,-\threepointsep) -- (1,-1);
  \end{tikzpicture}}
  \quad \text{or} \quad \mathord{\begin{tikzpicture}[baseline=-0.65ex]
  \draw[thick,black] (-1,1) -- (0,\threepointsep);
  \draw[thick] (0,-\threepointsep) -- (-1,-1);
  \draw[thick,dashed] (0,\threepointsep) -- (1,-1);
  \draw[thick] (0,\threepointsep) -- (0,-\threepointsep);
  \draw[thick,dashed
  ] (0,-\threepointsep) -- (1,1);
  \end{tikzpicture}} 
  &= 8
  \times
  \mathord{\begin{tikzpicture}[baseline=-0.65ex]
   \draw[thick,black] (-1,1) -- (-\currsep,0) -- (-1,-1);
   \draw[thick,black,dashed] (1,1) -- (\currsep,0) -- (1,-1);
   \currentmarker{\currsep,0}
   \currentmarker{-\currsep,0}
  \end{tikzpicture}}\\
    \mathord{\begin{tikzpicture}[baseline=-0.65ex]
    \draw[thick,black] (-1,1) -- (-\currsep,0) -- (-1,-1);
    \draw[thick,black] (1,1) -- (\currsep,0) -- (1,-1);
    \currentmarker{\currsep,0}
    \currentmarker{-\currsep,0}
     \node at (-\bracketsep,0) {$[$};
  \node at (\bracketsep,0) {$]$};
  \end{tikzpicture}}
  \cut
  \mathord{\begin{tikzpicture}[baseline=-0.65ex]
  \draw[thick,black] (-1,1) -- (0,\threepointsep);
  \draw[thick,black] (0,-\threepointsep) -- (-1,-1);
  \draw[thick, black] (0,\threepointsep) -- (1,1);
  \draw[thick,dashed] (0,\threepointsep) -- (0,-\threepointsep);
  \draw[thick] (0,-\threepointsep) -- (1,-1);
  \end{tikzpicture}}
  \quad \text{or} \quad \mathord{\begin{tikzpicture}[baseline=-0.65ex]
  \draw[thick,black] (-1,1) -- (0,\threepointsep);
  \draw[thick] (0,-\threepointsep) -- (-1,-1);
  \draw[thick] (0,\threepointsep) -- (1,-1);
  \draw[thick,dashed] (0,\threepointsep) -- (0,-\threepointsep);
  \draw[thick
  ] (0,-\threepointsep) -- (1,1);
  \end{tikzpicture}} 
  &= 4
  \times
  \mathord{\begin{tikzpicture}[baseline=-0.65ex]
   \draw[thick,black] (-1,1) -- (-\currsep,0) -- (-1,-1);
   \draw[thick,black] (1,1) -- (\currsep,0) -- (1,-1);
   \currentmarker{\currsep,0}
   \currentmarker{-\currsep,0}
  \end{tikzpicture}}
\end{align}


\noindent $\bullet$ $t/u$-channel Yukawa (double current cut):
  \begin{align}
\mathord{\begin{tikzpicture}[baseline=-0.65ex]
  \draw[thick] (-1,1) -- (0,\currsep) -- (1,1);
  \draw[thick] (-1,-1) -- (0,-\currsep) -- (1,-1);
  \currentmarker{0,\currsep}
  \currentmarker{0,-\currsep}
\end{tikzpicture}}
\cut
\mathord{\begin{tikzpicture}[baseline=-0.65ex]
  \draw[thick] (-1,1) -- (0,\threepointsep) -- (1,-1);
  \draw[thick] (-1,-1) -- (0,-\threepointsep) -- (1,1);
  \draw[thick,dashed] (0,\threepointsep) -- (0,-\threepointsep);
\end{tikzpicture}}
 &= -1
\mathord{\begin{tikzpicture}[baseline=-0.65ex]
  \draw[thick] (-1,1) -- (0,\currsep) -- (1,1);
  \draw[thick] (-1,-1) -- (0,-\currsep) -- (1,-1);
  \currentmarker{0,\currsep}
  \currentmarker{0,-\currsep}
\end{tikzpicture}}\\
\mathord{\begin{tikzpicture}[baseline=-0.65ex]
  \draw[thick] (-1,1) -- (0,\currsep) -- (1,1);
  \draw[thick] (-1,-1) -- (0,-\currsep) -- (1,-1);
  \currentmarker{0,\currsep}
  \currentmarker{0,-\currsep}
\end{tikzpicture}}
\cut
\mathord{\begin{tikzpicture}[baseline=-0.65ex]
  \draw[thick] (-1,1) -- (-\threepointsep,0) -- (-1,-1);
  \draw[thick] (1,-1) -- (\threepointsep,0) -- (1,1);
  \draw[thick,dashed] (\threepointsep,0) -- (-\threepointsep,0);
\end{tikzpicture}}
 &= -2
\mathord{\begin{tikzpicture}[baseline=-0.65ex]
  \draw[thick] (-1,1) -- (0,\currsep) -- (1,1);
  \draw[thick] (-1,-1) -- (0,-\currsep) -- (1,-1);
  \currentmarker{0,\currsep}
  \currentmarker{0,-\currsep}
\end{tikzpicture}}
\end{align}
\begin{align}
\mathord{\begin{tikzpicture}[baseline=-0.65ex]
  \draw[thick,dashed] (-1,1) -- (0,\currsep) -- (1,1);
  \draw[thick] (-1,-1) -- (0,-\currsep) -- (1,-1);
  \currentmarker{0,\currsep}
  \currentmarker{0,-\currsep}
\end{tikzpicture}}
\cut
\mathord{\begin{tikzpicture}[baseline=-0.65ex]
  \draw[thick,dashed] (-1,1) -- (0,\threepointsep);
  \draw[thick,dashed] (1,1) -- (0,-\threepointsep);
  \draw[thick] (-1,-1) -- (0,-\threepointsep) -- (0,\threepointsep) -- (1,-1);
\end{tikzpicture}}
 &= 2
\mathord{\begin{tikzpicture}[baseline=-0.65ex]
  \draw[thick,dashed] (-1,1) -- (0,\currsep) -- (1,1);
  \draw[thick] (-1,-1) -- (0,-\currsep) -- (1,-1);
  \currentmarker{0,\currsep}
  \currentmarker{0,-\currsep}
\end{tikzpicture}} \\
\mathord{\begin{tikzpicture}[baseline=-0.65ex]
  \draw[thick,dashed] (-1,1) -- (0,\currsep) -- (1,1);
  \draw[thick] (-1,-1) -- (0,-\currsep) -- (1,-1);
  \currentmarker{0,\currsep}
  \currentmarker{0,-\currsep}
\end{tikzpicture}}
\cut
\mathord{\begin{tikzpicture}[baseline=-0.65ex]
  \draw[thick] (-1,-1) -- (-\threepointsep,0) -- (\threepointsep,0) -- (1,-1);
  \draw[thick,dashed] (-1,1) -- (-\threepointsep,0);
  \draw[thick,dashed] (\threepointsep,0) -- (1,1);
\end{tikzpicture}}
 &= 1
\mathord{\begin{tikzpicture}[baseline=-0.65ex]
  \draw[thick,dashed] (-1,1) -- (0,\currsep) -- (1,1);
  \draw[thick] (-1,-1) -- (0,-\currsep) -- (1,-1);
  \currentmarker{0,\currsep}
  \currentmarker{0,-\currsep}
\end{tikzpicture}}
\end{align}

\subsection{IR divergent (soft)}
\label{app:IRdiv}
Here, the ``cut'' includes subtracting off the IR divergent piece due to triangle diagrams.
\begin{align}
\mathord{\begin{tikzpicture}[baseline=-0.65ex]
  \draw[thick,dotted,blue] (-1,1) -- (-\currsep,0) -- (-1,-1);
  \draw[thick] (1,1) -- (\currsep,0) -- (1,-1);
  \currentmarker{\currsep,0}
  \currentmarker{-\currsep,0}
\end{tikzpicture}}
\cutsub
\mathord{\begin{tikzpicture}[baseline=-0.65ex]
  \draw[thick] (-1,1) -- (0,\threepointsep) -- (1,1);
  \draw[thick] (-1,-1) -- (0,-\threepointsep) -- (1,-1);
  \draw[thick,decorate,decoration={snake}] (0,\threepointsep) -- (0,-\threepointsep);
\end{tikzpicture}}
 &=  6
\mathord{\begin{tikzpicture}[baseline=-0.65ex]
  \draw[thick,dotted,blue] (-1,1) -- (-\currsep,0) -- (-1,-1);
  \draw[thick] (1,1) -- (\currsep,0) -- (1,-1);
  \currentmarker{\currsep,0}
  \currentmarker{-\currsep,0}
\end{tikzpicture}} \\
\mathord{\begin{tikzpicture}[baseline=-0.65ex]
  \draw[thick,dotted,blue] (-1,1) -- (-\currsep,0) -- (-1,-1);
  \draw[thick,dashed] (1,1) -- (\currsep,0) -- (1,-1);
  \currentmarker{\currsep,0}
  \currentmarker{-\currsep,0}
\end{tikzpicture}}
\cutsub
\mathord{\begin{tikzpicture}[baseline=-0.65ex]
  \draw[thick,dashed] (-1,1) -- (0,\threepointsep) -- (1,1);
  \draw[thick,dashed] (-1,-1) -- (0,-\threepointsep) -- (1,-1);
  \draw[thick,decorate,decoration={snake}] (0,\threepointsep) -- (0,-\threepointsep);
\end{tikzpicture}}
 &=  8
\mathord{\begin{tikzpicture}[baseline=-0.65ex]
  \draw[thick,dotted,blue] (-1,1) -- (-\currsep,0) -- (-1,-1);
  \draw[thick,dashed] (1,1) -- (\currsep,0) -- (1,-1);
  \currentmarker{\currsep,0}
  \currentmarker{-\currsep,0}
\end{tikzpicture}} 
\end{align}
\begin{align}
\mathord{\begin{tikzpicture}[baseline=-0.65ex]
  \draw[thick] (-1,1) -- (-\currsep,0) -- (-1,-1);
  \draw[thick] (1,1) -- (\currsep,0) -- (1,-1);
  \currentmarker{\currsep,0}
  \currentmarker{-\currsep,0}
  \node at (-\bracketsep,0) {$[$};
  \node at (\bracketsep,0) {$]$};
\end{tikzpicture}}
\cutsub
\mathord{\begin{tikzpicture}[baseline=-0.65ex]
  \draw[thick] (-1,1) -- (0,\threepointsep) -- (1,1);
  \draw[thick] (-1,-1) -- (0,-\threepointsep) -- (1,-1);
  \draw[thick,decorate,decoration={snake}] (0,\threepointsep) -- (0,-\threepointsep);
\end{tikzpicture}}
 &=  24
 \mathord{\begin{tikzpicture}[baseline=-0.65ex]
   \draw[thick] (-1,1) -- (-\currsep,0) -- (-1,-1);
   \draw[thick] (1,1) -- (\currsep,0) -- (1,-1);
   \currentmarker{\currsep,0}
   \currentmarker{-\currsep,0}
   \node at (-\bracketsep,0) {$[$};
   \node at (\bracketsep,0) {$]$};
 \end{tikzpicture}} \\
\mathord{\begin{tikzpicture}[baseline=-0.65ex]
  \draw[thick,dashed] (-1,1) -- (-\currsep,0) -- (-1,-1);
  \draw[thick,dashed] (1,1) -- (\currsep,0) -- (1,-1);
  \currentmarker{\currsep,0}
  \currentmarker{-\currsep,0}
  \node at (-\bracketsep,0) {$[$};
  \node at (\bracketsep,0) {$]$};
\end{tikzpicture}}
\cutsub
\mathord{\begin{tikzpicture}[baseline=-0.65ex]
  \draw[thick,dashed] (-1,1) -- (0,\threepointsep) -- (1,1);
  \draw[thick,dashed] (-1,-1) -- (0,-\threepointsep) -- (1,-1);
  \draw[thick,decorate,decoration={snake}] (0,\threepointsep) -- (0,-\threepointsep);
\end{tikzpicture}}
 &= 14
 \mathord{\begin{tikzpicture}[baseline=-0.65ex]
   \draw[thick,dashed] (-1,1) -- (-\currsep,0) -- (-1,-1);
   \draw[thick,dashed] (1,1) -- (\currsep,0) -- (1,-1);
   \currentmarker{\currsep,0}
   \currentmarker{-\currsep,0}
   \node at (-\bracketsep,0) {$[$};
   \node at (\bracketsep,0) {$]$};
 \end{tikzpicture}} \\
 \mathord{\begin{tikzpicture}[baseline=-0.65ex]
   \draw[thick,dashed] (-1,1) -- (-\currsep,0) -- (-1,-1);
   \draw[thick,dashed] (1,1) -- (\currsep,0) -- (1,-1);
   \currentmarker{\currsep,0}
   \currentmarker{-\currsep,0}
   \node at (-\bracketsep,0) {$($};
   \node at (\bracketsep,0) {$)$};
 \end{tikzpicture}}
 \cutsub
 \mathord{\begin{tikzpicture}[baseline=-0.65ex]
   \draw[thick,dashed] (-1,1) -- (0,\threepointsep) -- (1,1);
   \draw[thick,dashed] (-1,-1) -- (0,-\threepointsep) -- (1,-1);
   \draw[thick,decorate,decoration={snake}] (0,\threepointsep) -- (0,-\threepointsep);
 \end{tikzpicture}}
  &=  26
  \mathord{\begin{tikzpicture}[baseline=-0.65ex]
    \draw[thick,dashed] (-1,1) -- (-\currsep,0) -- (-1,-1);
    \draw[thick,dashed] (1,1) -- (\currsep,0) -- (1,-1);
    \currentmarker{\currsep,0}
    \currentmarker{-\currsep,0}
    \node at (-\bracketsep,0) {$($};
    \node at (\bracketsep,0) {$)$};
  \end{tikzpicture}} 
\end{align}
Notice that the diagrams shown in (E.21), (E.22) and (E.23) are equivalent to the double-current cut diagrams in (E.29), (E.35) and (E.31), respectively. This is due to the inbuilt symmetry to exchange two legs in the LHS operator.

In the following, `$+$' on a Higgs leg denotes the Higgs particle, whereas `$-$' denotes the daggered Higgs anti-particle.
\begin{align}
  \mathord{\begin{tikzpicture}[baseline=-0.65ex]
    \draw[thick] (1,-1) node[right] {$\pm$} -- (0,-\currsep) -- (-1,-1);
    \draw[thick] (-1,1) -- (0,\currsep) -- (1,1) node[right] {$\pm$};
    \currentmarker{0,\currsep}
    \currentmarker{0,-\currsep}
  \end{tikzpicture}}
  \cutsub
  \mathord{\begin{tikzpicture}[baseline=-0.65ex]
    \draw[thick] (-1,1) -- (0,\threepointsep) -- (1,1);
    \draw[thick] (-1,-1) -- (0,-\threepointsep) -- (1,-1);
    \draw[thick,decorate,decoration={snake}] (0,\threepointsep) -- (0,-\threepointsep);
  \end{tikzpicture}}
   &= 0
  \mathord{\begin{tikzpicture}[baseline=-0.65ex]
    \draw[thick] (1,-1) -- (0,-\currsep) -- (-1,-1);
    \draw[thick] (-1,1) -- (0,\currsep) -- (1,1);
    \currentmarker{0,\currsep}
    \currentmarker{0,-\currsep}
  \end{tikzpicture}} \\
  \mathord{\begin{tikzpicture}[baseline=-0.65ex]
    \draw[thick] (1,-1) node[right] {$\pm$} -- (0,-\currsep) -- (-1,-1);
    \draw[thick] (-1,1) -- (0,\currsep) -- (1,1) node[right] {$\mp$};
    \currentmarker{0,\currsep}
    \currentmarker{0,-\currsep}
  \end{tikzpicture}}
  \cutsub
  \mathord{\begin{tikzpicture}[baseline=-0.65ex]
    \draw[thick] (-1,1) -- (0,\threepointsep) -- (1,1);
    \draw[thick] (-1,-1) -- (0,-\threepointsep) -- (1,-1);
    \draw[thick,decorate,decoration={snake}] (0,\threepointsep) -- (0,-\threepointsep);
  \end{tikzpicture}}
   &= (-6)
  \mathord{\begin{tikzpicture}[baseline=-0.65ex]
    \draw[thick] (1,-1) -- (0,-\currsep) -- (-1,-1);
    \draw[thick] (-1,1) -- (0,\currsep) -- (1,1);
    \currentmarker{0,\currsep}
    \currentmarker{0,-\currsep}
  \end{tikzpicture}} 
  \end{align}
  \begin{align}
  \mathord{\begin{tikzpicture}[baseline=-0.65ex]
    \draw[thick] (1,-1) node[right] {$\pm$} -- (0,-\currsep) -- (-1,-1);
    \draw[thick,dashed] (-1,1) -- (0,\currsep) -- (1,1) node[right] {$\pm$};
    \currentmarker{0,\currsep}
    \currentmarker{0,-\currsep}
  \end{tikzpicture}}
  \cutsub
  \mathord{\begin{tikzpicture}[baseline=-0.65ex]
    \draw[thick,dashed] (-1,1) -- (0,\threepointsep) -- (1,1);
    \draw[thick] (-1,-1) -- (0,-\threepointsep) -- (1,-1);
    \draw[thick,decorate,decoration={snake}] (0,\threepointsep) -- (0,-\threepointsep);
  \end{tikzpicture}}
   &=  (-4)
  \mathord{\begin{tikzpicture}[baseline=-0.65ex]
    \draw[thick] (1,-1) -- (0,-\currsep) -- (-1,-1);
    \draw[thick,dashed] (-1,1) -- (0,\currsep) -- (1,1);
    \currentmarker{0,\currsep}
    \currentmarker{0,-\currsep}
  \end{tikzpicture}} \\
  \mathord{\begin{tikzpicture}[baseline=-0.65ex]
    \draw[thick] (1,-1) node[right] {$\pm$} -- (0,-\currsep) -- (-1,-1);
    \draw[thick,dashed] (-1,1) -- (0,\currsep) -- (1,1) node[right] {$\mp$};
    \currentmarker{0,\currsep}
    \currentmarker{0,-\currsep}
  \end{tikzpicture}}
  \cutsub
  \mathord{\begin{tikzpicture}[baseline=-0.65ex]
    \draw[thick,dashed] (-1,1) -- (0,\threepointsep) -- (1,1);
    \draw[thick] (-1,-1) -- (0,-\threepointsep) -- (1,-1);
    \draw[thick,decorate,decoration={snake}] (0,\threepointsep) -- (0,-\threepointsep);
  \end{tikzpicture}}
   &= 4
  \mathord{\begin{tikzpicture}[baseline=-0.65ex]
    \draw[thick] (1,-1) -- (0,-\currsep) -- (-1,-1);
    \draw[thick,dashed] (-1,1) -- (0,\currsep) -- (1,1);
    \currentmarker{0,\currsep}
    \currentmarker{0,-\currsep}
  \end{tikzpicture}} 
  \end{align}
  \begin{align}
  \mathord{\begin{tikzpicture}[baseline=-0.65ex]
    \draw[thick] (1,-1) node[right] {$\pm$} -- (0,-\currsep) -- (-1,-1);
    \draw[thick] (-1,1) -- (0,\currsep) -- (1,1) node[right] {$\pm$};
    \currentmarker{0,\currsep}
    \currentmarker{0,-\currsep}
    \node[rotate=90] at (0,-\bracketsep) {$[$};
    \node[rotate=90] at (0,\bracketsep) {$]$};
  \end{tikzpicture}}
  \cutsub
  \mathord{\begin{tikzpicture}[baseline=-0.65ex]
    \draw[thick] (-1,1) -- (0,\threepointsep) -- (1,1);
    \draw[thick] (-1,-1) -- (0,-\threepointsep) -- (1,-1);
    \draw[thick,decorate,decoration={snake}] (0,\threepointsep) -- (0,-\threepointsep);
  \end{tikzpicture}}
   &=  0
  \mathord{\begin{tikzpicture}[baseline=-0.65ex]
    \draw[thick] (1,-1) -- (0,-\currsep) -- (-1,-1);
    \draw[thick] (-1,1) -- (0,\currsep) -- (1,1);
    \currentmarker{0,\currsep}
    \currentmarker{0,-\currsep}
    \node[rotate=90] at (0,-\bracketsep) {$[$};
    \node[rotate=90] at (0,\bracketsep) {$]$};
  \end{tikzpicture}} \\
  \mathord{\begin{tikzpicture}[baseline=-0.65ex]
    \draw[thick] (1,-1) node[right] {$\pm$} -- (0,-\currsep) -- (-1,-1);
    \draw[thick] (-1,1) -- (0,\currsep) -- (1,1) node[right] {$\mp$};
    \currentmarker{0,\currsep}
    \currentmarker{0,-\currsep}
    \node[rotate=90] at (0,-\bracketsep) {$[$};
    \node[rotate=90] at (0,\bracketsep) {$]$};
  \end{tikzpicture}}
  \cutsub
  \mathord{\begin{tikzpicture}[baseline=-0.65ex]
    \draw[thick] (-1,1) -- (0,\threepointsep) -- (1,1);
    \draw[thick] (-1,-1) -- (0,-\threepointsep) -- (1,-1);
    \draw[thick,decorate,decoration={snake}] (0,\threepointsep) -- (0,-\threepointsep);
  \end{tikzpicture}}
   &=  24
  \mathord{\begin{tikzpicture}[baseline=-0.65ex]
    \draw[thick] (1,-1) -- (0,-\currsep) -- (-1,-1);
    \draw[thick] (-1,1) -- (0,\currsep) -- (1,1);
    \currentmarker{0,\currsep}
    \currentmarker{0,-\currsep}
    \node[rotate=90] at (0,-\bracketsep) {$[$};
    \node[rotate=90] at (0,\bracketsep) {$]$};
  \end{tikzpicture}} 
  \end{align}
  \begin{align}
  \mathord{\begin{tikzpicture}[baseline=-0.65ex]
    \draw[thick,dashed] (1,-1)  node[right] {$\pm$} -- (0,-\currsep) -- (-1,-1);
    \draw[thick,dashed] (-1,1) -- (0,\currsep) -- (1,1) node[right] {$\pm$};
    \currentmarker{0,\currsep}
    \currentmarker{0,-\currsep}
    \node[rotate=90] at (0,-\bracketsep) {$($};
    \node[rotate=90] at (0,\bracketsep) {$)$};
  \end{tikzpicture}}
  \cutsub
  \mathord{\begin{tikzpicture}[baseline=-0.65ex]
    \draw[thick,dashed] (-1,1) -- (0,\threepointsep) -- (1,1);
    \draw[thick,dashed] (-1,-1) -- (0,-\threepointsep) -- (1,-1);
    \draw[thick,decorate,decoration={snake}] (0,\threepointsep) -- (0,-\threepointsep);
  \end{tikzpicture}}
   &=  (-2)
  \mathord{\begin{tikzpicture}[baseline=-0.65ex]
    \draw[thick,dashed] (1,-1) -- (0,-\currsep) -- (-1,-1);
    \draw[thick,dashed] (-1,1) -- (0,\currsep) -- (1,1);
    \currentmarker{0,\currsep}
    \currentmarker{0,-\currsep}
    \node[rotate=90] at (0,-\bracketsep) {$($};
    \node[rotate=90] at (0,\bracketsep) {$)$};
  \end{tikzpicture}} \\
  \mathord{\begin{tikzpicture}[baseline=-0.65ex]
    \draw[thick,dashed] (1,-1)  node[right] {$\pm$} -- (0,-\currsep) -- (-1,-1);
    \draw[thick,dashed] (-1,1) -- (0,\currsep) -- (1,1) node[right] {$\mp$};
    \currentmarker{0,\currsep}
    \currentmarker{0,-\currsep}
    \node[rotate=90] at (0,-\bracketsep) {$($};
    \node[rotate=90] at (0,\bracketsep) {$)$};
  \end{tikzpicture}}
  \cutsub
  \mathord{\begin{tikzpicture}[baseline=-0.65ex]
    \draw[thick,dashed] (-1,1) -- (0,\threepointsep) -- (1,1);
    \draw[thick,dashed] (-1,-1) -- (0,-\threepointsep) -- (1,-1);
    \draw[thick,decorate,decoration={snake}] (0,\threepointsep) -- (0,-\threepointsep);
  \end{tikzpicture}}
   &=  26
  \mathord{\begin{tikzpicture}[baseline=-0.65ex]
    \draw[thick,dashed] (1,-1) -- (0,-\currsep) -- (-1,-1);
    \draw[thick,dashed] (-1,1) -- (0,\currsep) -- (1,1);
    \currentmarker{0,\currsep}
    \currentmarker{0,-\currsep}
    \node[rotate=90] at (0,-\bracketsep) {$($};
    \node[rotate=90] at (0,\bracketsep) {$)$};
  \end{tikzpicture}} \\
  \mathord{\begin{tikzpicture}[baseline=-0.65ex]
    \draw[thick,dashed] (1,-1)  node[right] {$\pm$} -- (0,-\currsep) -- (-1,-1);
    \draw[thick,dashed] (-1,1) -- (0,\currsep) -- (1,1) node[right] {$\pm$};
    \currentmarker{0,\currsep}
    \currentmarker{0,-\currsep}
    \node[rotate=90] at (0,-\bracketsep) {$($};
    \node[rotate=90] at (0,\bracketsep) {$)$};
  \end{tikzpicture}}
  \cutsub
  \mathord{\begin{tikzpicture}[baseline=-0.65ex]
    \draw[thick,dashed] (-1,1) -- (0,\threepointsep) -- (1,1);
    \draw[thick,dashed] (-1,-1) -- (0,-\threepointsep) -- (1,-1);
    \draw[thick,decorate,decoration={snake}] (0,\threepointsep) -- (0,-\threepointsep);
  \end{tikzpicture}}
   &=  0
  \mathord{\begin{tikzpicture}[baseline=-0.65ex]
    \draw[thick,dashed] (1,-1) -- (0,-\currsep) -- (-1,-1);
    \draw[thick,dashed] (-1,1) -- (0,\currsep) -- (1,1);
    \currentmarker{0,\currsep}
    \currentmarker{0,-\currsep}
    \node[rotate=90] at (0,-\bracketsep) {$[$};
    \node[rotate=90] at (0,\bracketsep) {$]$};
  \end{tikzpicture}} \\
  \mathord{\begin{tikzpicture}[baseline=-0.65ex]
    \draw[thick,dashed] (1,-1)  node[right] {$\pm$} -- (0,-\currsep) -- (-1,-1);
    \draw[thick,dashed] (-1,1) -- (0,\currsep) -- (1,1) node[right] {$\mp$};
    \currentmarker{0,\currsep}
    \currentmarker{0,-\currsep}
    \node[rotate=90] at (0,-\bracketsep) {$($};
    \node[rotate=90] at (0,\bracketsep) {$)$};
  \end{tikzpicture}}
  \cutsub
  \mathord{\begin{tikzpicture}[baseline=-0.65ex]
    \draw[thick,dashed] (-1,1) -- (0,\threepointsep) -- (1,1);
    \draw[thick,dashed] (-1,-1) -- (0,-\threepointsep) -- (1,-1);
    \draw[thick,decorate,decoration={snake}] (0,\threepointsep) -- (0,-\threepointsep);
  \end{tikzpicture}}
   &=  (-18)
  \mathord{\begin{tikzpicture}[baseline=-0.65ex]
    \draw[thick,dashed] (1,-1) -- (0,-\currsep) -- (-1,-1);
    \draw[thick,dashed] (-1,1) -- (0,\currsep) -- (1,1);
    \currentmarker{0,\currsep}
    \currentmarker{0,-\currsep}
    \node[rotate=90] at (0,-\bracketsep) {$[$};
    \node[rotate=90] at (0,\bracketsep) {$]$};
  \end{tikzpicture}} \\
  \mathord{\begin{tikzpicture}[baseline=-0.65ex]
    \draw[thick,dashed] (1,-1) node[right] {$\pm$} -- (0,-\currsep) -- (-1,-1);
    \draw[thick,dashed] (-1,1) -- (0,\currsep) -- (1,1) node[right] {$\pm$};
    \currentmarker{0,\currsep}
    \currentmarker{0,-\currsep}
    \node[rotate=90] at (0,-\bracketsep) {$[$};
    \node[rotate=90] at (0,\bracketsep) {$]$};
  \end{tikzpicture}}
  \cutsub
  \mathord{\begin{tikzpicture}[baseline=-0.65ex]
    \draw[thick,dashed] (-1,1) -- (0,\threepointsep) -- (1,1);
    \draw[thick,dashed] (-1,-1) -- (0,-\threepointsep) -- (1,-1);
    \draw[thick,decorate,decoration={snake}] (0,\threepointsep) -- (0,-\threepointsep);
  \end{tikzpicture}}
   &=  (-8)
  \mathord{\begin{tikzpicture}[baseline=-0.65ex]
    \draw[thick,dashed] (1,-1) -- (0,-\currsep) -- (-1,-1);
    \draw[thick,dashed] (-1,1) -- (0,\currsep) -- (1,1);
    \currentmarker{0,\currsep}
    \currentmarker{0,-\currsep}
    \node[rotate=90] at (0,-\bracketsep) {$[$};
    \node[rotate=90] at (0,\bracketsep) {$]$};
  \end{tikzpicture}} \\
  \mathord{\begin{tikzpicture}[baseline=-0.65ex]
    \draw[thick,dashed] (1,-1) node[right] {$\pm$} -- (0,-\currsep) -- (-1,-1);
    \draw[thick,dashed] (-1,1) -- (0,\currsep) -- (1,1) node[right] {$\mp$};
    \currentmarker{0,\currsep}
    \currentmarker{0,-\currsep}
    \node[rotate=90] at (0,-\bracketsep) {$[$};
    \node[rotate=90] at (0,\bracketsep) {$]$};
  \end{tikzpicture}}
  \cutsub
  \mathord{\begin{tikzpicture}[baseline=-0.65ex]
    \draw[thick,dashed] (-1,1) -- (0,\threepointsep) -- (1,1);
    \draw[thick,dashed] (-1,-1) -- (0,-\threepointsep) -- (1,-1);
    \draw[thick,decorate,decoration={snake}] (0,\threepointsep) -- (0,-\threepointsep);
  \end{tikzpicture}}
   &= 14
  \mathord{\begin{tikzpicture}[baseline=-0.65ex]
    \draw[thick,dashed] (1,-1) -- (0,-\currsep) -- (-1,-1);
    \draw[thick,dashed] (-1,1) -- (0,\currsep) -- (1,1);
    \currentmarker{0,\currsep}
    \currentmarker{0,-\currsep}
    \node[rotate=90] at (0,-\bracketsep) {$[$};
    \node[rotate=90] at (0,\bracketsep) {$]$};
  \end{tikzpicture}} \\
  \mathord{\begin{tikzpicture}[baseline=-0.65ex]
    \draw[thick,dashed] (1,-1) node[right] {$\pm$} -- (0,-\currsep) -- (-1,-1);
    \draw[thick,dashed] (-1,1) -- (0,\currsep) -- (1,1) node[right] {$\pm$};
    \currentmarker{0,\currsep}
    \currentmarker{0,-\currsep}
    \node[rotate=90] at (0,-\bracketsep) {$[$};
    \node[rotate=90] at (0,\bracketsep) {$]$};
  \end{tikzpicture}}
  \cutsub
  \mathord{\begin{tikzpicture}[baseline=-0.65ex]
    \draw[thick,dashed] (-1,1) -- (0,\threepointsep) -- (1,1);
    \draw[thick,dashed] (-1,-1) -- (0,-\threepointsep) -- (1,-1);
    \draw[thick,decorate,decoration={snake}] (0,\threepointsep) -- (0,-\threepointsep);
  \end{tikzpicture}}
   &=  0
  \mathord{\begin{tikzpicture}[baseline=-0.65ex]
    \draw[thick,dashed] (1,-1) -- (0,-\currsep) -- (-1,-1);
    \draw[thick,dashed] (-1,1) -- (0,\currsep) -- (1,1);
    \currentmarker{0,\currsep}
    \currentmarker{0,-\currsep}
    \node[rotate=90] at (0,-\bracketsep) {$($};
    \node[rotate=90] at (0,\bracketsep) {$)$};
  \end{tikzpicture}} \\
  \mathord{\begin{tikzpicture}[baseline=-0.65ex]
    \draw[thick,dashed] (1,-1) node[right] {$\pm$} -- (0,-\currsep) -- (-1,-1);
    \draw[thick,dashed] (-1,1) -- (0,\currsep) -- (1,1) node[right] {$\mp$};
    \currentmarker{0,\currsep}
    \currentmarker{0,-\currsep}
    \node[rotate=90] at (0,-\bracketsep) {$[$};
    \node[rotate=90] at (0,\bracketsep) {$]$};
  \end{tikzpicture}}
  \cutsub
  \mathord{\begin{tikzpicture}[baseline=-0.65ex]
    \draw[thick,dashed] (-1,1) -- (0,\threepointsep) -- (1,1);
    \draw[thick,dashed] (-1,-1) -- (0,-\threepointsep) -- (1,-1);
    \draw[thick,decorate,decoration={snake}] (0,\threepointsep) -- (0,-\threepointsep);
  \end{tikzpicture}}
   &=  (-6)
  \mathord{\begin{tikzpicture}[baseline=-0.65ex]
    \draw[thick,dashed] (1,-1) -- (0,-\currsep) -- (-1,-1);
    \draw[thick,dashed] (-1,1) -- (0,\currsep) -- (1,1);
    \currentmarker{0,\currsep}
    \currentmarker{0,-\currsep}
    \node[rotate=90] at (0,-\bracketsep) {$($};
    \node[rotate=90] at (0,\bracketsep) {$)$};
  \end{tikzpicture}}
\end{align}

\subsection{Higgs quartic factors\label{sec:higgsquarticfac}}

\begin{table}
\begin{center}
\begin{tabular}{|p{10mm} || p{50mm} |p{30mm}|} \hline
& $HD+$ & $HD-$ \\ \hline \hline
$HD+$
& $(2\cdot\phs{-1}\cdot\gge{1}+\phs{-2}\cdot\gge{\frac54})\cdot\flv{-2\lambda c}$ & $\phs{2}\cdot\gge{\frac14}\cdot\flv{-2\lambda c}$ \\ \hline
$HD-$
& $\phs{6}\cdot\gge{\frac34}\cdot\flv{-2\lambda c}$ & $\phs{-6}\cdot\gge{\frac34}\cdot\flv{-2\lambda c}$\\ \hline
  \end{tabular}
    \caption{ Phase space factors, given in\phs{green}, can be found in \cref{sec:higgsquarticfac} and are explained in Sec.~\ref{sec:IRfinitephase}. Gauge factors, given in\gge{orange}, can be found in Tabs.~\ref{tab:colorfactors}, \ref{tab:colorfactorsnonsymmtosymm} and \ref{tab:colorfactorssymmtononsymm}, and are explained in Sec.~\ref{sec:gauge}.\label{tab:higgsquartic}}
    \end{center}
  \end{table}

For completeness, in Tab.~\ref{tab:higgsquartic} we list the terms of the anomalous dimension matrix which are dependent on the Higgs quartic coupling $\lambda$. Diagrammatically, the phase space pieces are as follows:

\begin{align}
\mathord{\begin{tikzpicture}[baseline=-0.65ex]
  \draw[thick,dotted,blue] (-1,1) -- (-\currsep,0) -- (-1,-1);
  \draw[thick,dashed] (1,1) -- (\currsep,0) -- (1,-1);
  \currentmarker{\currsep,0}
  \currentmarker{-\currsep,0}
\end{tikzpicture}}
\cut
\mathord{\begin{tikzpicture}[baseline=-0.65ex]
  \draw[thick,dashed] (-1,1) -- (0,0) -- (1,1);
  \draw[thick,dashed] (-1,-1) -- (0,0) -- (1,-1);
\end{tikzpicture}}
 &=  0
\mathord{\begin{tikzpicture}[baseline=-0.65ex]
  \draw[thick,dotted,blue] (-1,1) -- (-\currsep,0) -- (-1,-1);
  \draw[thick,dashed] (1,1) -- (\currsep,0) -- (1,-1);
  \currentmarker{\currsep,0}
  \currentmarker{-\currsep,0}
\end{tikzpicture}} \\
\mathord{\begin{tikzpicture}[baseline=-0.65ex]
  \draw[thick,dashed] (1,-1)  node[right] {$\pm$} -- (0,-\currsep) -- (-1,-1);
  \draw[thick,dashed] (-1,1) -- (0,\currsep) -- (1,1) node[right] {$\pm$};
  \currentmarker{0,\currsep}
  \currentmarker{0,-\currsep}
  \node[rotate=90] at (0,-\bracketsep) {$($};
  \node[rotate=90] at (0,\bracketsep) {$)$};
\end{tikzpicture}}
\cut
\mathord{\begin{tikzpicture}[baseline=-0.65ex]
  \draw[thick,dashed] (-1,1) -- (0,0) -- (1,1);
  \draw[thick,dashed] (-1,-1) -- (0,0) -- (1,-1);
\end{tikzpicture}}
 &=  (-1)
\mathord{\begin{tikzpicture}[baseline=-0.65ex]
  \draw[thick,dashed] (1,-1) -- (0,-\currsep) -- (-1,-1);
  \draw[thick,dashed] (-1,1) -- (0,\currsep) -- (1,1);
  \currentmarker{0,\currsep}
  \currentmarker{0,-\currsep}
  \node[rotate=90] at (0,-\bracketsep) {$($};
  \node[rotate=90] at (0,\bracketsep) {$)$};
\end{tikzpicture}} \\
\mathord{\begin{tikzpicture}[baseline=-0.65ex]
  \draw[thick,dashed] (1,-1)  node[right] {$\pm$} -- (0,-\currsep) -- (-1,-1);
  \draw[thick,dashed] (-1,1) -- (0,\currsep) -- (1,1) node[right] {$\pm$};
  \currentmarker{0,\currsep}
  \currentmarker{0,-\currsep}
  \node[rotate=90] at (0,-\bracketsep) {$[$};
  \node[rotate=90] at (0,\bracketsep) {$]$};
\end{tikzpicture}}
\cut
\mathord{\begin{tikzpicture}[baseline=-0.65ex]
  \draw[thick,dashed] (-1,1) -- (0,0) -- (1,1);
  \draw[thick,dashed] (-1,-1) -- (0,0) -- (1,-1);
\end{tikzpicture}}
 &=  0
\mathord{\begin{tikzpicture}[baseline=-0.65ex]
  \draw[thick,dashed] (1,-1) -- (0,-\currsep) -- (-1,-1);
  \draw[thick,dashed] (-1,1) -- (0,\currsep) -- (1,1);
  \currentmarker{0,\currsep}
  \currentmarker{0,-\currsep}
  \node[rotate=90] at (0,-\bracketsep) {$[$};
  \node[rotate=90] at (0,\bracketsep) {$]$};
\end{tikzpicture}} \\
\mathord{\begin{tikzpicture}[baseline=-0.65ex]
  \draw[thick,dashed] (1,-1)  node[right] {$\mp$} -- (0,-\currsep) -- (-1,-1);
  \draw[thick,dashed] (-1,1) -- (0,\currsep) -- (1,1) node[right] {$\pm$};
  \currentmarker{0,\currsep}
  \currentmarker{0,-\currsep}
  \node[rotate=90] at (0,-\bracketsep) {$($};
  \node[rotate=90] at (0,\bracketsep) {$)$};
\end{tikzpicture}}
\cut
\mathord{\begin{tikzpicture}[baseline=-0.65ex]
  \draw[thick,dashed] (-1,1) -- (0,0) -- (1,1);
  \draw[thick,dashed] (-1,-1) -- (0,0) -- (1,-1);
\end{tikzpicture}}
 &= (-2)
\mathord{\begin{tikzpicture}[baseline=-0.65ex]
  \draw[thick,dashed] (1,-1) -- (0,-\currsep) -- (-1,-1);
  \draw[thick,dashed] (-1,1) -- (0,\currsep) -- (1,1);
  \currentmarker{0,\currsep}
  \currentmarker{0,-\currsep}
  \node[rotate=90] at (0,-\bracketsep) {$($};
  \node[rotate=90] at (0,\bracketsep) {$)$};
\end{tikzpicture}} \\
\mathord{\begin{tikzpicture}[baseline=-0.65ex]
  \draw[thick,dashed] (1,-1)  node[right] {$\mp$} -- (0,-\currsep) -- (-1,-1);
  \draw[thick,dashed] (-1,1) -- (0,\currsep) -- (1,1) node[right] {$\pm$};
  \currentmarker{0,\currsep}
  \currentmarker{0,-\currsep}
  \node[rotate=90] at (0,-\bracketsep) {$($};
  \node[rotate=90] at (0,\bracketsep) {$)$};
\end{tikzpicture}}
\cut
\mathord{\begin{tikzpicture}[baseline=-0.65ex]
  \draw[thick,dashed] (-1,1) -- (0,0) -- (1,1);
  \draw[thick,dashed] (-1,-1) -- (0,0) -- (1,-1);
\end{tikzpicture}}
 &=  6
\mathord{\begin{tikzpicture}[baseline=-0.65ex]
  \draw[thick,dashed] (1,-1) -- (0,-\currsep) -- (-1,-1);
  \draw[thick,dashed] (-1,1) -- (0,\currsep) -- (1,1);
  \currentmarker{0,\currsep}
  \currentmarker{0,-\currsep}
  \node[rotate=90] at (0,-\bracketsep) {$[$};
  \node[rotate=90] at (0,\bracketsep) {$]$};
\end{tikzpicture}} \\
\mathord{\begin{tikzpicture}[baseline=-0.65ex]
  \draw[thick,dashed] (1,-1)  node[right] {$\mp$} -- (0,-\currsep) -- (-1,-1);
  \draw[thick,dashed] (-1,1) -- (0,\currsep) -- (1,1) node[right] {$\pm$};
  \currentmarker{0,\currsep}
  \currentmarker{0,-\currsep}
  \node[rotate=90] at (0,-\bracketsep) {$[$};
  \node[rotate=90] at (0,\bracketsep) {$]$};
\end{tikzpicture}}
\cut
\mathord{\begin{tikzpicture}[baseline=-0.65ex]
  \draw[thick,dashed] (-1,1) -- (0,0) -- (1,1);
  \draw[thick,dashed] (-1,-1) -- (0,0) -- (1,-1);
\end{tikzpicture}}
 &=  2
\mathord{\begin{tikzpicture}[baseline=-0.65ex]
  \draw[thick,dashed] (1,-1) -- (0,-\currsep) -- (-1,-1);
  \draw[thick,dashed] (-1,1) -- (0,\currsep) -- (1,1);
  \currentmarker{0,\currsep}
  \currentmarker{0,-\currsep}
  \node[rotate=90] at (0,-\bracketsep) {$($};
  \node[rotate=90] at (0,\bracketsep) {$)$};
\end{tikzpicture}} \\
\mathord{\begin{tikzpicture}[baseline=-0.65ex]
  \draw[thick,dashed] (1,-1)  node[right] {$\mp$} -- (0,-\currsep) -- (-1,-1);
  \draw[thick,dashed] (-1,1) -- (0,\currsep) -- (1,1) node[right] {$\pm$};
  \currentmarker{0,\currsep}
  \currentmarker{0,-\currsep}
  \node[rotate=90] at (0,-\bracketsep) {$[$};
  \node[rotate=90] at (0,\bracketsep) {$]$};
\end{tikzpicture}}
\cut
\mathord{\begin{tikzpicture}[baseline=-0.65ex]
  \draw[thick,dashed] (-1,1) -- (0,0) -- (1,1);
  \draw[thick,dashed] (-1,-1) -- (0,0) -- (1,-1);
\end{tikzpicture}}
 &=  (-6)
\mathord{\begin{tikzpicture}[baseline=-0.65ex]
  \draw[thick,dashed] (1,-1) -- (0,-\currsep) -- (-1,-1);
  \draw[thick,dashed] (-1,1) -- (0,\currsep) -- (1,1);
  \currentmarker{0,\currsep}
  \currentmarker{0,-\currsep}
  \node[rotate=90] at (0,-\bracketsep) {$[$};
  \node[rotate=90] at (0,\bracketsep) {$]$};
\end{tikzpicture}}
\end{align}
\section{(Anti)-symmetrised gauge contractions}
\label{app:gaugeantisymm}

Here we provide tables of gauge contractions, similar to Tab.~\ref{tab:colorfactors} in Sec.~\ref{sec:gauge}, but phrased in terms of the (anti)symmetrised tensor combinations $(\delta\,\delta)$ and $[\delta\, \delta]$. These give the gauge factors relevant for anomalous dimensions involving operators with indistinguishable currents. Tab.~\ref{tab:colorfactorsnonsymmtosymm} has the same contractions as Tab.~\ref{tab:colorfactors}, but the results are expressed in (anti)symmetrised form. These are relevant for anomalous dimensions from the running of distinguishable operators into operators with indistinguishable currents.  Tab.~\ref{tab:colorfactorssymmtononsymm} has contractions of (anti)symmetrised tensors with SM tensors, and the results are given in terms of non-symmetrised tensors. These are relevant for anomalous dimensions from the running of operators with indistinguishable currents into operators with distinguishable currents. In addition, Tab.~\ref{tab:colorfactorssymmtononsymm} has contractions of (anti)symmetrised tensors with SM tensors, and the results are given in terms of (anti)symmetrised tensors. These are relevant for anomalous dimensions from the running of operators with indistinguishable currents into operators with indistinguishable currents.

 \begin{table}
 \begin{center}
 \scalebox{0.78}{
 \begin{tabular}{| L | L | L | L |}
 \hline
 \mathcal{A}_6 \times \mathcal{A}_{\rm SM}~~{\rm (gauge ~part)}& {\rm General~} SU(N) & N=2 & N=3 \\
 \hline\hline
 \delta^a_b \delta^e_f \times \delta^f_e \delta^c_d & =N (\delta \, \delta) + N [\delta \, \delta] & = 2 (\delta \, \delta) + 2 [\delta \, \delta] & =3 (\delta \, \delta) + 3 [\delta \, \delta]\\
 \hline
 [\lambda^A]^a_b [\lambda^A]^e_f \times \delta^f_e \delta^c_d &= 0 &=0 &=0\\
 \hline
 \delta^a_b \delta^e_f \times \frac14 [\lambda^A]^f_e [\lambda^A]^c_d &  =0 & =0 & =0\\
 \hline
    [\lambda^A]^a_b [\lambda^A]^e_f \times \frac14 [\lambda^B]^f_e [\lambda^B]^c_d & = \left(\frac{N-1}{N}\right) (\delta \, \delta) - \left(\frac{N+1}{N}\right) [\delta \, \delta] & =\frac{1}{2}(\delta \, \delta) - \frac{3}{2} [\delta \, \delta] & =\frac{2}{3} (\delta \, \delta) - \frac{4}{3} [\delta \, \delta]  \\
   \hline  \hline
  \delta^a_b \delta^e_f \times \delta^f_d \delta^c_e & = (\delta \, \delta) + [\delta \,\delta] & =  (\delta \, \delta) + [\delta \,\delta] & = (\delta \, \delta) + [\delta \,\delta]\\
     \hline
     \delta^a_b \delta^e_f \times \frac{1}{4} [\lambda^A]^f_d [\lambda^A]^c_e & = \frac{N^2-1}{2N}\left((\delta \, \delta) + [\delta \,\delta]\right) & =  \frac{3}{4}\left((\delta \, \delta) + [\delta \,\delta]\right) & = \frac{4}{3}\left((\delta \, \delta) + [\delta \,\delta]\right)\\
 \hline
 [\lambda^A]^a_b [\lambda^A]^e_f \times \delta^f_d \delta^c_e &=  2 \left(\frac{N-1}{N}\right) (\delta \, \delta) - 2 \left(\frac{N+1}{N}\right) [\delta \, \delta] & =(\delta \, \delta) - 3 [\delta \, \delta] & =\frac{4}{3} (\delta \, \delta) - \frac{8}{3} [\delta \, \delta]\\
  \hline
   [\lambda^A]^a_b [\lambda^A]^e_f \times \frac{1}{4} [\lambda^B]^f_d [\lambda^B]^c_e &=  -\left(\frac{N-1}{N^2}\right) (\delta \, \delta) + \left(\frac{N+1}{N^2}\right) [\delta \, \delta] & =-\frac{1}{4}(\delta \, \delta) + \frac{3}{4} [\delta \, \delta] & =-\frac{2}{9} (\delta \, \delta) + \frac{4}{9} [\delta \, \delta]  \\
   \hline \hline
  [\lambda^A]^e_b [\lambda^A]^a_f \times  \frac{1}{4} [\lambda^B]^f_d [\lambda^B]^c_e & = \frac{N^3-2N+1}{N^2} (\delta \, \delta)-\frac{N^3-2N-1}{N^2} [\delta \, \delta]& = \frac{5}{4}(\delta \, \delta)-\frac{3}{4} [\delta \, \delta]  &  = \frac{22}{9}(\delta \, \delta)-\frac{20}{9} [\delta \, \delta]\\
  \hline
    [\lambda^A]^e_b [\lambda^A]^f_d \times  \frac{1}{4} [\lambda^B]^c_f [\lambda^B]^a_e &= \frac{(N-1)^2}{N^2} (\delta \, \delta)+\frac{(N+1)^2}{N^2} [\delta \, \delta]& = \frac{1}{4}(\delta \, \delta)+\frac{9}{4} [\delta \, \delta]  &  = \frac{4}{9}(\delta \, \delta)+\frac{16}{9} [\delta \, \delta]\\
    \hline
 \end{tabular}}
 \caption{\label{tab:colorfactorsnonsymmtosymm}Possible gauge contractions appearing in $\mathcal{A}_6 \times \mathcal{A}_{\rm SM}$, starting from non-symmetrised structures and expressing the result in terms of (anti)symmetrised structures. The identities are valid for $SU(N)$ generators $\lambda^A$ normalised such that $\Tr[\lambda^A \lambda^B] = 2 \delta^{AB}$, and which reduce to the Pauli (Gell-Mann) matrices when $N=2(3)$.}
 	\end{center}
 \end{table}
 
 \begin{table}
 \begin{center}
 \scalebox{0.78}{
 \begin{tabular}{| L | L | L | L |}
 \hline
 \mathcal{A}_6 \times \mathcal{A}_{\rm SM}~~{\rm (gauge ~part)}& {\rm General~} SU(N) & N=2 & N=3 \\
 \hline\hline
  \delta^{(a}_{(b} \delta^{e)}_{f)} \times \delta^f_e \delta^c_d & \makecell[l]{= \frac12(N+1) \delta^a_b \delta^c_d \\ =\frac12(N+1)(\delta \delta) + \frac12(N+1)[\delta\delta]} &\makecell[l]{=\frac{3}{2}\delta^a_b \delta^c_d  \\ = \frac32(\delta\delta) + \frac32 [\delta \delta]} & \makecell[l]{=2\delta^a_b \delta^c_d \\ = 2(\delta\delta) + 2 [\delta \delta]}\\
     \hline
   \delta^{(a}_{(b} \delta^{e)}_{f)} \times \frac14 [\lambda^A]^f_e [\lambda^A]^c_d  & \makecell[l]{= \frac18 [\lambda^B]^a_b [\lambda^B]^c_d\\ =\left(\frac14 \frac{N-1}{N}\right) (\delta \, \delta) - \left(\frac14 \frac{N+1}{N}\right) [\delta \, \delta] } & \makecell[l]{= \frac18 [\lambda^B]^a_b [\lambda^B]^c_d \\ = \frac{1}{8}(\delta \, \delta) - \frac{3}{8} [\delta \, \delta]} & \makecell[l]{= \frac18 [\lambda^B]^a_b [\lambda^B]^c_d \\  =\frac16(\delta\delta) -\frac13[\delta\delta]} \\
   \hline
   \delta^{[a}_{[b} \delta^{e]}_{f]} \times \delta^f_e \delta^c_d & \makecell[l]{= \frac12(N-1) \delta^a_b \delta^c_d \\ = \frac12(N-1) (\delta \, \delta) + \frac12(N-1) [\delta \, \delta]  } & \makecell[l]{= \frac{1}{2}\delta^a_b \delta^c_d \\ = \frac{1}{2}(\delta \, \delta) + \frac{1}{2} [\delta \, \delta]} & \makecell[l]{= \delta^a_b \delta^c_d \\ = (\delta \, \delta) + [\delta \, \delta]}\\
     \hline
   \delta^{[a}_{[b} \delta^{e]}_{f]} \times \frac14 [\lambda^A]^f_e [\lambda^A]^c_d & \makecell[l]{= -\frac18 [\lambda^B]^a_b [\lambda^B]^c_d \\ = -\left(\frac14 \frac{N-1}{N}\right) (\delta \, \delta) + \left(\frac14 \frac{N+1}{N}\right) [\delta \, \delta]  } & \makecell[l]{= -\frac18 [\lambda^B]^a_b [\lambda^B]^c_d \\ = \frac{1}{8}(\delta \, \delta) + \frac{3}{8} [\delta \, \delta]  } & \makecell[l]{= -\frac18 [\lambda^B]^a_b [\lambda^B]^c_d \\ =\frac{1}{6} (\delta \, \delta) + \frac{1}{3} [\delta \, \delta]} \\
   \hline \hline
   \delta^{(a}_{(b} \delta^{e)}_{f)} \times \delta^f_d \delta^c_e & \makecell[l]{ = \frac{N+1}{2 N} \delta^a_b \delta^c_d + \frac14 [\lambda^A]^a_b [\lambda^A]^c_d \\ =(\delta \, \delta)  }& \makecell[l]{=\frac{3}{4} \delta^a_b \delta^c_d + \frac14 [\lambda^A]^a_b [\lambda^A]^c_d \\ =(\delta \, \delta) } & \makecell[l]{=\frac{2}{3} \delta^a_b \delta^c_d + \frac14 [\lambda^A]^a_b [\lambda^A]^c_d \\ =(\delta \, \delta) } \\
  \hline
   \delta^{[a}_{[b} \delta^{e]}_{f]} \times \delta^f_d \delta^c_e & \makecell[l]{ = \frac{N-1}{2 N} \delta^a_b \delta^c_d - \frac14 [\lambda^A]^a_b [\lambda^A]^c_d \\= [\delta \, \delta] }& \makecell[l]{= \frac{1}{4} \delta^a_b \delta^c_d - \frac14 [\lambda^A]^a_b [\lambda^A]^c_d \\ = [\delta \, \delta]}& \makecell[l]{ = \frac{1}{3} \delta^a_b \delta^c_d - \frac14 [\lambda^A]^a_b [\lambda^A]^c_d \\= [\delta \, \delta]}\\
   \hline
    \delta^{(a}_{(b} \delta^{e)}_{f)} \times \frac{1}{4} [\lambda^A]^f_d [\lambda^A]^c_e & \makecell[l]{= \frac{(N-1)(N+1)^2}{4N^2}\delta^a_b \delta^c_d-\frac{1}{8N} [\lambda^B]^a_b [\lambda^B]^c_d \\ = \frac{N^2+N-2}{4N}(\delta \, \delta)+ \frac{N+1}{4}[\delta \, \delta]} & \makecell[l]{= \frac{9}{16}\delta^a_b \delta^c_d-\frac{1}{16} [\lambda^B]^a_b [\lambda^B]^c_d \\ = \frac{1}{2}(\delta \, \delta)+ \frac{3}{4}[\delta \, \delta] }& \makecell[l]{= \frac{8}{9}\delta^a_b \delta^c_d-\frac{1}{24} [\lambda^B]^a_b [\lambda^B]^c_d \\ =\frac{5}{6}(\delta \, \delta)+ [\delta \, \delta]} \\
  \hline
   \delta^{[a}_{[b} \delta^{e]}_{f]} \times \frac{1}{4} [\lambda^A]^f_d [\lambda^A]^c_e & \makecell[l]{= \frac{(N+1)(N-1)^2}{4N^2}\delta^a_b \delta^c_d+\frac{1}{8N} [\lambda^B]^a_b [\lambda^B]^c_d \\ = \frac{N-1}{4}(\delta \, \delta)+ \frac{N^2-N-2}{4N}[\delta \, \delta]} & \makecell[l]{= \frac{3}{16}\delta^a_b \delta^c_d+\frac{1}{16} [\lambda^B]^a_b [\lambda^B]^c_d \\ = \frac{1}{4}(\delta \, \delta) } & \makecell[l]{= \frac{4}{9}\delta^a_b \delta^c_d+\frac{1}{24} [\lambda^B]^a_b [\lambda^B]^c_d \\=\frac{1}{2}(\delta \, \delta)+ \frac{1}{3}[\delta \, \delta]}\\
   \hline
 \end{tabular}}
 \caption{\label{tab:colorfactorssymmtononsymm}Possible gauge contractions appearing in $\mathcal{A}_6 \times \mathcal{A}_{\rm SM}$, starting from (anti)symmetrised structures in $\mathcal{A}_6$ and expressing the result in terms of non-symmetrised structures and, equivalently, completely in terms of (anti)symmetrised structures for $\mathcal{A}_6$. The identities are valid for $SU(N)$ generators $\lambda^A$ normalised such that $\Tr[\lambda^A \lambda^B] = 2 \delta^{AB}$, and which reduce to the Pauli (Gell-Mann) matrices when $N=2(3)$.}
 	\end{center}
 \end{table}
 
\section{Clebsch-Gordan decomposition of Wilson coefficients}
\label{app:CGdecomp}
We use the conventions of \cite{deSwart:1963pdg,Kaeding:1995vq} developed for the $SU(3)$ of light flavours $u,d,s$, to decompose the Wilson coefficients in terms of irreps of their $SU(3)$ flavour symmetry. 
  \subsection{$c^p_q$}
  The $\mathbf{3} \otimes \overline{\mathbf{3}}$ of $c^p_q$ decomposes into the $\mathbf{1} \oplus \mathbf{8}$ of $c_{1,1}$ and $c_{8,1},\ldots,c_{8,8}$.
The Wilson coefficients are given in terms of the Clebsch-Gordan decompositions as:
\begin{gather}
 c_1^1=\frac{c_{1,1}}{\sqrt{3}}-\frac{c_{8,4}}{\sqrt{2}}-\frac{c_{8,6}}{\sqrt{6}},\quad c_2^1=c_{8,3}, \quad c_3^1=c_{8,1}, \\
 c_2^2=\frac{c_{1,1}}{\sqrt{3}}+\frac{c_{8,4}}{\sqrt{2}}-\frac{c_{8,6}}{\sqrt{6}}, \quad c_3^2=c_{8,2}, \quad c_3^3=\frac{c_{1,1}}{\sqrt{3}}+\sqrt{\frac{2}{3}} c_{8,6}.\nn
\end{gather}
And vice versa as:
\begin{gather}
\left(c_{8,8}\right)^*=-c_{8,1}=-c_3^1,\quad \left(c_{8,7}\right)^*=c_{8,2}=c_3^2,\quad
 \left(c_{8,5}\right)^*=-c_{8,3}=-c_2^1,\\
 c_{8,4}=\sqrt{\frac12} \left( c_2^2 - c_1^1 \right), \quad
 c_{8,6}=\sqrt{\frac16} \left( -c_1^1 - c_2^2 + 2 c_3^3 \right), \quad
 c_{1,1}=\sqrt{\frac13} \left( c_1^1 + c_2^2 + c_3^3 \right).\nn
\end{gather}
Recall that $(c^p_q)^* = c^q_p$.

\subsection{$c^{(pr)}_{(qs)}$}
The $(\mathbf{3} \otimes\mathbf{3})_\text{sym} \otimes (\overline{\mathbf{3}}\otimes \overline{\mathbf{3}})_\text{sym}=\mathbf{6} \otimes \overline{\mathbf{6}}$ of $c^{(pr)}_{(qs)}$ decomposes into the $\mathbf{1} \oplus \mathbf{8} \oplus \mathbf{27}$ of $c_{1,1}$ and $c_{8,1},\ldots,c_{8,8}$ and $c_{27,1},\ldots,c_{27,27}$. The Wilson coefficients are given in terms of the Clebsch-Gordan decompositions as:
\begin{gather}
c_{11}^{11}=\frac{c_{1,1}}{\sqrt{6}}+\frac{c_{27,12}}{\sqrt{6}}+\frac{c_{27,16}}{\sqrt{10}}+\frac{c_{27,18}}{\sqrt{30}}-\sqrt{\frac{2}{5}} c_{8,4}-\sqrt{\frac{2}{15}} c_{8,6},\\
 c_{12}^{11}=-\frac{1}{2} c_{27,11}-\frac{c_{27,15}}{2 \sqrt{5}}+\frac{c_{8,3}}{\sqrt{5}},\quad
 c_{13}^{11}=-\frac{c_{27,5}}{\sqrt{6}}-\sqrt{\frac{2}{15}} c_{27,8}+\frac{c_{8,1}}{\sqrt{5}},\nn\\
 c_{12}^{12}=\frac{c_{1,1}}{2 \sqrt{6}}-\frac{c_{27,12}}{\sqrt{6}}+\frac{c_{27,18}}{2 \sqrt{30}}-\frac{c_{8,6}}{\sqrt{30}},\quad
 c_{13}^{12}=-\frac{c_{27,6}}{\sqrt{6}}-\frac{c_{27,9}}{\sqrt{30}}+\frac{c_{8,2}}{2 \sqrt{5}},\nn \\
 c_{13}^{13}=\frac{c_{1,1}}{2 \sqrt{6}}-\frac{c_{27,16}}{\sqrt{10}}-\frac{1}{2} \sqrt{\frac{3}{10}} c_{27,18}-\frac{c_{8,4}}{2 \sqrt{10}}+\frac{c_{8,6}}{2 \sqrt{30}},\quad
 c_{22}^{11}=c_{27,10},\quad
 c_{23}^{11}=\frac{c_{27,4}}{\sqrt{2}},\nn \\
 c_{22}^{12}=\frac{1}{2} c_{27,11}-\frac{c_{27,15}}{2 \sqrt{5}}+\frac{c_{8,3}}{\sqrt{5}},\quad
 c_{23}^{12}=\frac{c_{27,5}}{\sqrt{6}}-\frac{c_{27,8}}{\sqrt{30}}+\frac{c_{8,1}}{2 \sqrt{5}},\quad
 c_{23}^{13}=\frac{c_{27,15}}{\sqrt{5}}+\frac{c_{8,3}}{2 \sqrt{5}},\nn\\
 c_{33}^{11}=c_{27,1},\quad
 c_{33}^{12}=\frac{c_{27,2}}{\sqrt{2}},\quad
 c_{33}^{13}=\sqrt{\frac{3}{10}} c_{27,8}+\frac{c_{8,1}}{\sqrt{5}},\quad
 c_{13}^{22}=-\frac{c_{27,7}}{\sqrt{2}},\nn\\
 c_{22}^{22}=\frac{c_{1,1}}{\sqrt{6}}+\frac{c_{27,12}}{\sqrt{6}}-\frac{c_{27,16}}{\sqrt{10}}+\frac{c_{27,18}}{\sqrt{30}}+\sqrt{\frac{2}{5}} c_{8,4}-\sqrt{\frac{2}{15}} c_{8,6},\nn\\
 c_{23}^{22}=\frac{c_{27,6}}{\sqrt{6}}-\sqrt{\frac{2}{15}} c_{27,9}+\frac{c_{8,2}}{\sqrt{5}},\quad
 c_{23}^{23}=\frac{c_{1,1}}{2 \sqrt{6}}+\frac{c_{27,16}}{\sqrt{10}}-\frac{1}{2} \sqrt{\frac{3}{10}} c_{27,18}+\frac{c_{8,4}}{2 \sqrt{10}}+\frac{c_{8,6}}{2 \sqrt{30}},\nn\\
 c_{33}^{22}=c_{27,3},\quad
 c_{33}^{23}=\sqrt{\frac{3}{10}} c_{27,9}+\frac{c_{8,2}}{\sqrt{5}},\quad
 c_{33}^{33}=\frac{c_{1,1}}{\sqrt{6}}+\sqrt{\frac{3}{10}} c_{27,18}+2 \sqrt{\frac{2}{15}} c_{8,6}. \nn
\end{gather}
And vice versa as:
\begin{gather}
  \left(c_{27,27}\right)^*=c_{27,1}=c_{33}^{11},\quad
    \left(c_{27,26}\right)^*=-c_{27,2}=-\sqrt{2} c_{33}^{(12)}, \quad
    \left(c_{27,25}\right)^*=c_{27,3}=c_{33}^{22},\\
    \left(c_{27,22}\right)^*=c_{27,4}= \sqrt{2} c_{(23)}^{11},\quad
  \left(c_{27,21}\right)^*=-c_{27,5}= -\sqrt{\frac23} \left( 2 c^{(12)}_{(23)} - c^{11}_{(13)} \right),\nn\\
  \left(c_{27,20}\right)^*=c_{27,6}= -\sqrt{\frac23} \left( 2 c^{(12)}_{(13)} - c^{22}_{(23)} \right), \quad
    \left(c_{27,19}\right)^*=-c_{27,7}= \sqrt{2} c_{(13)}^{22},\nn\\
    \left(c_{27,24}\right)^*=-c_{27,8}=-\sqrt{\frac{2}{15}} \left( 3 c_{33}^{(13)} - 2 c^{11}_{(13)} - 2 c_{(23)}^{(12)} \right), \nn\\
  \left(c_{27,23}\right)^*=c_{27,9}=\sqrt{\frac{2}{15}} \left( 3 c_{33}^{(23)} - 2 c^{22}_{(23)} - 2 c_{(13)}^{(12)} \right), \quad
    \left(c_{27,14}\right)^*=c_{27,10}=c_{22}^{11} \nn\\
\left(c_{27,13}\right)^*=-c_{27,11}=-c_{22}^{(12)} + c_{(12)}^{11},\quad c_{27,12}=\frac{1}{\sqrt{6}} \left( c_{11}^{11} -4c_{(12)}^{(12)} +c_{22}^{22}\right),\nn\\
 \left(c_{27,17}\right)^*=-c_{27,15}= -\sqrt{\frac15} \left( 4 c_{(23)}^{(13)} - c_{(12)}^{11} - c_{22}^{(12)} \right), \quad
    c_{27,16}=\sqrt{\frac{1}{10}} \left( c_{11}^{11} - 4 c_{(13)}^{(13)} - c_{22}^{22} + 4 c_{(23)}^{(23)} \right),\nn\\
c_{27,18}=\sqrt{\frac{1}{30}} \left( c_{11}^{11} + 2 c_{(12)}^{(12)} + c_{22}^{22}- 6 c_{(13)}^{(13)}  - 6 c_{(23)}^{(23)}+ 3 c_{33}^{33} \right),\nn \\ \left(c_{8,8}\right)^*=-c_{8,1}= -\sqrt{\frac45} \left( c_{(13)}^{11} + c_{(23)}^{(12)} + c_{33}^{(13)} \right) ,\quad
  \left(c_{8,7}\right)^*=c_{8,2}= \sqrt{\frac45} \left( c_{(13)}^{(12)} + c_{(23)}^{22} + c_{33}^{(23)} \right), \nn \\
  \left(c_{8,5}\right)^*=-c_{8,3}=  -\sqrt{\frac45} \left( c_{(12)}^{11}+ c_{22}^{(12)} + c_{(23)}^{(13)} \right), \quad
 c_{8,4}=\sqrt{\frac25} \left( -c_{11}^{11} - c_{(13)}^{(13)} + c_{22}^{22}+ c_{(23)}^{(23)} \right), \nn \\
 c_{8,6}=\sqrt{\frac{2}{15}} \left( - c_{11}^{11} - 2 c_{(12)}^{(12)} + c_{(13)}^{(13)} - c_{22}^{22}+ c_{(23)}^{(23)}+ 2 c_{33}^{33} \right), \nn \\
 c_{1,1}= \sqrt{\frac16} \left( c_{11}^{11} + 2 c_{12}^{12} + 2 c_{13}^{13} + c_{22}^{22} + 2 c_{23}^{23} + c_{33}^{33} \right). \nn
\end{gather}
Recall that $(c^{pr}_{qs})^* = c^{qs}_{pr}$.

\subsection{$c^{[pr]}_{[qs]}$}
The $(\mathbf{3} \otimes\mathbf{3})_\text{antisym} \otimes (\overline{\mathbf{3}}\otimes \overline{\mathbf{3}})_\text{antisym}=\overline{\mathbf{3}} \otimes \mathbf{3}$ of $c^{[pr]}_{[qs]}$ decomposes into the $\mathbf{1} \oplus \mathbf{8}$ of $c_{1,1}$ and $c_{8,1},\ldots,c_{8,8}$. The Wilson coefficients are given in terms of the Clebsch-Gordan decompositions as:
\begin{gather}
 c_{12}^{12}=-\frac{c_{1,1}}{2 \sqrt{3}}-\frac{c_{8,6}}{\sqrt{6}}\quad
 c_{13}^{12}=\frac{1}{2} c_{8,2},\quad
 c_{13}^{13}=-\frac{c_{1,1}}{2 \sqrt{3}}-\frac{c_{8,4}}{2 \sqrt{2}}+\frac{c_{8,6}}{2 \sqrt{6}},\\
 c_{23}^{12}=-\frac{1}{2} c_{8,1},\quad
 c_{23}^{13}=\frac{1}{2} c_{8,3},\quad
 c_{23}^{23}=-\frac{c_{1,1}}{2 \sqrt{3}}+\frac{c_{8,4}}{2 \sqrt{2}}+\frac{c_{8,6}}{2 \sqrt{6}}. \nn
\end{gather}
And vice versa as:
\begin{gather}
\left(c_{8,8}\right)^*=-c_{8,1}= 2 c_{[23]}^{[12]} ,\quad
 \left(c_{8,7}\right)^*=c_{8,2}= 2 c_{[13]}^{[12]} ,\quad
\left(c_{8,5}\right)^*= -c_{8,3}= -2 c_{[23]}^{[13]} ,\\
 c_{8,4}= \sqrt{2} \left( - c_{[13]}^{[13]} + c_{[23]}^{[23]} \right), \quad
 c_{8,6}= \sqrt{\frac23} \left( - 2 c_{[12]}^{[12]} + c_{[13]}^{[13]} + c_{[23]}^{[23]} \right), \nn\\
 c_{1,1}=- \sqrt{\frac43} \left( c_{[12]}^{[12]} + c_{[13]}^{[13]} + c_{[23]}^{[23]} \right).\nn
\end{gather}

\bibliographystyle{JHEP}
\bibliography{refs}

\end{document}